\documentclass[3p,times]{elsarticle}

\usepackage{ecrc}
\usepackage{amsmath}
\usepackage{color}
\def\beq{\begin{equation}}
\def\eeq{\end{equation}}

\volume{00}

\firstpage{1}

\journalname{Progress in Particle and Nuclear Physics}

\runauth{M. Baldo and G.F. Burgio}

\jid{}

\jnltitlelogo{Procedia Computer Science}

\CopyrightLine{2011}{Published by Elsevier Ltd.}

\usepackage{amssymb}
\usepackage{graphicx}

\usepackage[figuresright]{rotating}

\begin{document}

\begin{frontmatter}



\dochead{}

\title{The Nuclear Symmetry Energy}


\author{M. Baldo and G.F. Burgio}

\address{Istituto Nazionale di Fisica Nucleare, Sez. Catania, via S. Sofia 64, Catania Italy}

\begin{abstract}
The nuclear symmetry energy characterizes the variation of the binding energy as the neutron to proton ratio of a nuclear system is varied. This is one of the most important features of nuclear physics in general, since it is just related to the two component nature of the nuclear systems. As such it is one of the most relevant physical parameters that affect the physics of many phenomena and nuclear processes.  
This review paper presents a survey of the role and relevance of the nuclear symmetry energy in different fields of research and of the accuracy of its determination from the phenomenology and from the microscopic many-body theory. In recent years, a great interest was devoted not only to the Nuclear Matter symmetry energy at saturation density but also to its whole density dependence, which is an essential ingredient for our understanding of many phenomena. We analyze the nuclear symmetry energy in different realms of nuclear physics and astrophysics. In particular we consider the nuclear symmetry energy in relation to nuclear structure, astrophysics of Neutron Stars and supernovae, and heavy ion collision experiments, trying to elucidate the connections of these different fields on the basis of the symmetry energy peculiarities. The interplay between experimental and observational data and theoretical developments is stressed. The expected future developments and improvements are schematically addressed, together with  
most demanded experimental and theoretical advances for the next few years.    
\end{abstract}


\end{frontmatter}


\section{Introduction} 
\label{intro}
The nuclear matter (NM) Equation of State (EOS) is one of the central issue in Nuclear Physics.
It incorporates the fundamental properties of the nuclear medium, which
is present not only in terrestrial nuclei but in many astrophysical objects and 
phenomena. It plays an essential role in understanding and linking an extremely wide
set of data on different physical systems and processes, like     
%
nuclei in laboratory experiments, in particular exotic nuclei, heavy ion
collisions, the structure and evolution of compact astrophysical objects
as Neutron Stars (NS), supernovae and binary mergers, and so on.
On the other hand the laboratory experiments and the astrophysical observations
can put meaningful constraints on the nuclear EOS.
Unfortunately a direct connection between the phenomenology and the EOS is not possible,
and theoretical inputs are necessary for the interpretation of the data.
In particular the EOS above saturation density is much less constrained
than around or below saturation. \par 
In recent years a great attention has been payed to one of the main feature
of the EOS, i.e. the symmetry energy as a function of density \cite{EPJA2014}.
A distinctive aspect of the nuclear systems is the possibility of varying 
the relative contents of the two particles they are composed of, the neutrons and the protons.
The symmetry energy measures the change in binding of the system as the neutron to 
proton ratio is changed at a fixed value of the total number of particles. 
In nuclear matter one considers the energy per particle $E/A$, which is a
function of the total density $\rho$ and the asymmetry $\beta \,=\, (N - Z) / A $,
being $N,~ Z,~ A$ the neutron number, the proton number and the total particle number $A \,=\, N + Z$,
respectively. One then expands the energy per particle as a function of $\beta$ around $\beta = 0$
at a given density
\begin{equation}
\frac{E}{A}(\rho,\beta) \,=\, \frac{E}{A}(\rho,0) \,+\, S_{\!N}(\rho)\, \beta^2 \,+\, \cdots 
\label{eq:def}
\end{equation}
\noindent
The coefficient $S_{\!N}$ is indeed the nuclear ''symmetry energy". For a free Fermi gas of protons and neutrons 
the total energy and the symmetry energy read :  
\begin{eqnarray}
E& = & E_N + E_Z = \frac{3}{5} Z {E}_F^{(p)} +  \frac{3}{5} N {E}_F^{(n)} \nonumber \\
e& = &\frac{E}{A} = \frac{3}{10} \frac{\hbar^2}{2 m} 
 \left( {{3\pi^2}\over 2}\right)
^{{2\over 3}} \rho^{{2\over 3}} \left[ (1 + \beta)^{5\over 3}
\,+\, (1 - \beta)^{5\over 3} \right] \\
 & \approx & e(\beta = 0) \,+\, S_N \beta^2 + \cdots\cdots \nonumber \\
 S_N &=& \frac{1}{3} E_F  \nonumber
\label{eq:enas}
\end{eqnarray}
\noindent 
where $E_F$ denotes the Fermi energy for symmetric matter and $E_F^{(n)}, E_F^{(p)}$ are the neutron and proton Fermi energies.
At saturation density $\rho_0$, one finds $S_N \approx
12$ MeV. This value indicates a steep increase of the energy with asymmetry. We will see that the value in interacting nuclear matter is estimated to be more than twice larger. This steepness of the energy surface in the asymmetry direction is one of the main characteristic of $S_N$, and plays a fundamental role in a wide range of nuclear properties and phenomena.
\par
In general the symmetry energy is defined in nuclear matter, and its relation to
a similar quantity in finite nuclei requires a well defined theoretical framework.
However, the expansion around symmetry of Eq. (\ref{eq:def}) can be still used for finite nuclei to
define a symmetry energy, being now $S_N$ a function of the particle number $A$.
In general the main effort was focused on the determination of the nuclear matter
symmetry energy, that will be indicated from now on by $S$, and its density dependence.
Different strategies have been developed to extract from the experimental data the values of $S(\rho)$.
The simplest one is to use the liquid drop model (LDM) of nuclear binding to extract the bulk part
of $S_{\!N}$, which is identified with the value of $S$ at saturation. Alternatively the overall
$S_{\!N}$ is identified with the value of $S$ at some average sub-saturation density.
A different strategy relies on the use of energy density functionals (EDF), noticeably the ones generated by
Skyrme forces, to fit the nuclear binding throughout the nuclear mass table. The fit determines the
parameters of the best EDF, which is then used to calculate $S(\rho)$. In general, different EDF
can fit equally well the nuclear binding, and they usually agree on the value of $S$ at saturation, though 
its density dependence can be strongly dependent on the chosen EDF. However it is possible to extract correlations between
different physical parameters, like the values of $S$ at saturation and its slope.
In order to characterize its density dependence, $S(\rho)$ is usually expanded around the saturation density $\rho_0$
\begin{equation}
S(\rho) \,=\, S(\rho_0) \,+\, \left( \frac{d S}{d \rho} \right)_{\rho_0} (\rho \,-\, \rho_0) \,+\, 
                            \frac{1}{2} \left( \frac{d^2 S}{d \rho^2} \right)_{\rho_0} (\rho \,-\, \rho_0)^2 \,+\,
                            \frac{1}{6} \left( \frac{d^3 S}{d \rho^3} \right)_{\rho_0} (\rho \,-\, \rho_0)^3 \,+\, \cdot\cdot\cdot\cdot\cdot    
\end{equation}
\noindent and for convenience we define the following parameters
\begin{eqnarray}
S_0 &=& S(\rho_0) \nonumber \\
L &=& 3 \rho_0 \left( \frac{d S}{d \rho} \right)_{\rho_0} \nonumber \\
K_{sym} &=& 9 \rho_0^2 \left( \frac{d^2 S}{d \rho^2}\right)_{\rho_0}  \\
Q_{sym} &= & 27 \rho_0^3 \left( \frac{d^3 S}{d \rho^3}\right)_{\rho_0}  \nonumber 
\label{eq:expans}
\end{eqnarray}
\noindent 
having all an energy dimension (MeV). A similar strategy has been
followed in the analysis of other nuclear quantities that are believed to be related to $S(\rho)$,
like the neutron skin, the different isovector nuclear excitations, and the data on heavy ion collisions
like isospin diffusion and the isotopic distribution in multifragmentation processes.           
The large amount of novel exotic nuclei produced in laboratory and the development
of radioactive ion beams (RIB) have greatly stimulated new research projects on the symmetry 
energy as well as an intense theoretical activity on this fundamental subject.
In this case it is particularly evident the connection with the physics of
Neutron Stars, where very exotic nuclei can be present in their crust, even well beyond
the stability region of nuclei in laboratory conditions, since in NS the presence
of electrons hinders the beta decay of very asymmetric nuclei. Other observational data can
be used to constrain the value of $S(\rho)$, in particular at density above saturation. In this case
it is not easy to analyze the data on physical quantities, which are possibly sensitive to the symmetry
energy, and a certain amount of theoretical input is necessary. In other words,
it is difficult to eliminate the model dependence of the analysis.   
A constant characteristic of the field is the interplay between phenomenology
and theory, whose continuous reciprocal feedback is essential for its progress.   
In the last few years an enormous literature about this subject has developed, and 
numerous workshops have been organized throughout the world. 
It seems time to try a first summary 
of the established results and to indicate the possible future directions in the field. 
\par
There are already in the literature few excellent papers devoted to different aspects of the nuclear symmetry
energy and its relevance in Nuclear Physics and Astrophysics \cite{Tsang,Latt,Horo}. The emphasis of this
review will be on the comparison between the theoretical predictions, based on microscopic many-body
methods, and the phenomenological hints on different features of the symmetry energy. It is expected that
this detailed comparison will be helpful for the development of our knowledge not only on the symmetry
energy but in general on the microscopic structure of Nuclear Matter and Nuclei.  
\par 
The paper is organized as follows. In Section 2 we summarize the different methods, both phenomenological and more
microscopic, that link the nuclear structure data to symmetry energy. We distinguish between the semi-classical
approaches and the quantal approaches. Within the first category we discuss the Liquid Drop model, the Droplet model and
the Thomas-Fermi method. For the second category we describe in detail the Energy Density Functional method, and the way it has been developed in order to constrain the symmetry energy from the data on different aspects of nuclear structure. We mention here the Isobaric Analog State, the isovector excitations and the neutron skin width in highly asymmetric nuclei. Sec. 3 is devoted to the astrophysical role of the symmetry energy. The astrophysical objects, considered in some detail, are the Neutron Stars and the Supernovae. In particular in Neutron Stars we discuss separately the crust and the core and the role of the symmetry energy in each of them. Sec. 4 introduces the realm of the heavy ion collisions, where the presentation focus on the different signals that can be detected in the experiments, potentially useful for the study of the symmetry energy. Here one has to distinguish between the low density sub-saturation regime and the higher supra-saturation density. In Sec. 5 we discuss the microscopic many-body approaches that have been used to calculate he Equation of State of nuclear matter and the corresponding symmetry energy. We show that up to saturation density there is a substantial and satisfactory agreement between the microscopic theory and the phenomenological constraints for the density dependent symmetry energy. At supra-saturation density the different theoretical approaches start to diverge drastically,
but unfortunately the phenomenological constraints are here quite scarce. In Sec. 6 we collect a set of phenomenological constraints  on the symmetry energy at saturation and its slope, and we combine them in order to select the possible region of compatibility.
Comparison with the theoretical predictions is also performed. Finally in Sec. 7 we give a general survey on the subjects discussed in the different Sections, on the possible future developments and the most urgent quests.       


\section{Symmetry energy in nuclear structure.}
\label{sec2}
\subsection{Semi-classical methods.}
\label{semi}
\subsubsection{The Liquid Drop Model}
\label{LDM}
The Liquid Drop Model (LDM) assumes that nuclei can be treated as macroscopic drops
of an incompressible liquid, i.e. of nuclear matter. The assumption is based on
the short range nature of the nuclear interaction and the saturation property
of nuclear matter. This suggests that the nuclear binding energy of a nucleus of
mass $A$ and charge $Z$ can be written 
\begin{equation}
B(A,Z) \,=\, a_V A \,+\, a_S A^{2\over 3} \,+\,  S_N\,
 {(N - Z)^2\over A} \,+\, a_C {Z^2 \over A^{1\over 3}}
\label{eq:massf} 
\end{equation}
\par\noindent 
which contains the bulk contribution $a_V$, the surface correction $a_S$,the already introduced symmetry energy parameter 
$S_N$, and the Coulomb energy $a_C$.
The overall trend of the empirical binding energy of nuclei and how it can be reproduced by this simple formula, by adjusting the set of
parameters $a$, are discussed in basic books \cite{RS}, where the meaning and possible forms of the different
terms are discussed in more detail. The values of the parameters depend slightly on the particular experimental mass table used for the fit. 
In general the value of the bulk energy $ a_V$ is in all cases around $- 16 $ MeV, which is
interpreted as the single particle binding energy in nuclear matter. The formula \ref{eq:massf} provides an excellent
fit to the smooth part of the binding energy of nuclei throughout the nuclear mass table with a few parameters,
and allows to extract the finite nuclei symmetry energy $S_N$, at least averaged over
the mass table. It is worthwhile to mention a relatively recent fit \cite{Daniel2009} over 3100 nuclei with mass number A $>$ 10,
which was performed neglecting pairing and deformation contributions, as in Eq. (\ref{eq:massf}), and gives $ S_N\,=\, 22.5$ MeV. 
The average error for the total binding is 3.6 MeV, to be confronted with values of the binding energy that
can be as large as 2000 MeV. This value of $S_N$ cannot be directly related to the value $S$ of nuclear matter, because it is clearly affected by finite size effects. If the value of $S$, as expected, is an increasing function of density, at least up to
saturation, in a local density picture the value of $S_N$ can be viewed as the value of $S$ at some average
density of the nucleus. The value of $S_N$ is then expected to increase with the mass number $A$, since the surface contribution decreases and the average density
tends to the saturation value. At large value of the mass the value of $S_N$ is then expected to approach $S$, while at lower mass the total symmetry energy should include a negative term proportional to the surface area of the nucleus, and then for not too small mass a term for $S_N$ proportional to $A^{-1/3}$, in agreement with the leptodermous expansion \cite{lepto}. The value of $S_N$ as a function of $A$ should interpolate between these two trends. The simplest way to do so is to assume that $S_{\!N}$ is the sum of a bulk contribution, independent of $A$ and coinciding with $S$, and a (negative) surface term proportional to
$A^{-1/3}$. However a more refined method was proposed in ref. \cite{Daniel2009}, based on semi-classical considerations. In this case the mass-dependent formula for $S_{\!N}$ can be written
\begin{equation}
S_N \,=\, \frac{S}{1 \,+\, a_S/A^{1/3}}
\label{eq:int}\end{equation}
\par\noindent
which should be applicable also for relatively light nuclei.
In any case an additional parameter has to be introduced, along with a more refined fitting of the nuclear bindings to be used 
to extract the value of $S$. Unfortunately it turns out that the range of asymmetry values available from the nuclear data is not large enough to extract a value of $S$ with a satisfactorily small uncertainty. Indeed the results of most of the fits with the LDM model and its generalizations on the nuclear binding give a value of $S$ centered around 30 MeV, but with a spread of $\pm$ 3-4 MeV. The surface contribution to the symmetry energy is more uncertain, which can indicate that the separation between bulk and surface symmetry energy is difficult and not well defined, at least from the binding energy.
\subsubsection{The Droplet Model}
\label{droplet}
The form of the mass dependence of the symmetry energy as indicated in Eq. (\ref{eq:int}) is obtained by assuming implicitly 
that the proton and neutron surface are displaced between each other as the nuclear asymmetry increases \cite{Daniel2009}.
In the Droplet Model (DM) \cite{MS66,My69,MS69} this assumption is explicitly introduced and used to derive a more sophisticated mass formula. In addition the effects of the non-uniformity of the neutron and proton density distributions are included (within a perturbative expansion in the deviations from uniformity). Within this framework a formula of the form of Eq. (\ref{eq:int}) can be derived \cite{MS66}. The physical parameters that appear in the model can also be calculated \cite{My69} from the Thomas-Fermi approximation based on a simplified effective nucleon-nucleon interaction, which then will contain the only independent adjustable parameters. As a rule the fitting procedure of the droplet model has been performed with the inclusion of an additional microscopic term, usually indicated with the generic name of ''shell effects". In fact both LDM and Thomas-Fermi approaches do not take into account the well established presence of shells and the corresponding ''magic numbers", for which an additional binding is expected to occur. This is evident from the behaviour of the binding energy per particle along the mass table, where several enhancements of the binding against a smooth trend are observed at well defined values of neutron and proton numbers. This microscopic additional term is necessary also       
to describe deformations. In fact, for a purely semi-classical
model like LDM nuclei can be only spherical, because this is the shape that minimize the energy. The shell effect has been included already in ref. \cite{My69} in a schematic way, assuming that the effect vanishes at some critical deformation. A   
more microscopic method was introduced by Strutinski \cite{Strut}, and since then it was used systematically in the applications of the droplet model, which sometime is indicated as ''macroscopic-microscopic" method. The Strutinsky method is based on the fact that it is difficult to reconcile the LDM with the shell model (SM), since the shell model severely underestimates the binding.
One then assumes that only the increase of binding $\Delta E$ over the smooth trend obtained in the SM is indeed correct, while the smooth behaviour itself is better described by the LDM. One then subtracts from the shell model energies $E_i$ for each nucleus $i$ the smooth part $\overline{E}_i$ of the overall trend, as obtained with some smoothing procedure, which is then substituted by the LDM energy
\begin{equation}
 E_i \,= \overline{E}_i \,+\, \Delta E_i \ \rightarrow \ E_{LDM} \,+\, \Delta E_i
\end{equation}       
\noindent
Of course the choice of the smoothing procedure is critical in concrete applications. \par 
The microscopic part is calculated, for each deformation, by constructing a single particle potential and calculating the corresponding single particle energies. This can be done at different degrees of sophistication. The most advanced one is based on the effects of the finite range of the nucleon-nucleon interaction. A parametrized finite range interaction of Yukawa shape is used to get by folding the bulk energy and the single particle potential, which then acquires a diffuseness.  
 The procedure defines the finite range droplet model (FRDM). The single particle potential contains an isovector part and then one can define not only the symmetry energy but also the slope $L$ of its density dependence. In fact the model includes the compressibility of the neutron and proton components, which depend on the asymmetry and its slope with density . The sensitivity of the results to the parameters $S$ and $L$ can then be used to adjust them to their optimal values. Due to the limited range of asymmetry available from nuclear data it was found difficult to extract the value of $L$ without ambiguity. Only recently, by increasing the data set and its accuracy, it was possible \cite{NixPRL2012,2016ADNDT} to get from the fitting the values of both $S$ and $L$. The authors of ref. \cite{NixPRL2012,2016ADNDT} report the values $S \,=\, 32 \pm 0.5$ and $L \,=\, 70 \pm 15$ MeV. The errors correspond to the intrinsic uncertainty within the model, which were estimated by varying the data set and checking
the sensitivity of the results to different parametrizations. The reason of the difficulty to extract the value of $L$ from the fitting procedure is probably due to the indirect dependence of the calculated binding on this physical parameter. In fact $L$ appears essentially only in the calculation of the single particle potential for a given shape and asymmetry. The folding of the above mentioned Yukawa interaction is performed in the (sharp) volume of the nucleus whose spherical equivalent radius $R_{pot}$, is given by
\begin{equation}
R_{pot} \,=\, R_{den} \,+\, A_{den} \,-\, B_{den}/R_{den} 
\label{eq:Rpot}
\end{equation}                 
\noindent where $A_{den}$ and $B_{den}$ are two (fixed) parameters that take into account the diffuseness and the curvature of the density distribution, estimated from Thomas-Fermi calculations in semi-infinite nuclear matter. The radius $R_{den}$ is the average radius of the nucleus, which differs from the standard LDM one because the model assumes the matter as compressible
\begin{equation}
 R_{den} \,=\, r_0 A^{1/3} ( 1 \,+\, \overline{\epsilon} )
\label{eq:Rden} 
\end{equation}  
\noindent were $r_0$ is the standard radius parameter and $\overline{\epsilon}$ is the average deviation of the bulk density $\rho$ from a reference saturation value $\rho_0$ 
\begin{equation}
\epsilon \,=\, - \frac{1}{3} \frac{\rho - \rho_0}{\rho_0} 
\label{epsi}
\end{equation}
\noindent which gives the correct relation between the radius and the density. The value of $\overline{\epsilon}$ is determined from the minimization of the macroscopic energy and, neglecting for simplicity the Coulomb interaction, is given by
\begin{equation}
\overline{\epsilon} \,=\, \left(\, - \frac{2 a_2}{A_{1/3}} \,+\, L \overline{\delta}^2 \,\right)/ K
\label{eq:epsi1}
\end{equation}  
\noindent where $a_2$ is the parameter for the surface energy, $K$ the incompressibility at saturation, and $\overline{\delta}$
is the average bulk asymmetry. The latter is also obtained by minimization
\begin{equation}
\overline{\delta} \,=\, \beta\, /\, (\, 1 \,+\, \frac{9}{4} \frac{S}{Q} \frac{1}{A^{1/3}} \,)
\label{eq:delta}
\end{equation}
\noindent where $\beta$ is the usual overall asymmetry of the nucleus, as already introduced, and $Q$ is a parameter of the model that expresses the stiffness of the surface with respect to asymmetry. The parameter $L$ coincides with the one introduced
in Eq. (\ref{eq:expans}). \par 
In the fitting optimization of ref. \cite{NixPRL2012}, the standard deviation from experimental data on nuclear masses was reduced to 0.57 MeV. This is an astonishingly small value, which shows the exceptionally good performance of the model. With respect to the extraction of the nuclear symmetry energy parameters, the smallness of the overall deviation does not necessarily imply an accurate determination of these physical parameters. In fact the number of parameters included in the model is large enough to introduce possible compensations among different parameters in reaching this small deviation. It is then possible that the values of $S$ and $L$ so obtained are affected by the adjustment of some fitting parameters, which have not a direct physical meaning and are not directly related to them. In other words the set of fitting parameters could be redundant, and this would introduce spurious and unstable correlations among different physical quantities. A correlation analysis would be desirable, which however could be too time consuming for such a large set of parameters. However, the model extrapolates quite well \cite{NixPRL2012} to masses that have not been included in the fitting procedure, and this gives some confidence on the robustness of the results.     
\subsubsection{The Thomas-Fermi Model}
\label{TF}
Historically, the first method to relate the properties of finite quantal systems to the corresponding homogeneous
system was the semi-classical Thomas-Fermi scheme, which was devised to calculate the ground state properties of
large atoms and molecules in the mean field approximation. In fact, even in this approximation, the quantal
calculations can become quite complex. The original form of the scheme is equivalent to take the zero order term
in the expansion in $\hbar$ of the density matrix for the independent particle wave function, but it can be easily
derived by assuming that the system is locally equivalent to a free Fermi gas at the local density and local
potential. We remind here some features of the approximation that are useful for the development of the
presentation. In its simplest version it can be formulated within the Density Functional method, i.e. assuming
that the energy of the system can be written as a functional of the density $\rho(\mathbf{r})$
\beq
E_{TF} \,=\, T_F(\{\rho\}) \,+\, V_c(\{\rho\}) \,+\, V_{pp}(\{\rho\})
\label{eq:funTF}\eeq
\noindent where $T_F$ is the kinetic energy contribution, $V_c$ some external potential and $V_{pp}$ the
particle-particle interaction. In case of atoms, $V_c$ is the energy due to the central Coulomb potential and
$V_{pp}$ is the electron-electron Coulomb interaction energy
\begin{eqnarray} 
T_F(\{\rho\}) &=& \frac{3}{5} \int d^3r E_F(\rho(\mathbf{r}))\rho(\mathbf{r}) \ \ \ ; \ \
\ V_c(\{\rho\}) \,=\, -
Z\int d^3r' v(r)\rho(\mathbf{r'}) \nonumber \\
V_{pp}(\{\rho\}) &=&\, \int d^3r \int d^3r' v(|\mathbf{r} - \mathbf{r'}|)
\rho(\mathbf{r})\rho(\mathbf{r'}) \nonumber
\label{eq:termsTF}
\end{eqnarray}
\noindent with $Z$ the charge of the nucleus, $v = 4\pi e^2/r$ the Coulomb potential between two electrons (of
charge $e$) and $E_F$ the Fermi energy for a free gas at the density $\rho(\mathbf{r})$
\beq
E_F(\rho(\mathbf{r})) \,=\, \frac{\hbar^2}{2 m} \left(3\pi^2\right) \rho(\mathbf{r})^{2/3} 
\eeq 
\noindent 
The Euler-Lagrange equation corresponding to the minimization of this functional with the constraint of a fixed number of particles
is 
\beq
E_F(\rho(\mathbf{r})) \,-\, Z v(r) \,+\, \int d^3r' v(|\mathbf{r} - \mathbf{r'}|) \rho(\mathbf{r'}) \,=\, \mu
\label{eq:TF}
\eeq
\noindent where $\mu$ is the Lagrange multiplier, which has the meaning of chemical potential. Notice that $\mu$ is
a constant, independent of the position. The solution of this integral equation gives the density and then the energy of
the ground state. A way of solving this equation is to apply the Laplacian differential operator and use the
Poisson equation $\Delta V_C \,=\, - 4\pi e^2 \rho(\mathbf{r})$, where $V_C$ is the last term at the left hand
side of equation (\ref{eq:TF}), that is the potential produced by all electrons at $\mathbf{r}$. This gives the
familiar differential equation of the Thomas-Fermi scheme in its simplest form \cite{March}. We will not discuss
the many refinements of the approximation that have been developed, to include e.g. the exchange interaction, but
rather we consider the Thomas-Fermi approximation in the nuclear case. The physical situation is quite different.
There is no central interaction, and the binding must come from the NN interaction. The latter is not long range,
like the Coulomb potential, but it is short range, even zero range if we adopt a Skyrme effective interaction. Last, but not least, the system has two components, the protons and the neutrons.
For zero range interaction equation (\ref{eq:TF}) has no solution, except in the trivial case of an homogeneous system. To get any
sensible result for a nucleus one has to go to the next order in the expansion in $\hbar$ of the kinetic energy
term in the functional. This introduces gradient terms \cite{RS}. To order $\hbar$ only one term contributes,
proportional to $(\nabla \rho)^2$. With the inclusion of this term, a surface can develop and the nuclear
Thomas-Fermi approximation can describe the density profile of a nucleus. Alternatively one can introduce a phenomenological gradient term or a finite range effective interaction. The need of an effective interaction is another complication of the nuclear case, since one 
must assume its form with a set of parameters to be fitted to phenomenological data.  \par The discussion above makes also clear that in
the nuclear case the kinetic term must be treated at different order in the expansion in $\hbar$ and at a
different level of approximation. The full quantal treatment of the kinetic term must be considered to construct
any accurate nuclear energy functional. \par Despite that, the Thomas-Fermi approximation, or more advanced
expansions in $\hbar$ can be useful. In fact the Thomas Fermi approximation and its implementations are expected
to describe the smooth part e.g. of the density of states or the binding energy (with a suitable effective interaction), leaving outside of their possibility the description of
the shell effects, that are a pure quantal effect. Indeed the expansion in $\hbar$ of many quantities is actually
asymptotic, and the quantal effects depend non-analytically on $\hbar$. This property of the $\hbar$ expansions
can be used to evaluate the shell effects in the microscopic-macroscopic approach, introduced in section
\ref{droplet}.  Along these lines,
 in reference \cite{MarioPeter} the Kirkwood $\hbar$ expansion to fourth order has been used to estimate
the shell effects by comparing the results with the quantal calculations. This method and similar ones based on
$\hbar$ expansions are methods alternative to the Strutinski procedure \cite{Strut}.\par
For a nucleus the TF integral equation (\ref{eq:TF}) becomes a system of two coupled integral equations for the proton and neutron density profiles. A characteristic of the TF approximations is that the density distributions go to zero at a finite distance from the center of the nucleus. Despite that, the nucleus displays a well defined surface region. Because all properties of the system are derived from the assumed effective interaction, or equivalently from the TF functional, one can relate in a unique way the nucleus to nuclear matter. In particular, given an effective interaction, the symmetry energy and its density dependence in nuclear matter is determined together with the corresponding symmetry energy in nuclei along the nuclear mass table.      
\par 
In ref. \cite{MS1996} extensive TF calculations were performed based on an updated version of the Seyler-Blanchard two-body interaction \cite{SB1961}, which is a schematic effective potential that contains both a momentum and a density dependence  
\beq
v(r,p) \,=\, v_0 Y(r) \left[ -a_1 \,+\, a_2\, \left(\frac{p}{P_0}\right)^2 \, -a_3\, \left(\frac{P_0}{p}\right) \,+\, \sigma\, \left(\frac{2\rho}{\rho_0}\right)^{2/3}  \right]
\eeq
\noindent where $\sigma$ and the $a_i$ are parameters are fitted to nuclear data within the TF approach, while $v_0, P_0$ fix the energy and momentum scales, respectively, just for convenience. The use of both the coordinate $r$ and momentum $p$ as independent classical variables is related to the semi-classical character of the approach. 
The form factor $Y(r)$ is chosen as the standard Yukawa interaction
\beq
Y(r) \,=\, \frac{1}{4\pi a^3}\, \frac{e^{-r/a}}{r/a}
\eeq 
The interaction is purely central and no explicit spin-isospin dependence was considered, but the parameters are distinct for "like nucleons" (neutron-neutron and proton-proton) and "unlike particle" (neutron-proton), which affects the interaction contribution to the symmetry energy. 
\begin{figure}[hbt]
\centering
\includegraphics[scale=0.7,clip]{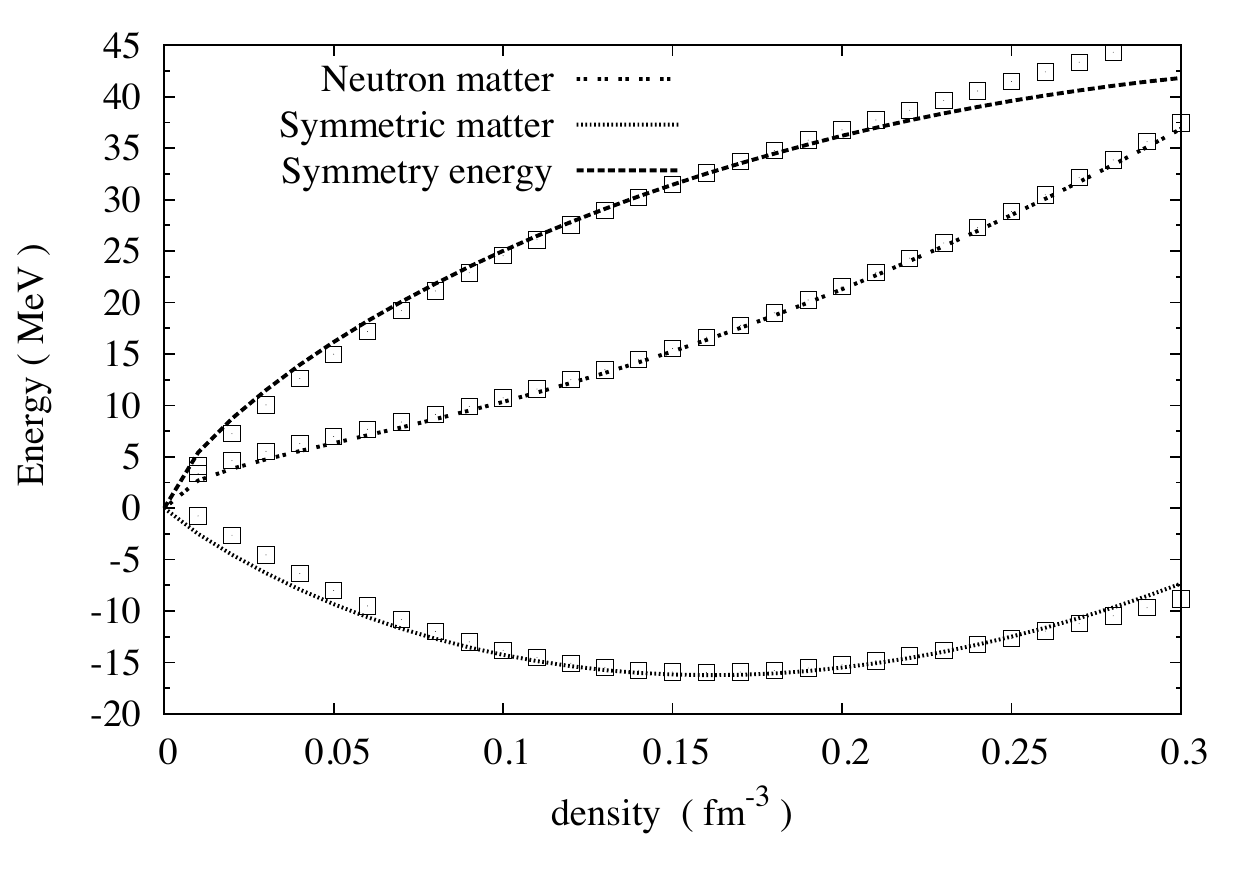}
\caption{Equation of state in the Thomas-Fermi approximation, as reported in ref. \cite{MS1998}. The open square correspond to the microscopic many-body calculation of ref. \cite{BCPM}}
\label{fig:EOS_TF}
\end{figure}
\noindent The fitting was performed on the smooth part of the data set on the masses of 1654 nuclei with $N, Z \geq 8$. To get the smooth behaviour of the data, shell corrections to the experimental masses were introduced according to the Strutinski method. This is in line with the main assumption of the model, i.e. the TF approximation should be able to fit the smooth part of the masses data with a suitable effective interaction. The resulting fit is able to reproduce this data set on binding energy with an average error 
(root mean square deviation) of 0.655 MeV. The interaction was also constrained to reproduce the phenomenological optical potential and a set of fission barriers. Although the data set on the masses is slightly more restricted than in more recent applications of e.g. the droplet model \cite{NixPRL2012}, this is a remarkable result, in view of the schematic form of the effective interaction and the limited number of free parameters (essentially 6).\par
Once the parameters of the effective interaction have been fixed, the EOS of nuclear matter can be determined in a straightforward way. In this case the TF calculation can be performed analytically. The resulting EOS for arbitrary asymmetry can be written as a polynomial form  in the Fermi momentum or in $\rho^{1/3}$
\beq
\frac{E}{A}\,(\rho,\beta) \,=\, b_1(\beta)\, \Omega^2 \,+\, b_2(\beta)\, \Omega^3 \,+\, b_3(\beta)\, \Omega^5 
\label{eq:TFEOS}
\eeq  
\noindent with $\Omega \,=\, \rho^{1/3}$. In any case the functions $b_i(\beta)$ are completely determined by the parameters 
fixed by the fitting procedure. In the following we report them to show explicitly the dependence on asymmetry $\beta$ and for later discussion.
Putting $q \,=\, (1-\beta)^{1/3}$ and $p \,=\, (1+\beta)^{1/3}$, one has
\begin{eqnarray}
b_1(\beta) &=& \frac{3}{10} \left( 1 \,+\, a_{3l} \right) \left( p^5 \,+\, q^5 \right) \,+\, \frac{3}{20}  a_{3u} 
\left( 5 p^2 q^3 - q^5 \right) \nonumber \\
b_2(\beta) &=& -\, \frac{1}{4} a_{1l} \left( p^6 \,+\, q^6 \right) \,-\, a_{1u} q^3 p^3  \\
b_3(\beta) & = & \frac{3}{10} \left( a_{2l} \,+\, \frac{5}{6} \sigma_{l} \right) \left( p^8 \,+\, q^8 \right) \,+\, \frac{3}{10} \left( a_{2u} \,+\, \frac{5}{6} \sigma_{u} \right) p^3 q^3 \left( p^2 \,+\, q^2 \right)  \nonumber
\label{eq:bTF}
\end{eqnarray} 
\noindent where the subscripts $l, u$ stand for like and unlike particles, respectively.
The symmetric matter EOS, the pure neutron matter EOS and the symmetry energy have then the same polynomial form of Eq. (\ref{eq:TFEOS}). They are all reported in Fig. \ref{fig:EOS_TF}. In particular the symmetry energy at saturation turns out to be $S \,=\, 32.65$ MeV, and the corresponding slope parameter $L \,=\, 49.9$ MeV. 
It is interesting to notice that the value of the symmetry energy would be equal to the difference between the pure neutron EOS and the symmetric matter EOS at the saturation density if the dependence on asymmetry is quadratic. This difference turns out to be $32.72$ MeV, very close to the just reported value obtained from the expansion around symmetry, see Eq. (\ref{eq:def}), despite the analytical form of the EOS is not quadratic in the asymmetry $\beta$.\par  
One has to stress that the symmetric matter EOS is characterized by a incompressibility $K \,=\, 234$ MeV, which is within the phenomenological range, and a saturation point well compatible with the experimental estimate. This gives further support to the method followed in this analysis, based on the Thomas-Fermi approach.  \par
A detailed report on the EOS obtained in this application of the nuclear TF can be found in ref. \cite{MS1998}.\par
For comparison in Fig. \ref{fig:EOS_TF} we also report (open squares) the EOS and symmetry energy obtained within an {\it ab initio} microscopic calculation \cite{BCPM}. Up to density of 0.2 fm$^{-3}$ there is a close agreement. Above this density some discrepancy starts to appear, which is not surprising since the fitting of nuclear data on binding can fix the EOS and the symmetry energy only up to saturation or slightly above. The neutron matter EOS was shown in ref. \cite{MS1996} to agree also with the variational calculation of ref. \cite{FP}.

\subsubsection{Summarizing.}
\label{semi_survey}
Semi-classical methods have been extensively applied along the years to the study and analysis of the bulk properties of nuclei throughout the mass table, in particular the nuclear binding, the radius and possibly the density profile. The main purpose, common to the different semi-classical methods, is to get an overall picture of known nuclei as their mass and charge are changing.
Quantum corrections are considered as embodied mainly in the shell effects (including the small pairing contribution). It is then assumed that once the fluctuations in the binding due to the presence of shells are subtracted, following e.g. the Strutinsky method, what remains is a smooth behavior that can be reproduced by a semi-classical approach. To characterize the properties of the nuclear medium one can try to connect the results of the fitting, which are embodied in the determination of the parameters of the model, to the macroscopic properties of nuclear matter. In this view the difference between the properties of nuclei and nuclear matter can be referred as "finite size effects". It is of particular relevance the extraction from the model and the fitting of the nuclear matter symmetry energy and possibly its density dependence. This is not so simple in the case of the pure Liquid Drop Model,
since the symmetry energy of a finite nucleus is strongly affected by the presence of the surface, and a single parameter for all fitted nuclei can give only an average value. In other words the connection to the nuclear matter symmetry energy is expected to be dependent on the mass number of the nucleus. This connection can be obtained again by semi-classical methods which include a surface contribution. The fitting of the additional parameters turns out to be difficult because of the limited range of the asymmetry available from the known nuclei \cite{Daniel2009}. To test this approach it is necessary to go beyond the simple LDM. \par 
The extraction of the nuclear matter symmetry energy is easier in the Droplet Model and in Thomas-Fermi model. In the DM the NM symmetry energy is embodied in the parameters of the single particle potential used to calculate the shell effects.
In the TF approach the symmetry energy can be calculated once the parameters of the effective nucleon-nucleon interaction have been fitted on the nuclear data. Actually the whole explicit density dependence of the EOS can be obtained. Notice that both models are extremely accurate in reproducing the total binding energies of nuclei, with a root mean square deviation
close to 600 KeV. 
\par
In any case, if one considers the latest calculations available \cite{NixPRL2012,MS1996,MS1998}, the resulting value of the symmetry energy at saturation is quite close in the two models, both centered around $32$ MeV. This is not the case for the slope parameter $L$, which is around 70 MeV for DM and around 50 MeV for the TF model. The disagreement on $L$ points out again the difficulty of extracting the slope of the symmetry energy. The agreement on $S$ does not necessarily indicate that this is the correct value. Only the comparison with reliable microscopic NM calculations can validate this value.

\subsection{Density functional methods.}
\label{EDF}
 A step further
towards a microscopic approach within the same method of fitting nuclear data is the construction of a general Energy Density functional that is assumed to include
all the correlations in an effective way and without any $\hbar$ expansion. In general these functionals are devised to be used at the mean field level. They are necessarily partly
phenomenological, i.e. they contain a certain number of parameters. One of the main goals is to relate and fix the parameters by fitting not only the binding energies but also other experimental quantities. In fact, once the functional has been chosen, it can be applied to a wide range of nuclear phenomena,  
 On the other hand, if these functionals are treated at purely
phenomenological level in fitting the binding energies, they turn out to be quite accurate. In this case both the smooth part as well as shell effects are included in the calculations, which therefore can be considered microscopic to a certain extent.\par
The energy density functional (EDF) can be constructed directly or derived from a effective forces, which can contain a two-body as well as a three-body interaction. In general, apart from a few exceptions, these forces are density dependent. The justification of the density dependence stems on the Hohenberg and Kohn theorem \cite{HK}, which states that the ground state energy of a system is a unique functional only of the density profile. This gives in general a foundation to the use of a functional for determining the energy of the ground state. One can assume in fact that a chosen functional is a good approximation of the exact, but unknown, functional.   
The most used effective forces 
are the zero range Skyrme forces and the Gogny forces. The latter include a finite range part \cite{Gogny}, but recently a fully finite range Gogny force has been proposed \cite{Gogny_nl}. Notice that the Hoenberg and Kohn exact functional is expected to be non-local,
so that the zero range assumption is an approximation. However many Skyrme forces introduce gradient terms, which can be viewed as the result of an expansion in the non-locality. For a review of the method see ref. \cite{RMPreview}. A related approach is based on the relativistic formulation, where a relativistic mean field (RMF) is introduced \cite{RMPreview}, which is generated by the coupling with a set of mesons. Both the nucleon and meson fields are treated at the mean field level. The couplings can be density dependent and self-interactions in the meson fields can be also introduced.\par The Skyrme functionals, as well as the RMF functional, as a rule include the pairing interaction, which is an essential ingredient to get a good accuracy. 
Hundreds of Skyrme forces have been devised along the years and in general the parameters that they contain were fixed by fitting a large set of nuclear binding throughout the mass table, with the possible addition of nuclear charge radii, fission barriers, a selected set of single particle levels and in some cases also by reproducing some physical quantities calculated by microscopic many-body theory, that are considered well established. \par 
From the point of view of reproducing the nuclear binding, the Skyrme functional with the smallest root mean square deviation appears to be the one by the Bruxelles group \cite{HFB17}, which includes pairing, with a deviation of $0.581$ MeV. This deviation is competitive with the droplet model \cite{NixPRL2012} (and the TF model \cite{MS1998}, but with a smaller set of nuclei). Also in this case this remarkable precision does not necessarily imply that the corresponding symmetry energy is accurate. The latter can be deduced from the functional once the force parameters have been fixed, as explained below.    
The simplest form of a generic Skyrme force can be written
\beq
\begin{array}{rl}
f(\mathbf{r}_1,\mathbf{r}_2)\!\!\!\! &\,=\, t_0 \left( 1 \,=\, x_0 P_\sigma \right) \delta(\mathbf{r}) \\
 &\ \\
                       \     &\,+\, \frac{1}{2} t_1 \left( 1 \,+\, x_1 P_\sigma \right) \left[ \delta(\mathbf{r}) \mathbf{k}'^2 \,+\,
                       \mathbf{k}^2 \delta(\mathbf{r}) \right] \\
  &\ \\                \     &\,+\, t_2 \left( 1 \,+\, x_2 P_\sigma \right) \mathbf{k}' \delta(\mathbf{r}) \mathbf{k}  \\
  &\ \\
                       \     &\,+\,  t_3 \rho(\mathbf{r})^\alpha \left( 1 \,+\, x_3 P_\sigma \right) \delta(\mathbf{r})  \\
  &\ \\                \     &\,+\, iW_0 \mathbf{\sigma} \cdot \left[ \mathbf{k}' \times \delta(\mathbf{r}) \mathbf{k} \right]                                                                
\end{array}
\label{eq:Skyrme}\eeq 
\noindent
where
\beq
\begin{array}{rl}
\mathbf{r}\!\!\! &=\, \mathbf{r}_1 \,-\, \mathbf{r}_2 \\
\mathbf{k}\!\!\! &=\, \frac{1}{2i} \left( \nabla_1 \,-\, \nabla_2 \right) \\
\sigma\!\!\! &=\, \sigma_1 \,+\, \sigma_2 \\
P_\sigma\!\!\! &=\, \frac{1}{2} \left( 1 \,+\, \sigma_1 \cdot \sigma_2 \right)
\end{array}
\eeq
\noindent and a prime indicates that the differential operator acts on the left.\par
Additional density dependent terms, with integer or fractional exponent, have been introduced by several authors \cite{Agra}, in order to improve the performance of the force or to avoid possible collapse of nuclear matter at density above saturation \cite{HFB17}. Also tensor force contribution has been considered \cite{Colo_t,Lesi,Otsu}. Despite the main part of this form is common to most Skyrme forces,
the values of the parameters can be different because of the different set of physical quantities that have been fitted and different fitting procedure.
The energy mean value corresponding to this force for a generic Slater determinant many-body wave function has then to be minimized with respect to the single particle wave functions, at a fixed value of N and Z (i.e. a definite nucleus). For a density independent force this results in the well known Hartree-Fock (HF) self-consistent equations \cite{RS} for the single particle wave functions. For  
density dependent forces, as the Skyrme one of Eq. (\ref{eq:Skyrme}), one still gets equations of the HF type, but the effective force to be used is different from the original (\ref{eq:Skyrme}) because the variation of the energy involves also the density appearing in the two-body force. Indeed, if we indicate with $i, j, k, l .....$ the single particle states of a given basis set,
the Skyrme functional is just the mean value of the Skyrme hamiltonian in a generic Slater determinant
\beq
E \,=\, T \,+\,  \frac{1}{2} \sum_{iklj} \rho_{ik}\, f_{il ; kj}^A(\rho) 
\, \rho_{lj} \,=\, T \,+\, V(\rho), 
\label{eq:fun}\eeq
\noindent where $T$ is the kinetic energy, $\rho_{ik}$ is the density matrix and the effective force $f$ can be written in general as a functional of the density matrix. The matrix elements of the effective force are anti-symmetrized 
\beq
f_{il ; kj}^A  \,=\, <\, i l\, |\, f\, |\, k j\, >  
\,-\,  <\, i l\, |\, f\, |\, j k\, >.  
\label{eq:antif}\eeq
The density matrix can be written as the expectation value
\beq
\rho_{ik} \,=\, < \psi_i\,^\dag \psi_k >
\label{eq:ro}\eeq
\noindent where $\psi^\dag$, $\psi$ are the single particle creation and annihilation operators and the mean value is on the Slater determinant. In minimizing the functional with respect to the single particle states, one has to take the functional derivative of the potential part with respect to the density matrix
\begin{eqnarray} 
 < k\,\, |\,\, {\cal V}\,\, |\,\, k' > & = & \big( \frac{\delta V}{\delta \rho_{kk'}}\big) \nonumber \\
  & = &  \frac{1}{2} \sum_{lj} \big[ f_{kl ; k'j}^A\, \rho_{lj}  \,+\, \rho_{lj}\, f_{lk ; jk'}^A \,+\,
 \sum_{ip} \rho_{ip}\, \big( \frac{\delta f_{il ;pj}^A }{ \delta \rho_{kk'} }\big)\, \rho_{lj} \big] 
\label{eq:VHF}
\end{eqnarray}
\noindent The functional derivative defines the single particle potential ${\cal V}$, which coincides with the Hartree-Fock single particle potential once the HF equations have been solved for the HF single particle orbitals (and the corresponding density matrix). For a density independent force one gets only the usual first two (equal) terms inside the square brackets. We will not discuss the different technical details and problems \cite{Egi,Bend,Lac,Wash} that are present in the applications, in particular for deformed nuclei \cite{Satu}, but we will focus on the comparison among the different Skyrme forces in connection with the symmetry energy and its density dependence. \par
On the basis of the expression of Eq. (\ref{eq:Skyrme}) for the Skyrme force, it is straightforward to calculate the EOS for asymmetric nuclear matter
\begin{eqnarray}
\frac{E}{A} (\rho,\beta) &=& \frac{3 \hbar^2}{10 m} \left( \frac{3 \pi^2}{2} \right)^{2/3}\, G_5(\beta) \,+\, \frac{t_0}{8}\, 
\rho \left[ 2 (x_0 \,+\, 2) \,-\, (2 x_0 \,+\, 1)\, G_6 \right]  \nonumber \\
& + & \frac{1}{48} t_3\, \rho^{\alpha + 1} \left[ 2 (x_3 \,+\, 2) \,-\, (2 x_3 \,+\, 1)\, G_6(\beta) \right] \nonumber \\
& + &  \frac{3}{40} \left(\frac{3\pi^2}{2}\right) \rho^{5/3} \left[ c_1\, G_5(\beta) \,+\, c_2\, G_8(\beta) \right],  \\
c_1 &=& t_1 ( x_1 \,+\, 2 ) \,+\, t_2 ( x_2 \,+\, 2 ) \nonumber \\
c_2 &=& \frac{1}{2} \left[ t_2 ( 2 x_2 \,+\, 1 ) \,-\, t_1 ( 2 x_1 \,+\, 1 ) \right]  \nonumber \\
G_n(\beta) &=& \frac{1}{2} \left[ q^n \,+\, p^n \right]   \nonumber 
\label{eq:EOS_Skyrme}
\end{eqnarray}
\noindent where $p$ and $q$ have the same meaning as in Eq. (\ref{eq:bTF}). The expansion of the EOS at $\beta = 0 $ gives the symmetry energy as a function of density
\begin{eqnarray}
 S(\rho)&=& \frac{\hbar^2}{6m}\, \left(\frac{3 \pi^2}{2}\right)^{2/3}\, \rho^{2/3} \,+\, \frac{t_0}{8}\, ( 2 x_0 \,+\, 1)\, 
 \rho \,-\, \frac{1}{48}\, t_3\, ( 2 x_3 \,+\, 1 )\, \rho^{\alpha + 1} \nonumber \\
& + & \frac{1}{24}\, \left(\frac{3 \pi^2}{2} \right)^{2/3}\, \left[ c_1 \,+\, 4\, c_2\, \right]\, \rho^{5/3}       
\label{eq:SE_Skyrme}
\end{eqnarray}
\noindent The expression for the slope $L$ of $S(\rho)$ contains of course the same terms, with only a different set of coefficients.
Analogous expressions can be derived for other physical quantities of NM, like the incompressibility and its derivative.
The same parameters appear in all these expressions. It is then unavoidable that a certain degree of correlation exists among the different physical quantities. This is a consequence of the simplified form of the force and of the NM properties derived from the force. These correlations ca be considered in general as spurious, since in a fully microscopic approaches they cannot be defined or they are much more complex. However particular correlations can have a physical meaning, since particular physical quantities are surely determined by the properties of homogeneous and non-homogeneous NM.\par
\begin{figure}[h]
\vskip -3 cm
\centering
\includegraphics[scale=0.5]{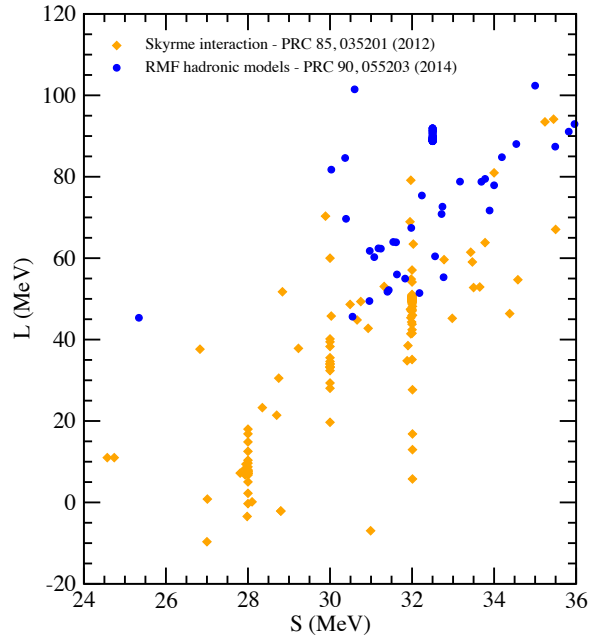}
\vskip -1.5 cm
\caption{(Color on line) Values of the symmetry energy $S$ and the slope parameter $L$ for a set of 142 Skyrme forces from ref.\cite{Jirina}, (orange diamonds), and 87 relativistic mean fields from ref. \cite{Jirina2},, (blue full dots). 
They have been selected according to the criteria described in the text.} 
\label{fig:LS}
\end{figure}

A compilation of a set of 240 Skyrme forces was presented in ref. \cite{Jirina}. From this set we selected 142 Skyrme forces that satisfy the following requirements : i) They give an acceptable saturation point, which means a saturation density $\rho_0$ in the interval $\, 0.15 < \rho_0 < 0.17$ fm$^{-3}\, $ and the corresponding energy per particle $e_0$ in the interval $\, -17 < e_0 < -15\, $ MeV ; ii) The calculated incompressibility $ K_0 $ at saturation has a value in the interval $\, 200 < K_0 < 260$ MeV, a restriction that is suggested by the phenomenological analysis on the Giant Monopole Resonance, as originally discussed in ref. \cite{Blaizot};
iii) They are compatible with the constraints from the heavy ion phenomenology on the flux and kaon production, as it will be discussed in Sec. \ref{HI}. In Fig. \ref{fig:LS} we report in a two-dimensional plot the value of $S$ and $L$ for each Skyrme
force (orange diamonds) and 87 relativistic mean field \cite{Jirina2} (blue full dots) satisfying these criteria. 
\noindent
Despite the strong restrictions we have imposed, the values of these two quantities spread all over the plot, with values ranging from $30$ to $100$ MeV for $L$ and from $24$ to $36$ MeV for $S$. The only gross trend that is barely visible is an increase of $L$ as $S$ increases. The conclusion that can be obtained from this plot is that there is no evidence of
 even weak correlation between the values of $S$ and $L$ when only data on binding energy are used in the parameters fitting. They are physical parameters mainly independent from each other and from other physical quantities like the incompressibility and the specific EOS which characterize the Skyrme forces or RMF models. 
 The spread of the values of $S$ and $L$ means also that the density dependence of the symmetry energy is different in different Skyrme forces. This is exemplified in Fig.1 of ref. \cite{Tsang}, where $S$ as a function of the density $\rho$ is reported for a selected set of Skyrme force. A considerable spread in the symmetry energy is apparent, but even Skyrme forces that have a comparable value of $S$ at saturation can have a quite different slope at saturation. 
\par      
The lesson, that one can get by inspection on this overview of results, is that it could be possible to extract a correlation, or even the separate values of $S$ and $L$, only if other constraints are introduced, which are more sensitive to the symmetry energy.    
To this respect the correlations between different physical quantities, that can be established following the same procedure, can be of great help.   
\par
An alternative method to construct an EDF is based on the Kohn-Sham scheme \cite{KS}. In this case the functional is constructed directly from the EOS of the homogeneous system (nuclear matter), which is taken from a reliable microscopic many-body calculation.
In addition surface terms are introduced, in order to include the surface energy \cite{BCPM_KS,Fay,Stei}. The surface terms contain the main fitting parameters. The symmetry energy is therefore fixed from the adopted EOS. The EDF so devised cannot compete with the most performing Skyrme functionals, since they use a much more limited set of parameters.

\subsubsection{Isovector excitations.}
\label{IVex}
The natural choice for finding physical quantities that could be sensitive to the symmetry energy is to look at the isovector elementary nuclear excitations, because in this case it is expected that at least part of the neutrons and of the protons moves out of phase.
These excitations should produce local or overall oscillations of the asymmetry in the nucleus.
Excitations of different multipolarities have been analyzed with the aim of finding correlations between observable quantities and the symmetry energy. An analysis of all these collective vibrations has been presented in ref. \cite{multipol}.\par 
The prototype of the isovector excitation is the Isovector Giant Dipole Resonance (IVGDR), historically the first one that has been discovered. The excitation can be viewed macroscopically as a collective oscillation of the protons against the neutrons, which
necessarily involves a variation of the asymmetry in the bulk or surface part of the nucleus. The main uncertainty is on the relevance of the surface, where the local symmetry energy is surely decreasing with density. This indicates that the effective symmetry energy involved in the excitation must be smaller than the one in NM at saturation. \par
\begin{figure}[hbt]
\centering
\includegraphics[scale=0.4,clip]{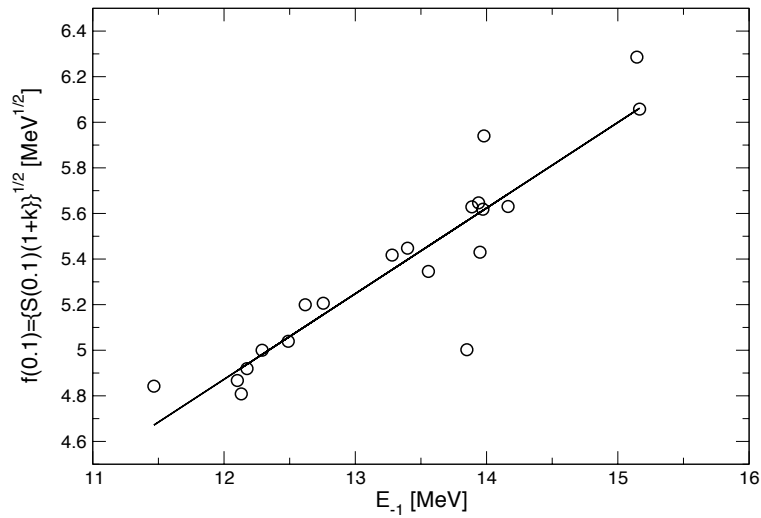}
\caption{Correlation between symmetry energy at $ 0.1~\rm fm^{-3}$, in the indicated combination,  and the dipole excitation energy. Figure taken from ref. \cite{Trippa}.} 
\label{fig:dipole}
\end{figure}

The connection between symmetry energy and the dipole excitation energy can be elucidated with the help of macroscopic models.
In general one can use sum rules, derived within specific macroscopic models, to estimate the excitation energy for each multipolarity. We will focus on the energy weighted sum rule $m_1$ and inverse energy weighted sum rule $m_{-1}$, from which the excitation energy $E_{coll}$ of the collective vibration can be estimated as
\beq
E_{coll} \,=\, \sqrt{m_1/m_{-1}}
\label{eq:estimate}
\eeq    
\noindent The generic sum rule $m_k$ is defined according to
\beq
m_k \,=\, \sum_n \, E_n^k \, | <\, n\, |\, Q\, |\, 0\, >|^2  
\label{eq:sum}
\eeq
\noindent where $E_n$ is the excitation energy of the state $n$, $0$ indicates the ground state and $Q$ is the operator associated to the excitation (of a given multipolarity). The $m_1$ sum rule can be calculated in an almost model independent way. One gets
\beq
m_1 \,=\, \frac{2 \pi^2 e^2 \hbar}{m c}\, \frac{NZ}{A}\, ( 1 \,+\, \kappa )
\eeq 
\noindent where $\kappa$ is a measure of the non-locality of the effective force (for a local zero range force $\kappa \,=\, 0$).
The other sum rules present different degrees of model dependence. For the IVGDR excitation the general form of Eq. (\ref{eq:estimate}), that is obtained from different models, can be written
\beq
E_{-1} \,=\, \sqrt{\frac{6 \hbar^2}{m <\, r^2\, >} \frac{S(\rho_0)}{1 \,+\, C_S}\, ( 1 \,+\, \kappa\ ) }
\label{eq:ED}
\eeq
\noindent where $<\, r^2\, >$ is the mean value of $ r^2 $ in the ground state, which is an estimate of the square of the nuclear radius. The quantity $C_S$ takes into account the surface correction to the symmetry energy $\, S(\rho_0)\, $ of NM.
This parameter is model dependent, and has different expressions according to the method used to estimate the surface contribution. One can mention the leptodermous expansion \cite{RS}, the hydrodynamical model \cite{LS}, semi-infinite matter calculations \cite{siNM} and the semi-classical method of ref. \cite{Daniel2009}, see Eq. (\ref{eq:int}). This expresses the uncertainty, mentioned above, on the surface contribution to the symmetry energy for finite nuclei. To circumvent this uncertainty one can 
introduce \cite{multipol} an effective symmetry energy as the one in NM calculated at an average density $\overline{\rho}$
\beq
\frac{S(\rho_0)}{1 \,+\, C_S} \,\approx\, S(\overline{\rho})
\label{eq:roav}
\eeq  
\noindent This is in line with ref. \cite{XavierPRL}, where it was found a strong correlation between the symmetry energy for a given nuclear mass number and the symmetry energy of NM calculated at a suitable density $ \overline{\rho} $. This was shown by considering a set of Skyrme forces with different behaviour of the symmetry energy as a function of density. The value of $ \overline{\rho} $ is smoothly dependent on the mass region, but in any case it turns out to be close to $ 0.1 $fm$^{-3}$. This is also the density where the correlation ( "Pearson correlation" ) between the nuclear symmetry parameter and the NM symmetry energy $ S(\rho) $ has its maximum \cite{DanielIAS}, close to 1 (tight correlation), with only a slight increase with the mass number. As an illustration of the general procedure that is usually followed to extract the symmetry energy from the isovector nuclear collective vibration,
we follow ref. \cite{Trippa}. In Fig. \ref{fig:dipole}, taken form ref. \cite{Trippa}, for a set of Skyrme forces the approximation of Eq. (\ref{eq:roav}) is used, with $ \overline{\rho} \,=\, 0.1 ~$fm$^{-3} $, to show the existing correlation between the symmetry energy, in the combination of Eq. (\ref{eq:ED}), and the IVGDR excitation energy. 
The latter has been calculated within the Hartree-Fock + RPA scheme for each Skyrme force to get the estimate of Eq. (\ref{eq:ED}). The non-locality parameter $ \kappa $ was obtained by imposing the fulfillment of the sum rule $ m_1 $ (with a large enough energy cut-off). All calculations are for $^{208}$Pb. Notice that the Skyrme forces were selected among the ones that produce a reasonable incompressibility value ($\rm 210~ < K <  270~ MeV$) and reproduce the IVGDR energy within 2 MeV. The straight line is a linear best fit, which turns out to have a correlation coefficient close to 0.9 . 
\noindent The observed correlation is surely related to the mentioned selection of the Skyrme forces, because it includes a constraint that should be sensitive to the symmetry energy. In principle, the linear correlation of Fig. \ref{fig:dipole}, or similar, can be used to extract the symmetry energy, at least at $ \overline{\rho} = ~0.1~\rm fm^{-3}$. To this purpose one considers the experimental IVGDR excitation energy to select the value of $ S(0.1) $  consistent with the data. The authors find, with an estimate of errors
\beq
 23.3\,\, {\rm MeV} \, < \, S(0.1) \, < \, 24.9\,\, {\rm MeV}
\label{eq:estdipole} 
\eeq
\par
\begin{figure}[hbt]
\vskip -3cm
\centering
\includegraphics[scale=0.6,clip]{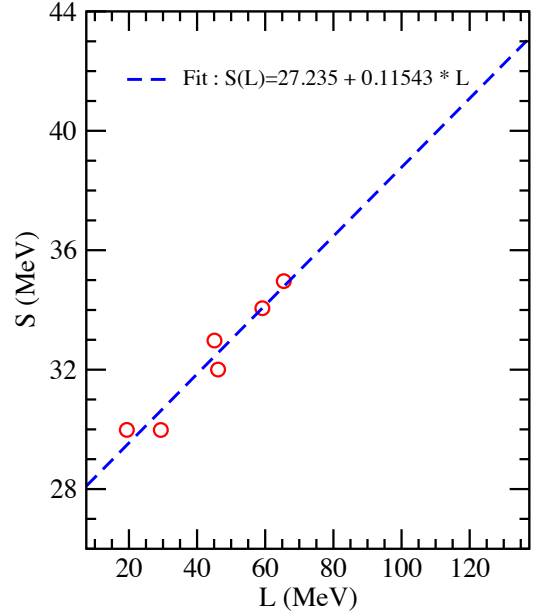}
\vskip -3cm
\caption{Correlation between symmetry energy and its slope at saturation for a selected set of Skyrme functionals (open circles) that reproduce the data on the dipole polarization. Adapted from ref. \cite{polar}.} 
\label{fig:alpha}
\end{figure}

 
A related quantity is the so-called (static) polarizability  $ \alpha_D $ of a nucleus. It is the linear response to an external static dipole field. If $\lambda\, D$ is the external field, where $D$ is the dipole operator, one has 
\beq
\alpha_D \,=\, \frac{\partial <\, D\, >}{\partial \lambda}
\label{eq:alpha}\eeq
\noindent where $ <\, D\, > $ is the dipole mean value in the perturbed nuclear state. This is just the static (zero frequency) response function \cite{RS} $ R(0) $
\beq
\alpha_D \,=\, R(0) \,=\, 2 \sum_{n} \, \frac{| <\, n\, |\,\, D\,\, |\, 0\, > |^2}{E_n \,-\, E_0} \,=\, 2\, m_{-1}
\label{eq:resp}\eeq 
\noindent where $ E_n $ is the energy of the $ n^{th} $ excited state, with $ 0 $ the ground state. The relevance of this quantity is that it can be measured from the photoabsorption cross section $ \sigma (\omega) $, where the dipole contribution is dominant
\beq
\alpha_D \,=\, \frac{\hbar c}{2 \pi^2 e^2}\, \int \frac{\sigma(\omega)}{\omega^2}\, d\omega 
\label{eq:cross}\eeq  
\noindent Notice the enhancement at low energy due to the $ 1/\omega^2 $ factor. The polarizability can be obtained also from Coulomb excitation and polarized proton inelastic scattering at forward angles \cite{Tamii}. If $ \alpha_D $ is fixed, one can expect that this implies some correlations between suitable physical parameters connected with symmetry energy. The experimental value of $ \alpha_D $, although affected by uncertainty, can provide relevant information on the nuclear symmetry energy. In Fig. \ref{fig:alpha}, adapted from ref. \cite{polar}, are reported the values of $ S $
and $ L $ for a set of Skyrme energy functionals which are able to reproduce the data on $ \alpha_D $ for the nuclei $^{68}$Ni, $^{120}$Sn, $^{208}$Pb . Because of this constraint the set of functionals is of reduced size, but a correlation between the two parameters is evident, at variance with the results shown in Fig. \ref{fig:LS}. One should notice, however, that experimentally the regions well above and well below the IVGDR are not easily accessible, so that the data are affected by a systematic uncertainty, see ref. \cite{polar} for a discussion on this point. It is not straightforward to explain the correlation induced by the $ \alpha_D $ data. A first insight can be obtained by looking at the droplet model. In ref. \cite{dropol} the expression for the polarizability was derived following the same minimization procedure in the standard droplet model, see Sec. \ref{droplet}, but in presence of a static dipole external field
\beq
\alpha_D \,=\, \frac{A r^2}{24 S}\, \left( 1 \,+\, \frac{15}{4 A^{1/3}}\, \frac{S}{Q}\right) 
\label{eq:dropalpha}\eeq   
\noindent where $ r $ is the radius of the nucleus. Taken literally together with the correlation shown in Fig.\ref{fig:alpha}, this relation would imply a connection between $ S/Q $ and the slope parameter $ L $ appearing in Fig. \ref{fig:alpha}. It turns out that this is the case, as we will discuss in Section \ref{skin}.   
\par 

\par 
As already mentioned, the low energy part of the dipole excitation strength is particularly relevant for the dipole polarizability experiments. The contribution of the low energy region is further enhanced in the case of neutron rich exotic nuclei by the observed presence of an
appreciable concentration of dipole strength at few MeV of excitation. This strength has been classified as '' Pygmy "
resonance, since its size is much smaller than the IVGDR one, in general few percents of the total sum rule for isovector dipole electric transitions. An overview of the experimental data and methods, up to the year 2012, can be found in ref. \cite{Savran}. 
In heavy asymmetric nuclei a neutron skin appears, i.e. a neutron radius substantially larger than the proton one. This is expected also in non-exotic nuclei like $ ^{208}$Pb, but it is particularly pronounced in exotic nuclei, especially unstable ones.
It has been then argued \cite{Mohan} that the Pygmy strength could be produced by a collective motion of the neutron skin against the symmetric neutron-proton core, and therefore decoupled by the dipole giant resonance. This picture has some support from the strong isoscalar component of the mode. However this macroscopic picture of a collective  low frequency mode could be misleading since at the quantal level this dipole strength could be due to a certain number of non-collective excitations, mainly single particle-hole excitations. This is a controversial issue and accordingly the connection of the Pygmy strength with the symmetry energy is also controversial.        
In refs. \cite{Klimkiewicz,Carbon} it was argued that the Pygmy strength can be related to the symmetry energy and the size of the neutron skin. This conclusion was questioned in ref. \cite{Daoutidis} on the basis of an analysis of the experimental and theoretical uncertainty, and in refs. \cite{Reinhard1,Reinhard2} on the basis of a covariance analysis of the correlations. From this discussion one can conclude that the issue of the relation between Pygmy strength and symmetry energy (or neutron skin width) is at least controversial. Additional theoretical analysis and experimental data are needed to get any firm conclusion.  
\begin{figure}[hbt]
\vskip 1 cm
\centering
\includegraphics[scale=0.7]{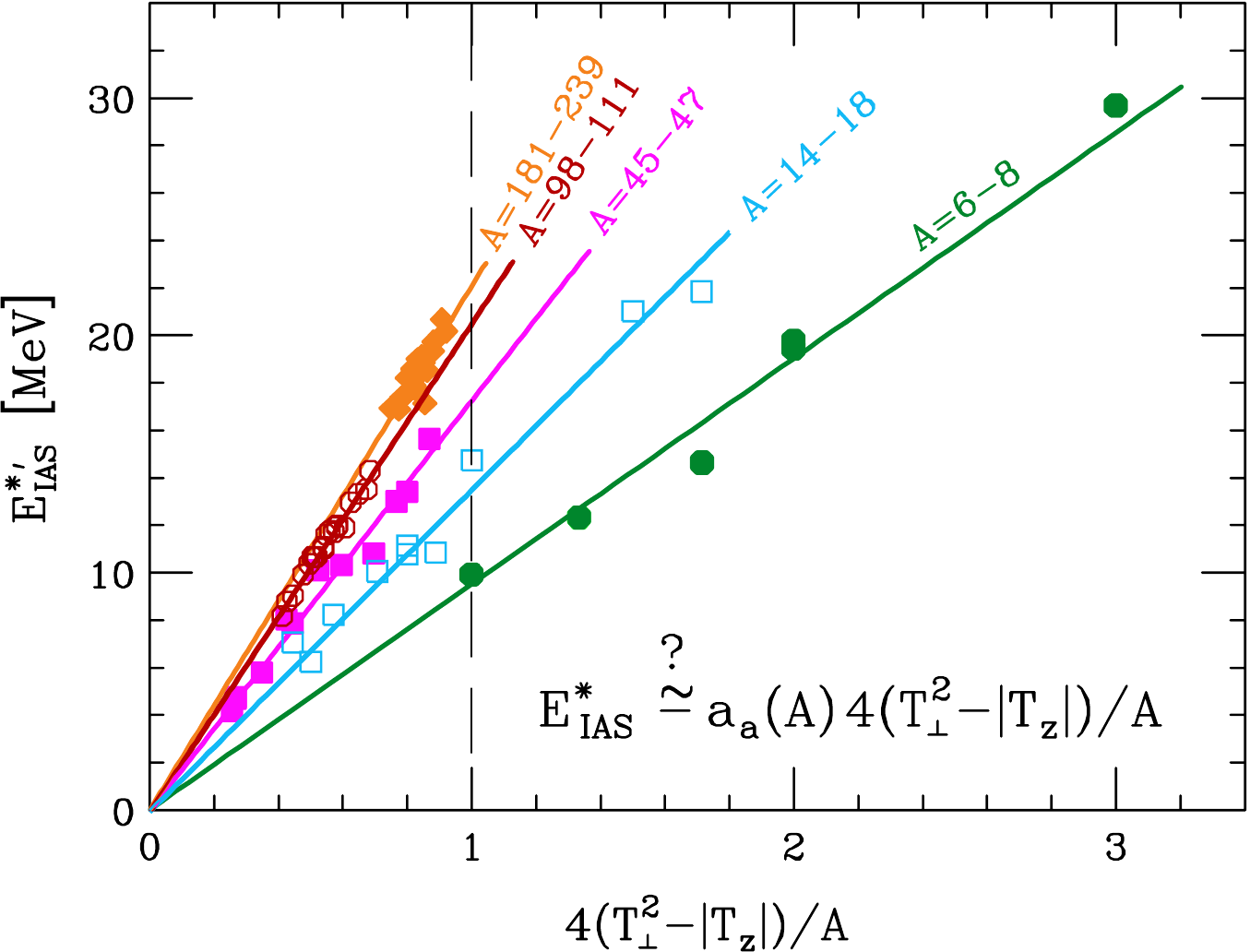}
\caption{(Color on line) Linear fits of the IAS energies as a function of the indicated isospin variable, for each of the $ A $ regions, as labeled. Taken from \cite{DanielIAS}.} 
\label{fig:fitIAS}
\end{figure}

\begin{figure}[hbt]
\vskip 1cm
\centering
\includegraphics[scale=0.7]{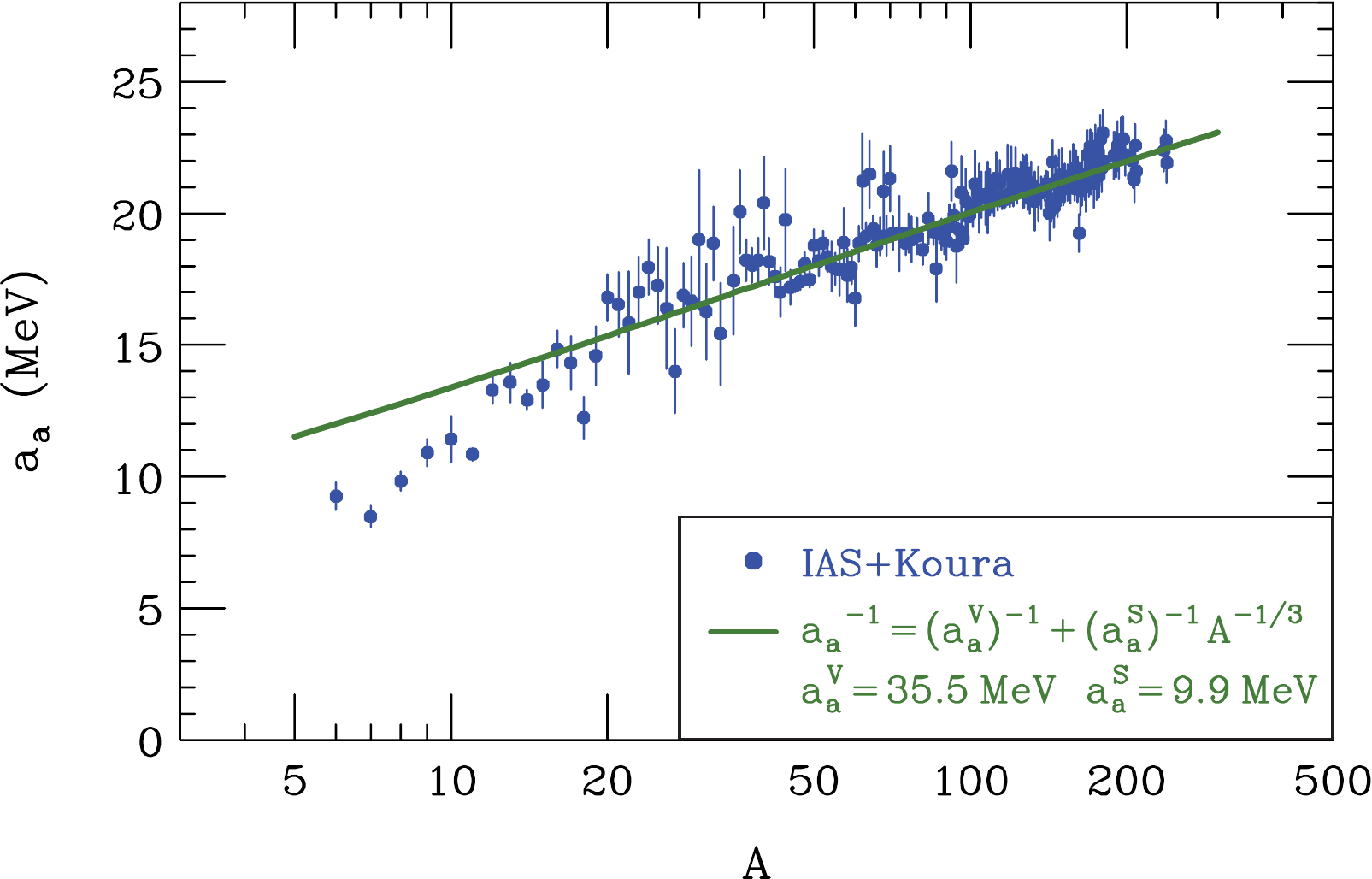}
\caption{(Color on line) Fit of the nuclear symmetry energy $ S_N(A) $ as a function of $ A $ with the form of Eq. \ref{eq:int}. Taken from \cite{DanielIAS}.} 
\label{fig:aVIAS}
\end{figure}
\par 

The isovector giant quadrupole resonance (IVGQR) has been also analyzed in connection with the nuclear symmetry energy.
Macroscopically the IVGQR should correspond to a quadrupole oscillation of the neutron and of the proton components out of phase between each other. This creates regions of the nucleus, in particular close to the surface, where the nucleonic matter asymmetry
oscillates around the equilibrium value. The restoring force of the vibration is therefore not only the surface tension, like in the isoscalar giant quadrupole resonance (ISGQR), but also the symmetry energy. For this reason the excitation energy of the IVGQR should be above the ISGQR one. The difference in excitation energy is therefore connected with the nuclear symmetry energy. The experimental observation of the IVGQR is more difficult than the well known ISGQR, and only in the last few years detailed data have been presented in the literature \cite{IVGQRpre}. 
For the same reasons, theoretical studies of the connection between symmetry energy and IVGQR are scarce. In ref. \cite{IVGQRskin} the harmonic oscillator model \cite{BM}, supplemented with the droplet model, was used to relate the IVGQR and ISGQR energy difference directly to the symmetry energy 
\beq
S(\overline{\rho}) \,\approx\, S_0 \left( \frac{A^{2/3}}{8 E_F^2}\, \left[ E_V^2 \,-\, 2 E_S^2 \right] \,+\, 1 \right)
\label{eq:diff}\eeq    
\noindent where $E_V$ and $E_S$ are the excitation energies of the IVGQR and ISVQR, respectively, and $ S_0 \,=\, E_F/3 $ is the symmetry energy of a free Fermi gas at saturation density and $ E_F $ the corresponding Fermi energy. In Eq. (\ref{eq:diff}) $\overline{\rho} $ is an average nucleus density, close, as usual, to 0.1 fm$^{-3}$. Using the experimental values of the excitation energies, properly averaged over different experiments \cite{IVGQRskin}, one gets
\beq
S(\overline{\rho}) \,=\, 23.3 \,\pm\, 0.6 \ \ {\rm MeV}
\label{eq:SIV}
\eeq      
\noindent which is compatible with the estimate (\ref{eq:estdipole}) from the IVGDR. Given a functional, one can calculate both $ E_V $ and $ E_S $ and from the relation (\ref{eq:diff}) estimate $ S(\overline{\rho}) $. Comparing this value to the direct calculation from the functional gives a test of the simplified relation (\ref{eq:diff}). The results \cite{IVGQRskin} obtained by considering families of functionals indicate a fair agreement, but in a few cases it appears that some characteristics of the force, other than the symmetry energy, are relevant.
\subsubsection{The Isobaric Analog State.}
\label{IAS}
Special consideration has to be devoted to the issue of the Isobaric Analog States (IAS) in connection with the nuclear symmetry energy, mainly developed by P. Danielewicz and J. Lee \cite{DanielIAS}. In a nucleus with a given neutron number N and proton number Z the ground state is characterized by a value of the total isotopic number projection $ T_Z \,=\, ( N \,-\,Z )/2 $ and total isospin number $ T \,=\, | T_Z | $. Let us imagine to keep the same wave function except for the change of a neutron into a proton, or equivalently to act with the operator $ T_{-} $, which lowers the value of $ T_Z $ by one unit. The new total wave function will correspond in general to the excited state of a neighboring isobar, with $ T \,=\, T_Z \,-\, 1 $
\beq
 T_{-} |\, \{n\},\, T,\,  T_Z = T\, >  \,\,\,=\, \sqrt{2 T}\, |\, \{n\},\, T,\,  T_Z = T\, -\, 1\, >
\label{eq:Tminus}\eeq
\noindent where $ n $ indicates all the other quantum numbers. The only correction is due to the change in Coulomb energy, because
the Coulomb potential is the only term of the nuclear Hamiltonian that does not commute with the isospin operator. This isobar analog state is expected to be the lowest state for this value of the total isospin $ T $ for the '' daughter nucleus ", and therefore it corresponds to a sharp resonance, since isospin is a good quantum number for the nuclear interaction. The charge invariance does not imply only that the energy depends on the $ T_Z^2 $, but more generally that it is invariant on rotation in isospin space, and therefore it should depend only on the modulus square of the isospin
\beq
 E_T \,=\, 4\, S_N(A)\, \mathbf{T}^2
\label{eq:T2}\eeq   
\noindent where $ E_T $ is the nuclear energy term that depends on isospin and $ S_N(A) $ is the symmetry energy for the nucleus of atomic number $ A $. This means that for a given $ T_Z $ a multiplet of states should be present with an energy separation dictated by Eq. (\ref{eq:T2}). The excitation energy of each member of the multiplet with respect to the ground state is then
\beq
E_x(T) \,=\, 4 S_N\, \left[\, T( T \,+\, 1 ) \,-\, | T_Z |\, (\, | T_Z | \,+\, 1\, )\, \right] \,=\, 4 S_N \left[\, \mathbf{T}_{\bot}^2 \,-\, | T_Z |\, \right]
\label{eq:EX}\eeq
\noindent Of course this conclusion holds if the other contributions to the energy are the same in the ground and excited states.
This sounds plausible if the member of the multiplet are the lowest ones for each value of $ T $. In this case one can assume that the energy can be derived from the LDM or DM, with possible microscopic corrections, like shell effects and deformations. After the ( differences of ) microscopic contributions has been subtracted, the IAS energies should be then proportional to $ \mathbf{T}_{\bot}^2 \,-\, | T_Z | $. A fitting procedure can be used to verify to what extent such a behavior is valid and to extract the symmetry energy on the nucleus-by-nucleus basis. In practice the number of IAS for a given isobaric chain at the atomic number $ A $ turns out to be too small, and the IAS were grouped in different mass region, which is justified by the expected smooth dependence of the nuclear symmetry energy $ S_N $ on the atomic number $ A $. In Fig. \ref{fig:fitIAS}, taken from ref. \cite{DanielIAS}, the corrected IAS energies are reported as function of $ 4\, ( \mathbf{T}_{\bot}^2 \,-\, | T_Z | )/A $, grouped in different atomic mass regions. The microscopic corrections in this calculations were taken from the ones tabulated in ref. \cite{Koura}.  The linear fits can be used to extract the average nuclear symmetry energy within the indicated $ A $ interval, by reading off the slope of the linear fit or equivalently taking the intersection with the vertical line at "1 " (dashed line).  
\begin{figure}[hbt]
\vskip -3.9cm
\centering
\includegraphics[scale=0.55,clip]{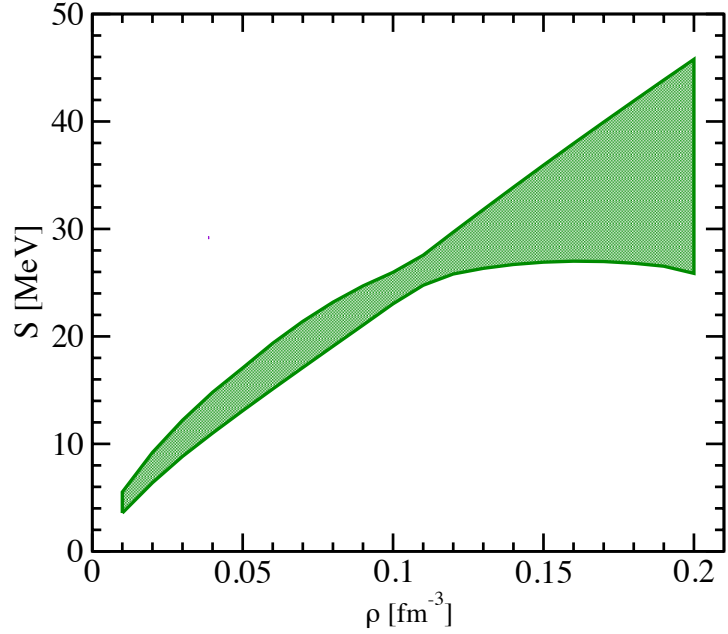}
\vskip -1 cm
\caption{(Color on line) Region (filled area) that indicates the boundary for the values of the symmetry energy, once the Skyrme functional are constrained to reproduce the results on the IAS analysis. Adapted from \cite{DanielIAS}.} 
\label{fig:Sro}
\end{figure}
\par 
The plots of Fig. \ref{fig:fitIAS} show that Eq. (\ref{eq:EX}) is fulfilled to a good accuracy. Once the nuclear symmetry energies have been extracted, one can look if the dependence on the atomic number follows some smooth behaviour. In particular one can verify to what extent equation (\ref{eq:int}) is fulfilled. It turns out that for large enough atomic number $(A  > 30)$ the trend of the equation is followed with a fair accuracy. This is shown in Fig. \ref{fig:aVIAS}, taken again from ref. \cite{DanielIAS}  (the parameter $a_a$ of the figure corresponds to the nuclear symmetry energy $S_N(A)$ at a given atomic mass A). 
\noindent
 If we interpret the parameter $ a_V $ as the nuclear matter symmetry energy $S$, the extracted value, reported in the figure, looks substantially higher than the values estimated from other methods. However the uncertainty associated with each point, as indicated by the error bars in the figure, and the fluctuations with respect to the average smooth trend are large enough to accomodate a substantial spread of the fitted value. The indicated error bars have been estimated by a statistical analysis of the IAS fit for each individual nucleus \cite{DanielIAS}. \par 
This extensive analysis of the IAS data aims to find not only the NM symmetry energy $ S $ at saturation, but also its whole density dependence, at least at sub-saturation density. This is not an easy task, since nuclear data correspond to overall properties of nuclei. One procedure, which is often used also in other context, is the use of a representative set of functionals that is compatible with the given constraints. In the present case of the data on IAS, one should select the functionals that reproduce the nuclear symmetry energy $S_N(A)$ extracted from the IAS, as described above. One can then check to what extent the selected functionals
constrain the density dependent symmetry energy $S(\rho)$ of NM. In order to be viable, the method requires the possibility of extracting from the calculations, which are mainly within the Hartree-Fock (HF) scheme, the symmetry energy for each individual nucleus. The obvious method would be to extract $ S_N(A) $ from the ground state energy differences with respect to the neighbouring nuclei, i.e. by considering the energy change with respect to the close isobars (in the ground state). It was found in ref. \cite{DanielIAS} that this procedure is strongly biased by the shell effects and it is then not possible to extract any trend as a function of $ A $.               
Another possibility is to treat the variables $ (A,Z) $, which identify each nucleus, as continuous, i.e. to adopt a macroscopic picture of the nucleus, where only the smooth part of the energy is present. It is then possible to define the symmetry energy as the usual derivative with respect to asymmetry. Under the assumption that the symmetry energy part of the binding can be expressed as a local functional of the density profile, one can relate \cite{DanielIAS} the nuclear symmetry energy $ S_N(A) $ to the NM symmetry energy $ S(\rho) $ and to the difference
between the neutron profile $ \rho_n(r) $ and proton profiles $ \rho_p(r) $. The latter can be calculate directly from the functional once the HF calculation for the ground state is performed. We report here the final result for convenience
\beq
 \frac{A}{S_N(A)} \,\approx\, \frac{N \,-\, Z}{\int_0^{r_c} d^3 r \left[ \rho_n(r) \,-\, \rho_p(r) \right] } \, \int_0^{r_c} \, 
d^3 r \frac{\rho(r)}{S(\rho(r))}  
\label{eq:SNA}\eeq  
\noindent The interested reader can resume to the original paper for the detailed derivation and discussion. 
In Eq.\ref{eq:SNA}) $r_c$ is some cutoff that defines the ''interior" part of the density distribution. Therefore the factor in front gives
a measure of the contribution of the surface region. Eq. (\ref{eq:SNA}) expresses the inverse of the nuclear symmetry energy as a suitable average over the inverse of the local symmetry energy $ S(\rho) $, which is reminiscent of the law for the addition of the capacitance for capacitors. The formula is actually valid only if the Coulomb interaction is switched off. In practice one has to correct this formula for the interference to the symmetry energy produced by the Coulomb energy. We notice that this result has been derived assuming a local functional form for the symmetry energy term in the binding energy expression. This can sound plausible if only local Skyrme functionals are considered, but it is not necessarily true. 
\par 
For each density functional one can then derive the nuclear symmetry energy $S_N$ for each nucleus with given $ (A,Z)$
and selecting the ones that are compatible with the results obtained from the analysis of the isobaric analog states, as described above. This strongly restricts the region where the symmetry energy $ S(\rho) $ can pass in the ($ S, \rho $) plane.
The result for the allowed region, adapted from ref. \cite{DanielIAS}, is illustrated in Fig. \ref{fig:Sro}. The filled region is where the symmetry energy of all the selected Skyrme functionals is passing through. Actually, to avoid a too small sample of functionals, some limited deviations from the IAS constraints were allowed in selecting the functionals. More details can be found in the original paper. The limited region of Fig. \ref{fig:Sro} gives also a constraint on the slope parameter L, whose possible values
appear more restricted with respect to  the wide spread region spanned by the generic and unrestricted set of Skyrme functionals reported in Fig. \ref{fig:LS}.  
\noindent

\subsubsection{The neutron skin and the PREX experiment.}
\label{skin}
\begin{figure}[htb]
\centering
\includegraphics[scale=0.4,clip]{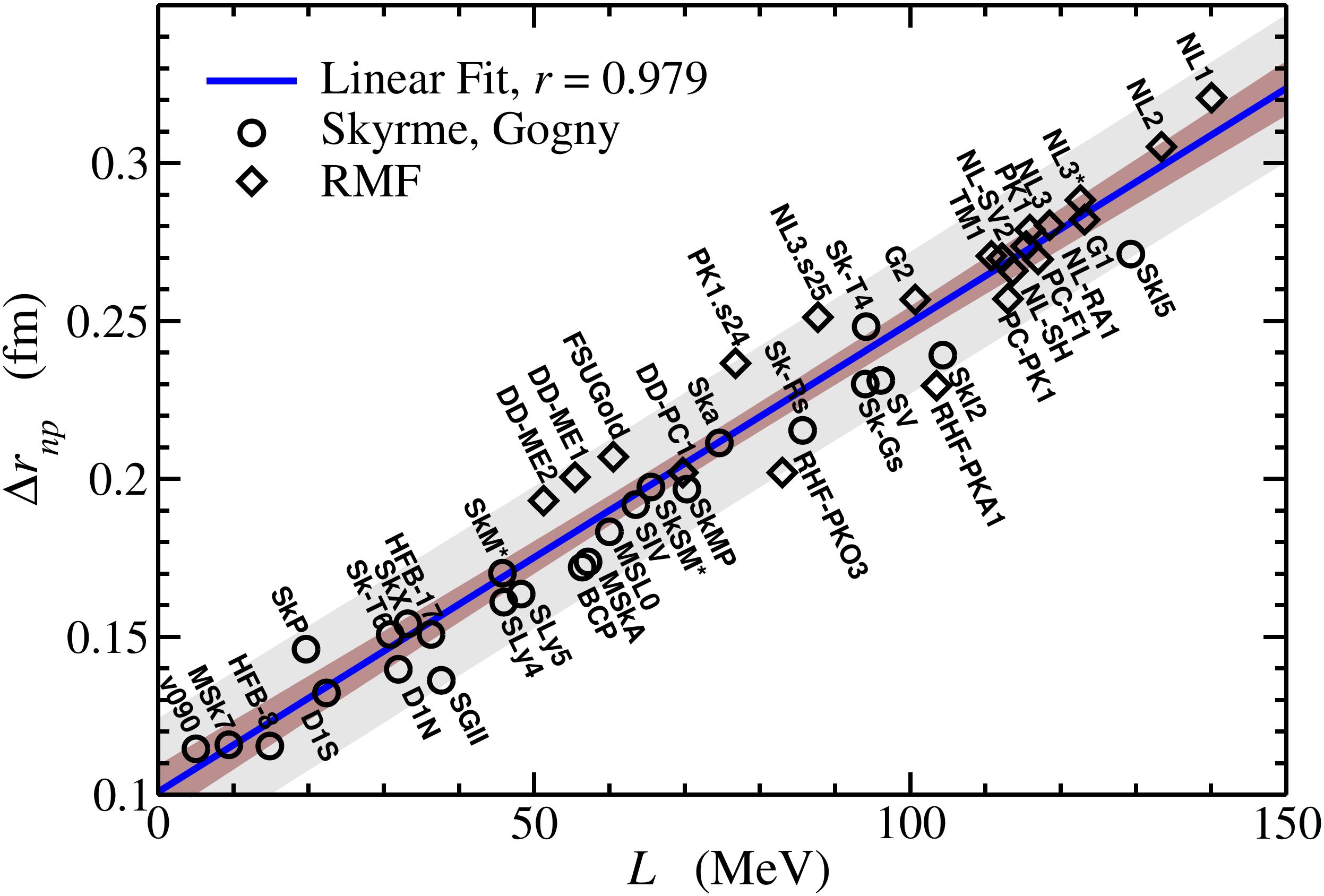}
\caption{(Color on line) Correlation between the neutron skin width of $^{208}$Pb and the slope parameter $ L $, calculated from different Skyrme and Gogny forces, and from relativistic mean field models. Taken from \cite{XavierEPJA}, where the meaning of the labels can be found.} 
\label{fig:skinL}
\end{figure}

In a nucleus with a neutron excess it can be expected that the neutron component spills out from the average radius since the corresponding pressure is larger than for the proton component. The spill out is contrasted by the symmetry energy, because a larger asymmetry produces an increase of the energy. The competition between these two effects determines the width of the surface region where the neutron component is dominant well above the $ N/Z $ ratio. This region is usually called ''skin" region. A paradigmatic example is $^{208}$Pb, which is not exotic but possesses a well defined neutron skin. From these simple considerations it is clear that the width of the neutron skin is sensitive to the symmetry energy, at least at sub-saturation density. In particular it can be expected that the skin is sensitive not only to the strength of the symmetry energy but also to its slope, because this determines how fast the symmetry energy decreases with density. A larger slope, that is the value of $ L $, makes the symmetry energy weaker at low density, thus increasing the skin width. To see this effect, one can define the skin size as the difference between the neutron and proton rms radius
\beq
 \Delta\, r_{np} \,=\, <\, r_n\,^{2}\, >^{1/2} \,-\, <\, r_p\,^{2}\, >^{1/2}
\label{eq:rnrp}\eeq   
\noindent and perform ground state calculations with a set of functionals which display different values of the slope parameter $ L $. The skin width extracted from these calculations, according to Eq. (\ref{eq:rnrp}), can be called "quantal skin width", according to ref. \cite{XavierPRL}.
As can be seen in Fig. \ref{fig:skinL}, taken from ref. \cite{XavierEPJA}, the quantal skin width (for $^{208}$Pb) is linearly correlated with $ L $, especially for Skyrme and Gogny functionals. The meaning of the labels for the functionals can be found in the original paper. In principle also in this case an experimental measure of the skin width would pin down the value of the slope parameter $L$.  

\par
\begin{figure}[htb]
\centering
\vskip -8cm
\includegraphics[scale=0.55,clip]{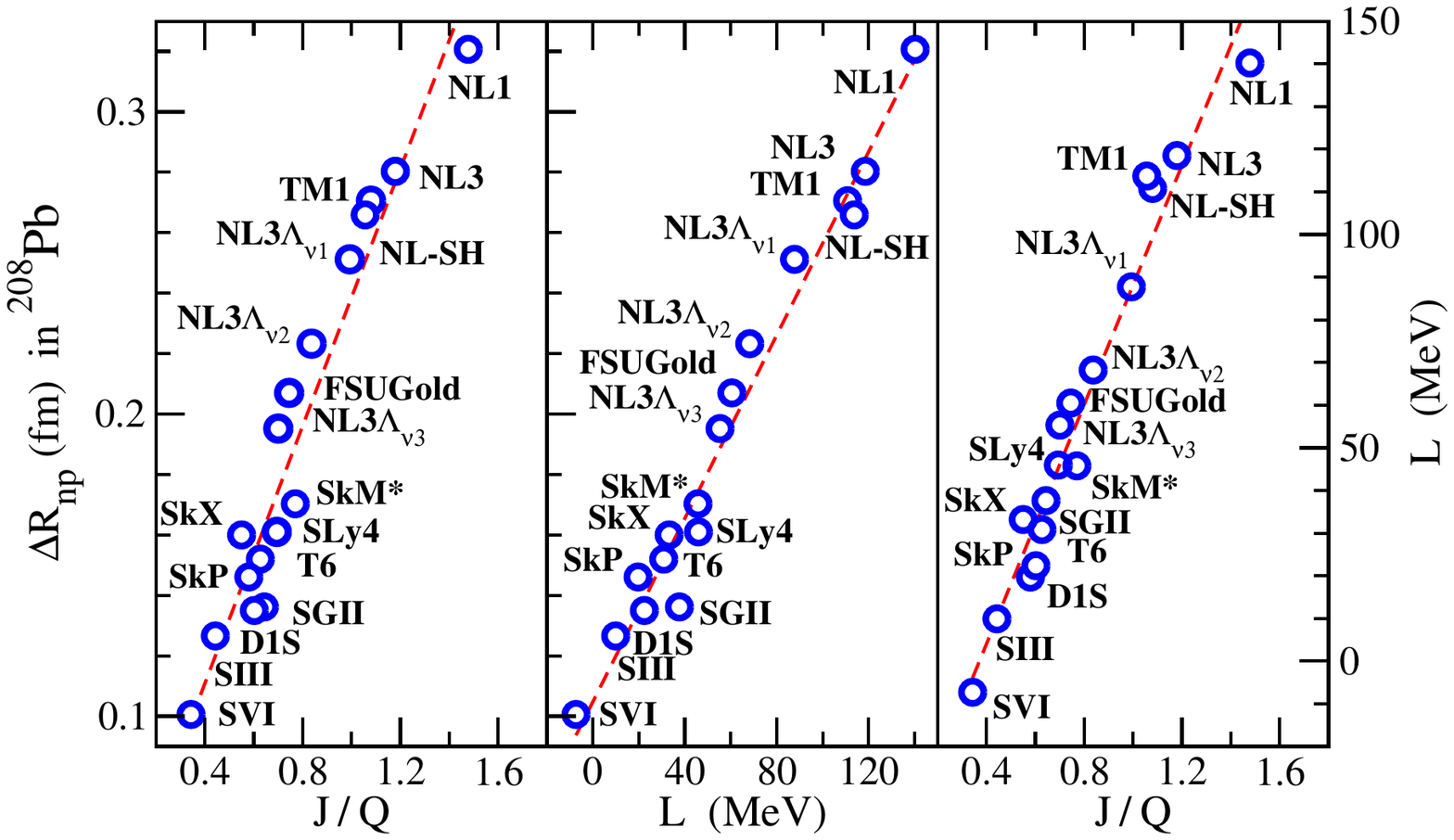}
\vskip -0.8 cm
\caption{(Color on line). Correlations between the skin width and the ratio $ J/Q $, and between the skin width and the slope parameter $ L $. On the right panel is displayed the resulting correlation between $ L $ and the ratio $ J/Q $. Here $ J $ is the symmetry energy at saturation ($ S $ in our notation) and $ Q $ is the parameter of the DM appearing in e.g. Eq. \ref{eq:delta}.  Figure from ref. \cite{XQ2009}.} 
\label{fig:LQ}
\end{figure}

 It turns out that several other physical parameters are correlated with the skin width, and therefore they are correlated among each other. A first example is the relation between the slope parameter $ L $ and the stiffness parameter $ Q $ of the DM. In fact in the DM the neutron skin is directly related to the ratio $ S/Q $ \cite{MS66,MS69}, according to (neglecting Coulomb for simplicity)
\beq
\Delta r_{np} \,=\, \frac{3}{2} r_0 \frac{S}{Q} \frac{\beta}{1 \,+\, \frac{9}{4} \frac{S}{Q} A^{-1/3}}
\label{eq:skinwidth}
\eeq 
\noindent where $ r_0 $ is the parameter of the nuclear radius, $ <\, r^2\, >^{1/2} \,=\, r_0\, A^{1/3}\, \equiv r\, $. In this expression we neglect also an additional term related to the possible difference between neutron and proton surface widths. Each functional is characterized by a different value of the parameter $ Q $. It can be extracted \cite{XQ2009} by comparing
Eq. (\ref{eq:skinwidth}) with Thomas-Fermi calculations in semi-infinite nuclear matter ( A $\rightarrow \infty $ ) at different asymmetry, where one can check the proportionality of $ \Delta r_{np} $ and $ S/Q $. With the same functional, one can then calculate in a given nucleus directly the quantal value of the skin width. For $^{208}$Pb results are displayed in Fig.\ref{fig:LQ}, taken from ref. \cite{XQ2009},
for a set of functionals with different values of $S$ and the corresponding correlations between the skin width and $ S/Q $ ($J \,\equiv\,S$), and between the skin width and the slope parameter $L$. This double correlations entails the correlation between $L$ and $ S/Q $, as anticipated in Sec. \ref{IVex}, displayed on the right panel of the figure.  
Notice that there is no ground for a correlation between the skin width and the symmetry energy $S$ (at saturation) if a set of unrestricted functionals is considered. Only for a restricted set of functionals, like the ones used for Fig. \ref{fig:alpha} or for Fig. \ref{fig:Sro}, the resulting correlation between $L$ and $S$ would produce of course a correlation between $\Delta r_{np}$ and $S$. 

\par
\begin{figure*}[t]                    
\includegraphics[width=0.475\linewidth,clip=true]{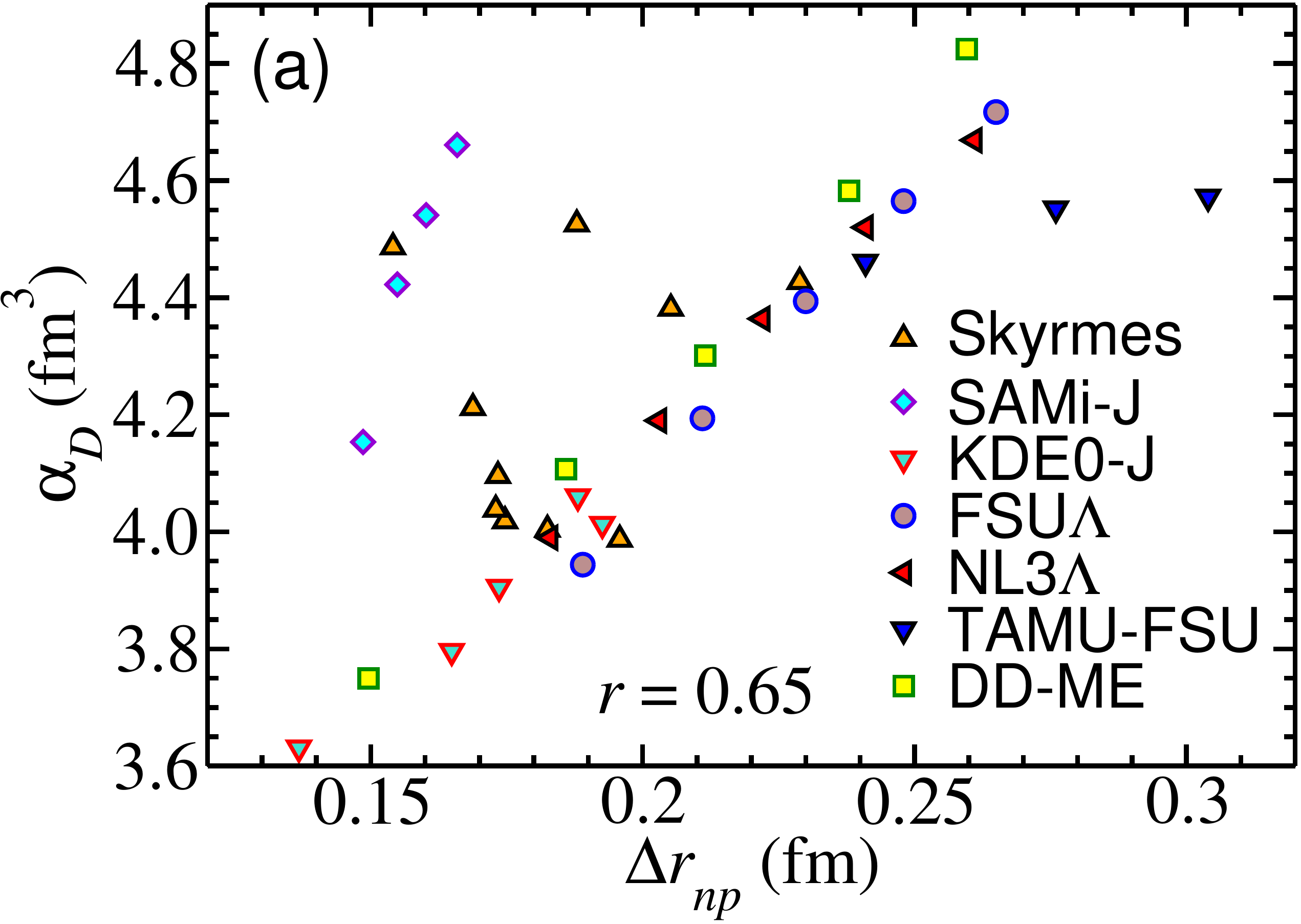}
\hspace{10pt}
\includegraphics[width=0.475\linewidth,clip=true]{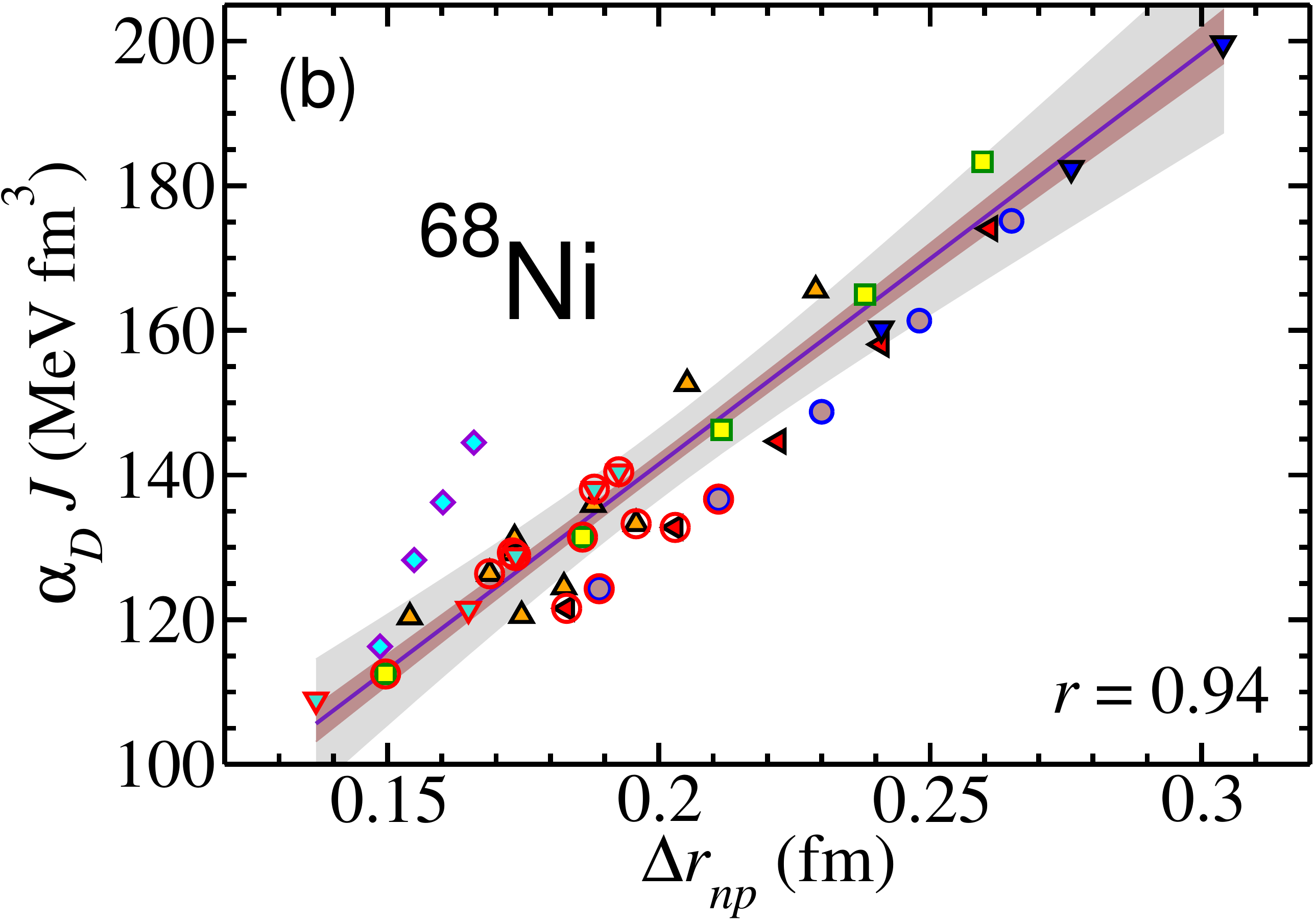}
\caption{(Color online) Correlation between the polarizability $ \alpha_D $ of $^{68}$Ni and the skin width for different EDF (panel (a)) and between the product $ \alpha_D S $ of the polarizability and the symmetry energy and the skin width (panel (b)). The quantity $ r $ is the linear correlation coefficient. Figure taken from ref. \cite{alphaskin} .} 
\label{fig:alphaskin} 
\end{figure*}

For analogous reasons a correlation should exist between the polarizability and the neutron skin width. Indeed, combining Eq. (\ref{eq:dropalpha}) and Eq. (\ref{eq:skinwidth}) of the droplet model, neglecting the small term proportional to $ A^{-1/3} $ in the denominator, one gets
\beq
\alpha_D \,=\, \frac{A r^2}{24 S} \left(\, 1 \,+\, \frac{5}{2} \frac{\Delta r_{np}}{\beta\, r}\, \right)  
\label{eq:alphaskin}\eeq  
\noindent Actually this relationship suggests rather a correlation between $ \alpha_D\, S $ and $ \Delta r_{np} $. This correlation has been recently verified \cite{alphaskin} for the nuclei $^{68}$Ni, $^{120}$Sn and $^{208}$Pb, within the EDF formalism and procedure described above. In Fig. \ref{fig:alphaskin}, taken from ref. \cite{alphaskin}, the case of $^{68}$Ni is considered and it is shown that there is a loose correlation between $ \alpha_D $ and $ \Delta_{np} $, while a much tighter correlation is present between 
$\alpha_D S$ and $\Delta_{np}$ ($J\, \equiv S $).
\par Following a similar procedure, it was found that the skin width correlates also with the excitation energy or strength of most of the isovector nuclear excitations, like the Pygmy resonance \cite{skinpyg} and the IVGQR \cite{IVGQRskin}, but in general such correlations do not look so tight. 
\par
\begin{figure}[hbt]
\vskip 0.8cm
\centering
\includegraphics[scale=0.4,clip]{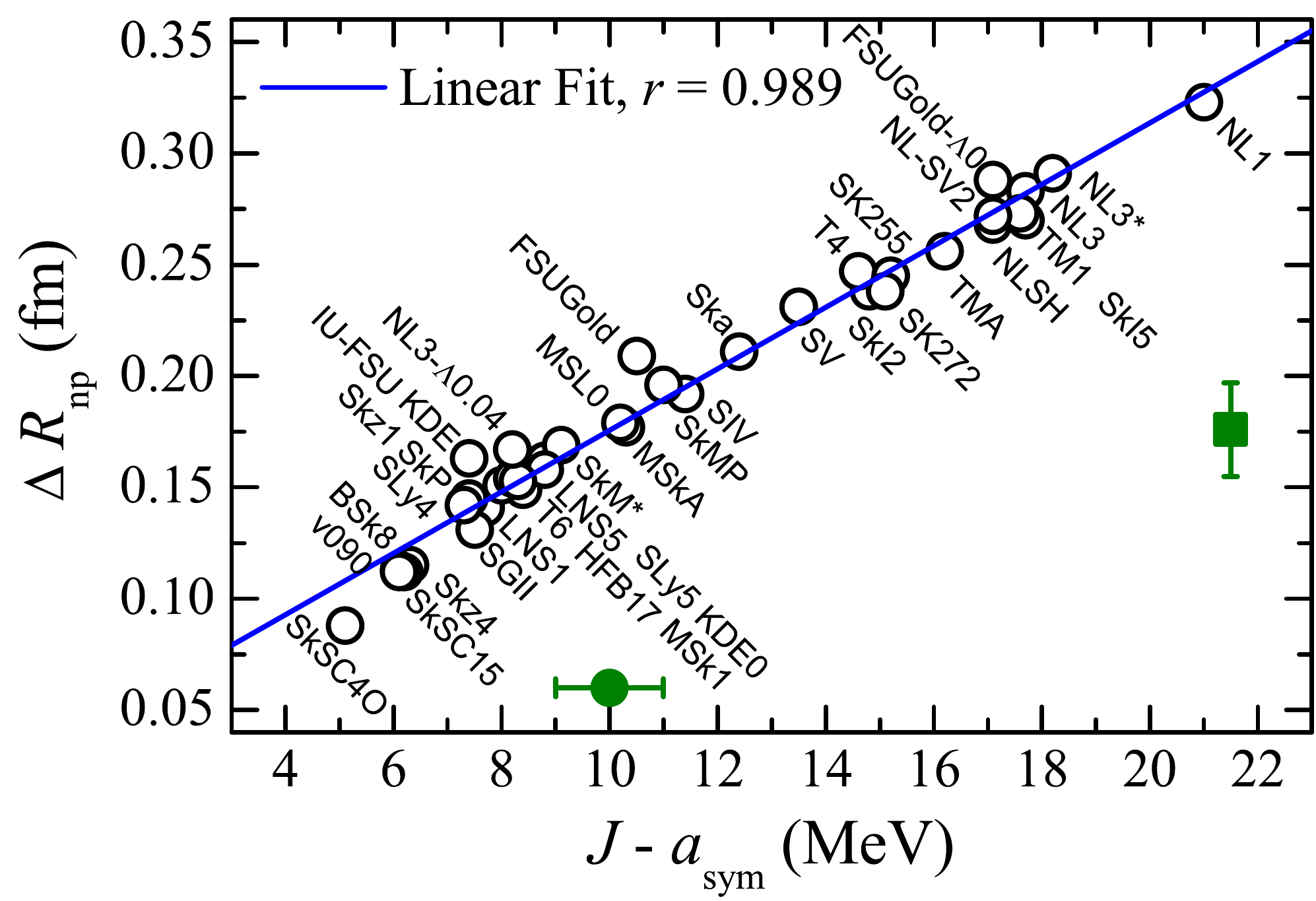}
\caption{(Color on line) Correlation between the skin width and the difference $ J \,-\, a_{sym} $ between the symmetry energy at saturation and the symmetry energy at a specific atomic number $ A $, taken from ref. \cite{Zuo} .} 
\label{fig:skinS_a}
\end{figure}

We have seen that the neutron skin does not correlate with the symmetry energy $S$. However it has been found \cite{XavierPRL,Zuo} that
it correlates with the difference between the NM symmetry energy (at saturation) and the nuclear symmetry energy  of a given nucleus of atomic mass $ A $,  $S \,-\, S_N(A)$. This correlation is displayed in Fig. \ref{fig:skinS_a}, taken from ref. \cite{Zuo}
($a_{sym} \,\equiv\, S_N(A)$).
\noindent
The meaning of the labelling of the different functionals can be found in the original paper. The filled (green) dots with error bars
indicate the estimated values of the skin values from beta decay data (dot on the right) and the corresponding value for $S \,-\, S_N(A)$. 
\par 
This correlation can again be understood from the Droplet Model, where the nuclear symmetry energy $ S_N(A) $ can be written
\beq
S_N(A) \,=\, \frac{S}{1 \,+\, C_S} \ \ \ \ \ \ \ \ \  ; \ \ \ \ \ \ \ \ \ C_S \,=\, \frac{9}{4} \frac{S}{Q} A^{-1/3}
\label{eq:SN}\eeq 
\noindent Then, from Eq. (\ref{eq:skinwidth}), one gets (neglecting Coulomb)
\beq
\Delta r_{np} \,=\, \frac{2}{3} \frac{r}{S} \left( S \,-\, S_N(A) \right) \beta
\label{eq:skinDM}\eeq 
\noindent In the case of a functional, the nuclear symmetry energy $S_N(A)$ must be calculated, since no simple expression is available. In ref. \cite{XavierPRL} semi-infinite nuclear matter calculations were employed, while in ref. \cite{Zuo} the Thomas-Fermi expression for the isospin dependent part of the functional was integrated over the nuclear volume using the calculated density profile. The correlations appear quite stringent in both cases. 
\par 
A general discussion on the correlation of the skin width with other observables can be found in ref. \cite{general}.  
\par
\begin{figure}[t]
\vskip -3.5 cm
\centering
\includegraphics[angle=0,scale=0.4]{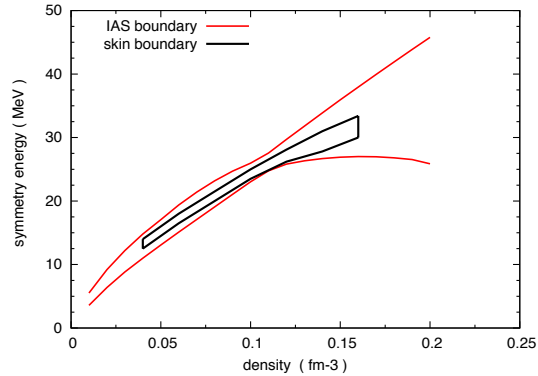}
\vskip -2 cm
\caption{(Color on line) Boundary of the constrained region for the symmetry energy as a function of density coming from the IAS analysis ( red outer boundary ) and on the skin width (black inner boundary), adapted from ref. \cite{DanielIAS} .} 
\label{fig:2box}
\end{figure}

 
From all previous considerations it is evident that an experimental estimate of the skin width would put strong constraints on several other nuclear physical parameters. Furthermore, combined to other constraints, it would further restrict the symmetry energy as a function of density. This line was followed in ref. \cite{DanielIAS}, where the constraints obtained from the analysis of the IAS data were implemented with the data on the neutron skin width. The latter were obtained by a statistical analysis of the data on pionic atoms, pion elastic scattering, polarized and unpolarized proton elastic scattering and alpha scattering. If the functional
set is restricted to the ones that, besides the IAS constraints, are able to reproduce within the uncertainty the values obtained for the skin width, then their overall density dependence is restricted to a much smaller region in the ($S, \rho$) plane.
The result of this analysis in shown in Fig. \ref{fig:2box}, adapted from ref. \cite{DanielIAS}. In the figure the region bounded by the grey (red) full line corresponds to the area depicted in Fig. \ref{fig:Sro}, while the internal smaller region bounded by the thick black line is the one where the symmetry energy $ S(\rho) $ of the selected Skyrme functionals pass through. This looks indeed a tight constraint, which almost fixes the symmetry energy up to saturation. 

\par 
Generally speaking the data from hadronic probes that test the skin width are affected by considerable errors because of the uncertainty on the reaction mechanism and the role of absorption. A direct measurement of the neutron density distribution would allow to extract the neutron skin thickness since the proton density distribution is known to a very good accuracy \cite{Xavier_el}
from electron scattering data. To this purpose one needs a probe sensitive only to neutrons. The only possibility that has been proposed up to now is the neutral current weak interaction \cite{Donn}, mediated by the $ Z_0 $ meson according to the standard model. In fact the weak charge for neutron is much larger than for the proton, even if radiative corrections are included \cite{Erl,Data}, by more than two orders of magnitude. However one needs to discriminate the weak interaction contribution from the much larger Coulomb scattering. A distinctive feature of the weak interaction is the parity violation, and therefore one needs an observable quantity that is sensitive to parity violation. The simplest quantity is the scattering asymmetry between electrons with opposite elicities, which is absent in Coulomb scattering
\beq
A_{PV} \,=\, \frac{\sigma_R \,-\, \sigma_L}{\sigma_R \,+\, \sigma_L} 
\label{eq:APV}\eeq 
\noindent where $ \sigma_{R,L} $ is the differential cross section of electron elastic scattering for right (R) and left (L) elicity, respectively. Unfortunately this quantity is very small, about few units of 10$^{-7}$, and many experimental difficulties hinder
its accurate determination. Among them, a prominent one is the accurate control of the electron elicity during the whole period of the experiment.  For a review of the experimental problems to be overcome in this type of measurements, see ref. \cite{Hor2001}, where the theoretical predictions and uncertainty are also discussed. Notice that the cross sections in Eq. (\ref{eq:APV}) are of course dependent on the four momentum transfer square $ Q^2 $, and therefore the    
determination of the form factor of the neutron distribution is in principle possible. The electron scattering amplitudes is the sum of the Coulomb part and the weak part, corresponding to photon and $ Z_0 $ exchange processes. The former is much larger than the latter, and therefore in the cross section the weak interaction contribution is dominated by the interference term in the modulus square. This determines the structure of the asymmetry of Eq.(\ref{eq:APV}). In the plane wave Born approximation it can be written  \cite{Hor2001} 
\beq
A_{PV} \,=\, \frac{G_F Q^2}{4 \pi \alpha \sqrt{2}}\, \frac{F_n(Q^2)}{F_p(Q^2)}
\label{eq:APVex}\eeq     
\noindent where $ G_F $ is the Fermi constant, $ \alpha $ the fine structure constant and $ F_n, F_p $ the neutron and proton form factors, respectively. They are connected to the density distributions in the standard form (elastic scattering) if one assumes point nucleons
\beq
F_{n,p} \,=\, N \int r^2 dr\, j_0(qr)\, \rho_{n,p}(r) 
\label{eq:FF}\eeq    
\noindent being $ j_0 $ the Bessel function of order $ 0 $. In general for finite size nucleons, the density has to be folded with the single nucleon form factor. The ratio $ G_F/\alpha $ is of the order of $ 10^{-6} $, which determines the smallness of the effect. 
For an accurate determination of $A_{PV}$, one has to perform distorted wave relativistic calculations in the Coulomb field,
and Eq.(\ref{eq:FF}) must be properly modified.   
\par 
\begin{figure}[hbt]
\centering
\vskip -8.5 cm
\includegraphics[scale=0.6,clip]{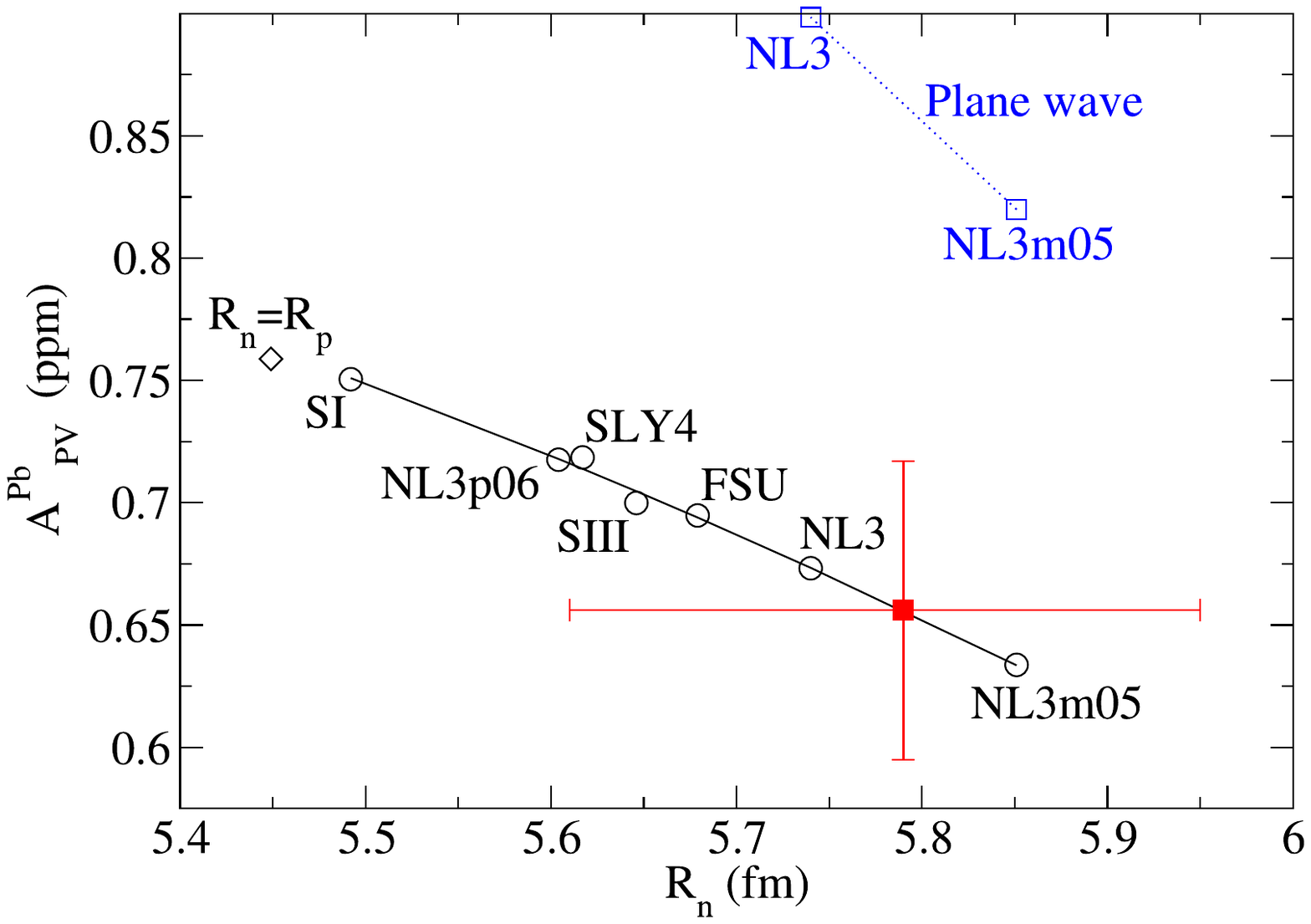}                    
\caption{(Color online) Correlation between the electron scattering asymmetry and the skin width for $^{208}$Pb and the skin width for different functionals. The cross indicates the experimental value with errors. Figure from ref. \cite{PREXI}} 
\label{fig:PREX_exp}
\end{figure} 
\begin{figure}[hbt]
\centering
\includegraphics[scale=0.4,clip]{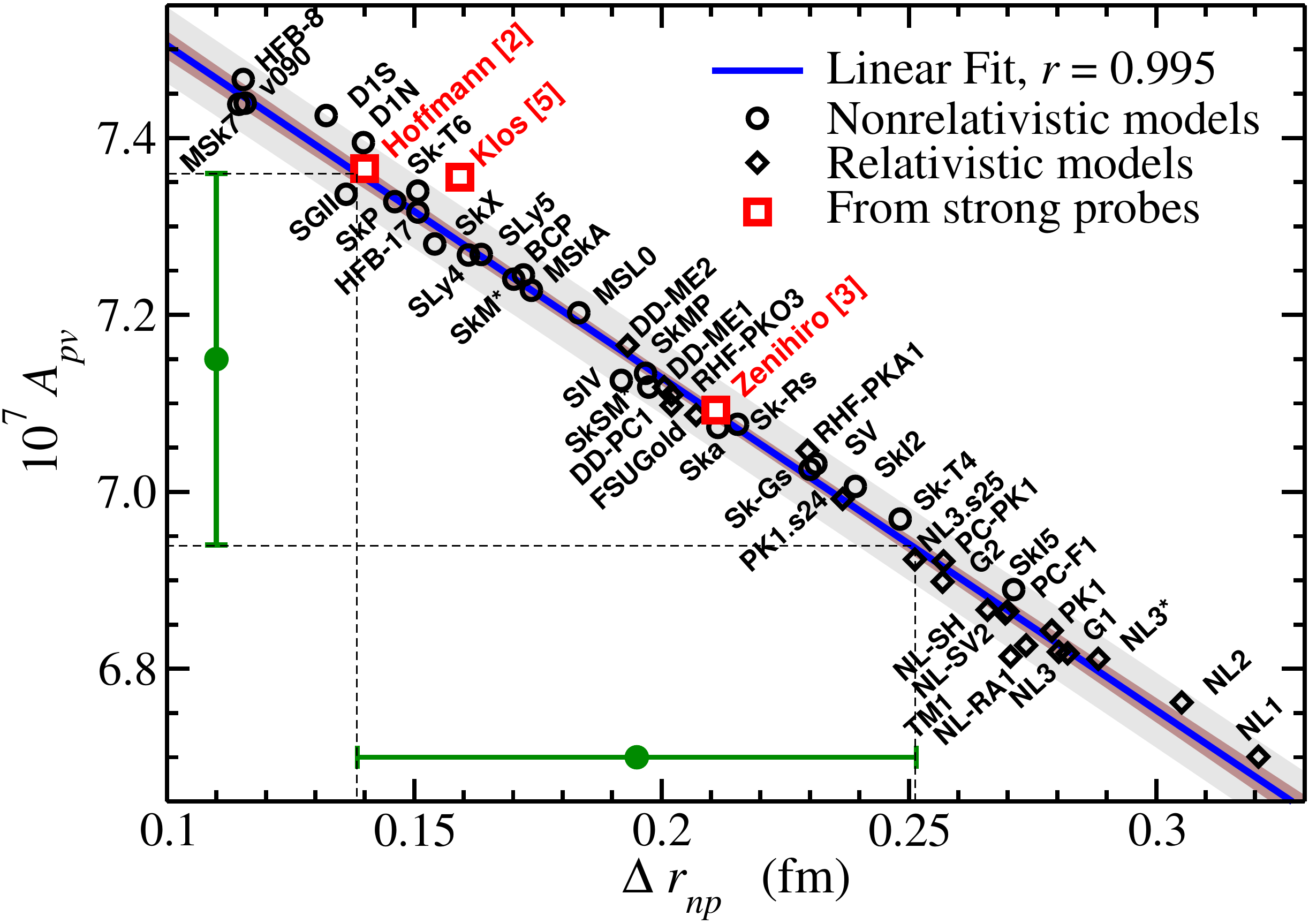}                    
\caption{(Color online) Linear correlation between the electron scattering asymmetry and the the skin width for a wide set of functionals. Taken from ref. \cite{XavierPREX}. } 
\label{fig:PREX_Xavier}
\end{figure} 
\par
On the experimental side, thanks to the continuous progress along the years, this method has become very promising towards an accurate determination of the asymmetry. The experimental status updated to the year 2014 can be found in ref. \cite{Horo}. The efforts have been concentrated on the nucleus $^{208}$Pb (PREX experiment) and $^{48}$Ca (CREX experiment) at the Jefferson Lab. The two nuclei are both neutron rich enough to expect a substantial skin thickness. The PREX experiment has been performed in the preliminary configuration (PREX-I), while CREX experiment has been approved. The updated version of PREX (PREX-II) has also been approved.\par
 The result of PREX-I experiment is reported in Fig. \ref{fig:PREX_exp} from ref. \cite{PREXI}. Here is indicated the value of the measured asymmetry (the square at the center of the cross) with the estimated error (vertical bar). For comparison are reported the predictions based on a set of functionals, which show a correlation between asymmetry and neutron radius $ r_n $. Assuming this linear correlation (the  straight line passing through the open circles) to hold, one can get the center value of $ r_n $ and the corresponding uncertainty (horizontal bar). Notice that the asymmetry is measured at one scattering angle (5 degrees in the lab frame), which is taken as small as possible. In fact according to Eq. (\ref{eq:FF}) at $ Q \,=\, 0 $ the form factor is proportional to the mean square radius, and therefore at small $ Q^2 $ the correlation should be stronger. The nominal error for $ r_n $ is about 3$\%$. Unfortunately this is too large for the estimate of the neutron skin thickness, which could be used to estimate e.g. the slope parameter $ L $, see Fig. \ref{fig:LQ}. The PREX-II is expected to reach a precision of about $ 1\% $. \par
In ref. \cite{XavierPREX} a larger set of functionals was considered, and it was shown that the correlation is less stringent even if the functionals are selected among the ones that fit well the nuclear binding and charge radii. In particular functionals that 
give the same asymmetry can produce a different neutron radius. It was shown that this difficulty can be overcome by calculating directly the skin width $ \Delta r_{np} $, since then the uncertainty in the neutron radius is compensated by the fluctuations in the charge radius and a much tighter correlation is present, see Fig. \ref{fig:PREX_Xavier}. In the figure is indicated (vertical and horizontal bars) that an error of 5$\%$ in $ A_{PV} $ should result in an error of 25$\%$ in the skin thickness. This is approximately the precision that PREX-II should obtain.

\subsubsection{Critical discussion on the density functionals method.}
\label{critical}
We have repeatedly reported results based on the density functionals method. It consists in considering a set, as wide as possible, eventually constrained to fit well some relevant physical characteristics of nuclei, and look for correlations between a measurable quantity and physical parameters connected with the symmetry energy. The unavoidable question that arise when applying this method is mainly twofold. Is the sample of functionals large enough or the correlation is biased by the choice of the set and a larger or different set would make the correlation weaker ? Is the correlation meaningful or rather it is an artifact of the necessarily limited complexity of a generic functional ? 
The first question could be, at least partly, answered by considering the cross-correlation (covariance) between the parameters of the functional that have been fitted without reference to the symmetry energy and the ones directly related to the symmetry energy.
If the co-variance is small, one could confirm that the correlation is a genuine effective one between the measured quantity and the physical parameters related to the symmetry energy. The co-variance analysis has been developed in general in the last few years \cite{Modal,GL_cov,Naz}, and it would be desirable that more and more such analysis will be performed to clarify the meaning of a generic correlation based on this method. \par 
The second question cannot be easily answered. Microscopic calculations cannot be clarifying, since they cannot be tuned to produce a correlation. Therefore the correlations that have been found are necessarily due to the parametrization of the functionals, and as such they are biased by their general simplified structure. Despite that, the correlations can have a genuine physical meaning and can be validated by cross checking them in different combinations to verify their mutual consistency.      

\section{The symmetry energy in Astrophysics}
\label{Astro}
Nuclear Physics plays a fundamental role in the structure and dynamics of astrophysical compact objects, like Supernovae (SN) and Neutron Stars (NS). In particular the behaviour of the nuclear symmetry energy as a function of density and temperature is crucial
for interpreting many astrophysical observations, and for understanding many phenomena related to compact objects. Besides the structure of NS, the evolution of supernovae explosions, the NS oscillations and gravitational waves emission, the transient phenomena in NS, the binary mergers final stage, are all strongly influenced by the features of the nuclear symmetry energy. These phenomena occur in very different physical contexts, and accordingly they are sensitive to different aspects of the symmetry energy. In the following we analyze the relevance of the symmetry energy in astrophysical objects and phenomena, and to what extent observational data can constrain the symmetry energy.

\subsection{The crust of Neutron Stars.}
\label{crust}
\begin{figure}[t]
\centering
\vskip -5cm
\includegraphics[scale=0.5,clip]{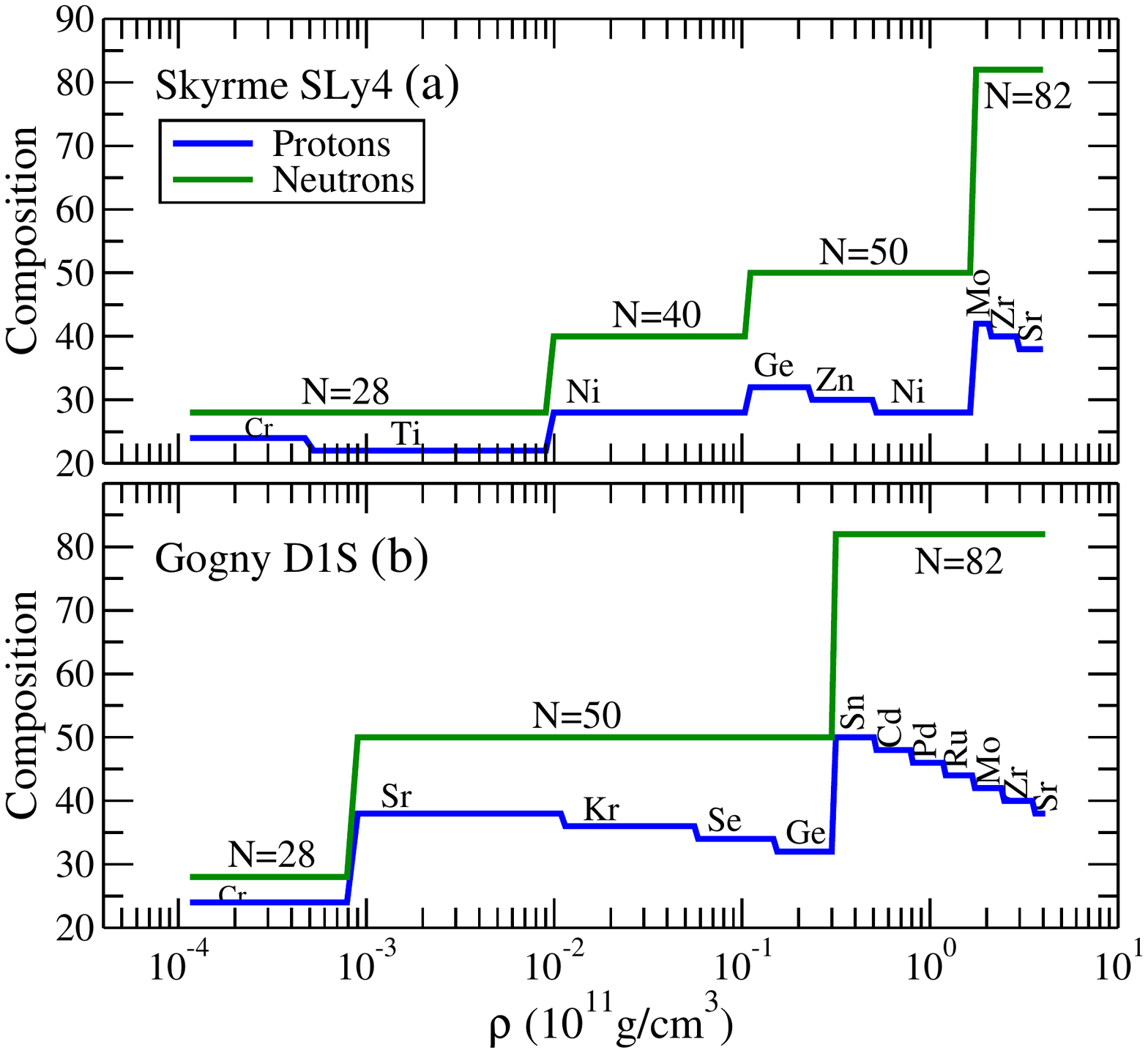}         
\vskip -0.3cm           
\caption{(Color online) Composition of the Neutron Star outer crust for two different functionals, from ref. \cite{book_Bert}.} 
\label{fig:outer_crust}
\end{figure}

 In the standard physical model NS contain nuclear matter at very different density, from few times saturation density down to values
 pertinent to terrestrial materials. We are referring here to cold non-accreting NS. In the more external part nuclear matter is non-homogeneous and arranged in a lattice of different shapes. This part is solid and it is denominated as ''crust" of the NS. The most external part  of the crust, the so-called outer crust, is a Coulomb lattice of nuclei immersed in a gas of relativistic electrons. The atomic and charge numbers (A,Z) of these nuclei change at increasing density towards the core of a NS, starting from $ ^{56}$Fe  ( the most bound nucleus in nature ) and becoming more and more neutron rich. The (A,Z) composition is determined by beta equilibrium and charge neutrality,
 and therefore it is directly affected by the symmetry energy of nuclei, mainly through the chemical potentials of neutrons and protons. This is equivalent to minimize the energy at a given crust density, and this  can be determined if the binding energy for each (A,Z) is known. At the lowest density the nuclei are the ones available in the laboratories, and therefore the binding energy can be obtained from the well established data on nuclear masses. When proceeding downwards, the asymmetry of nuclei is increasing and one starts to encounter nuclei that are so beta unstable that are not available in laboratory, i.e. they are very ''exotic". They do not decay in the crust because the electron gas prevents the electron emission due to the Pauli blocking. Then It is clear that an extrapolation of the nuclear models is mandatory, and no experimental data in laboratory can be helpful. The extrapolation can be performed using all the methods that we have considered in Section \ref{sec2}, including the LDM and DM. If one follows the density functionals scheme, this extrapolation can be dependent on the functional, and the main reason for that is a different behavior of the nuclear symmetry energy $ S_N $ in this exotic region, which cannot be experimentally tested. We have seen that the value of $ S_N $ is directly connected with the symmetry energy at sub-saturation density. The relevance of $ S(\rho) $ is illustrated in Fig. \ref{fig:outer_crust}, from ref. \cite{book_Bert}, where the sequence of nuclei at increasing density is reported for two different functionals. The Gogny functional D1S has a lower symmetry energy than the Skyrme functional Sly4 and can support more neutron rich nuclei, so that for the former there is a larger plateau at neutron number N $\,=\,$50, whereas it is located 
 at N $\,=\,$40 for the latter. For the same reason the plateau at N $\,=\,$82 is reached at lower density for D1S than for Sly4. \par 
The increase of neutron excess with density can be understood looking at LDM for the mass of nuclei, see Eq. (\ref{eq:massf}).
Once the neutron and proton masses, $ m_n , m_p $, are included, one can get the chemical potentials for neutrons and protons, $ \mu_n , \mu_p $,
by differentiating with respect to the neutron number $ N $ and proton number $ Z $, i.e.
\beq
\mu_n \,-\, \mu_p \,=\, m_n \,-\, m_p \,-\, \frac{1}{3}\, a_c\, \frac{x_p}{A^{1/3}} \,+\, 4 S_N ( 1 \,-\, 2 x_p )
\label{eq:chem}\eeq       
\noindent where $ x_p \,=\, Z/A $ is the proton fraction. At increasing density the electron density and the corresponding chemical potential $ \mu_e $ must increase. Since by chemical equilibrium $ \mu_n \,-\, \mu_p \,=\, \mu_e $, according to Eq. (\ref{eq:chem})
the proton fraction $ x_p $ must decrease. Then the role of the symmetry energy $ S_N $ is evident. \par
As we proceed downwards inside the NS, the neutron excess becomes so large that the neutron chemical potential becomes positive and the neutrons start to drip from nuclei. This is the region of the so-called inner crust, where nuclei are surrounded by a neutron gas. This means that nuclei are unstable also with respect to the strong interaction. Energy density functionals are in principle unable to predict the drip point since it is far away from the region of mass and charge where they have been tested. However microscopic calculations can be reliable to this respect, and it turns out that most of them agree on the value of crust density where the drip starts to occur \cite{Bert}, 
$\rho_{drip} \approx \rm 2.6 \times 10^{-4}~fm^{-3}$, even if some discrepancy exists on the first nucleus that starts to drip. In any case it is clear that the drip point does not depend only on the symmetry energy, but it is determined by more subtle nuclear structure effects. In particular shell and single particle effects are essential and only at the Thomas-Fermi level some correlations with the symmetry energy can be found \cite{Bao}. This was confirmed recently in ref. \cite{dripacc}, where the drip point was analyzed in the case of the presence of a strong magnetic field (e.g. magnetars) and for accreting and non-accreting NS. Although some correlations were found between symmetry energy and drip point, which turns out to be quite different in presence of accretion, the symmetry energy has only a minor relevance for the determination of the drip point, which is more influenced by the nuclear structure features of the nuclei in the crust. \par

At increasing density the nuclei at the center of the lattice sites can become deformed, and finally the nuclear matter can be arranged in more extended structures, like slabs, rods, tubes, bubbles and more complicated structures or even in a disordered structure, similar to an amorphous material. Finally the matter becomes homogeneous at a well defined density, corresponding to the edge of the crust. This is 
the point of the so-called spinodal instability of homogeneous asymmetric NS matter towards the formation of non homogeneous structures. The region between the regular lattice of nuclei and homogeneous matter is called ''pasta phase".   
\begin{figure}[t]
\centering
\vskip -17cm
\includegraphics[scale=0.8,clip]{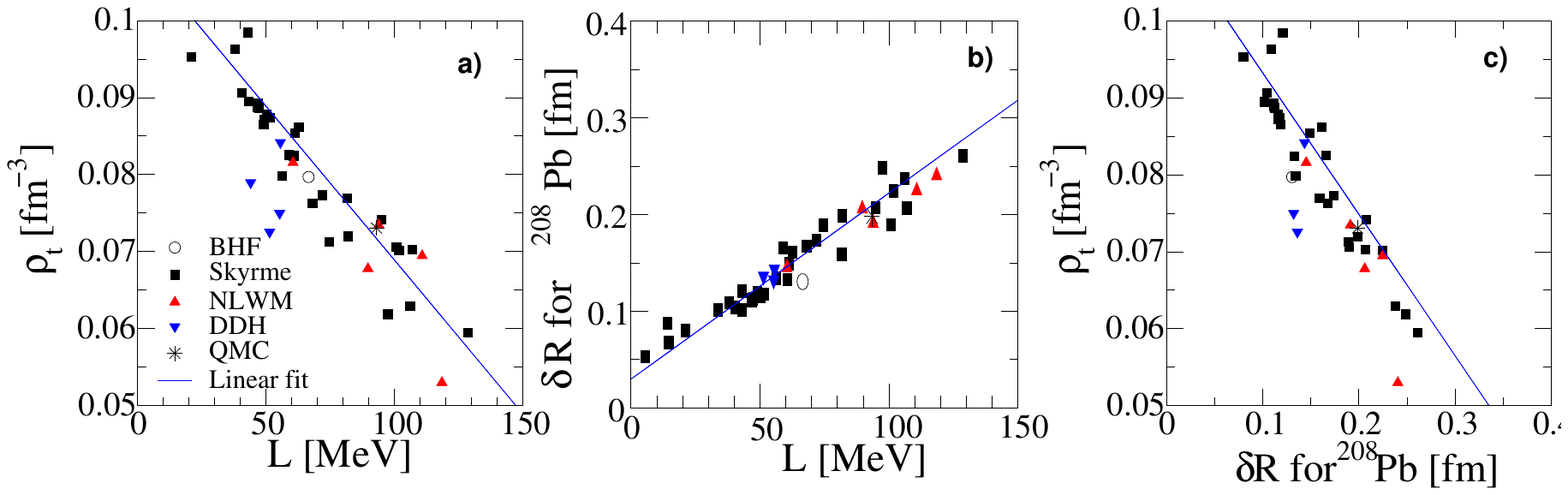}   
\vskip -1cm                 
\caption{(Color online) (Anti-)Correlation between the density $ \rho_t $ for the crust-core transition and the slope parameter $ L $,
(panel a)). Correlation between the skin width $ \delta R $ and $ L $ (panel b)). Resulting correlation between $ \rho_t $ and $ L $ (panel c)). Figure from ref. \cite{rhot,Bert_t}.  } 
\label{fig:trans_den}
\end{figure} 
If the symmetry of the inner crust is restricted to a lattice with spherical nuclei (that would be better called ''nuclear matter blobs") the transition from the lattice to the homogeneous matter is sharp. If the pasta phase occupies a restricted
zone of the NS, the density at which the spinodal instability occurs will identify approximately the boundary of the inner crust.
The spinodal instability can be found by looking for the point in the $ ( \rho_n,\rho_p )$ plane  where the total curvature of the energy surface $ E(\rho_n,\rho_p) $ vanishes \cite{Avan}. Here $ \rho_n , \rho_p $ are the neutron and proton density, respectively.
For a given functional, the transition density can be identified following the evolution of the energy and composition of the NS matter as the total density decreases from the homogeneous phase. It turns out that the transition density $ \rho_t $ is  (anti-)correlated with the slope parameter $ L $ that characterizes the functional \cite{rhot}. In panel a) of Fig. \ref{fig:trans_den},
taken from ref. \cite{rhot,Bert_t}, this (anti-)correlation is illustrated for a set of Skyrme functionals and relativistic mean fields.
The linear correlation looks stronger for the former than for the latter. Two more microscopic calculations (BHF, QMC), to be discussed in Section \ref{theory}, seem to follow the linear correlation determined mainly by the Skyrme functionals. Since the skin thickness is also correlated with $ L $, panel b) ($\delta R \,\equiv\, \Delta r_{np}$) and Fig. \ref{fig:LQ}, this results in a direct correlation between $ \rho_t $ and skin thickness, as shown in panel c). The reason of the correlation between $ \rho_t $ and $ L $ is unclear, since NS matter is extremely asymmetric ($ x_p $ is of the order of few percents), while $ L $ is defined for symmetric matter.  \par 
The smallness or even non-existence of the pasta phase for NS matter in beta equilibrium  was observed for Skyrme forces \cite{Bert_t} or other functionals \cite{BCPM}. The pasta phase within the functional method is obtained by considering different periodic structure of one dimensional character (rods, tubes), and of two-dimensional character (slabs), or three-dimensional, i.e the ''ordinary" spheres or bubbles (empty spheres), all calculated in the mean field scheme (Hartree or Hartree-Fock calculations), and then finding the lowest energy configuration. If such a configuration does not correspond to a lattice of nuclei (spheres), one concludes that the considered density is inside the pasta phase. Even if the functional method can allow for more complex periodic configurations \cite{Jirina_pasta}, it is unfit to describe disordered configurations, which however cannot be excluded. The origin of a disordered phase in the inner crust is the mechanism of ''frustration" due to the competition between the repulsive long range Coulomb interaction and the attractive short range nuclear force. The effect of frustration is the appearance of 
numerous local minima, energetically very close. A brief discussion on frustration from the point of view of a nuclear physicist can be found in ref. \cite{Hor69}. Semi-classical Molecular Dynamics was developed in refs. \cite{Hor69,Hor70,Hor78,Hor88} for handling this problem. Due to the complexity of the problem, even special computer chips were developed for the simulations. Disordered configurations in realistic physical situations were indeed found \cite{Hor78}. However the size of the disordered phase could not be firmly established, also in view of the semi-classical approximation. The size of the disordered phase, or even its mere existence, can be considered still an open question. A possible extension of the method to the quantal treatment was presented in ref. \cite{HorMAD} (the MADNESS project), which probably will give the ultimate solution of this intriguing problem. \par 
The relation between the pasta phase and the symmetry energy is difficult to establish, even at qualitative level, but in any case the possible detection of the disordered phase can put further constraints on the functionals.  \par
Once the possible connection between the characteristics of NS crust and nuclear symmetry energy is established, the natural question that arises is
if the properties of the crust can have observational counterparts. First of all, as a general remark, one must stress that what is observed is usually filtered through the atmosphere of NS. Then one has to look for observational data that are not dependent on the composition of the atmosphere, or to rely on cases where such a composition can be considered established. \par 
We have seen that the solid crust of a NS is formed by a Coulomb lattice. The shear modulus $ \mu $ of such a lattice depends only on the charge of the ions, as it is well known in solid state \cite{Fuchs}, and it is given by
\beq
\mu \,=\, 0.1194\, \frac{n (Ze)^2}{a}
\label{eq:shear}\eeq 
\noindent where $ n $ is the ion density and $ a \,=\, ( \frac{3}{4\pi n} )^{1/3} $ is the average lattice spacing. The formula is for body-centered cubic lattice, which is the lowest in energy, and it is valid also for NS crust since the possible neutron component does not contribute to the shear modulus. Shear oscillations are associated to a non-zero shear modulus, and if the NS crust is sufficiently decoupled from the core, these oscillations can produce observable phenomena sensitive to the symmetry energy. The oscillation frequency depends on the $ Z $ value and lattice spacing $ a $, according to Eq. (\ref{eq:shear}), or more precisely to the average value of $ Z^2/a $.
The best candidates that could be related to the crust oscillations are the Quasi Periodic Oscillations that were observed in giant flares of highly magnetized NS, and in ref. \cite{Steiner} it was suggested that the frequencies of these oscillations could be associated with the fundamental shear oscillation of the crust and its overtones. The observation of these oscillations is obtained by the Fourier analysis of the X-ray signal coming from the NS, which besides the pulsar rotational frequency shows additional low frequency peaks. Different crust models were examined, and variations even of factors 2-3 in frequency were found, and those are related to the different values of $ Z/a $. Since the crust composition and structure depend mainly on the symmetry energy, this could be an indirect method to extract the symmetry energy from the observational data. References to the observational data can be found in the original paper \cite{Steiner}. Unfortunately up to now there is no general agreement on the interpretation of the observed oscillation frequencies. \par 
Also transport properties of the crust depend on its structure and play a major role in many NS processes, since any phenomena that occur
in the core is filtered by the crust before it can be observed. In particular neutrino opacity is affected by the presence of the pasta phase \cite{Hor69}. The composition of the inner crust has influence on the deep crustal heating in accreting NS \cite{accr}. Thermal conductivity is also affected by the crust structure, but it is a quite complex process, involving many features of the crust \cite{thermal}. 
All these phenomena can in principle provide information on the nuclear symmetry energy, but unfortunately they are quite indirectly related to it and the results obtained from their analysis are necessarily model dependent.

\subsection{The Core and the overall Structure of Neutron Stars.}
\label{core}
The NS crust extends up to a density that is usually between 1/2 and 2/3 the saturation density, depending on the model.
According to the NS standard model, below the crust there is a liquid core of homogeneous nuclear matter, where neutrons, protons, electrons and muons are in beta equilibrium, and the fraction of each one of these components is determined by imposing charge neutrality and thermodynamic equilibrium. These two conditions correspond to the relations
\beq
\begin{array}{rl}           
n_p &\!\!=\, n_e \,+\, n_{\mu} \\
&\ \\
\mu_n &\!\!=\, \mu_p \,+\, \mu_e \\
&\ \\
\mu_{\mu} &\!\!=\, \mu_e
\end{array}
\label{eq:equi}\eeq   
\noindent where the $ n's $ indicate number densities, the $ \mu's $ the chemical potentials, and $n, p, e, \mu $ stand for neutrons, protons, electrons and muons, respectively. The electrons are ultra-relativistic, and, neglecting their (weak) electromagnetic interaction their chemical potential can be calculated from a free Fermi gas, i.e. $ \mu_e \,=\, \hbar c k_{Fe} $, where $ k_{Fe} $
is the electron Fermi momentum. To calculate the nucleon chemical potentials one must know the asymmetric nuclear matter EOS. Let be $ E(N,Z) $ the total energy as a function of the neutron numbers $ N $ and the proton number $ Z $, and $ e(\rho,\beta) \,=\, E/A $ the corresponding energy per nucleon. One gets
\beq
\mu_n \,-\, \mu_p \,=\, \left( \frac{\partial E(N,A-N}{\partial N} \right)_A \,=\, -\left( \frac{\partial E(A-Z,Z}{\partial Z} \right)_A \,=\, 2 \left( \frac{ \partial e (\rho,\beta}{\partial \beta} \right)_{\rho} 
\label{eq:mudiff}\eeq          
\noindent where the subscripts label the quantity that is held fixed in the differentiation. If we assume that the dependence of the EOS on the asymmetry is quadratic, i.e. we stop the expansion of Eq. (\ref{eq:def}) at the first non-vanishing term (which turns out to be a good approximation), then Eq. (\ref{eq:mudiff}) simplifies to
\beq
 \mu_n \,-\, \mu_p \,=\, 4 S(\rho) \beta
\label{eq:quadr} 
\eeq
\noindent Then the condition for the chemical potentials in equation (\ref{eq:equi}) becomes an explicit equation for the proton fraction $ x_p \,=\, Z/A $ 
\beq
4 S(\rho) \beta \,=\, \mu_e \,\equiv\, \hbar c k_{Fe} \,=\, \hbar c ( 3\pi^2 n_e )^{1/3} \,=\, \hbar c ( 3\pi^2 \rho x_p )^{1/3}
\label{eq:xp}\eeq
\noindent This equation shows that the proton fraction is determined by the nuclear matter symmetry energy. For $ x_p \,<<\, 1 $, one gets an explicit expression for $ x_p $
\beq
x_p(\rho) \,=\, \left( \frac{(4 S(\rho))^3}{(\hbar c)^3 (3 \pi^2 \rho)} \right) 
\label{eq:xp1}\eeq
\noindent We have seen that the symmetry energy at small enough density increases at least linearly with density, and therefore $ x_p $ is indeed very small and increases also with density. At higher density, the behaviour of $ x_p $ depends on the EOS, but in any case the symmetry energy is the basic quantity. It follows that if we could determine $ x_p $ in the core we could extract $ S(\rho) $. Of course this is not possible in a direct way, and one has to rely on indirect methods, i.e. look for observable phenomena or physical parameters that are sensitive to the values of $ x_p $. \par
If one assumes a quadratic dependence of the symmetry energy on the asymmetry parameter $ \beta $, one can relate in a simple way the pressure of NS matter $ P(\rho) $ to the pressure $ P_S(\rho) $ of symmetric nuclear matter at a given density $ \rho $
\beq
P(\rho) \,=\, P_S(\rho) \,+\, \rho^2\, \frac{d\, ( S(\rho)\beta^2 )}{d \rho}
\label{eq:press}\eeq
\noindent which shows explicitly that the behavior of the symmetry energy determines the stiffness of the NS matter EOS in the core with respect to symmetric matter. Once the EOS is known as a function of density and asymmetry, one can calculate the density profile of NS, which fixes the relation between its mass $ M $ and its radius $ R $ \cite{Shap}. This must be done in the framework of General Relativity, due to the extremely intense gravitational field, and the basic equation to be solved is the celebrated Tolman-Oppenheimer-Volkoff equation \cite{Shap}, where the pressure as a function of energy density is the only input. In principle the observation of masses and radii of a certain number of NS could tightly constrain the EOS, even if not determine it uniquely, and therefore also the symmetry energy as a function of density. In ref. \cite{MREOS} the data on six compact objects were analyzed, three Type-I bursters and three transient low-mass X-ray binaries and constraints on their masses and radii were obtained. These constraints were then used to restrict the range of possible EOS, that were parametrized by a flexible analytical form. 
The main uncertainty is the behavior of the EOS above saturation, and in refs. \cite{MREOS,MREOS1} it was shown that it is possible in this way to put constraints on the symmetry energy at saturation. The following range of values were obtained
\beq 
 28 \,< S \,< 34 \ {\rm MeV}  \ \ \ \ \ \ \ \ \ \ \ \ \ \ , \ \ \ \ \ \ \ \ \ \ \ \ \ \ \  43.3 \,< L \,< 66.5 \ {\rm MeV}
\eeq
On the other hand the EOS for symmetric matter at high density has been constrained from the data on heavy ion collisions \cite{DanielHI}, as it will be discussed in Sec.\ref{HI}. Combining the latter results with the ones of ref. \cite{MREOS}, it is possible to restrict the symmetry energy as a function of density. In principle from Eq. (\ref{eq:press}) one could extract $ S(\rho) $ by simple integration if $ P(\rho)$ and $ P_S(\rho) $ are known. However, since the constraints are not strong enough to avoid large uncertainties, a more viable method is to consider a large enough sample of functionals, already optimized on other data, e.g on nuclear binding, and select the ones that are compatible with the two EOS, for symmetric matter and NS matter respectively. The restricted set of functionals would provide the range of possible symmetry energy.
 Alternatively, as we will discuss in Sec. \ref{theory}, one can select the theoretical microscopic EOS that satisfy the same criterion and extract the corresponding symmetry energy. 

\par 
A process that is strongly dependent on the symmetry energy is the cooling of NS, which can last million years. We are referring here to isolated non-accreting NS. The simplest process for cooling is the beta decay of neutron or the electron capture by protons, which ultimately consists in the loss of thermal energy by emission of neutrinos and anti-neutrinos. Neutrino emission dominate the cooling process after thermal relaxation of the star and for a long period until it decreases so much that the photon emission starts to dominate. The neutrino emission occurs through the following reactions  
\beq
\begin{array}{rl}
n & \rightarrow \ \,\, p \,+\, e \,+\, \overline{\nu} \\
p & \!\!\!+\,\ \, e \,\,\ \rightarrow \,\,\  n \,\,+\,\, \nu 
\end{array}
\label{eq:cool}\eeq                   
\noindent which defines the so called direct URCA (DURCA) process \cite{Shap}. For the NS matter in the core, due to energy and momentum conservation, these processes can occur only if the fraction of protons \cite{Shap} exceeds some minimal value. In absence of muons this threshold fraction is $ x_p \,\approx\, 12 \% $, and $ x_p \,\approx\, 14 \% $ when muons are included. Since the proton fraction is determined by the density dependent symmetry energy, this threshold percentage implies also a density threshold. The latter is however dependent on the EOS, which fixes the symmetry energy as a function of density. It can also happen that this threshold proton fraction is actually not reached inside the NS and the DURCA process does not occur. Then neutrino emission can take place ins the core through the so-called modified URCA (MURCA) processes, where the energy and momentum conservation can be fulfilled by the interaction with an additional nucleon. The URCA process is the fastest process for cooling, and its occurrence without any hindrance, even in a restricted region of NS, cools down the star in a few hundreds of years. The MURCA process is several orders of magnitude slower, as other neutrino emission processes, e.g. nucleon-nucleon bremsstrahlung. An extensive discussion on neutrino emission processes from NS can be found in ref. \cite{n_emis}. The observational data are usually summarized by reporting in a plot the temperature and age of those NS in which they have been approximately determined. The cooling curve of each NS of a given mass can be calculated once the EOS of NS matter and its composition are known. These simulations are quite complex, and we cannot go into details, thus limiting ourselves to a schematic discussion. The calculated cooling profiles must pass through one or few of the observational data for some choice of the NS mass. For small enough masses the central density is so small that DURCA cannot occur and the cooling is slow. At increasing mass the onset of DURCA can be possible, according to the EOS. Then the cooling becomes fast. Therefore, at first sight, small NS should cool slowly, while heavier NS should cool fast. As the mass is varied the calculated cooling curves should cover the region of the plot where the observational data are present. Since the onset of DURCA depends on the proton fraction reached inside the NS, this comparison with the data should give relevant indications about the symmetry energy of nuclear matter, at least at the high asymmetry expected in the NS core. Unfortunately the analysis of the observational data is much more complicated. \par 
First of all the NS that cool fast can reach in a few hundred years a temperature so low that they cannot be observed, which means that  
they cannot appear in the cooling plot of the data. Then different density threshold for DURCA could simply result, at least to a certain extent, to just a different mass assignment to each observational data. \par
\begin{figure}[hbt]
\centering  
\includegraphics[scale=0.4,clip]{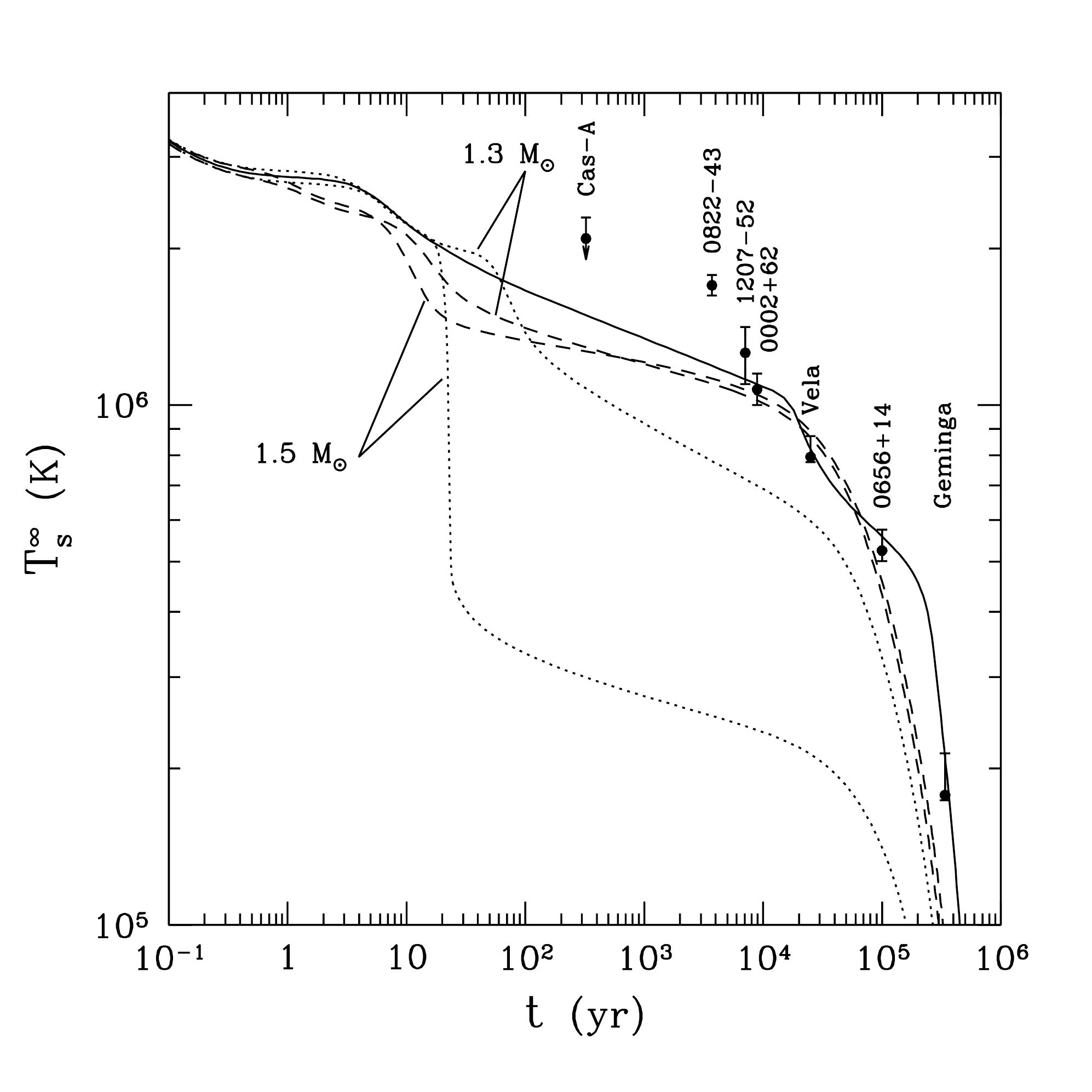}                    
\caption{(Color online) Superfluidity effects on the cooling curve of Neutron Stars. The dotted lines correspond to Neutron Stars without superfluidity. From ref. \cite{n_emis}} 
\label{fig:pairURCA}
\end{figure}

 
The second complication is the possible presence of superfluidity in NS matter, both for neutrons and for protons. In superfluid matter DURCA process can be strongly reduced, and for very strong superfluidity it can become comparable to MURCA \cite{n_emis,Kamin,Usp}. Actually the superfluid matter can cool faster in the initial stage, when the temperature is still a fraction of the superfluid critical temperature, due to the so-called pair breaking process \cite{n_emis}. However all observed NS are in the later stage (with the possible exception of the noticeable case of the NS in Cassiopeia A ), where the temperature is much smaller then the critical temperature and the superfluid hindrance is in operation. The reason of such suppression of DURCA is the reduction of the available phase space when pairing is present. In fact pairing is characterized by an energy gap in the single particle spectrum, which is a strong hindrance to the low energy neutrino emission process. The DURCA process is then exponentially reduced by a factor $ \exp \,(\, -\, T/T_c\, )\, $, where $ T $ is the matter temperature and $ T_c $ the critical temperature for pairing. The effect of pairing is illustrated in Fig. \ref{fig:pairURCA}, taken from ref. \cite{n_emis}. The EOS included in the simulations can support DURCA process at sufficiently high density. If one excludes pairing (dotted lines) the small mass NS cooling is relatively slow (1.3 solar mass in the figure), while more massive NS cool fast (mass 1.5 in the figure).
With the introduction of superfluidity (dashed lines) the distinction between fast and slow cooling basically disappears. The simulations have been performed with a strong pairing gaps constant all throughout the NS core. Such an extensive presence of pairing is crucial to suppress the DURCA process. Therefore pairing can be considered the principal regulator of the NS cooling. Unfortunately the theoretical estimates of the nuclear pairing gaps in nuclear matter is quite uncertain \cite{UJ,BCS50}.\par 
Roughly speaking one can say that the data on cooling can be explained or in a scenario with no DURCA and no pairing, or in a scenario where DURCA is possible but suppressed by strong pairing \cite{Noimnras}. The assumption that the DURCA process is absent or strongly quenched belongs to the so-called "Minimal Cooling Paradigm" \cite{stellar}.                 
\par In conclusion the possibility to connect directly cooling and the nuclear matter symmetry energy is strongly reduced by the complexity of the problem, but in modeling the cooling process the symmetry energy plays a decisive role.\par 
Another possible relevance of the symmetry energy in the NS physics is in the determination of the NS maximum mass. Recent observations have established the existence of NS of 2 solar masses, in PSR J348+0432 (2.1 $\pm$ 0.4 M$_{\odot}$) \cite{Anto} and PSR J1614-2230
(1.97$\pm$0.4 M$_{\odot}$) \cite{Demo}. These observational data put stringent constraints on the nuclear matter EOS, which must be able to support such NS mass, or larger. This means that the NS matter EOS must be relatively stiff. Since the symmetric nuclear matter seems to be relatively soft \cite{DanielHI}, this poses constraints on the symmetry energy as a function of density, see 
Eq. (\ref{eq:press}). A few theoretical EOS \cite{Tara2013,Ken,BCPM} are able to fulfill both the NS maximum mass condition and at the same time to be compatible with the constraints of ref. \cite{DanielHI}, which indicate that these EOS are characterized by a symmetry energy compatible with phenomenology and observational data. The sensitivity of NS mass (and radius) to the symmetry energy at saturation was studied in ref. \cite{SETBF}, where a flexible form of three-neutrons interaction was used to estimate the uncertainty on the NS mass. Even if the results are only upper limits on the NS mass, they indicate the relevance and the role of the symmetry energy in the problem of the maximum mass. \par 

Another possible physical property of NS that could be sensitive to symmetry energy is their radius. In refs. \cite{Rad2001} it was indeed shown that, for the parametrized PAL EOS \cite{PAL}, the variation of the value of the symmetry energy at saturation alters the NS radius, especially for NS of smaller mass. Later works have confirmed the correlation \cite{RadApJ,BaoPLB}. However the radius is sensitive to the crust EOS \cite{Rcrust}. To have a clear conclusion is then necessary to be consistent between the EOS for the crust and the core. In any case the possible constraints on the symmetry energy from the data on the radius are necessarily mixed with the crust structure. From the observational point of view recent analyses of data on quiescent low-mass X-ray binaries (QLMXRB) \cite{RadGR} and X-ray busters \cite{RadGO} seem to indicate that the radius could be as small as about 10 Km. Although more studies could be needed \cite{RadLS2014}, these results, if confirmed, would pose serious constraints on the overall behaviour of the symmetry energy, pointing in the direction of a moderate increase of the symmetry energy as a function of density. Notice that a steep increase of the symmetry energy does not necessarily imply a stiff EOS for NS matter, since a large symmetry energy produces a large proton fraction, which will reduce more strongly the pressure.    
\par
The structure of NS could be more complex than described above. In fact exotic matter, like hyperonic or quark matter could appear in the core. A review on this problem can be found in ref. \cite{maxmass}. If this possibility is indeed realized inside NS, then a different prospect should be taken, since the inclusion of exotic matter would require a revision of the EOS and other properties of the NS matter. The discussion on this subject is outside the scope of the present paper.           
\subsection{The role of Symmetry energy in Supernovae.}
\label{supernovae}
The supernovae (SN) explosion triggered by gravitational collapse \cite{Janka} is one of the most spectacular astrophysical phenomenon. When a star evolves along the principal sequence and has a mass large enough (roughly larger than 8 M$_\odot$)
it reaches a stage where the growing central iron core becomes unstable against collapse (Chandrasekhar instability). The contraction of the core reduces the electron fraction by beta capture on nuclei, so the electron pressure is further reduced and the collapse can stop only when the central density is close to the nuclear matter saturation density or slightly higher. At this point the repulsive action of nuclear matter dominates and the collapse stops (provided that the mass is not so high that star enters the stage of black hole). Then the nuclear matter bounces and a shock wave is formed which expands against the neutrino sphere and the still in-falling matter. If the shock wave is energetic enough an explosion occurs, whose luminosity can exceed the luminosity of the entire galaxy where the star resides. This schematic description of the supernovae explosion already suggests that the nuclear matter EOS, in particular its symmetry energy, should play a relevant role. A review of the connection between the general EOS and the supernovae characteristics and on the numerical simulations can be found in ref. \cite{FishEPJA}.\par 
The nuclear physics involved in supernovae is quite different from the one in NS, in particular the nuclear matter EOS that has to be considered under rather different physical conditions. The temperature reached during the collapse and after the bounce is high enough to change the behaviour of the EOS. One must then consider the free energy as the basic thermodynamic potential, rather than the internal energy, and correspondingly the symmetry free energy. Furthermore the typical asymmetry which develops in the supernovae process is much smaller than in NS matter. \par 
We do not enter into details about the supernovae mechanism, but for the later discussion it is important to notice that after many years of sophisticated simulations the most reliable model for the explosion is the one of the delayed neutrino revival. The initial shock wave in its outward expansion is expected to loose its energy by the dissociation that produces on the nuclei of the falling material, and by neutrino emission from the electron capture on free protons. The shock wave 
produced in simulations turns out to be not large enough to overcome the total loss of energy, and the shock stalls at a certain distance (100-150 Km) from the compact object that is forming at the center, the proto-neutron star (PNS). The mostly accepted solution of this problem is that the flux of neutrinos emitted by the hot PNS, and that were trapped there just before bounce, is so large that it can revive the shock, which leads finally to an explosion. The picture that emerges is a delayed explosion mechanism. Notice that the trapping of neutrinos in the PNS can last for several seconds.  
Furthermore it has been found that it is essential to perform the simulations in two or, better, three dimensions (rather than in spherical symmetry) to obtain a successful explosion, at least for the larger masses \cite{AAMarek,ApJSuwa,ApJCouch}. This is due to the relevance of the convective processes and hydrodynamic instabilities, that cannot occur in spherical symmetry. They increase the neutrino opacity of the matter behind the shock, thus increasing the energy deposition of the neutrinos, and introduce mechanical instability in the supernovae evolution. 
\par                  
Generally speaking the connection of micro-physics, e.g. the nuclear EOS, and the evolution and properties of supernovae faces 
the difficulties summarized in the so-called "Mazurek law" \cite{Janka}. According to this "law" the supernovae event is affected by many factors that conspire to keep the final results quite stable with respect to changes in the micro-physics. This is probably a consequence of the complexity of the phenomenon. This trend was confirmed in ref. \cite{SNNS}, where it was found that the central density at bounce is weakly correlated with the slope parameter $ L $ of the symmetry energy, and the symmetry energy itself has only a moderate influence on the evolution of the electron fraction. \par 
The matter which is present along the supernovae evolution is in an extremely wide range of density and temperature. In the outer part of the supernovae, both during the collapse and just after the bounce, the SN matter is at density low enough that it is composed of nuclei and unbound protons and neutrons. At bounce and in the PNS the matter is expected to reach a density a few times the saturation value and a temperature of few tens of MeV. It is therefore rather challenging to device a single EOS appropriate for such different physical conditions. There are three EOS which indeed describe the overall supernovae matter. One is based on the compressible liquid drop model \cite{LS}, where the low density matter is assumed to contain nucleons, alpha particle and a representative heavy nucleus for mimicking the statistical distributions of nuclei. The uniform nuclear matter EOS is parametrized
phenomenologically with few possible value of the incompressibility. The smooth transition from low to high density is achieved by imposing statistical equilibrium among the different species, and introducing different matter shapes in the (finite temperature) pasta phase region. The excluded volume approximation is used to include the interaction between the different species in the low density regime. Similarly in the EOS of ref. \cite{Shen} the relativistic finite temperature Thomas-Fermi approximation is used
to describe the non-homogeneous matter with a representative nucleus and the gas of nucleons and alpha particles. Finally the EOS of ref. \cite{HS} introduces a statistical ensemble of nuclei in the low density part together with a relativistic mean field for nucleons and alpha particles, as well as for the homogeneous matter. In the low density regime these EOS give different N and Z for the heavy nuclei as a function of density and temperature. Since the collapse evolution is mainly determined by the electron capture process on nuclei, these EOS give different \cite{FishEPJA} rate of deleptonization (i.e. neutrino emission) and therefore different lepton fraction $ Y_L $ at the trapping stage in the PNS. These differences are related to the nuclear symmetry energy only very indirectly through the predicted binding of nuclei, but many other effects are relevant and no imprint of the nuclear matter symmetry energy can be identified. However in the collapse phase up to bounce also the (inhomogeneous) matter symmetry free energy is relevant.  
The contraction rate is driven by the pressure of the matter, which depends also on the total asymmetry. One can derive the contribution $ P_B $ to the total pressure of the baryonic matter from the free energy $ F $
\beq
P_B \,=\, \rho_B^2 \left( \frac{\partial F}{\partial \rho_B} \right)_{T,Y_e} 
\eeq
\noindent where the subscript indicates the quantity to be held fixed in the partial derivative. The dependence of the free energy on the asymmetry can be expanded as for the nuclear matter symmetry at zero temperature
\beq
F(T,\rho,\beta) \,=\, F(T,\rho,0) \,+\, S_B(T,\rho) \beta^2 \,\,\,\, \cdots\cdots\cdots 
\eeq
\noindent where  $ S_B $ can be identified as the baryonic matter symmetry energy at finite temperature, or better symmetry free energy. It has to be stressed that this symmetry energy must be distinguished from the symmetry energy for nuclear matter, since it is defined for baryonic matter that contains heavy nuclei and eventually light clusters. It can have a quite different value with respect to the one for uniform nuclear matter at the
same density. Even if the latter is actually unstable towards cluster formation at these low densities, it can be still defined. The symmetry energy $ S_B $ can have some connection with the symmetry energy that appears in heavy ion collisions in the multifragmentation regime, as it will be discussed in Section \ref{HI}. \par
The PNS cannot be observed since in the post-bounce period the mantle, which eventually will be expelled by the explosion, will obscure all possible signals coming from the PNS. A noticeable exception is the neutrino signal, that, despite its elusive character, was observed by the Kamiokande apparatus, which detected a few neutrinos coming from the supernova 1987a \cite{nsuper}. A galactic supernova is expected to give a neutrino signal coming from the PNS strong enough to enable a detailed study of the PNS evolution. The connection of the neutrino signal and the nuclear matter symmetry energy was studied in ref. \cite{nsignal}, where the neutrino luminosity of the PNS as a function of time was studied by simulations that include convection. The convection was shown to modulate the neutrino luminosity in a characteristic way, that looks dependent on the symmetry energy of the matter. In fact
the convection affects the neutrino transport and the neutrino luminosity of the matter.  \par
\begin{figure}[hbt]
\vskip -5 cm
\centering
\includegraphics[scale=0.4,clip]{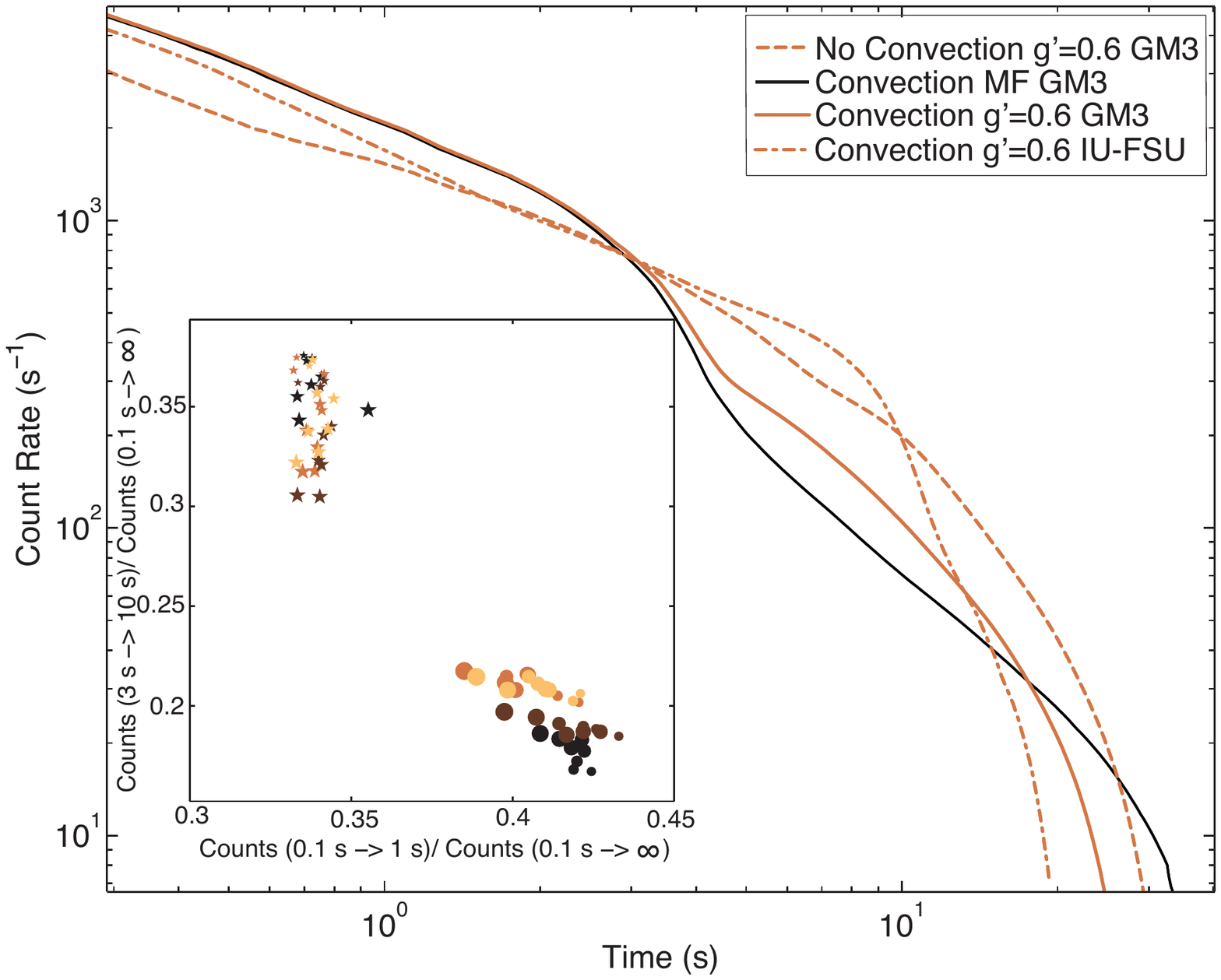}                    
\caption{(Color online) Neutrino luminosity from a proto-neutron star as a function of time. Three cases are considered : no convection with the functional GM3 with short range correction, convection regime from the functional GM3 (with and without short range correction), and convection from the functional IU-FSU with short range correction. The drops of luminosity along the different curves at different times correspond to the onset of convection. In the insert additional characterization of the different luminosity curves are displayed, see the text. Figure from ref. \cite{nsignal}.} 
\label{fig:lumin}
\end{figure}

In general the convection phenomenon is due to buoyancy instability. In order to see the connection with symmetry energy, we will describe briefly the mechanism of convection, following refs. \cite{Wilson,nsignal}. Let us consider a certain amount of PNS matter in equilibrium under the influence of the gravitational field. For simplicity we can assume spherical symmetry. Each small portion of matter (a blob, say) centered at a given radial distance $ r $ is characterized by the local matter density $ \rho $,
the pressure $ P $, the entropy $ s $ and the composition. i.e. the lepton fraction $ Y_L $. As a consequence of the matter EOS we can assume that the density is a function of the other physical variables, 
\beq
 \rho(r) \,=\, \rho(P(r),s(r),Y_L(r)) 
\label{eq:convec1}\eeq
 If the blob is moved at large radial distance slowly enough, it will adjust to be in equilibrium with the surrounding medium, i.e. its thermodynamic variables will change and acquire the values of the matter around it. If the motion is not slow enough, thermodynamic equilibrium will not be reached and some of the variables can be assumed to keep their initial value. Which ones of the thermodynamic variables will equilibrate and which ones will keep their initial value will depend on the physical conditions. In any case the density of the blob at its shifted position will be different from the one of the surrounding matter, according to Eq. (\ref{eq:convec1}). If this density is smaller than the one of the surrounding matter, the blob will be pushed up by buoyancy. This means that the medium is unstable for an ascending motion of the lower part, which entails convection. It is clear from the discussion that different types of convection are possible, according to which variables are kept adiabatic and which ones equilibrate with the medium. The possible convection that seems to be appropriate, if possible, in the matter of a PNS is the so-called Ledoux convection, where the motion is locally adiabatic, i.e. at constant entropy, and the composition also does not change, while the pressure equilibrates. Therefore the  condition for the onset of convection is in this case
\beq
 \rho(P(r+\lambda),s(r+\lambda),Y_L(r+\lambda)) \,-\, \rho(P(r+\lambda),s(r),Y_L(r)) \,=\, \lambda \left(\ \left(\frac{\partial \rho}{\partial s}\, \right)_{P,Y_L} \frac{d s}{d r} \,+\, \left(\, \frac{\partial \rho}{\partial Y_L}\, \right)_{P,s} \frac{d Y_L}{d r}\ \right) \,\geq\, 0
\label{eq:convec2}\eeq 
\noindent where the density difference is expanded to first order in $ \lambda $. This condition can be satisfied or not during the evolution of the PNS and the corresponding mantle. It can be used in the simulations to introduce convection in the proper  regions of the PNS. The connection with the symmetry energy can be seen considering the second term (inside the brackets) in
 the inequality (\ref{eq:convec2}). Taking into account the condition of the equilibration of the pressure, one gets
\beq
\left( \frac{\partial \rho}{\partial Y_L} \right)_{P,s} \,=\, - \left(\ \frac{\partial P}{\partial Y_L}\ \right)_{\rho,s} \, 
\left( \ \frac{\partial \rho}{\partial P}\ \right)_{Y_L,s}
\label{eq:convec3}\eeq                
\noindent For simplicity we will further neglect the neutrino component, i.e. $ Y_L \,\equiv\, Y_e $, and the effect of temperature.
Then, using Eq. (\ref{eq:press}), one gets for the contribution $ P_B $ of the baryons to the pressure $ P $
\beq
\left( \frac{\partial P_B}{\partial Y_e} \right)_{P,s} \,=\, - 4 \rho_B^2 \left( \frac{\partial S}{\partial \rho_B} \right)\, (\, 1 \,-\, 2 Y_e \,) 
\label{eq:convec4}\eeq 
\noindent where $ S $ is the nuclear matter symmetry energy taken at the local density. One finally gets
\beq
\left( \frac{\partial \rho_B}{\partial Y_e} \right)\, \left( \frac{d Y_e}{d r} \right) \,=\, 4 \rho_B^2 \left( \frac{\partial S}{\partial \rho_B} \right)\, (\, 1 \,-\, 2 Y_e \,)\, \left( \frac{\partial \rho_B}{\partial P_B} \right)_{Y_e,s} \, \left( \frac{d Y_e}{d r} \right) 
\label{eq:convec5}\eeq
\noindent The factor in front of the radial derivative of $ Y_e $ is positive for any reasonable NS matter EOS. Therefore a positive gradient of the electron fraction $ Y_e $ will move the matter toward a possible convection instability, a negative one has a stabilizing effect. In particular for a reasonable EOS a positive gradient both for entropy and electron fraction implies the onset of Ledoux convection. The effect of the electron fraction gradient is larger for larger slope of the nuclear symmetry energy at the local density. Following similar considerations one can obtain an estimate of the growing time of the convection inside the instability region. In general the faster is the developing of the convection more robust is its presence with respect to the physical conditions of the PNS matter. This means that convection will persists for longer time and again this depends also on the average slope of the symmetry energy, according to Eq. (\ref{eq:convec5}). Since the presence of convection increases neutrino luminosity,
the time profile of the neutrino emission is affected by the features of the symmetry energy of the EOS of the PNS matter. This was indeed checked in ref. \cite{nsignal}, where two functionals with a different symmetry energy were employed in the simulations of the PNS
after bounce. From this paper we report here Fig. \ref{fig:lumin}, where the corresponding neutrino counting as a function of time is drawn for the two functionals. A galactic supernova was assumed and the efficiency of the detector, similar to Kamiokande, was properly included. The drop of the counting at about 3 sec. (time of the simulation) is due to the end of the mantle convection in the simulation for the functional GM3 (full black line). This is only slightly modified if the neutrino opacity is calculated in the RPA scheme (g' = 0.6) rather than in mean field (MF), see full red line. For the other functional IU-FSU, convection lasts for a longer time and the drop (red dash-dot line) accordingly occurs at later time, around 12 sec. This is in line with the larger symmetry energy slope, at least at supra-saturation density, which produces a stronger stabilizing effect. In any case the shape of the counting rates as a function of time are different, and they reflect the features of the symmetry energy in the two functionals. More quantitatively this is shown in the insert, where the integrated luminosity at short time and at higher time, with respect to the total one, are reported in a two-dimensional plot. A clear separation of the two functional is evident, see ref. \cite{nsignal} for details..
Without convection the neutrino luminosity has again a different behaviour. 

\subsection{Imprint of the Symmetry Energy on Gravitational Wave emission.}

The field of "gravitational-wave astronomy" has recently become reality thanks to the first ever detected gravitational wave (GW) by the Laser-Interferometer-Gravitational-Wave Observatory (LIGO)  \cite{GW}, successively confirmed by an additional observation \cite{GW2}.  In both
cases, the signal seems to be originated by a binary black hole merger, but it can be expected that similar signals could come also from NS binary mergers. In this case the EOS and the symmetry energy can be strongly involved \cite{GWPRD90}. The GW astrophysics is in its infancy, but its development will be of enormous relevance for our knowledge of the Universe and the phenomena that can occur, as well as of their fundamental laws.
This is an exciting prospect, because gravitational-wave observations
have the potential to probe several aspects of neutron star physics \cite{2009astro}.
Moreover, the information gleaned will be complementary to
electromagnetic observations, thus providing constraints on the state of matter at extreme densities.
\begin{figure}[hbt]
\centering
\includegraphics[scale=0.4,angle=-90,clip]{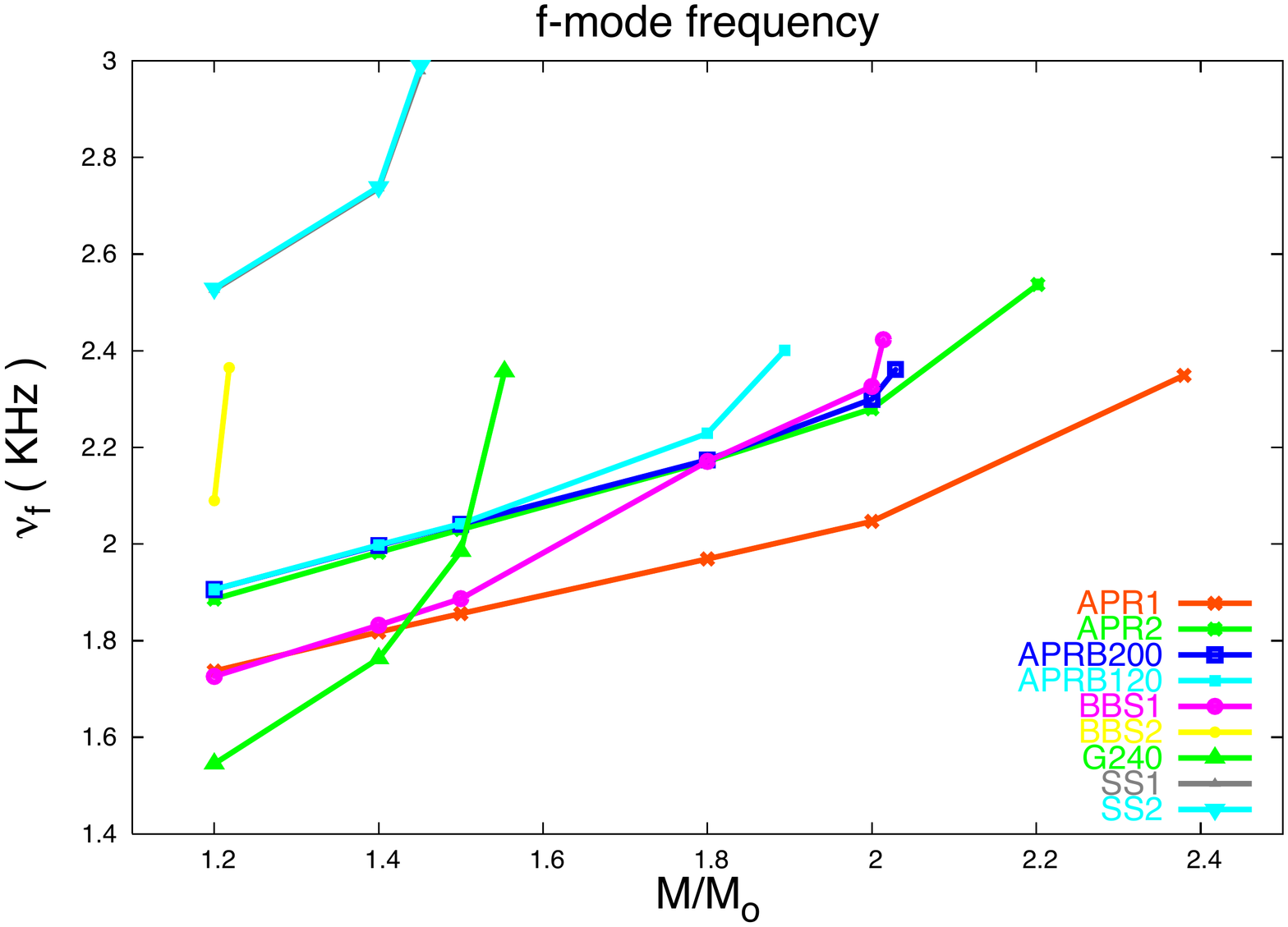}                    
\caption{(Color online) The frequency of the fundamental mode is displayed as a function of the mass of the star for several choices of the EOS. For the the meaning of the labels and corresponding references, see the text. Figure taken from ref. \cite{2004GW_Roma}.} 
\label{fig:freq}
\end{figure} 
Neutron stars radiate gravitationally in a number of ways. The most promising scenarios involve oscillations and instabilities,
binary inspiral and merger,  core-collapse supernovae and hot remnants, rotating deformed neutron stars \cite{2011GW_Rev}.
They are briefly discussed below.

Neutron stars have rich oscillation spectra which, if detected,
could allow us to probe the internal composition. The basic strategy for such ``gravitational-wave
asteroseismology'' has been set out in ref.\cite{1998MNRAS.299}, but neutron stars models need to be made much 
more realistic if the method has to be used in practice. An extensive discussion of this issue is presented in ref.\cite{2004GW_Roma},
where using several different EoS for neutron star matter the frequencies and damping times of the quasinormal  modes 
are calculated. As an example, we show in Fig.\ref{fig:freq}
the frequency of the fundamental mode as a function of the neutron star mass, for different choices
of the equation of state \cite{2004GW_Roma}. Purely nucleonic EOS (labelled APR2\cite{apr}, BBS1\cite{BBS1}), including hyperons (BBS2\cite{BBS1}, G240\cite{Glen})
or a hadron-quark phase transition (APRB200\cite{aprb}, APRB120\cite{Rubino}) were considered, along with models of strange stars (SS1, SS2\cite{DeyBomb}).

We notice that the BBS1 and APR2 EOS, which are based on different many-body technique, yield appreciably different f-mode frequencies,
likely due to the  different treatment of three-nucleon interactions \cite{RepPP}.
The transition to hyperonic matter, predicted by the BBS2 model, produces a sizeable softening of the EOS, thus leading to stable NS configurations of  mass smaller than 1.4 $\rm M_\odot$. As a consequence, the corresponding f-mode frequency is significantly higher than those obtained with the other EOS. So much higher, in fact, that its detection would provide evidence of the presence of hyperons in the NS core. It is also interesting to compare the f-mode frequencies corresponding to models BBS2 and G240 , as they both predict the occurrence of hyperons but are obtained from different theoretical approaches. The behavior of  the oscillation frequency $\nu_f$ displayed in Fig.\ref{fig:freq} directly reflects the
relations between mass and central density obtained from the two EOS, larger frequencies being always associated with larger densities. 
Finally, we notice that the presence of quark matter in the star inner core (EOS APRB200 and APRB120) does not seem to significantly affect the pulsation properties of the star. This is a generic feature, which we observe also in the p- and w- modes behaviour.
On the other hand, strange stars models (SS1 and SS2)  correspond to values of  frequencies well above those obtained from the other models. The peculiar properties of these stars largely depend upon the self-bound nature of strange quark matter. \par
In any case it has to be noticed that the connection with the symmetry energy at high density is indirect,
since the EOS for the NS matter largely depends on the behavior of the symmetry energy.
\par
A further scenario for GW emission is the birth of a proto-neutron star (PNS) in a core-collapse supernova. This  is a very difficult phenomenon to model, since it requires not only accurate descriptions of the microphysics of the collapsing matter, but also of the violent dynamical processes occurring in the contracting-exploding star, which need to be treated in the framework of general relativity.
The description of the subsequent PNS evolution is also challenging, because a PNS is a hot and rapidly evolving object. Thus, most simulations of gravitational core collapse to a PNS end shortly after the core bounce and the launch of the supernova explosion-typically after a few hundreds of milliseconds-and only a few dynamical simulations extend to the first minute of the PNS life. In this latest phase of the PNS life, the evolution is quasistationary and  the frequencies of the PNS quasinormal modes of oscillation are associated to gravitational wave signals with sizeable amplitudes. It turns out \cite{pns_hot} that those frequencies are lower than those typical of mature neutron stars, and this would favour their detection by ground-based interferometers LIGO/Virgo and their future version, the Einstein Telescope \cite{fermin}.
\begin{figure}[hbt]
\centering
\includegraphics[scale=0.25,clip]{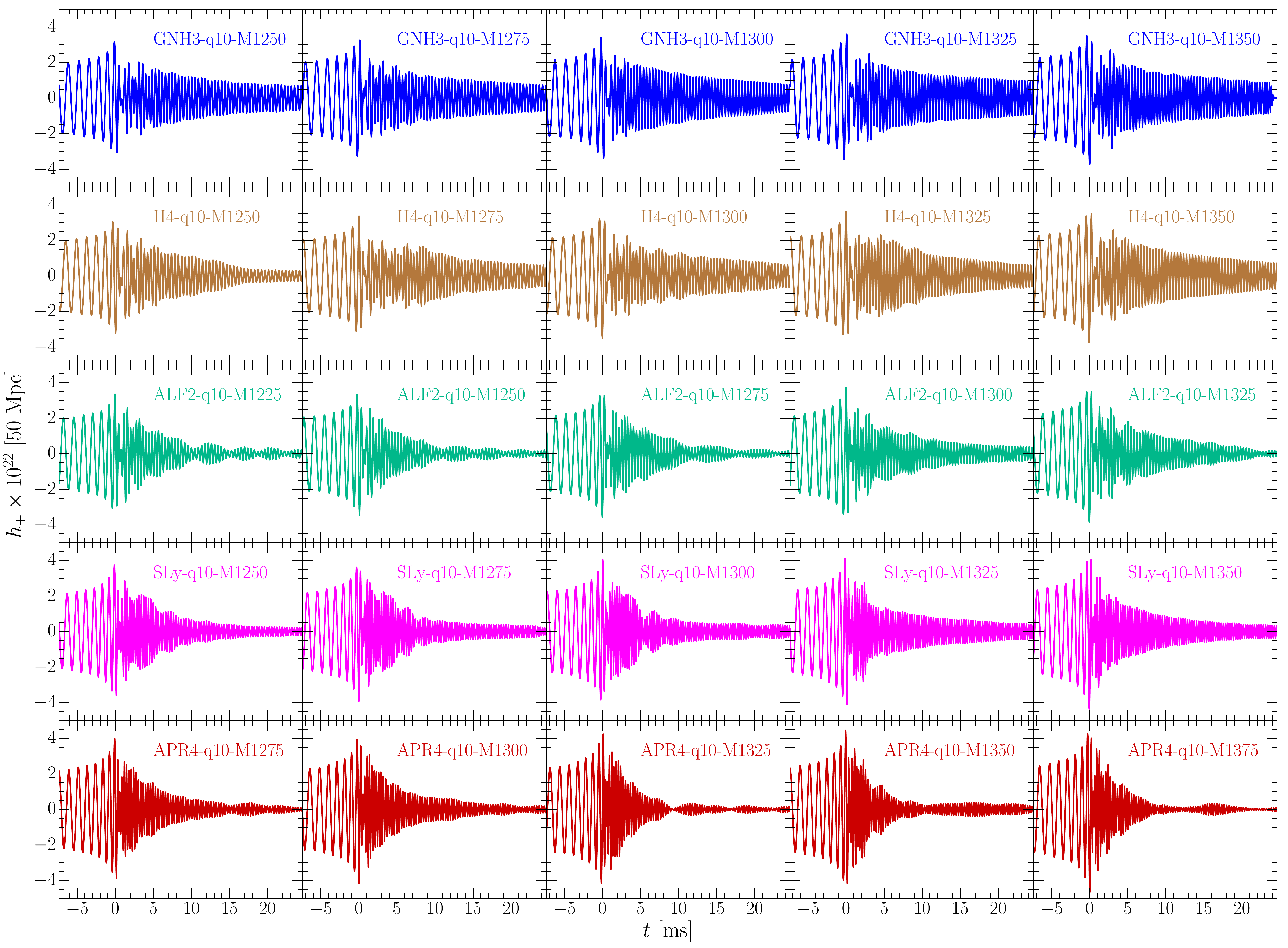}                    
\caption{(Color online)  Gravitational waveforms for all the binaries with equal masses and nuclear-physics EOSs. Each row refers to a given EOS, while each column corresponds to a given initial mass. Different colors indicate different EOS. Taken from ref.\cite{2015Rez}.} 
\label{fig:merg_EOS}
\end{figure} 

\begin{figure}[hbt]
\centering
\includegraphics[scale=0.4,angle=90,clip]{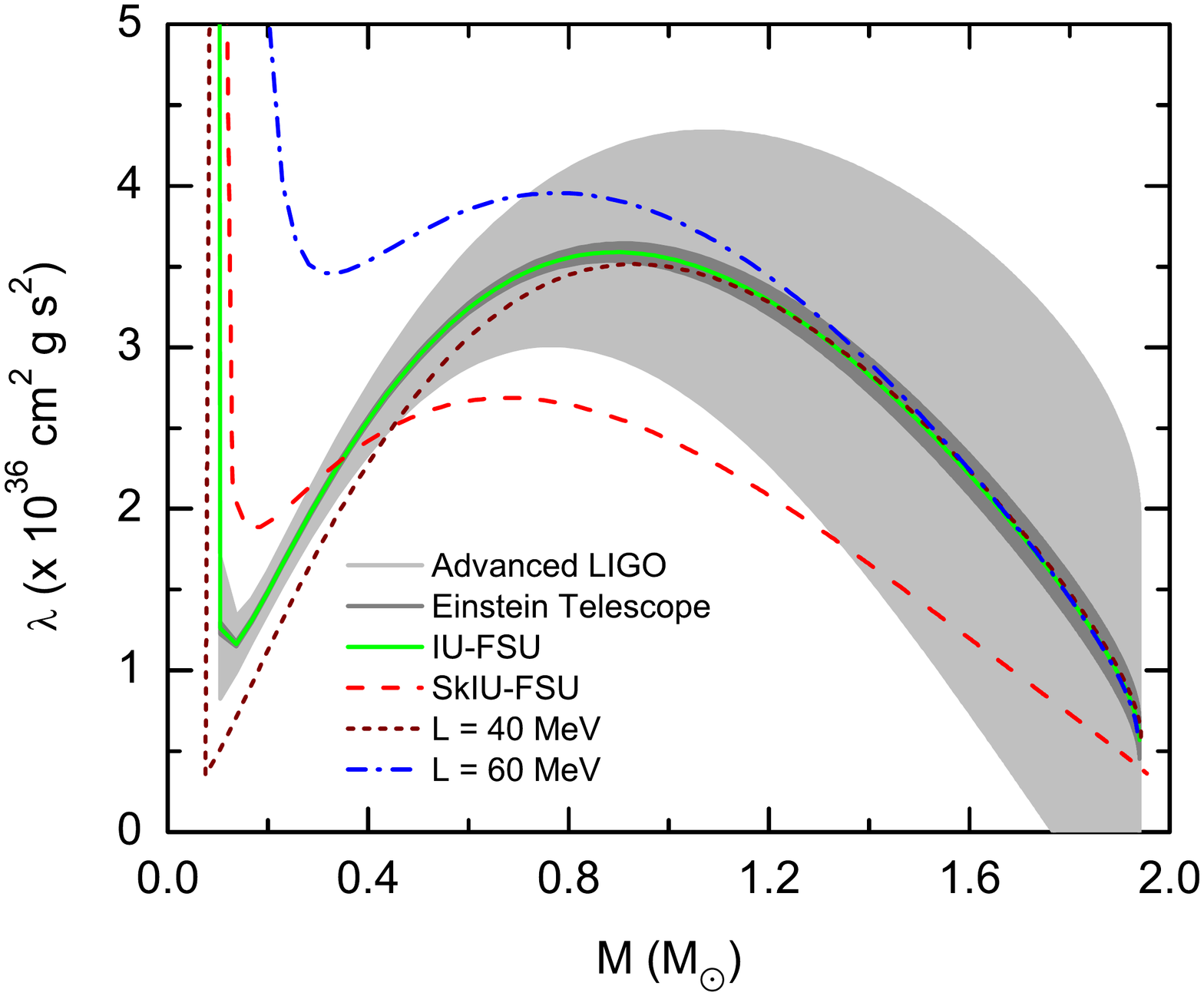}                    
\caption{(Color online) Tidal polarizability $\lambda$ of a single neutron star as a function of neutron-star mass for several EOS which differ in the $L$. The shaded light-grey (dark-grey) area represents a crude estimate of uncertainties in measuring $\lambda$ for equal mass binaries at a distance of D = 100 Mpc  for the Advanced LIGO detector (Einstein Telescope). Taken from ref.\cite{2013Fatto}.} 
\label{fig:tidal}
\end{figure}

Also NS mergers may  yield information about the nuclear EoS, because the dynamics of the coalescence depend
sensitively on the behaviour of high-density matter \cite{2010Duez,2009Faber}. 
The merger of NS is a consequence of gravitational wave emission, 
which extracts energy and angular momentum from the binary and thus forces the binary components on inspiraling trajectories. 
The sources that are more likely to be detected are, in particular, the inspiral and post- merger of NS
binaries or NS-black hole binaries, and binary black holes. In particular, binary neutron star (BNS) mergers are widely 
considered to be the most common source, with an expected detection rate of $\approx \rm 40~yr^{-1}$ \cite{2010Abadie} .

In their pioneering works, Bauswein and Janka \cite{2012PRLBaus,2012PRDBaus} showed that it is possible to extract 
information about the EOS of nuclear matter from 
the analysis of the spectral properties of the post-merger signal. Using a set of many different equations of state, they pointed out the presence of a peak at high-frequency in the spectrum (labelled $f_{peak}$), and showed a tight correlation with the radius of the maximum-mass non-rotating configuration.  It was later recognized that $f_{peak}$ corresponds to a fundamental fluid mode with $m=2$ of the hypermassive NS, formed by the merger, and that the information from this frequency could also be used to set constraints on the maximum mass of the system and hence on the EOS. 
Recently, the authors of ref.\cite{2015Rez} found that the post-merger emission is characterized also by a low-frequency peak, 
which is related to the total compactness of the stars in the binary. An example of the expected signals is displayed in Fig.\ref{fig:merg_EOS}, which collects all  waveforms for the equal-mass models with nuclear-physics EOSs.
Each row refers to a given EOS, while each column indicates a given initial mass, and the different EOSs are distinguished by different colors. 
Some features are common among the different signals, e.g., the pre-merger signal and the post-merger one are in frequency ranges that are considerably different. This is not surprising given the significant difference in compactness in the system before and after the merger.

We notice that the analysis illustrated above aims at exploring the dependence of the GW signal on the equation of state.
To our best knowledge, the dependence on the symmetry energy has been studied only in ref.\cite{2013Fatto,2014Fatto}.
Using a set of model EOSs satisfying the latest constraints from terrestrial nuclear experiments, modern nuclear many-body calculations of the pure neutron matter EOS, and astrophysical observations consistently, the authors of ref.\cite{2013Fatto,2014Fatto} study various GW signatures of the high-density behaviour of the nuclear symmetry energy. In particular, they find the tidal polarizability of NS, potentially measurable in binary systems just prior to merger, is a very sensitive probe of the high-density component of the nuclear symmetry energy, more than the symmetry energy at nuclear saturation density. We remind the reader that the
tidal deformation NS undergo as they approach each other prior to merger, may be effectively described through the tidal polarizability parameter $\lambda$ \cite{2008Hinder} defined via $Q_{ij}=-\lambda\cal E_{\rm ij}$ , where $Q_{ij}$ is the induced quadrupole moment of a star in binary, and  $\cal E_{\rm ij}$ is the static external tidal field of the companion star. The tidal polarizability can be expressed in terms of the dimensionless tidal Love number $k_2$ and the neutron star radius R as $\lambda=2 k_2 R^5/(3G)$, being the Love number solution of a 
system of first-order differential equations where the EoS is the only input \cite{2008ApJHind}.
Using two classes of nuclear EOS within the relativistic mean field (RMF) model and the Skyrme Hartree-Fock (SHF) approach,
the authors of ref.\cite{2013Fatto} found that for coalescing binaries consisting of two massive neutron stars with equal masses, the tidal 
polarizability $\lambda$ is rather insensitive to variations of $L$ within the constrained range, but it changes up to
$\pm 10\%$  with the nuclear compressibility and the nucleon effective mass. On the other hand, the $L$ parameter affects significantly the tidal polarizability of neutron stars with $M \leq 1.2 M_\odot$.
In Fig.\ref{fig:tidal} the tidal polarizability $\lambda$ is reported as a function of neutron-star mass for the adopted EOSs,
which are different only in their predictions for the nuclear symmetry energy above about $1.5\rho_0$, and this lead to significantly 
different $\lambda$ values in a broad mass range from $0.5 M_\odot$ to $2 M_\odot$.
The question comes up whether the Advanced LIGO-Virgo detector may potentially measure the tidal polarizability of binary neutron stars. 
As an example, we report uncertainties in measuring $\lambda$ for equal-mass binaries at an optimally oriented distance of
$D=100 Mpc$  \cite{2010Abadie}, shown in Fig.\ref{fig:tidal} for the Advanced LIGO-Virgo detector (shaded light-grey area) and the Einstein Telescope (shaded dark-grey area). We see that the Advanced LIGO-Virgo detectors' sensitivity for stars of mass 
$1.4 M_\odot$ and below is at the limit for discerning between high-density symmetry energy behaviours.
On the other hand, the narrow uncertain range for the proposed Einstein Telescope will enable  to tightly constrain the symmetry energy especially at high densities.

\subsection{Summarizing and Prospects.}
In this section we have considered the relevance of the nuclear symmetry energy in the astrophysical context. We have not touched the wide area of the nucleosynthesis problem. In this field the role of symmetry energy is only quite indirect. It is involved in the predictions of exotic nuclei binding energy, not reachable in laboratory, and the reaction rate of slow and fast processes.  
\par 
More direct is the influence of the symmetry energy on the structure and dynamics of compact objects like Neutron Stars and Supernovae. We analyzed from this point of view the whole structure of NS, from the crust to the core. Transient phenomena,
like deep crustal heating and data on QLMXRB and X-ray busters will surely shade further light on the structure of the crust and therefore on the symmetry energy. Despite the model dependence, data analysis on NS oscillations can give access to the composition of the crust and then on the symmetry energy. The maximum mass of NS is a basic question that can be decisive for the constraints one can put on the nuclear matter EOS and the corresponding symmetry energy. In Supernovae the most spectacular advances can come from the neutrino observatories. There can be little doubt that future improvements on the sensitivity of the detectors and on the accuracy of the simulations will be able to provide basic and detailed insights on the initial evolution of supernovae and the properties of the nuclear EOS and it symmetry energy.\par 
 We have also considered the field of gravitational waves (GW). The first detected GW \cite{GW} seem to be originated by binary black hole mergers, but it can be expected that similar signal could come from NS binary mergers. In this case the EOS and the symmetry energy can be strongly involved \cite{GWPRD90}. The GW astrophysics is in its infancy, but its development will be of enormous relevance and opens a new window in the observational astrophysics, complementary to the other methods based on different signals, like electromagnetic waves and neutrinos.           
\section{Symmetry energy in heavy ion collisions.}
\label{HI}
The density dependence of the symmetry energy can be explored in laboratory with heavy ion collisions.
Its advantage is that by choosing different collision systems, incident energies and impact parameters,
one can access different densities and asymmetries of nuclear matter, and study the momentum dependence of
the nuclear interaction. In the energy range from the Fermi energy to about 150 MeV per nucleon, the collision dynamics
proceeds through the formation of a hot, compressed participant region, followed by an expansion
where the decomposition into fragments with the emission of light clusters is the main decay mode.
This allows to obtain information on the properties of the nuclear symmetry energy, since the forces resulting from such a compression
and the following expansion determine the motion of ejected matter.  A number of observables have been identified as sensitive to the nuclear symmetry energy. One is the pre-equilibrium emission of nucleons and light fragments in the initial stages of the collision, which depends directly on the neutron/proton potentials. To enhance the sensitivity to the nuclear symmetry energy one considers ratios or 
differences of observables of isobaric pairs of particles, such as yields of neutrons to protons, tritons to $^3$He, or
$\pi^-$ to $\pi^+$.
\par
In this section we first give a sketch of the main features of heavy ion collisions and the constraints on the nuclear EOS that is possible to device from the experimental data. Then we will consider different aspects of the symmetry energy that appear in heavy ion collision experiments and analysis.

\subsection{Constraints on the EoS.}
\label{sec:eos}

The asymmetric nuclear matter EOS can be expanded around saturation according to 
\begin{eqnarray}
\frac{E}{A}(\rho, \beta) &=& \frac{E}{A}(\rho, 0) + S_N (\rho) \beta^2  \\ 
& = &  \frac{E}{A}(\rho_0) + \frac{1}{18} K_0 \, e^2 + \Big[S_0 - \frac{1}{3} Le + \frac {1}{8} K_{sym} e^2 \Big ] \beta^2 \nonumber
\end{eqnarray}
where $E(\rho, 0)$ is the binding energy for symmetric nuclear matter, $e = (\rho - \rho_0)/\rho_0$, $K_0$ is the incompressibility at the saturation point
\begin{equation}
K_0 = k_F^2 \left(  \frac{d^2E/A}{dk_F^2} \right),
\end{equation}
\noindent
$ \beta $ is the asymmetry parameter, see Eq.(\ref{eq:def}), and the other parameters have been introduced in Eq.(\ref{eq:expans}).
Significant constraints for the term  $E(\rho, 0)$ in the range $1 \leq \rho/\rho_0 \leq 4.5$  have been obtained from 
measurements of collective flow \cite{DanielHI} in heavy ion collisions (HIC) at energies
ranging from few tens to several hundreds MeV per nucleon (hereafter indicated as MeV/A). 
The main goal has been the extraction from the data of the gross properties of the nuclear EoS. It can be expected that in heavy ion collisions at large enough energy nuclear matter is compressed and that, at the same time, the two partners of the collisions produce flows of matter. In principle the dynamics of the collisions should be connected with the properties of the nuclear medium EoS and its viscosity. In the so called ''multifragmentation" regime,  after the collision numerous nucleons and fragments of different sizes are emitted, and the transverse flow, which is strongly affected by the matter compression during the collision, can be measured.

\begin{figure}[t]
\centering
\vspace{-3cm}
\includegraphics[scale=0.5,clip]{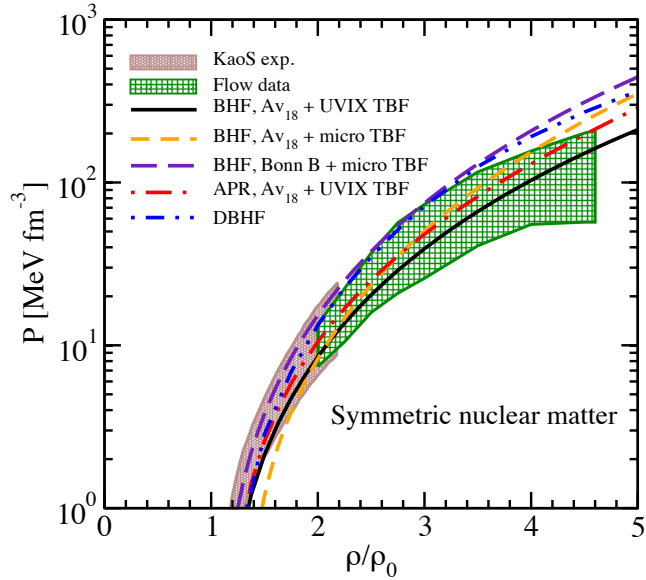}
\vspace{-1cm}
\caption{(Color online)
Pressure as a function of baryon density for symmetric nuclear matter.
The green dashed box represents the boundary obtained from
the numerical simulations for the experimental data discussed in Ref.\cite{DanielHI}. 
The brown filled region represents  the experimental data on sub-threshold kaon production 
obtained by  the KaoS \cite{kaos} and FOPI \cite{fopi} collaborations. The EOS of nuclear matter
is constrained to pass through both filled regions. Some theoretical EOS are reported, following ref.
\cite{Tara2013}, from where the figure was taken. The meaning of the labels for each EOS can be found
in that reference. }
\label{fig:p_sym}
\end{figure}

Based on numerical simulations, in reference \cite{DanielHI} a phenomenological range of densities
was proposed where any reasonable EoS for symmetric nuclear matter should pass through in the pressure vs. density plane. 
The plot is reproduced in Fig.\ref{fig:p_sym}, where the green dashed box represents the
results of the numerical simulations of the experimental data discussed in Ref.\cite{DanielHI}. In ref.\cite{Tara2013} it was found that 
most of the currently used EoS derived within microscopic approaches are compatible with the experimental data.
The meaning of the labels can be found in the original paper.

In Fig.\ref{fig:p_sym} the brown filled region represents  the experimental data on sub-threshold kaon production 
obtained by  the KaoS \cite{kaos} and FOPI \cite{fopi} collaborations.
The optimal energy for this type of investigation is close or even below two-body threshold, since then the only way to produce the kaons is by compression of the matter. Since at threshold the production rate increases steeply, there is a strong sensitivity to the value of the maximum density reached during the collision, and this is an ideal situation for studying the EoS and its incompressibility. The comparison of the simulations with the experimental data on $K^+$ production  points in the direction of a soft EoS \cite{Fuchs}. However it has to be kept in mind that, in the simulations, kaon production occurs at density  $\rho \geq 2-3\rho_0$, and therefore this set cannot be directly compared to the one of the monopole oscillations. In
any case, a stiff EoS above saturation seems to be excluded from this analysis, as it is evident in Ref. \cite{Fuchs}.
This is in agreement with the values extracted from the phenomenology of monopole oscillations \cite{Blaizot}. One finds indeed that a correlation exists between incompressibility and
position of the monopole Giant Resonance, so that, in principle it is possible to extract from the experimental data the value of the incompressibility in nuclear matter. 
At present, the constraints on the value of the nuclear matter incompressibility from the
monopole excitation are not so tight. It can be approximately constrained between 210 and 250 MeV
\cite{Jirina}, though a more refined value can be expected to come out in the near future from additional analysis of phenomenological data.
\par

\subsection{Symmetry term in heavy ion collisions} \label{sym_eos}

The knowledge of the equation of state for asymmetric nuclear matter beyond
saturation density still remains poor. The symmetry energy 
reflects this uncertainty, as found in  the different predictions of modern microscopic many-body theories \cite{Tara2013}.

In the transport theories, widely used for interpreting HIC, 
effective interactions, associated with a given EOS, are usually employed as an input of the transport code, and
from the comparison with experimental data one can get some hints on nuclear matter properties.
The usual way of treating the symmetry term is to split it into two contributions, i.e. a kinetic contribution 
coming directly from the basic Pauli correlations, and a potential contribution from the properties of
the isovector part of the effective nuclear interactions in the medium.
Since the kinetic part can be exactly evaluated, one can separate
the two contributions, reducing the discussion just to a function $F(u)$
of the density $u \equiv \rho/ \rho_0$ linked to the interaction:

\begin{equation}
\epsilon_{sym}(\rho) \equiv {E_{sym} \over A} (\rho) = {\epsilon_F(\rho) \over 3} 
+ {C \over 2} ~ F(u),
\label{kipot}
\end{equation}
\noindent
with  $F(1)=1$,  $\epsilon_F(\rho)$ is the Fermi energy, $\epsilon_F=\frac{\hbar^2}{2m} \left( \frac{3\pi^2\rho}{2} \right )^{2/3}$,
and the parameter $C$ is of the order $C \simeq 32$ MeV, chosen
in order to reproduce the symmetry term of the Bethe-Weisz\"acker mass formula.
\noindent
The choices of the $F(u)$ behavior are not arbitrary, and reflect the wide spectrum of theory predictions for effective forces
in the isovector channel. For instance, $F(u)=const=1$  (around saturation density) is
typical of Skyrme-like forces, whereas  $F(u)=\sqrt{u}$ behavior is obtained in
variational approaches with realistic $NN$ interactions 
\cite{apr}. Moreover,  the linear dependence $F(u)=u$ is found in 
non-relativistic Brueckner-Hartree-Fock (BHF) \cite{BombaciPRC55}
as well as in Relativistic Mean Field (RMF) \cite{yoshidaPRC58} and
Dirac-Brueckner-Hartree-Fock (DBHF) \cite{LeePRC57} approaches. 
A stiffer symmetry term in general enhances the pressure
gradient of asymmetric matter, and therefore one can expect direct effects on the
nucleon emissions in the reaction dynamics, fast particles and collective
flows. 

In the following we elaborate more on the isospin dependence of the Skyrme-like  effective interactions
\cite{vau3}, since in this report we will  often show results on heavy-ion collisions obtained from non-relativistic kinetic equations 
with Skyrme forces. For instance, let us consider a simplified Skyrme-like effective interaction \cite{Col1998}, see also Eq. (\ref{eq:EOS_Skyrme}) :
\begin{equation}
V_{ij} = t_0 (1+x_0 P_\sigma) \delta({\bf r}_i - {\bf r}_j) + \frac{1}{6} t_3 (1+x_3 P_\sigma) 
\left [ \rho \left( \frac{{\bf r}_i + {\bf r}_j}{2} \right) \right ]^{\sigma-1} \delta({\bf r}_i - {\bf r}_j) 
\end{equation}
\noindent 
where $P_\sigma = \frac{1}{2} (1 - { \sigma}_1 \cdot { \sigma}_2 )$ is the  spin exchange operator. 
If we impose the correct value of the saturation density and binding energy, and an incompressibility value 
equal to $K=200$ MeV, we obtain the following values for the coefficients, i.e.,
$t_0 = \rm -2973 ~ MeV ~ fm^{3}$, $t_3 = \rm 19034 ~MeV ~fm^{3\sigma}$,
$x_0 = 0.025$, $x_3=0$, and $\sigma=\frac{7}{6}$. 
This parametrization can be considered as a simplified Skyrme SkM* force with effective mass $m^*=m$.
The interaction contribution to the energy density functional can be expressed under the form :
\begin{equation}
\cal E\rm_{pot} (\rho, \beta) = \frac{A}{2} \frac{\rho^2}{\rho_0} + \frac{B}{\sigma+1} \frac{\rho^{\sigma+1}}{\rho_0^\sigma} + 
\frac{C(\rho)}{2} \frac{\beta^2\rho^2}{\rho_0} 
\label{eq:func}
\end{equation}
\noindent
where $\rho_0=0.16~ \rm fm^{-3}$ is the saturation density of nuclear matter and the coefficients
$A,B$ and $C$ in the energy density functional (\ref{eq:func}) are related to the Skyrme parameters as :
\begin{equation}
A= \frac{3}{4} t_0 \rho_0, \quad B= \frac{\sigma+1}{16} t_3 \rho_0^\sigma, \quad C(\rho) = -\frac{1}{2} \rho_0 \left [t_0(x_0+\frac{1}{2}) +
\frac{t_3}{6} (x_3+\frac{1}{2}) \rho^{\sigma-1} \right ]
\label{eq:coeff}
\end{equation}
\noindent 
The coefficient $C$ in eqs.(\ref{eq:coeff}) at saturation density is related to the symmetry coefficient in the Weisz\"acker mass formula :
\begin{equation}
a_{sym} = {{1\over 2} {\partial^2(E/A) \over \partial \beta^2}} {\Big |}_{\beta=0}  = \frac{\epsilon_F}{3} + \frac{C(\rho_0)}{2}
\label{eq:asym}
\end{equation}
where $\epsilon_F$ is the Fermi energy.
With the considered parametrization $\epsilon_F= 37$ MeV  and $C(\rho_0) = 31.3$ MeV, Eq.(\ref{eq:asym}) gives the standard value of the symmetry coefficient in the mass formula $a_{sym}= 28 $ MeV.
From $a_{sym}$ provided by the Bethe-Weisz\"acker formula we notice that the kinetic contribution represents less than one half, the remaining part being a consequence of the properties of the nucleon-nucleon interaction.

At the same time, we can calculate the mean-field potentials for protons and neutrons as 
functional derivatives of $\cal E\rm_{pot}$ with respect to the proton (neutron) density :
 \begin{eqnarray}
 U_p &=& \frac{\delta\cal E\rm_{pot}}{\delta\rho_p} = A \frac{\rho}{\rho_0} + B \left(\frac{\rho}{\rho_0} \right)^{\sigma} - C(\rho) \frac{\rho_i}{\rho_0}  + \frac{1}{2} \frac{\rm dC(\rho)}{\rm d\rho} \frac{\rho_i^2}{\rho_0} \\
 U_n &=& \frac{\delta\cal E\rm_{pot}}{\delta\rho_n} = A \frac{\rho}{\rho_0} + B \left(\frac{\rho}{\rho_0} \right)^{\sigma} + C(\rho) \frac{\rho_i}{\rho_0}  + \frac{1}{2} \frac{\rm dC(\rho)}{\rm d\rho} \frac{\rho_i^2}{\rho_0} 
 \label{eq:U}
 \end{eqnarray}

The mean fields are one of the main ingredients of the transport theory widely used to interpret HIC.
 The transport theory describes the temporal
evolution of the one-body phase-space distribution function $f($\rm {\bf  r}$, $\rm {\bf p}$, t)$ under the
action of a mean field potential $U($\rm {\bf r}$, $\rm {\bf p}$)$, possibly momentum dependent, and 2-body
collisions with the in-medium cross section $\sigma(\Omega)$. In a non-relativistic approach it reads
\begin{eqnarray}
\frac{df_i}{dt} &=& \frac{\partial f_i}{\partial t} + \frac{\bf p_i}{m} \nabla^{(r)} f_i -
\nabla^{(r)} U_i ({\bf r}, {\bf p}) \nabla^{(p)} f_i  - \nabla^{(p)} U_i ({\bf r}, {\bf p}) \nabla^{(r)} f_i  = I \\
I & = & \sum_{j,i',j'}\int d{\bf p_j} d {\bf p_{i'}} d{\bf p_{j'}} v_{ij}  \sigma_{{i,j}\rightarrow{i'j'}} (\Omega) \, 
\delta({\bf p_i} +  {\bf p_j} - {\bf p_{i'}} - {\bf p_{j'}} ) \nonumber \\
& \times & [(1-f_i) (1-f_j) f_{i'} f_{j'} - f_i f_j (1-f_{i'}) (1-f_{j'}) ]
\label{eq:lv}
\end{eqnarray}
\noindent
In this expression, $f($\rm {\bf  r}$, $\rm {\bf p}$, t)$  can be viewed semi-classically as the probability of finding a
particle, at time t, with momentum $\bf p$ at position $\bf r$. The indexes (i, j, i',j') run over neutrons and protons, such that 
these are coupled equations via the collision term $I$ and indirectly via the potentials. The average (mean field) potential $U$ 
is computed self-consistently using  the distribution functions $f($\rm {\bf  r}$, $\rm {\bf p}$, t)$ that satisfy Eq.(\ref{eq:lv}).  
The collision integral $I$  governs the modifications of $f($\rm {\bf  r}$, $\rm {\bf p}$, t)$
by elastic and inelastic two body collisions caused by short-range residual
interactions \cite{ber_88,dan_00}. Therefore the motions of particles reflect a complex interplay between such
collisions and the density and momentum dependence of the mean fields. 
 If the production of other particles is considered, like mesons, these have their own transport equations coupled through the corresponding inelastic cross sections. We notice that mean fields and cross sections should be related through a
theory for the in-medium effective interaction, like e.g. Brueckner theory, but
this is not necessarily done in many applications. The isospin effects enter via the
differences in neutron and proton potentials and the isospin dependent cross sections,
but they are always small relative to the dominant isoscalar effects. Thus one often
resorts to differences or ratios of observables between isospin partners, in order to
eliminate as much as possible the uncertainties in the isoscalar part. 

Several transport models have been developed in the past years, and most of them can account reasonably well for many characteristic properties experimentally observed. We mention FMD \cite{fmd}, AMD \cite{amd_1,amd_2,amd_3}, 
CoMD\cite{comd},  ImQMD \cite{ImQMD}, QMD \cite{QMD}, BNV \cite{BNV}, SMF \cite{SMF}, BUU \cite{BUU} among others. 

\subsection{Constraints on the symmetry energy at sub-saturation density.}
\label{sub}

Since large variations in nuclear density can be attained momentarily in nuclear collisions, constraints
on the EoS can be obtained by comparing measurements to transport calculations
of such collisions. The symmetry energy has been recently probed at sub-saturation densities via
double ratios involving neutron and proton energy spectra \cite{fami} and isospin diffusion 
\cite{tsang_2004,tsang_2009,liu_2007}. These two
observables largely reflect the transport of nucleons under the combined influence of the mean fields and
the collisions induced by residual interactions.  Additional experimental observables used to constrain the  symmetry energy
at sub-saturation density are : transverse collective flow of light charged particles \cite{kohley_11}, 
ratio of fragments yields and isoscaling \cite{xu_00,tsang_01}, cluster formation at very low densities \cite{nato_10,wada_12}. They will be discussed in more details in the following subsections.

\subsubsection{Neutron/proton double ratio}
\label{npratio}
\begin{figure}[t]
\centering
\includegraphics[scale=0.5,clip]{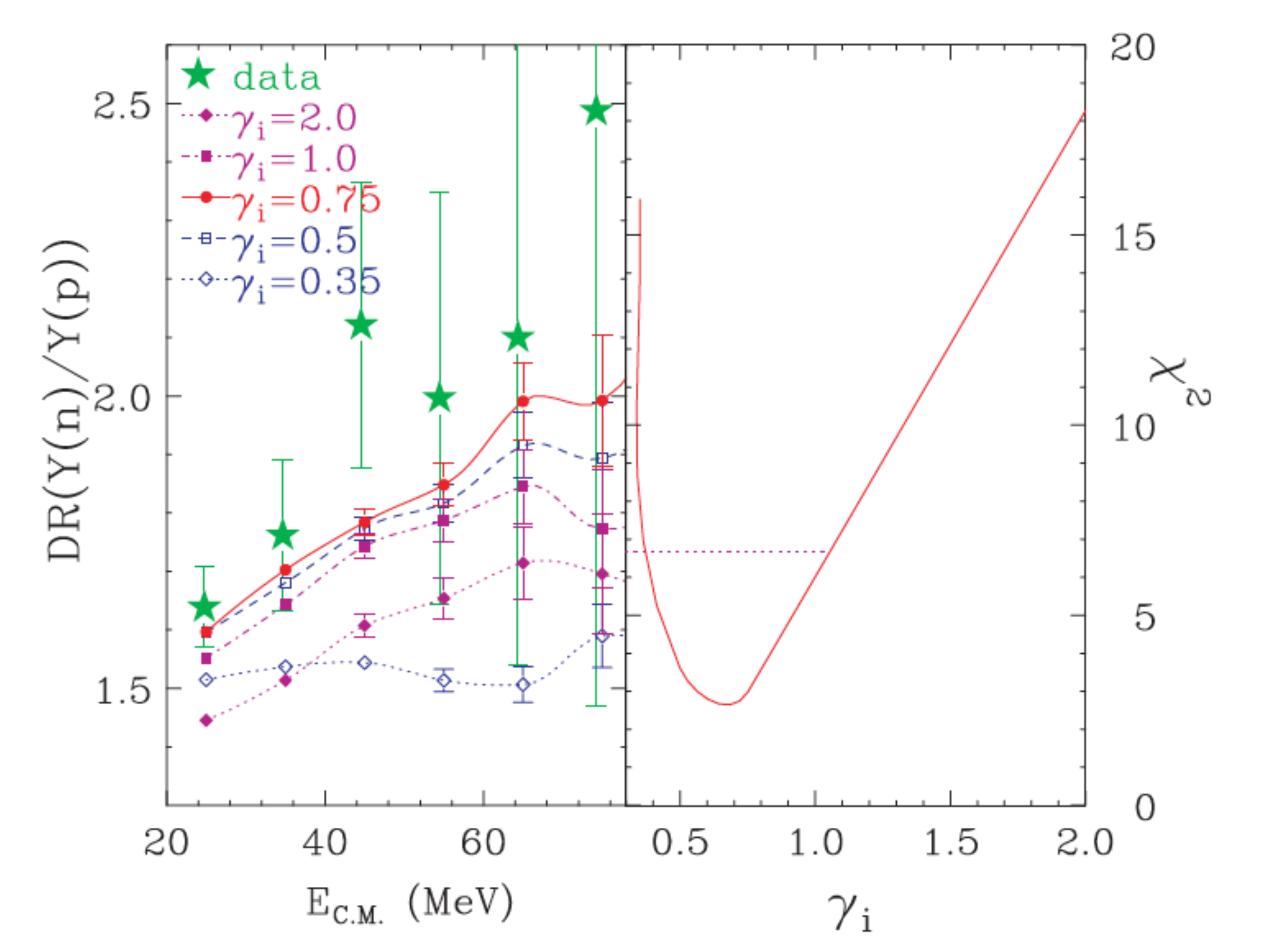}
\caption{(Color online)
Left panel: Comparison of experimental neutron-proton ratios (green stars) vs. the nucleon center-of-mass
 energy with calculations performed with the improved quantum molecular dynamic model (lines) for the impact
parameter $1 fm \leq b\leq 5 fm$
and several values of $\gamma_i$. The parameter $\gamma_i$ is the exponent in Eq. (\ref{eq:s_rho}), 
which describes the density dependence of the symmetry energy. Right panel: $\chi^2$ analysis 
 as a function of $\gamma_i$. Figure taken from ref. \cite{tsang_2009}. }
\label{fig:dr}
\end{figure}

We first concentrate on the interpretation of neutron/proton double ratio data, which derives
its sensitivity to the symmetry energy from the opposite sign of the symmetry force for neutrons as
compared to protons \cite{bali_97}. First experimental comparisons of neutron and proton spectra in
ref.\cite{fami} used a double ratio in order to reduce sensitivity to uncertainties in the neutron detection efficiency and to
relative uncertainties in energy calibrations of neutrons and protons. The double ratio is defined as
\begin{eqnarray}
DR (Y(n)/Y(p)) & = & R_{n/p}(A) / R_{n/p}(B)  \nonumber \\ 
& = & \frac{dM_n(A)/dE_{c.m.}} {dM_p(A)/dE_{c.m.}}  \cdot \frac{dM_p(B)/dE_{c.m.}} {dM_n(B)/dE_{c.m.}}
\end{eqnarray}
\noindent
and was constructed by measuring the energy spectra, $dM/dE_{c.m.}$, of neutrons and protons for two systems
A and B characterized by different isospin asymmetries. 
The green stars in Fig.\ref{fig:dr}, left panel, represent the measured double ratios at $\rm 70^o\leq\theta_{c.m.}\leq 110^o$  
as a function of center-of-mass (c.m.) energy of nucleons emitted from the central collisions of
$\rm ^{124}Sn +^{124}Sn$  and $\rm ^{112}Sn +^{112}Sn$ at E/A=50 MeV \cite{tsang_2009}.
That data set had an impact parameter over a range $1 fm \leq b\leq 5 fm$.
The lines show the results of the simulations performed within the improved quantum molecular dynamics (ImQMD) model,
using different values of $\gamma_i$. We remind the reader that the symmetry energy used in the numerical simulations is
usually parametrized as
\begin{equation}
S(\rho) = \frac{C_{s,k}}{2} \left( \frac{\rho}{\rho_0} \right)^{2/3} + \frac{C_{s,p}}{2} \left( \frac{\rho}{\rho_0} \right)^{\gamma_i}
\label{eq:s_rho}
\end{equation}
with $\rm C_{s,k}=25~MeV$, $\rm C_{s,p}=35.2~MeV$, and the symmetry energy at saturation $\rm S_0= 30.1~MeV$.
The exponent $\gamma_i$ indicates the stiffness of the potential symmetry energy, which is then termed as asy-soft or asy-stiff.
Several calculations are shown for different values of $\gamma_i$ in the range $0.35 \leq \gamma_i \leq 2$, but the optimal value 
turns out to be around $\gamma_i = 0.7$, as shown also in the right panel of Fig.\ref{fig:dr} where the $\chi^2$ analysis computed from
the difference between predicted and measured double ratios is shown. 

\begin{figure}[t]
\centering
\includegraphics[scale=0.5,clip]{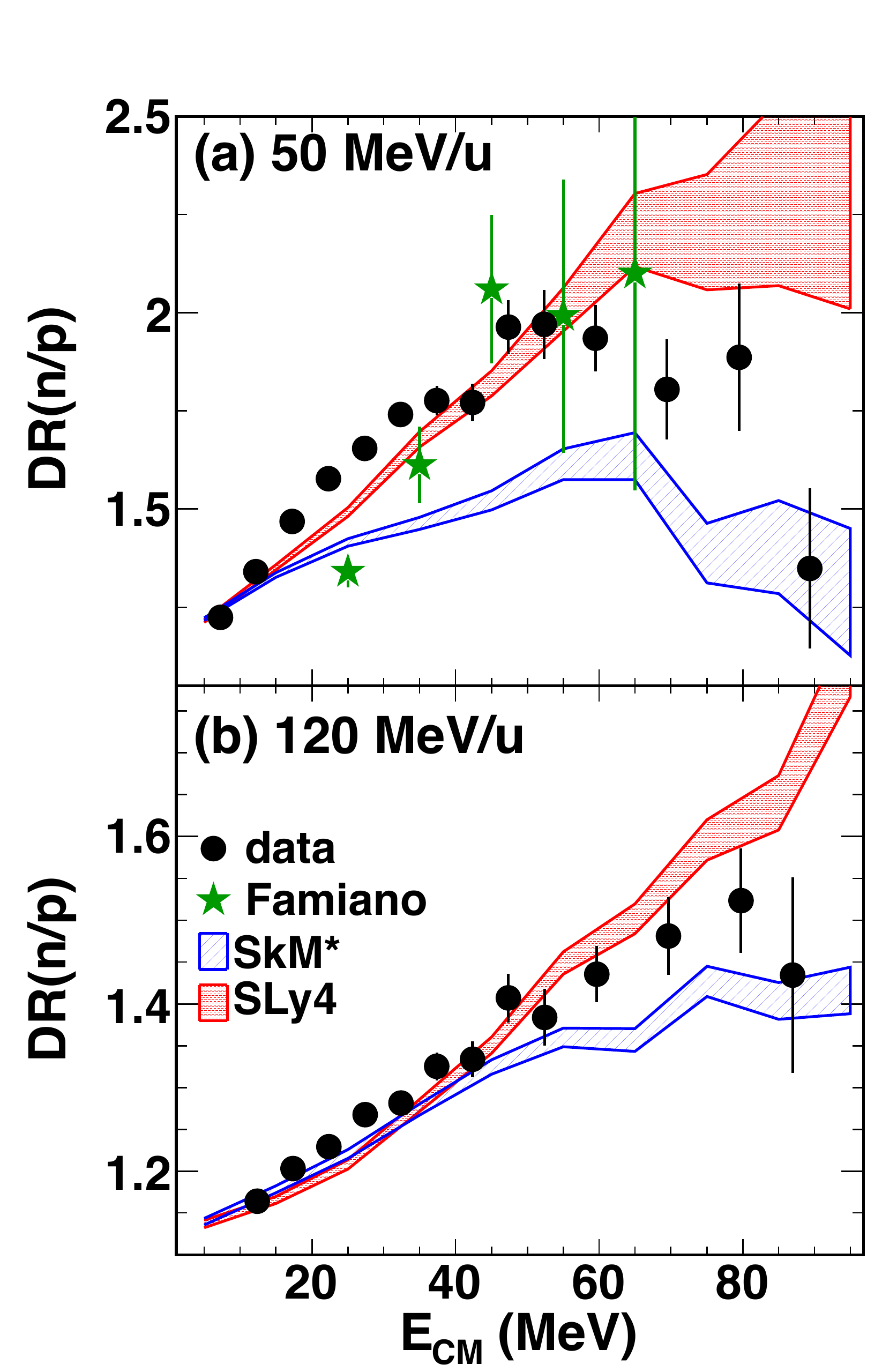}
\caption{(Color online)
Neutron to proton double ratio for $Sn+Sn$ collisions at 50 MeV/A (a) and 120 MeV/A (b).
Taken from ref.\cite{coup}.}
\label{fig:dr_new}
\end{figure}

Additional data have been recently published in ref.\cite{coup}, and they are  shown in Fig.\ref{fig:dr_new}.
The black dots in Fig.\ref{fig:dr_new} represent the recently measured double ratios at $\rm 70^o\leq\theta_{c.m.}\leq 110^o$  
as a function of center-of-mass (c.m.) energy of nucleons emitted from the central collisions of
$\rm ^{124}Sn +^{124}Sn$  and $\rm ^{112}Sn +^{112}Sn$ at E/A=50 MeV (upper panel) and E/A=120 MeV (lower panel).
Previous double ratio data from ref.\cite{tsang_2009} are displayed as green stars in the upper panel.
Considering both statistical and systematic uncertainties, the new data set is statistically more precise than the previous one,
and is consistent with ref.\cite{tsang_2009}, except at the lowest energy data point.  Detailed comparison of the neutron and proton data of the two experiments seems to indicate that the difference lies in the free neutron data, where hardware problems in the previous experiment required large systematic corrections to the neutron spectrum. The filled regions indicate the ranges predicted
by the ImQMD simulations performed with two different Skyrme forces, i.e. the $\rm SkM^*$ \cite{SkM*} and SLy4 \cite{SLy4} parametrizations.
The data at 120 MeV/u lie between the SLy4 and SkM* calculations, whereas the data at 50 MeV/u generally agree with the SLy4 calculations better than with the SkM* calculations. 

The adopted Skyrme forces are characterized by a different momentum dependence of the nucleonic mean-field potential,
and this can be approximated by replacing bare nucleon masses by effective masses. 
Differences between the neutron and proton effective masses strongly influence the isovector contributions contained in the 
symmetry mean-field potentials. Microscopic calculations performed in the non-relativistic Brueckner-Hartree-Fock approach 
\cite{zuo_1999,zuo_2002} have predicted that $m_n^* > m_p^*$  in neutron rich matter while relativistic
mean field (RMF) \cite{chen_07,baran_05} and other calculations using relativistic Dirac-Brueckner approach\cite{Dalen_2005,Rong_2006,Ma_2004,Sam_2005,Ulr_97} 
predict that $m_p^* > m_n^*$ . Large uncertainties do exist in the analyses of nucleon-nucleus elastic scattering,
which prefers $m_n^* > m_p^*$ \cite{bauge_2001}. Hence, the sign and magnitude of the effective mass splitting are not well 
constrained.

Central collisions of neutron-rich nuclei represent the ideal environment where to study the effective mass splitting.
In fact, transport model calculations predict that fast neutrons from the compressed participant region will experience a more 
repulsive potential and a higher acceleration for $m_n^* < m_p^*$ than do fast protons at the same momentum, resulting
in an enhanced ratio of neutron over proton (n/p) spectra at high energies. In contrast, calculations for
$m_n^* > m_p^*$  predict that the effective masses enhance the acceleration of protons relative to neutrons resulting in a
lower n/p spectral ratio \cite{zhang_2014}.  In the case reported in Fig.\ref{fig:dr_new} the data lie between the SLy4 and SkM*
calculations at high center-of-mass energy, but they are generally well reproduced by SLy4 EoS at energies below 50 MeV. 
We remind that SLy4 and SkM* Skyrme potentials have similar $S_0$, $L$ but opposite mass splitting at saturation density.
Therefore, this suggests that the sign of the effective mass splitting is the one proposed by the Sly4 EoS, i.e. $m_n^* < m_p^*$
but its magnitude can be even smaller than for the SLy4 mean field. Moreover, such a large reduction $m_p^* < m_n^*$  
as proposed in the SkM* can be ruled out. As shown in Fig.\ref{fig:dr_new}, however, these calculations do not accurately reproduce all aspects of the data, indicating the need for a throughout evaluation of the theoretical uncertainties in these tentative constraints.

This conclusion is in contradiction with the analyses reported in ref.\cite{bao_2013}, where the current data on 
$E_{sym}(\rho_0)$ and $L(\rho_0)$, as derived from various terrestrial nuclear laboratory experiments
and astrophysical observations, were used to put constraints on the effective mass splitting. However, in ref.\cite{bao_2013}
it was shown that while the mean values in most analyses are rather consistent and point toward $m_n^* >  m_p^*$ at saturation
density, it is currently not possible to scientifically state surely that $m_n^* >  m_p^*$ within the present knowledge of the uncertainties.
Another analysis \cite{kong_2015} of the same data using the IBUU11 transport model found that indeed the assumption of 
$m_n^* < m_p^*$ leads to a higher neutron/proton ratio although the underlying symmetry potential disagrees with the constraints
from optical model analyses of nucleon-nucleus scattering data. This situation clearly calls for more detailed theoretical studies of the free
neutron/proton ratio with different transport models and consider possibly new
mechanisms, such as effects of the short-range nucleon-nucleon correlations on the
symmetry energy \cite{bao_2015_1}. A complete discussion on this issue has been recently published in \cite{bao_2015_2}.

\subsubsection{Isospin diffusion}
\label{isdiff}

In a heavy-ion collision involving a projectile and a target with different proton fractions, Z/A, the symmetry energy
tends to propel the system toward isospin equilibrium so that the difference between neutron and proton densities is minimized
\cite{shi_2003}. The isospin asymmetry $\delta = \frac{N-Z}{A}$ of a projectile-like residue produced in a peripheral collision reflects the exchange of nucleons with the target; significant diffusion rates should lead to residues with larger isospin asymmetries for collisions
with neutron-rich targets and smaller isospin asymmetries for collisions with proton-rich targets.
To isolate the isospin diffusion effects from similar effects caused by pre-equilibrium emission, Coulomb, or sequential
decays, relative comparisons involving different targets are important. In recent studies, isospin diffusion has been measured
by ¡Ècomparing¡É $A+B$ collisions of a neutron-rich ($A$) nucleus and a proton-rich ($B$) nucleus to symmetric collisions
involving two neutron-rich nuclei ($A+A$) and two proton-rich ($B+B$) nuclei under the same experimental conditions \cite{tsang_2004}.
Non-isospin diffusion effects such as pre-equilibrium emission from a neutron-rich projectile should be approximately
the same for asymmetric $A+B$ collisions as for symmetric $A + A$ collisions. Similarly, non-isospin diffusion effects from
a proton-rich projectile in $B+A$ collisions and $B+B$ collisions should be the same.
The degree of isospin equilibration can be quantified by
rescaling the isospin asymmetry $\delta$  of a projectile-like residue from a specific collision according to the isospin transport ratio
$R_i(\delta)$ 
\begin{equation}
R_i(\delta) = 2 \frac{\delta-(\delta_{A+A} + \delta_{B+B})/2}{\delta_{A+A} - \delta_{B+B}}
\label{eq:isoratio1}
\end{equation}
\noindent
In the absence of isospin diffusion, the ratios are  $R_i(\delta_{A+B}) = R_i(\delta_{A+A})=1$ 
and $R_i(\delta_{B+A}) = R_i(\delta_{B+B})=-1$. If isospin equilibrium is achieved, then $R_i(\delta_{A+B}) = R_i(\delta_{B+A}) \approx 0$ 
for the mixed system. By focusing on the differences in isospin observables between
mixed and symmetric systems, $R_i (\delta)$ largely removes the sensitivity to pre-equilibrium emission and enhances the
sensitivity to isospin diffusion.

\begin{figure}[h]
\centering
\includegraphics[scale=0.6,clip]{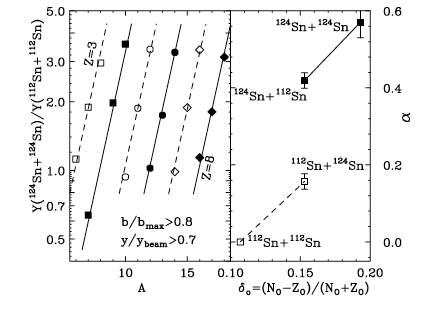}
\caption{Left panel: Measured values for the ratio $R_{21}(N,Z)=Y_{124+124}(N,Z)/Y_{112+112}(N,Z)$ 
(points) of the isotopic yields for the indicated reaction, and corresponding fits with Eq.(\ref{eq:r21}) 
(lines). The solid line and points represent even Z=4, 6, 8
isotopes while the dashed lines and open points represent odd
Z=3, 5, 7 isotopes. Right panel: Best fit values for the parameter $\alpha$ of Eq. (\ref{eq:r21}) as a
function of the isospin asymmetry $\delta_o$. The reactions are labeled next to the data points. Solid (open) points denote 
$^{124}Sn$ ($^{112}Sn$) as the projectile. The lines  guide the eye. Figure taken from ref.\cite{tsang_2004}. }
\label{fig:alpha_iso}
\end{figure}

\begin{figure}[h]
\centering
\includegraphics[scale=0.6,clip]{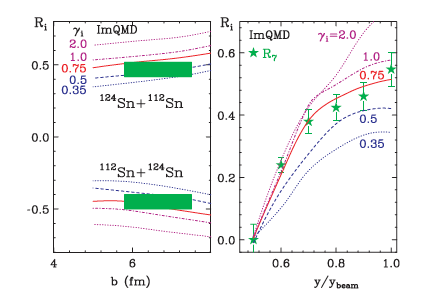}
\caption{(Color online)
 Left panel: Comparison of experimental isospin transport ratios (green boxes) 
to ImQMD results (lines), as a function of impact parameter for different values of $\gamma_i$.
Right panel: Comparison of experimental isospin transport ratios obtained from the yield ratios of
A=7 isotopes (green stars), as a function of the rapidity to ImQMD calculations (lines) at $b \rm =6$ fm.
Taken from ref.\cite{tsang_2009}.}
\label{fig:Ri}
\end{figure}

Ideally, one would like to know the asymmetry of the projectile-like residue immediately after the collision and
prior to secondary decay because this is the quantity that is calculated in transport theory \cite{tsang_2004}. To do this, one can
measure an observable $X$ that is linearly dependent on the residue asymmetry, i.e., $X = a \cdot \delta+b$, and construct the
corresponding isospin transport ratio $R_i (X)$
\begin{equation}
R_i(X) = 2 ~\frac{X-(X_{A+A} + X_{B+B})/2}{X_{A+A} - X_{B+B}}
\label{eq:isoratio2}
\end{equation}
\noindent
where $X$ is the isospin observable. After some algebra, one gets $R_i(\delta) = R_i(X)$.
As experimental observables, the authors of ref.\cite{tsang_2004} focus on features of the isotopic yields $Y_j(N, Z)$ of 
particles measured for reaction $"j"$  at $y/y_{beam} \geq 0.7$. Here, $N$ and $Z$ are the
neutron and proton numbers for the detected particles. They found that ratios of isotopic yields $R_{21}(N,Z) = Y_2(N,Z)/
Y_1(N, Z)$ for a specific pair of reactions, with a different total isotopic composition, follow the isoscaling relationship
\begin{equation}
R_{21}(N,Z)=C ~ {\rm exp}(\alpha N+ \beta Z)
\label{eq:r21}
\end{equation}
where $\alpha$, $\beta$ and $C$ are the isoscaling parameters, obtained by fitting the isotope yield ratios to Eq.(\ref{eq:r21}). 
The above idea has been adopted to study the isospin diffusion involving two
asymmetric collisions $\rm^{124}Sn + ^{112}Sn$ ($A+B$) and $\rm^{112}Sn+ ^{124}Sn$ ($B+A$), and two symmetric collisions, i.e.
$\rm^{124}Sn+^{124}Sn$  ($A + A$) and $\rm^{112}Sn + ^{112}Sn$ ($B+B$). 
Because there are no isospin differences between identical projectiles and targets, the symmetric 
collisions are used to establish diffusion-free baseline values for the measured and predicted
observables. The asymmetric collisions, on the other hand, have the large isospin differences needed to explore the isospin diffusion.
In Fig.\ref{fig:alpha_iso} the left panel displays the measured values for $R_{21}(N,Z)$ when using 
$\rm^{124}Sn + ^{124}Sn$ collisions as reaction 2 and $\rm^{112}Sn + ^{112}Sn$ collisions as reaction 1 in Eq.(\ref{eq:r21}). 
The fits to Eq.(\ref{eq:r21}) are represented by the solid and dashed lines. In the right panel of Fig.\ref{fig:alpha_iso}, 
we plot the best fit values for the isoscaling parameter $\alpha$ versus the overall isospin asymmetry of the
colliding system $\delta_o = \rm (N_o - Z_o)/(N_o + Z_o)$ where $\rm N_o$ and $\rm Z_o$ are the corresponding total neutron and proton
numbers. The solid and open points represent data for $\rm^{124}Sn$ and $\rm^{112}Sn$ projectiles, respectively. In general, the
isoscaling parameter increases with the overall isospin asymmetry $\delta_o$, while the overall trend indicates that isospin equilibrium
is not achieved in the asymmetric reaction systems. A complete discussion on this point is reported in ref.\cite{tsang_2004}.
In the same reference, using $\alpha$ for $X$ in eq.(\ref{eq:isoratio2}), $|R_i(\alpha)| \approx 0.5$ was found.
This result was reproduced by simulations performed with the BUU transport model with a stiff asymmetry term, but without
momentum dependence, thus requiring additional analysis.
In Fig.\ref{fig:Ri} ImQMD calculations of $R_i(\alpha)$ performed at impact parameters of b=5, 6, 7, and 8 fm are shown as lines,
whereas the experimental isospin transport ratios are displayed as shaded green boxes. We remind the reader that the interaction component of the asymmetry term provides a contribution to the symmetry energy per nucleon of the form 
$E_{sym,int}/A = C_{sym} (\rho/\rho_0)^{\gamma_i}$, where $C_{sym}$ is set to 12.125 MeV.
The experimental boxes put serious constraints on the exponent $\gamma_i$, which however depend on the choice of the symmetry energy at saturation density, and on  L.  

In the right panel of Fig.\ref{fig:Ri}  it is shown that experimental isospin transport ratios obtained from isoscaling parameters, $\alpha$, 
and from yield ratios of $A=7$ mirror nuclei, $\rm R_7= R_i(X_7 = ln(Y(^7Li)/Y(^7Be)))$ are consistent, i.e.
$R_i(\alpha)  \approx R_7$, reflecting linear relationships between $\alpha$, $X_7$, and the asymmetry $\delta$ of the 
emitting source. Experimental isospin diffusion transport ratios, $R_7$, are plotted as green stars as a function of rapidity,
and a $\chi^2$ analysis for the impact parameters favors the region $0.45 \leq \gamma_i \leq 0.95$ at $b=$ 6 and 7 fm,
thus favoring a moderately soft behavior of the symmetry energy.
\begin{figure}
\centering
\includegraphics[scale=0.4,clip]{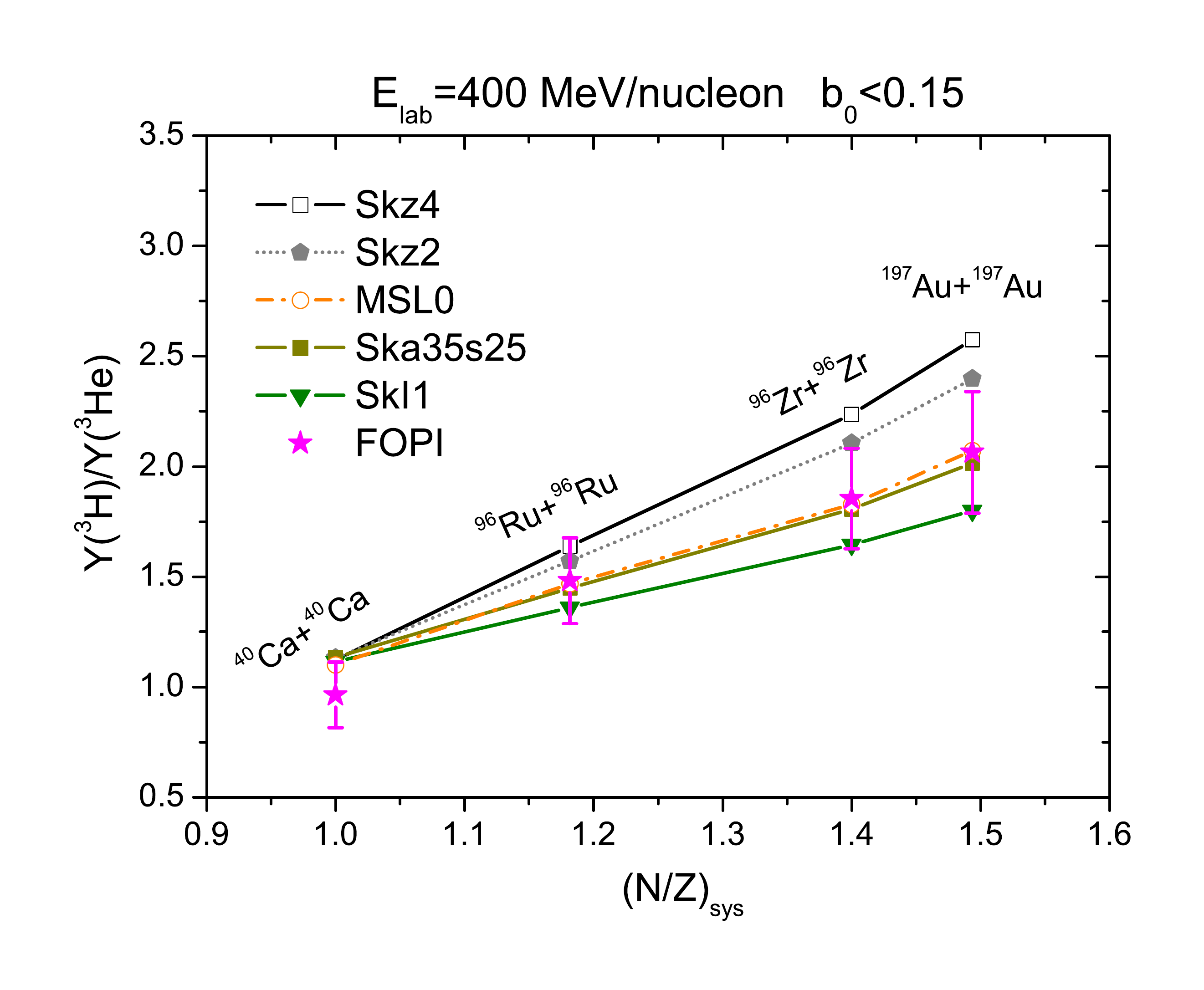}
\caption{(Color online)
$\rm^3H/^3He$ ratio from central ($b_0 < 0.15$) collisions at $E_{lab}$ = 400 MeV/nucleon as a function of neutron/
proton ratio of the colliding system. The FOPI data are indicated as stars. Figure taken from ref. \cite{wang_15}.}
\label{fig:trizi}
\end{figure}
\begin{figure}[t]
\centering
\includegraphics[scale=0.4,clip]{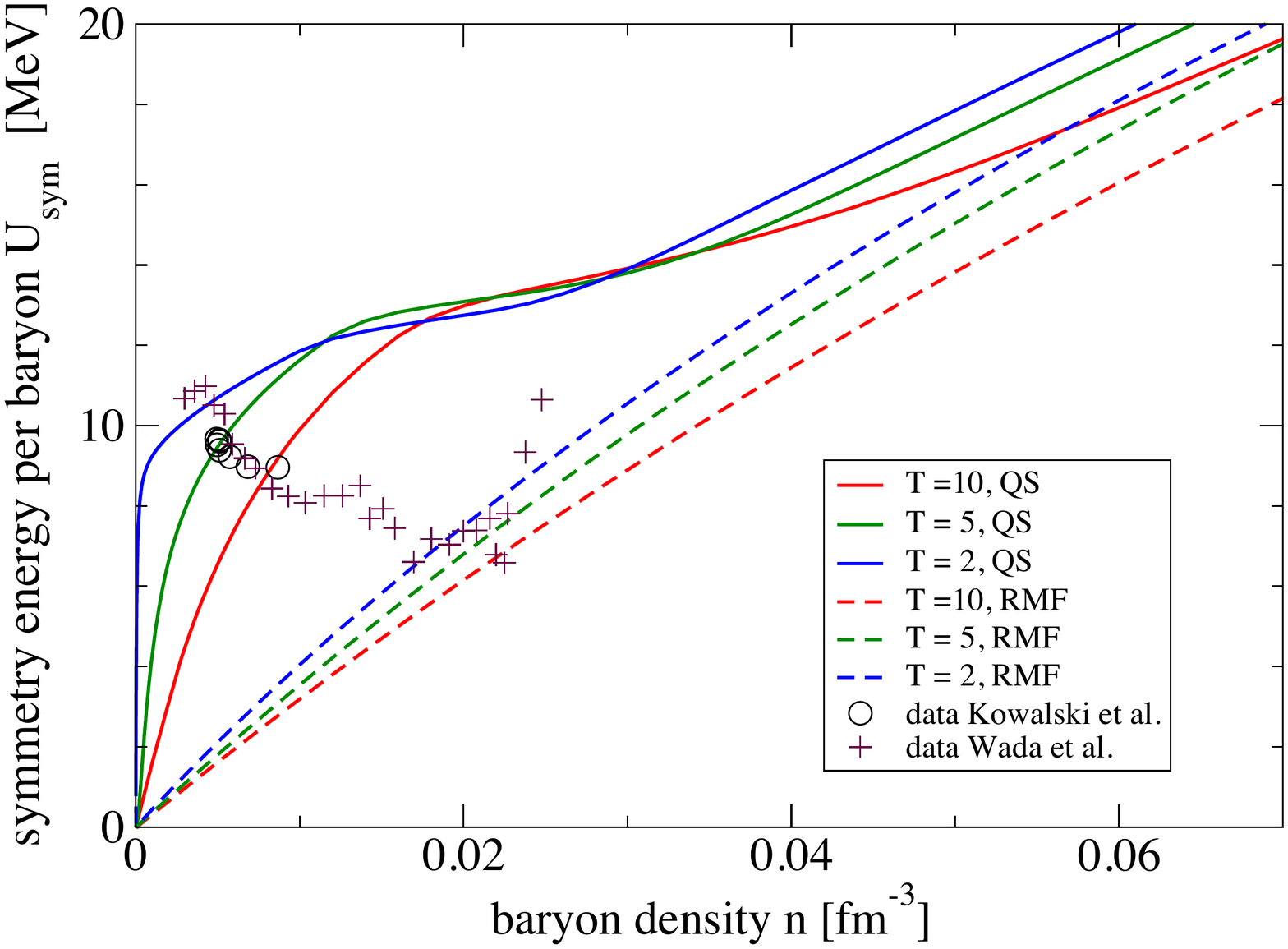}
\caption{(Color online)
Internal symmetry energy coefficients as a function
of baryon density. Experimental results \cite{kowa_07,wada_12} are compared to results of QS model calculations.
Taken from ref.\cite{hagel_14}.}
\label{fig:esymt}
\end{figure}
\par
This result has been confirmed \cite{wang_15} by the simulations obtained 
within the newly updated version of the Ultra-relativistic quantum molecular dynamics
(UrQMD) model \cite{urqmd} where the yield ratio between $\rm^{3}H$ and $\rm^{3}He$ clusters emitted from central 
$\rm^{40}Ca+^{40}Ca$, $\rm^{96}Zr+^{96}Zr$, $\rm^{96}Ru+^{96}Ru$, and
$\rm^{197}Au+^{197}Au$ collisions in the beam energy range from 0.12 to 1 GeV/nucleon is studied, and
compared with the recent FOPI data \cite{fopi_1,fopi_2}. In the UrQMD calculations 13 different Skyrme
interactions have been used,  all characterized by similar values of isoscalar incompressibility but very different density
dependences of the symmetry energy. It is found that the $\rm^3H/^3He$ ratio is sensitive to the nuclear
symmetry energy at sub-saturation densities. Model calculations with moderately soft to linear
symmetry energies are in agreement with the experimental FOPI data, thus confirming the findings 
discussed above. The comparison between simulations and FOPI data is shown in Fig.\ref{fig:trizi}, where 
the ratio $\rm^3H/^3He$ is shown for several colliding systems and different Skyrme forces adopted in the simulations.
The calculated results are well separated due to an increasingly stronger effect of the symmetry energy
at sub-normal densities. We notice that  calculations using both MSL0 and Ska35s25, which represent a moderately
soft to linear symmetry energy, reproduce the data fairly well. Although a desirable tighter constraint to the density
dependence of the symmetry energy is still not achieved here, partly due to the large experimental uncertainties,
a very satisfactory consistency among the presented comparisons is achieved.
This result is also consistent with previous results based on the elliptic flow of free nucleons
(and hydrogen isotopes) as a probe \cite{wang_14,russ_11,cozma_13}  which mainly provide information on the symmetry energy at
supra-normal densities. However, we remind the reader that there are still some puzzling inconsistencies among
different simulation codes.  For example, in refs.\cite{li_05,zhang_05} two QMD-type model calculations showed that the yield of 
$\rm^3H$ calculated with a soft symmetry energy is larger than that with a stiff one, while in refs.\cite{chen_03_1,chen_03_2}
the isospin-dependent BUU (IBUU) model calculations showed the opposite trend. In ref.\cite{chen_04}, 
using the Gogny effective interaction in the IBUU model, it was found that both, $\rm^3H$ and $\rm^3He$ yields, 
did not exhibit significant differences between the results for a soft and a stiff symmetry energy. Given that, 
the sensitivity of the nuclear symmetry energy to the $\rm^3H/^3He$ ratio is a subject which needs to be
further explored. \par
In any case, the value of the exponent $\gamma_i$ is not well constrained. In ref.\cite{chim1} a new method to fix the timescale
of intermediate mass fragments emission allowed to detect a process of  isospin "migration" towards the low density region.
The result of the data analysis favors a value of $\gamma_i \approx 1$. On the other side, the yields of the mirror nuclei 
$^7$Li and $^7$Be in collisions between Sn isotopes, with different neutron excess, was used \cite{chim2} to study isospin diffusion.
The result favors a value $\gamma_i \approx 0.5$.

\subsubsection{Symmetry energy at subsaturation density and finite T}
\label{ST}

The symmetry energy of nuclear matter at finite temperature is a fundamental ingredient in the investigation 
of astrophysical phenomena like supernovae explosions.
As shown in ref.\cite{sumy_08}, the calculated density, electron fraction and temperature profiles of a post core 
bounce collapse supernova display values accessible in near Fermi energy heavy ion collisions \cite{kowa_07}.
The role of the cluster formation in the neutrino-sphere region (the region of last neutrino interaction) is of particular
interest \cite{Oco_10,Lentz_12}, and is accessible in collisions of heavy ions at intermediate energies.

The experimental information is derived from heavy-ion collisions of charge asymmetric nuclei, where transient
states of different density can be reached, depending on the incident energy and the centrality of the collision.
The experimental investigations of low-density nuclear matter have utilized near Fermi energy heavy-ion collisions
of $\rm^{64}Zn$ on $\rm^{92}Mo$ and $\rm^{197}Au$ at 35 MeV per nucleon,
to produce heated and expanded matter \cite{kowa_07,wang_07,qin_12,nato_10,wada_12}.
Cluster production was studied using the 4$\pi$ multi-detector, NIMROD, at the Cyclotron Institute at Texas
A\&M University.  Yields of light particles produced in the collisions of 47A MeV
$\rm^{40}Ar$ with $\rm^{112}Sn$, $\rm^{124}Sn$ and $\rm^{64}Zn$ with $\rm^{112}Sn$, $\rm^{124}Sn$ were
employed in thermal coalescence model analyses to derive densities and temperatures of the evolving emitting
systems.
Experimental data were used to extract the symmetry free energy and the symmetry internal energy at subsaturation
densities and temperatures below 10 MeV.  An extensive discussion on the methods to determine
the temperature and density regions actually sampled in the collision is reported in ref.\cite{hagel_14}.
In Fig.\ref{fig:esymt} the internal symmetry energy derived from the experimental data \cite{kowa_07,wada_12,nato_10}
in an expanded low-density region is compared with the predictions of the relativistic mean field
(RMF without clusters)  and a quantum statistical (QS) approach  \cite{roep_82,roep_84,roep_90}. 
The temperature T varies in the interval 3-11 MeV. 
It is clearly seen that the quasiparticle mean-field approach disagrees strongly with the experimentally
deduced symmetry energy, while the QS approach gives a rather good agreement with the experimental data,
We remind the reader that the QS approach takes the formation of clusters into account, at variance with the mean-field 
approach, and this leads to a significant increase of the symmetry energy in the low-density region in contrast to the linear increase 
with density predicted by many mean-field motivated models. In the low-density region, the symmetry
energy turns out to be strongly depending on temperature.

We stress that the symmetry free energies of these systems were determined through isoscaling parameters 
deduced from isotopic yields measured in two similar reactions with different isotopic composition.
Recently, it has been shown that symmetry free energy coefficients can be extracted from
fragment yield data produced in Fermi-energy heavy-ion collisions by employing quantum fluctuation analysis technique
based on the Landau's free energy approach \cite{mabi_15}.
The temperature- and density-dependent symmetry free energies turn out to be consistent with those
derived from isoscaling analyses. 

\subsection{Constraints on the symmetry energy at high density}
\label{high}

The high-density behavior of nuclear symmetry energy is among the most uncertain properties of dense neutron-rich matter. 
Its accurate determination has significant consequences in understanding not only the reaction dynamics of heavy-ion 
reactions but also many interesting phenomena in astrophysics, such as the explosion mechanism of supernova
and the properties of neutron stars.  

A big experimental effort has been devoted during the last few years to constrain the high-density symmetry energy
using various probes in heavy-ion collisions at relativistic energies.
It still remains a challenge, however, to find observables suitable for extracting 
information on the properties of the equation of state during the brief compression phase.
During the past decades an intense field of research has rapidly concentrated
on collective flows and meson production. 
Studies of flow and kaon production within the framework of transport theory
have indeed demonstrated that a soft EoS with
compressibility $K \approx 230$~MeV and momentum dependent interactions best describes the
response of symmetric nuclear matter to compression (see section above).

In the density regime exceeding saturation, the symmetry energy is still largely unknown.
In fact phenomenological forces are well constrained near or just below
saturation but lead to largely diverging results if they are extrapolated to higher 
densities \cite{Tsang,ABrown}. Microscopic many-body calculations with realistic potentials face the difficulty
that three-body forces and short-range correlations are not sufficiently well known
at higher densities at which their importance increases \cite{RepPP}. 
The need for experimental high-density probes is thus obvious, with the above-mentioned 
collective flows and sub-threshold particle production as the main candidates. 

A strong motivation for exploring the information contained in isotopic flows was provided by Bao-An Li \cite{bao_02},
who pointed to the parallels in the density-dependent isotopic compositions of neutron stars and of the transient systems 
formed in collisions of neutron-rich nuclei as a function of the EoS input used in the calculation.
A close comparison allows  to infer properties of these exotic astrophysical
objects from data obtained in laboratory experiments.  However, the main difficulty is the
comparatively small asymmetry of available nuclei. Symmetry effects are, therefore,
always small relative to the dominating isoscalar forces which one hopes will
cancel in differences or ratios of observables between isotopic partners. 
The observable proposed by Li is the so-called differential directed flow which is the difference
of the multiplicity-weighted directed flows of neutrons and protons, and 
describes the rapidity dependence of the mean in-plane transverse momenta of observed
reaction products. Besides that, also the elliptic flow has been studied with model 
calculations to test its usefulness as a probe of the stiffness of the symmetry energy~\cite{lipr08,lisustich01,baran05}. 
Elliptic flow relates to the azimuthal anisotropy of particle emissions, mainly differentiating
between predominantly in-plane emissions as recently observed in ultrarelativistic 
heavy-ion collisions~\cite{abelev07,aamodt10,adare12,aad12} and the out-of-plane emissions 
or squeeze-out observed in the present regime of lower energies as a
consequence of the pressure build-up in the collision zone~\cite{gutbrod90}.  

A data set to test these predictions has been available from earlier experiments of the 
FOPI/LAND Collaboration. It was originally collected and shown to provide evidence for the  
squeeze-out of neutrons emitted in $^{197}$Au + $^{197}$Au collisions at 400 MeV per
nucleon~\cite{leif93,lamb94}. The capability of the Large Area Neutron Detector 
LAND~\cite{LAND} used in these experiments
of detecting neutrons as well as charged particles permitted the differential analysis of the
observed flow patterns in the form of flow ratios~\cite{russotto11} or flow 
differences~\cite{cozma11}. Both analyses favor a density dependence between moderately soft 
and moderately stiff,  consistent with the experiments made at sub-saturation density.

Densities of two to three times the saturation density may be reached on time scales of $\approx$ 10-20 fm/c in 
the central zone of heavy-ion collisions at relativistic energies of up to $\approx$ 1 GeV/nucleon.
The resulting pressure produces, besides  a collective outward motion of the compressed material, 
the excitation of $\Delta$ resonances in hard nucleon-nucleon scatterings, which leads to the production and
subsequent emission of charged and neutral $\pi$ and K mesons. The relative intensities
of isovectors pairs of mesons depend directly on the proton-neutron content of the
site where they are produced, suggesting them as sensitive probes for the high density symmetry energy.

Whether meson production yields will become similarly useful for the same purpose is not so 
clear at present.  Measurements of $\rm K^+/ K^0$ production ratios have been performed by the FOPI
Collaboration for the neutron-rich and neutron-poor A = 96 systems $\rm^{96}Zr + ^{96}Zr$
and $\rm^{96}Ru + ^{96}Ru$ \cite{hi_96} at 1.53 GeV per nucleon.
A significant sensitivity of this observable to the chosen 
stiffness of the asymmetric EOS was expected from the calculations for infinite nuclear matter. It
diminished, however, by one order of magnitude when the calculations were performed for 
the actual heavy-ion collisions studied in the experiment~\cite{xlopez07}. The measured 
double ratio is satisfactorily reproduced irrespective of the choice made for the asymmetric EoS.
An even more puzzling situation is  
encountered in the case of the $\pi^-/\pi^+$ yield ratios measured by the FOPI
Collaboration at several energies up to 1.5 GeV per nucleon for  $^{40}$Ca + $^{40}$Ca, $^{96}$Zr + $^{96}$Zr,  $^{96}$Ru + $^{96}$Ru,
and $^{197}$Au + $^{197}$Au~\cite{reisdorf07} collisions.
Theoretical analyses of this data set came to
rather conflicting conclusions, suggesting everything from a rather stiff to a
super-soft behavior of the symmetry energy~\cite{xiao09,feng10,xie13}. 
The hard interpretation of the meson data emphasizes the urgent need to improve the statistical accuracy beyond 
that of the existing FOPI/LAND data set.  The dedicated ASY-EOS experiment \cite{russo_14}, aiming at measurements of collective flows in collisions of $^{197}$Au + $^{197}$Au as well 
as of the $^{96}$Zr + $^{96}$Zr and $^{96}$Ru + $^{96}$Ru, was started in 2011 at the GSI laboratory. 
The LAND~\cite{LAND} detector operated together with a subset of the CHIMERA~\cite{chimera} detector array complemented with the KraTTA \cite{kratta} additional detector, aiding in the measurement of the reaction plane orientation and of the flow of 
light fragments. 

\begin{figure}[t]
\centering
\includegraphics[scale=0.4,clip]{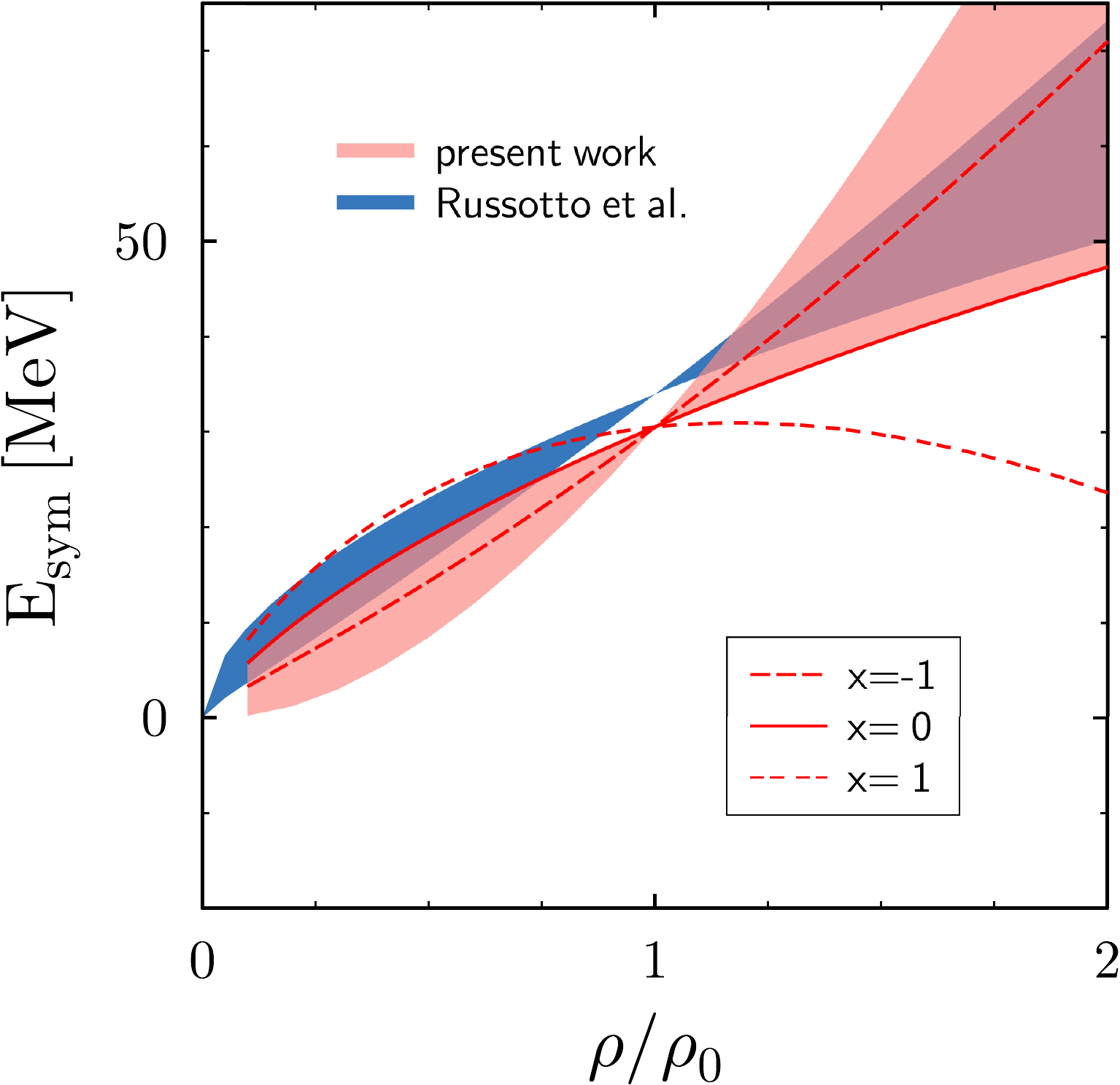}
\caption{(Color online)
Constraints on the density dependence of the symmetry energy obtained
from comparing theoretical predictions for npEFD and npEFR to FOPI-LAND experimental data (pink region).
The result of ref.\cite{russotto11} is also shown (blue band) together with the Gogny inspired 
symmetry energy  parametrization
for three values of the stiffness parameter: x=-1 (stiff), x=0 and x=1 (soft).}
\label{fig:esymhigh}
\end{figure}

Experimental data for the neutron-proton elliptic-flow difference (npEFD) and ratio (npEFR) 
have been obtained and the corresponding  model parameter dependence has recently been investigated in 
detail~\cite{cozma_13}. The effects of the selected microscopic nucleon-nucleon cross-sections, 
of the compressibility of nuclear matter, of the optical potential, and of the parameterization 
of the symmetry energy were thoroughly studied. For a detailed report see ref.\cite{russo_14}.
In  Fig.~\ref{fig:esymhigh}, the explicit constraints on the density dependence
of the symmetry energy obtained in this study from the comparison of theoretical and 
experimental values  of npEFD and npEFR are displayed (pink coloured band). 
For comparison, the result of Ref.~\cite{russotto11} is added (blue band). 
The two studies employ independent versions of the QMD transport model (T\"{u}bingen
QMD vs. UrQMD), characterized respectively by parameterizations of isovector EoS that differ : 
Gogny inspired (momentum dependent) vs. power-law (momentum independent).
The constraints on the density dependence of 
the symmetry energy obtained with these different ingredients are in agreement with
each other. By combining the two estimates, a moderately stiff to linear density dependence 
corresponding to a symmetry energy parametrization with $x=-1.0 \pm 1.0$ as stiffness parameter, 
is obtained. It indicates a somewhat faster increase of the symmetry energy with density than what is extracted 
from nuclear structure and reactions for sub-saturation densities. 

\section{Microscopic theory.}
\label{theory}
 The phenomenological constraints on the symmetry energy and its features are based on fitting procedures on experimental data.
In semi-classical methods (LDM, DM, ETF) the symmetry energy and its slope are parameters that are extracted by fitting an extensive set of nuclear binding energy and other nuclear structure data, see Sec. \ref{semi}. Similarly, within the energy density functional method the parameters of the effective force or of the functional are extracted by fitting mainly the binding energy throughout the mass table, see Sec. \ref{EDF}. Once the parameters are fixed, one can calculate the whole density dependence of the symmetry energy.
The same procedure is followed in the analysis of data on heavy ion collisions, see Sec. \ref{HI}, and on the astrophysical observational data, Sec. \ref{Astro}.
Comparing the results of different fits, with different forces or data, one can get an estimate of the uncertainty on the extracted values of the symmetry energy. One could try to reduce the uncertainty relying on the microscopic many-body theory of nuclear matter and finite nuclei. Unfortunately the theory is not so advanced to reduce the uncertainty in a conclusive way. On the other hand the comparison of the theoretical predictions with the phenomenological constraints can be of great value in establishing the microscopic structure of nuclear matter and nuclei. However the phenomenological constraints could be biased by the limited flexibility of the considered models or forces, and therefore it is important to keep the comparison only to the constraints that can be considered less model dependent as much as it is possible. After the selection of the microscopic theories, the symmetry energy, with its density dependence, is well defined for each theoretical model and it can be then extracted with some uncertainty.
The aim of this analysis is to try to answer to the following questions
\par\noindent
1. How much the symmetry energy depends on the (realistic) NN force ?
\par\noindent
2. What is the role of three-body forces ?
\par\noindent
3. How relevant are the many-body correlations ?
\par\noindent
4. How large is the resulting theoretical uncertainty ?
\par                
Before going to the selection procedure, we will give a short overview of the theoretical many-body schemes included in the analysis, mainly for nuclear matter.
\subsection{Theoretical model overview.}
The goal of a theoretical many-body scheme is the calculation, as accurately as possible, of the ground state of (asymmetric) nuclear matter starting from a realistic nuclear interaction. A realistic two-body force is assumed to be able to reproduce the experimental phase shifts and deuteron data. Such a realistic force must contain a large enough set of terms for the possible two-body operators, like tensor, spin-orbit, angular momentum square and so on, with the proper isospin dependence. Several two-body forces have been developed along the years, with different operatorial complexity and different theoretical background. 
One can distinguish mainly three categories of NN interaction. One is based on the meson exchange model for the NN interaction, which suggests, in a direct or indirect way, the form factors of each operatorial term. The family of chiral interactions is based on the expansion of the interaction, besides the explicit one pion exchange term, in a series of point-like interactions whose form is suggested by the chiral symmetry of the underling QCD theory. The third class of interaction is quite restricted and includes the models for the NN interaction which use explicitly the quark structure of the nucleons. Some of these interactions, covering all three categories, will be mentioned in the sequel in connection with the nuclear matter EOS and the corresponding symmetry energy.         \par
Besides the two-body forces, it has been established, at least in a non-relativistic approach, that three-body forces are necessary
to reach agreement with basic phenomenological data like the saturation point of symmetric nuclear matter. Later we will briefly discuss
some of the three-body forces that have been introduced in the theory of nuclear matter EOS.    
\par 
The microscopic many-body methods that will be considered are the Bethe-Brueckner-Goldstone (BBG) expansion, non-relativistic or relativistic, the variational method and the effective interaction approach.  \par
In the BBG method the ground state energy is expanded in a series of diagrams, where the original interaction is replaced by the so-called G-matrix, obtained by the explicit summation of the ladder diagrams. In this way the effect of the strong short range repulsive core of the NN interaction is embodied in the G-matrix, which is expected to have a much softer behavior at short distance or large momentum. In this way the possible convergence of the expansion is facilitated. The diagrams in terms of the G-matrix are then ordered according to the number of hole-line that they contain ("hole expansion"), which should be in agreement with the size of their contribution. The diagrams containing n hole-lines should take into account the correlations of order n, i.e. the contribution of the interaction processes that involve n particles in an irreducible way.  
Furthermore a scheme is adopted to introduce a single particle potential which should include part of the correlations and is calculated following the Brueckner self-consistent procedure.  A pedagogical introduction of the BBG approach can be found in ref. \cite{book}. A variant of the approach is the Coupled Cluster approach, where the ordering of the diagrams is slightly different \cite{Martino}. The relativistic extension of the method is the so-called Dirac-Brueckner-Hartree-Fock (DBHF) approach \cite{Machlei1989}. A fully relativistic many-body theory for nuclear matter is quite challenging, and in DBHF usually the relativistic four dimensional covariance is approximated by a proper three dimensional reduction, but still keeping the spinor structure of the nucleon field. The variational approach is developed in coordinate space. The correlated wave function is obtained by multiplying
the uncorrelated (free particles) wave function by a set of correlation functions, which introduce two-body correlations, and eventually correlations of higher order (three-body and so on). The correlation functions are obtained minimizing the energy, and
solving the corresponding Euler-Lagrange equations. To simplify the procedure one can parametrize the correlation functions and extract the parameters by minimization. 
For nuclear matter the complex structure of the NN interaction makes the procedure more involved, because the correlation functions
are then actually operators with the same structure as the NN interaction \cite{apr}. A related approach is based on the effective interaction scheme, where the correlation functions are used to substitute the original bare interaction by an effective interaction. The advantage of such an approach is the possibility of using the same effective interaction for the calculations of other physical quantities, in particular of astrophysical interest, and to get an unified description of e.g. NS physical properties \cite{OM1,OM2}. The effective interaction so constructed has some resemblance with the G-matrix of the BBG expansion.  \par 
Another approach that has been developed is mainly connected with the chiral interactions. This is the Renormalization Group method (RG).
In this case an effective interaction in the medium is constructed by the RG procedure, where the momentum cut-off is lowered until the interaction is weak enough to be treated perturbatively \cite{RG}. \par 
A brief survey of all these microscopic many-body methods can be found in ref. \cite{RepPP}.\par    
\subsection{Selecting the EOS.}
\label{EOSsec}
A microscopic many-body approach is expected to be able to calculate within the same framework the whole EOS both for symmetric and asymmetric nuclear matter, in particular pure neutron matter. The symmetry energy and its density dependence can be therefore
obtained directly from the calculations. The EOS should be compatible with all the constraints coming from phenomenology. Some of these constraints are well established and, within the uncertainty, model independent. The first basic requirement is the compatibility with the phenomenological saturation point. We will consider the density and energy $ (\rho_0,e_0) $ of the nuclear matter saturation point constrained according to
\beq
0.14\,\, {\rm fm}^{-3}\, \leq \,\, \rho_0 \, \leq \, 0.18\,\, {\rm fm}^{-3} \ \ \ \ \ \ \ \ ; \ \ \ \ \ \ \ \ \
- 17.\,\,\, {\rm MeV}\, \leq \,\, e_0 \, \leq \, - 15.\,\,\, {\rm MeV}   
\label{sat}\eeq     
Most of the mass formula, based on phenomenological (LDM, DM) or more microscopic (ETF, EDF) approaches, correspond to saturation points well inside these boundaries. Another quantity that can be considered relatively  constrained, within the uncertainty, is the incompressibility $ K_0 $, mainly from the monopole excitation, where the relevance of the effective mass is minimal. We will assume
\beq
 200\,\, {\rm MeV} \,\, \leq \,\, K_0 \,\, \leq \,\, 260 \,\, {\rm MeV}
\label{compress}\eeq 
\noindent All these boundaries are in line with the ones adopted in ref. \cite{Jirina}. Although these constraints look not so stringent, they turn out to be quite selective, and in fact they rule out several microscopic many-body approaches. In particular we extended the incompressibility constraint down to $200$ MeV to avoid a too restricted set of EOS, although the preferred value of $ K_0 $ is usually larger \cite{compr_Stone}. Notice however that the value of $ K_0 $ extracted from the Isoscalar Giant Moopole Resonance is quite model dependent and it can be affected by other features of the
effective force, like the symmetry energy \cite{compr1,compr2}, and the connection between the incompressibility in nuclei and nuclear matter is not so straightforward \cite{compr_Stone}.      
We finally arrive to select a few EOS from different methods. We list them here, together with some of their features.
\par\noindent
1. EOS derived from Brueckner-Hartree-Fock (BHF) calculations reported in refs. \cite{Tara2013,BCPM}. Two-body interaction $ Av_{18} $ \cite{v18}, three-body forces (TBF) of the Urbana model \cite{UIX,UIX_1,UIX_2}.
\par\noindent
2. EOS from BHF, reported in \cite{Li1,Li2}. Two-body interaction Bonn B \cite{Bonn}, TBF derived from two-body.  
\par\noindent
3. EOS from variational method, reported in ref. \cite{apr}. Two-body interaction $ Av_{18} $, Urbana model TBF.
\par\noindent
4. EOS from the effective interaction approach \cite{OM1,OM2}.
\par\noindent
5. EOS from Dirac-Brueckner, reported in ref. \cite{Dalen}, two-body interaction Bonn A \cite{Machlei1989}.
\par\noindent
6. EOS from renormalized chiral forces, reported in ref. \cite{Drisch}.  
\par\noindent
7. EOS from BBG up to three hole-line contributions, reported in refs. \cite{Kenji1,Ken}, two-body interaction from meson-quark model.
\par
We summarize in Table 1 the properies at saturation for each one of these EOS.
\par 
\begin{table*}[h]
\begin{center}
\begin{tabular}{|l|c|c|c|c|c|}
\hline
 EoS &  $\rho_0 \rm (fm^{-3})$ & $e_0$ (MeV)& $K_0$ (MeV)& $S_0 $ (MeV) & $L$ (MeV)\\
\hline\hline
 BHF, \, Av$_{18}$ + Urbana~TBF  & 0.16 & -15.98 & 212.4 & 31.9  & 52.9  \\
\hline
BHF, \, Bonn~ B + TBF from NN  & 0.17 & -16. & 254. & 30.3 & 59.2  \\
\hline
 Variational,  \, Av$_{18}$ + Urbana~TBF & 0.16 & -16. & 247.3 & 33.9 & 53.8  \\
\hline
Effective interaction & 0.16 & -16.0 & 234.0 & 37.25 & 65.8  \\
\hline
 DBHF,  \, Bonn~ A  & 0.18 & -16.15 & 230. & 34.4 & 69.4  \\
\hline
 Chiral model  \,     &  --   &  --    &  --   & 30.08$\pm$0.8   &  --   \\
\hline
 Quark model    \,    & 0.157 & -16.3  & 219.  & 31.8 & 52.0   \\ 
\hline
\end{tabular}
\end{center}
\caption{Calculated properties of symmetric nuclear matter of different EOS.}
\label{t:sat}
\end{table*}  
\noindent     
The EOS of refs. \cite{Li1,Li2} (second line) is calculated in the BHF scheme but with a three-body force derived consistently from the two-body one by considering a set of processes involving three nucleons and meson exchange with the same coupling constants. This is a quite challenging procedure, since in principle no additional parameter is introduced for the construction of the TBF. The data for the effective interaction approach were extracted from Fig. 2 of ref. \cite{OM1}, where the reported EOS has the correct saturation point. They were obtained by fitting the EOS around saturation.  
The EOS in the DBHF scheme was obtained without the contribution of TBF. The relativistic formulation of the Dirac-Brueckner method gives a saturation point within the phenomenological boundaries with only two-body NN interaction (the Bonn A in our case). It has been argued \cite{DBHF_TBF} that the reason for this saturation effect is that the relativistic formulation implicitly introduces a particular TBF if re-formulated in the non-relativistic framework. The main origin of the TBF is the use of the Dirac spinors, that in the medium are "rotated" with respect to the free ones, which introduces an antiparticle component. 
\par   
The lack of necessity for a TBF holds true also for the quark-meson model for the NN interaction. In this case the reason is different, it stems on the non-local short range interaction that originates from the quark structure of the nucleons. In this case the Bethe-Brueckner-Goldstone expansion has been extended beyond the Brueckner approximation, i.e. up to the three hole-line level of approximation.\par
The symmetry energy form the chiral model was obtained by expanding the EOS from the pure neutron matter one in the proton fraction, and the full density dependence of the symmetry energy was obtained.
Since we are indeed interested on the symmetry energy, we choose this result, even if the symmetric matter EOS was not calculated in that paper (this is the reason of the lack of data for the other physical parameters in Table 1).    

\begin{figure}[htb]
\centering
\vskip -5 cm
\includegraphics[scale=0.5]{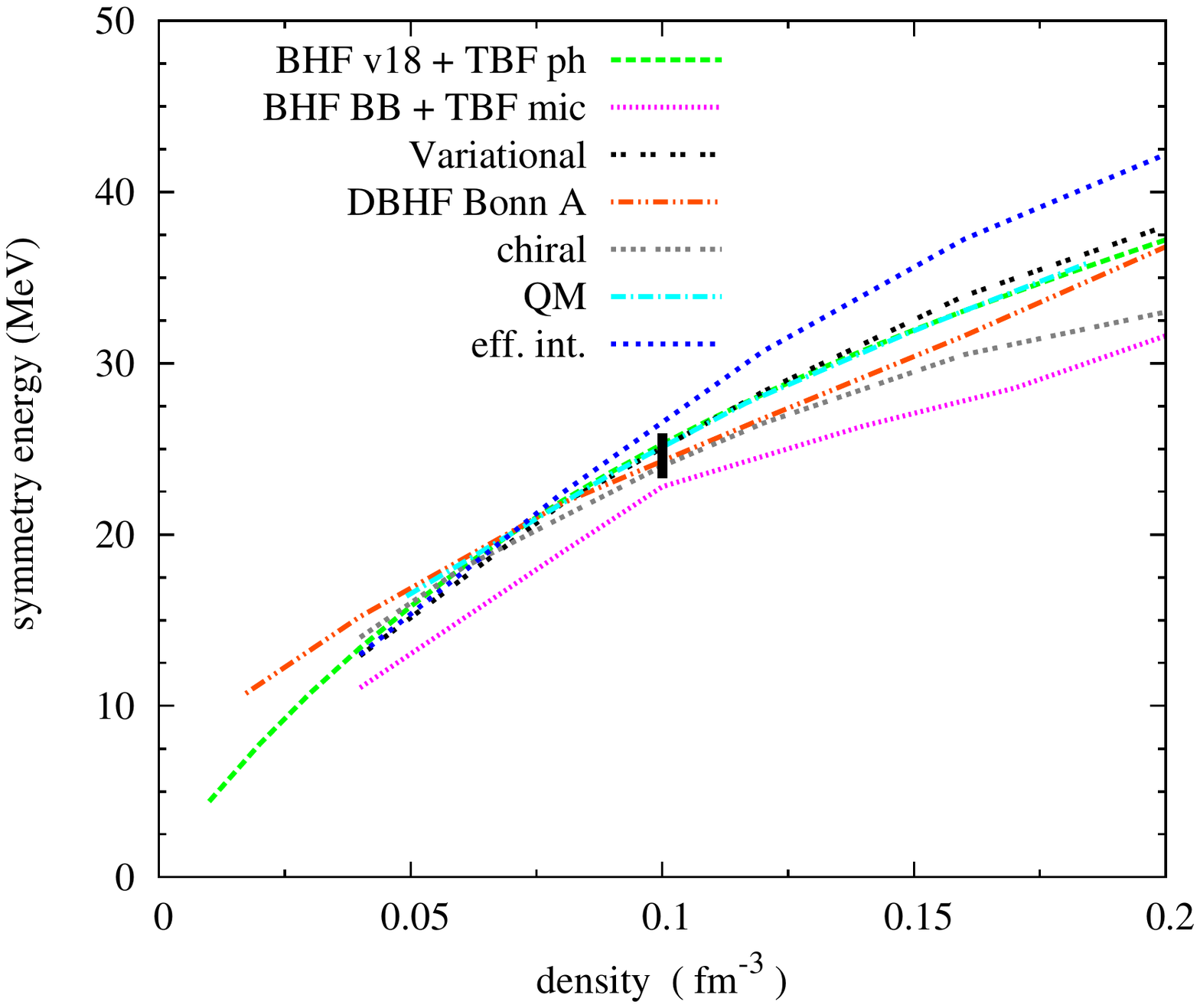}
\vskip -2.5 cm
\caption{(Color on line) The symmetry energy up to saturation for the selected EOS of Table \ref{t:sat}. The label BHF v18 + TBF indicates the Brueckner-Hartree-Fock calculation reported in ref. \cite{Tara2013,BCPM}, with the Argonne v$_{18}$
two-body interaction and the Urbana IX three-body force. BHF BB labels the Brueckner-Hartree-Fock calculation of ref. \cite{Li1,Li2} with the Bonn B two-body interaction and consistent ('microscopic') three-body force. 'Variational' labels the variational calculation of ref. \cite{apr}, with the Argonne v$_{18}$ two-body interaction and the Urbana IX three-body force. DBHF labels the Dirac-Brueckner calculation of ref. \cite{Dalen} with the Bonn A two-body interaction. 'chiral' indicates the calculation of ref. \cite{Drisch}, based on the renormalized chiral forces. QM indicated the calculation of ref. \cite{Kenji1}, based on the quark model for the two-body force. 'eff. int.' labels the calculation of ref. \cite{OM1,OM2}, based on the effective forces derived from the correlation function. 
The (black) vertical bar indicates the constraint of ref. \cite{Trippa}.} 
\label{fig:SEbelowa}
\end{figure}

\subsection{Symmetry energy up to saturation density.}
\label{symsub}
We report in Fig.\ref{fig:SEbelowa} the symmetry energy as a function of density for the selected EOS up to the saturation density.
At very low density the plots extend down to where the data are available. For the chiral results in ref. \cite{Drisch} also an estimate of the theoretical uncertainty is given. For simplicity of the presentation we have taken the average values inside the error range.   
As it is obvious, these values of the symmetry energy cannot be tuned in any way, and therefore the relative agreement among the different EOS below saturation cannot be considered trivial or expected, since they have been calculated within quite different theoretical schemes and different forces. It seems then that the restrictions we have used in selecting the EOS are physically meaningful, and that the different physical parameters that characterize the EOS are strongly inter-related in a generic microscopic approach. 
For comparison we report (vertical bar) the range of values constrained by the analysis on the GDR of ref. \cite{Trippa}. 
Around saturation there is some spread of values for the symmetry energy and its slope, as it can be seen also in Table \ref{t:sat}.
\begin{figure}[htb]
\centering
\vskip -5 cm
\includegraphics[angle=0,scale=0.5]{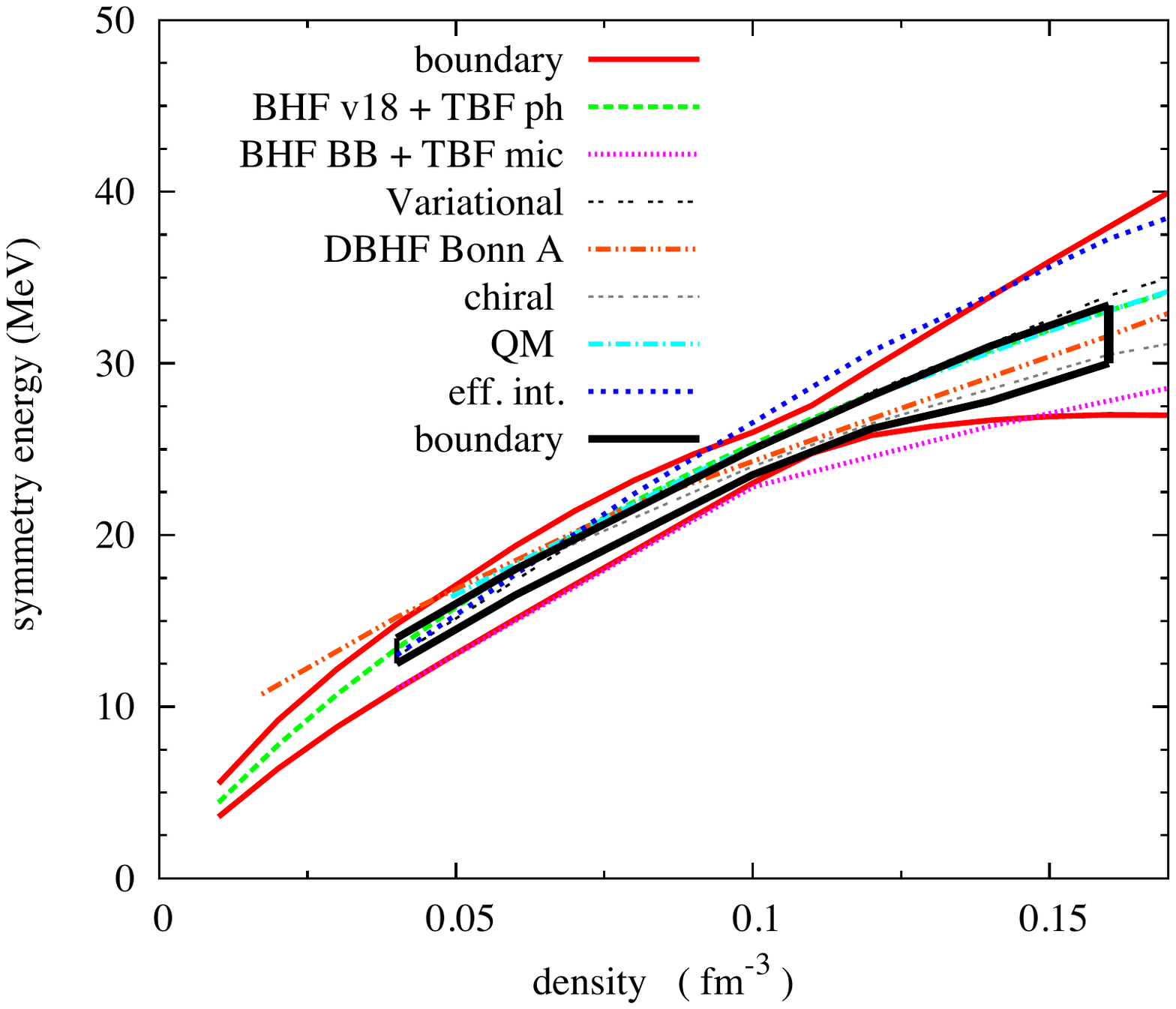}
\vskip -2.5 cm
\caption{(Color on line) The symmetry energy up to saturation for the selected EOS of Table \ref{t:sat} in comparison with the constraints from ref. \cite{DanielIAS}. Labels as in Fig. \ref{fig:SEbelowa}}. 
\label{fig:SEbelowb}
\end{figure}

In Fig. \ref{fig:SEbelowb} we compare the different symmetry energies with the phenomenological boundaries discussed in Section \ref{IAS}.
The smaller restricted area, as we have discussed in Sec. \ref{IAS}, is coming from the analysis on the neutron skin width within the energy functional method. The selected symmetry energies are not all inside this area, but the five EOS which agree substantially among each other stay quite close to it in the low density region, with the possible exception of DBHF at very low density, where it could have problem. In view of the uncertainty in the phenomenological boundaries, this looks a noticeable result. Notice that the agreement is still valid at saturation. The agreement strengthen the relevance of the phenomenological constraints, obtained in the EDF method, and gives a sound basis to the extracted values of the symmetry energy below saturation with an estimate of an uncertainty throughout the sub-saturation density range, where five of the seven considered EOS are still pretty close. The symmetry energy for the BHF with Bonn B potential of ref. \cite{Li1,Li2} is definitely below these five EOS, while the one from the effective interaction approach of ref. \cite{OM1} is substantially above. If one
consider only the five EOS and one assumes a power law for the symmetry energy
\beq
S(\rho) \,\approx\, S(\rho_0)\, \left(\, \frac{\rho}{\rho_0}\, \right)^\gamma
\label{eq:gamma}\eeq
\noindent one finds that the overall trend both of the different theoretical predictions and of the constraint (small box) is compatible with
\beq
 S(\rho_0) \,\approx\, 32 \,\pm\, \\ \rm 1~ MeV \ \ \ \ \ \ \ \ ; \ \ \ \ \ \ \ \ \gamma \,\approx\, 0.65 - 0.70 
\label{Sestimate}\eeq  
\noindent in agreement with the estimate of ref. \cite{DanielIAS}, where it is noticed that the value for $ \gamma\ $ is close to the one
for a free Fermi gas. This is a little surprising, since this exponent is expected to be reached only at very low density.
This value is also compatible with the one extracted from heavy-ion collisions, see Sec. \ref{npratio}.
 It has to be stressed that this simple behaviour is referring only to the overall trend of the symmetry energy and should not be used to extract more detailed properties, like the slope parameter $ L $,
 which depends on the precise functional dependence of $ S(\rho) $. Indeed the values reported in Table \ref{t:sat} have been obtained by an accurate fit of $ S(\rho) $ around saturation. To this respect Table \ref{t:sat} shows that the values of $ L $ for the different EOS stay close, with some deviation for the DBHF one. A more detailed comparison with phenomenology will be reported in Section \ref{LSplot}.

\subsection{Symmetry energy above saturation density.}
\label{symsupra}
As we have discussed in Sections \ref{Astro},\ref{HI} the symmetry energy above saturation is of paramount relevance for the physics of astrophysical compact objects and heavy ion collisions. It is therefore meaningful to extend the comparison of the predictions of different theoretical schemes above saturation density. 

\begin{figure}[hbt]
\centering
\vskip -5 cm
\includegraphics[scale=0.55]{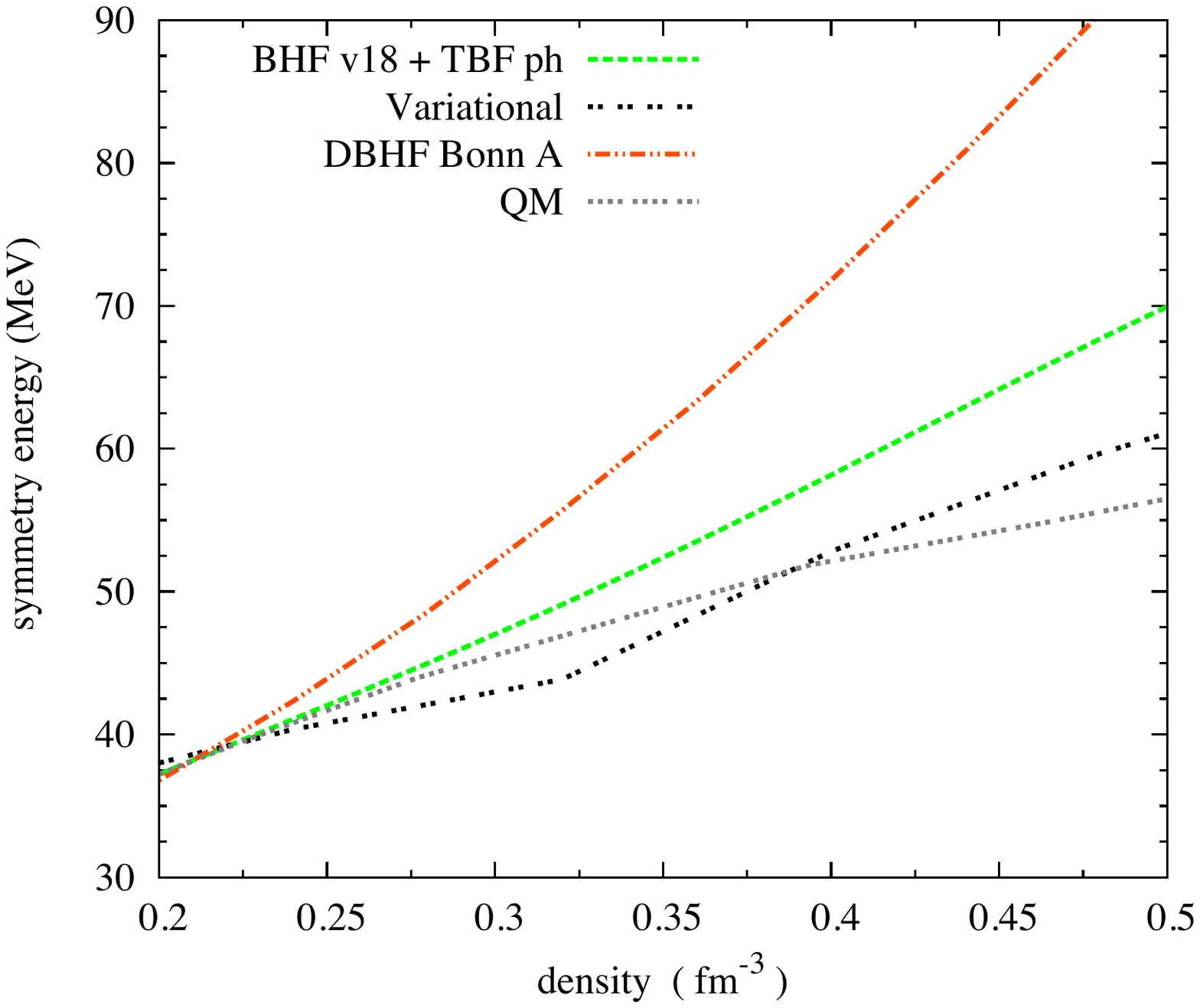}
\vskip -2.5 cm
\caption{(Color on line) The symmetry energy above saturation for the EOS of Table \ref{t:sat} which are compatible with the boundary of ref. \cite{DanielIAS}, see Fig. \ref{fig:SEbelowb}. The labels have the same meaning as in Fig. \ref{fig:SEbelowa}.} 
\label{fig:SEabove}
\end{figure}
\begin{figure}[hbt]
\centering
\vskip -5 cm
\includegraphics[scale=0.5,clip]{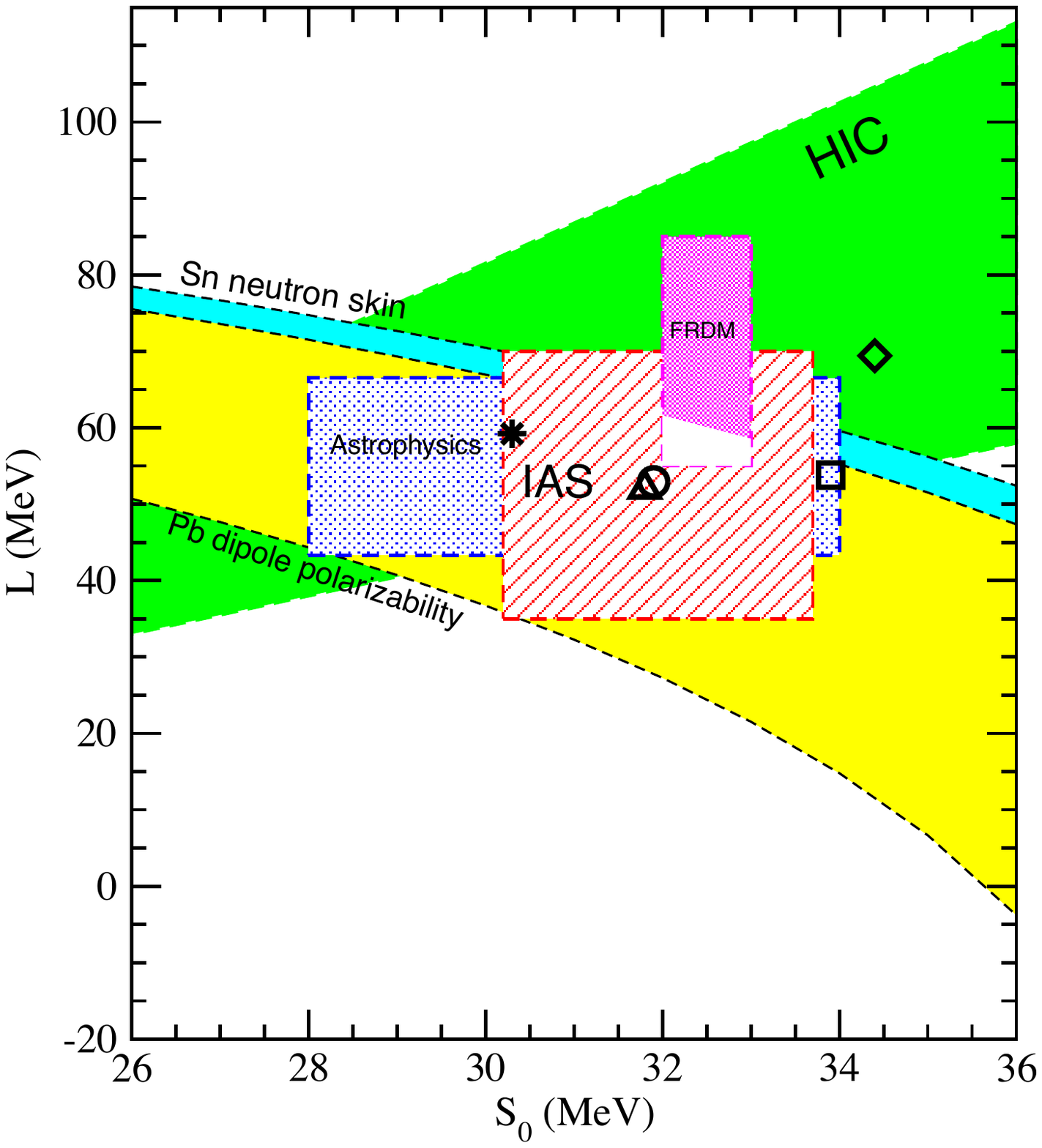}
\vskip -2.5 cm
\caption{(Color online)
Boundaries of the constraints on the values of the symmetry energy $ S $ and slope parameter $ L $ at saturation as derived by different methods. The (pink) rectangle labeled FRDM corresponds to the boundary obtained from the Finite Range Droplet Model in ref. \cite{NixPRL2012} The (orange) rectangle labeled IAS is from the analysis on the Isobaric Analog States of ref. \cite{DanielIAS}. The long (blue) rectangle labeled 'Astrophysics' is the boundary from the NS data analysis discussed in Sec. \ref{core}. The (green) band going upwards at increasing value of $ L $ corresponds to the boundary obtained from isospin diffusion in ref. \cite{tsang_2009}. The (blue) band going downwards at increasing   $ L $, labeled 'Sn neutron skin', was obtained from different data analysis on heavy ion collision and on the neutron skin thickness, as elaborated in ref. \cite{Chen_2010} and discussed in Sec. \ref{HI}. This band partially overlaps with the (yellow) band, going downwards, corresponding to the boundary from dipole polarizability (label 'PB dipole polarizability') as discussed in ref. \cite{Latt} and Sec. \ref{IVex}. The central (white) small open region is the area where all the boundaries overlap. The symbols in the figure correspond to five of the EOS reported in Table 1. Open circle :  BHF, \, Av$_{18}$ + Urbana~TBF. Star : BHF, \, Bonn~ B + TBF from NN. Open square : Variational,  \, Av$_{18}$ + Urbana~TBF. Open diamond :  DBHF,  \, Bonn~ A. Open triangle : Quark model.  
 }
\label{fig:L_S}
\end{figure}
In Fig. \ref{fig:SEabove} are reported the symmetry energies calculated with those theoretical schemes of Section \ref{symsub} which provide symmetry energy values that agree throughout the density range up to saturation.  The theoretical prediction from the chiral approach is missing, since the scheme cannot be extended to higher density. The figure shows that even the nice agreement obtained up to saturation density for these EOS rapidly deteriorates as the density increases. This is somehow a disappointing result since the theoretical predictions are very much demanded in astrophysics and heavy ion physics. Indeed the constraints from phenomenology in these fields are quite indirect and not so stringent, see Fig. \ref{fig:esymhigh} and corresponding references. On the other hand any phenomenological firmly established constraint would be of great value in restricting the selection of the acceptable EOS and the underlying theoretical approaches. This looks as a great opportunity for the advancement in
our knowledge of that fundamental physical quantity which is the nuclear EOS.   

\section{Constraints on the symmetry energy at saturation.}
\label{LSplot}

The constraints on the symmetry energy are abundant at the saturation density. As in recent papers \cite{Tsang,Latt} we collect a set of constraints on the values of $ S $ and $ L $ to see if they are compatible, and if an overlap region exists where the constraints are all fulfilled. We will report in the $ (L,S) $ plot of Fig. \ref{fig:L_S} the following constraints.
\par\noindent
i) The boundaries for $ S $ and $ L $ obtained from the DM fit of nuclear binding in ref. \cite{NixPRL2012}, discussed in Sec. \ref{droplet}. The size of these boundaries were derived by varying the conditions of the fitting, e.g. different sets of data
and different refinements of the model. Since the uncertainties were obtained within the model, they can be biased by the peculiarities of the approach and probably they are underestimated. Nevertheless we report as they are in the plot, keeping in mind that the constraints are expected to be too tight.  
\par\noindent
ii) As discussed in Sec. \ref{IAS}, the isobaric analog state phenomenology and the skin width data can put tight constraints on the density dependence of the symmetry energy up to saturation. These constraints can be translated to ranges of possible values for $ S $ and $ L $. The reported boundaries in ref. \cite{DanielIAS} are
\beq
 30.2 \,<\, S \,<\, 33.7 \  {\rm MeV} \ \ \ \ \ \ \ \ \ \ \ \ \ \ \ \ , \ \ \ \ \ \ \ \ \ \ \ \ \ \ \ \ \ 35 \,<\, L \,<\, 70 \ {\rm MeV}
\eeq   
\noindent which is represented by a slashed box in the figure.
\par\noindent
iii) The astrophysical constraint from NS data analysis, as discussed in Sec. \ref{core}. This is indicated by the dotted box.
\par\noindent
iv) The isospin diffusion constraints is represented by the (green) band going upwards, taken from ref. \cite{tsang_2009}.
\par\noindent
v) We finally add the constraints from the skin width data, combined with isospin diffusion and neutron to proton double ratio in heavy ion collision, as elaborated in ref. \cite{Chen_2010}, see also Sec. \ref{HI}. This is represented by the (blue) band going downwards. We report also the (yellow) band elaborated in ref. \cite{Latt} on the basis of data on nuclear dipole polarizability, see also Sec. \ref{IVex}, indicated by the label. 
\par\noindent

\par\noindent
If we take literally all the constraints, only a small region of the plane is compatible with all of them. It is indicated by the white area at the crossing of all the boundaries, i.e. inside all of them. This is indeed a narrow area, restricted mainly by the constraint from the droplet model (FRDM). Let us stress again that the size of this boundary is expected to be underestimated, and therefore the white area can be appreciably larger. \par 
The theoretical predictions of Table \ref{t:sat} are reported in the same plot by different symbols. The result from the effective interaction method lies outside the considered range for $ S $. All the others are close to the allowed region, two of them lie even on its boundary. However the values from the Dirac-Brueckner approach appears to be a little far. To be fair one has to notice that all the constraints and calculations were obtained within a non-relativistic approach. This could in principle bias the relativistic scheme, but further investigation is needed. In any case the closeness of the results of the majority of the microscopic
approaches to the allowed region should be appreciated since it is not an obvious or trivial result.     
\section{Overall Survey and Prospects.}
\label{sum}
In this review paper we presented an overall survey of the crucial relevance of the nuclear symmetry energy in the numerous contexts where it is involved. On the other hand we tried to summarize what is the present status of our knowledge about the symmetry energy and its density dependence. The symmetry energy is one of the main properties of the nuclear Equation of State and the constraints on its density dependence can be established from the phenomenological analysis of very different systems and phenomena, which are affected by the symmetry energy.\par 
The original field where symmetry energy started to be studied was the nuclear binding throughout the mass table. Here the fitting of the nuclear masses by semi-classical models, like the Liquid Drop Model or Droplet model, provides constraints on the symmetry energy in nuclei, including its possible parametrization as a function of the mass number  A  and proton number  Z. This symmetry energy has to be distinguished from the one defined in nuclear matter, since it is strongly affected by finite size effects, i.e. the presence of the surface. It has been found that its mass dependence can be parametrized by a form as in Eq. (\ref{eq:int}), obtained by introducing the coupling between the bulk and surface parts \cite{Daniel2009}, and within the LDM and DM, as well as by comparing with the nuclear mass data. In the DM it is also possible to fit the slope parameters $ L $. The extrapolation to nuclear matter can be then obtained by taking the limit $ A \,\rightarrow\, \infty $. These approaches converge overall to values that overlap approximately in the range $ S \,=\, 32 \pm 2$ MeV, and $ L \,=\, 70 \pm 20$ MeV.\par A more microscopic procedure is based on the density functional method, again from the nuclear binding fit. In this case, once fixed the parameters of the functional the whole density dependence of the symmetry energy can be deduced,
at least up to saturation density. This is evident in the Thomas-Fermi approximation for the functional. It has been shown that in this approximation one can device an effective interaction which reproduces fairly well a wide set of nuclear binding, see Section \ref{TF}. From the fit one gets the nuclear matter EOS and therefore the density dependent symmetry energy up to saturation.
The results are in agreement with the latest LDM and DM calculations, and, surprising enough, with microscopic calculations, see
Fig. \ref{fig:EOS_TF}. Despite the reduced set of nuclear binding considered in this analysis, with respect to the most recent nuclear mass compilations, this result is remarkable.\par  
However if one goes beyond the TF approximation, extending the analysis to general functionals, in particular Skyrme functionals, the fitting procedure of the nuclear bindings turns out to be not enough to constrain in a meaningful way the symmetry energy. Even restricting the set of functionals to the ones that satisfy the phenomenolgical constraints on the saturation point and the incompressibility, the symmetry energy $ S $ and its slope parameter $ L $ are widely unconstrained, see Fig \ref{fig:LS}. This means that the symmetry energy is mainly uncorrelated to other physical parameters, and other constraints on the functional are needed.   
Improvements along this direction, i.e. the fitting of nuclear mass, can be expected to come from new data on exotic nuclei, where the asymmetry is pushed to an extreme. However it is doubtful that more extended mass fitting could reduce drastically the uncertainty on the symmetry energy coming from this approach. 
\par 
A natural choice for additional constraints extracted from nuclear structure data concerns the set of isovector collective excitations in nuclei,
since they involve the motion of protons against neutrons, see Sec. \ref{IVex}. The IVGDR excitation energy looks indeed correlated with the symmetry energy, see Fig. \ref{fig:dipole}, as obtained within the EDF method. What appears explicitly in this case is the symmetry energy pertinent to a finite nucleus, which can be translated to constraints on the symmetry energy in nuclear matter at a fixed value $ \overline{\rho} $ of the (sub-saturation) density in nuclear matter (mainly $ 0.1$ fm$^{-3}$). Similar correlations have been found for the low density part of the dipole strength, the so-called Pygmy resonance, but these correlations look less stringent. Another dipole physical parameter is the dipole polarizability $ \alpha_D $. It is just the value of the static response function of the nucleus to an
external dipole field. It can be obtained from the photoabsorption cross section, see Eq. (\ref{eq:cross}). Again within the EDF method the value of $ \alpha_D $ entails a correlation between the symmetry energy $ S $ at saturation and the slope parameter $ L $, see Fig. \ref{fig:alpha}. For the IVGQR the analysis is a little simplified, since, in a macroscopic picture, the comparison of the excitation energies of the IVGQR and of the ISGQR can be related directly to the symmetry energy at the reference density  
$ \overline{\rho} $, see Eq.(\ref{eq:diff}). However a check of this relation within the EDF method reveals that it could be affected by other characteristics of the force, besides the symmetry energy.\par

Among the additional constraints coming from nuclear structure, special mention has to be given to the ones related to the Isobaric Analog States and to the neutron skin width in strongly asymmetric nuclei. Both possibilities were analyzed in ref. \cite{DanielIAS}. The method followed for the IAS is discussed in Section \ref{IAS}. It is based on the nice observation that charge invariance of the nuclear forces implies not only the symmetry for the interchange of protons and neutrons, i.e. dependence of the binding only on the modulus of the third component $ | T_z | $ of the total isospin (neglecting Coulomb interaction), but also invariance under rotations in isospin space. This implies a linear relationship between the IAS energy, with respect to the ground state, and a simple combination of isospin quantum numbers, see Eq. (\ref{eq:EX}). The coefficient of the linear relationship is the nuclear symmetry energy $ S_N(A) $ at a given mass number A. If one separates different intervals of A-values, this relationship is expected to be still valid, provided the coefficient is interpreted as its average value within each interval. This is due to the smooth dependence of the symmetry energy. Despite the relation is approximate, it is confirmed by phenomenology, see Fig.(\ref{fig:fitIAS}),
at least for medium-heavy nuclei. One can then even check the mass dependence of $ S_N $, and it is found that the form of Eq. (\ref{eq:int}) is confirmed, see Fig.\ref{fig:aVIAS}. Going to the EDF analysis, one can select the functionals that are compatible with these relationships and extract the constraints on the density dependence of the symmetry energy that come out from this restriction. The result is summarized in Fig. \ref{fig:Sro}, where it is depicted the region of the $ (S,\rho) $ plane where the symmetry energy should pass through. Despite all the uncertainties inherent to the method, discussed in Section \ref{IAS}, this is a constraint tight enough to exclude a quite large fraction of EDF. An even more tight constraint can be obtained by including also the phenomenology on the neutron skin thickness in strongly asymmetric nuclei, see Sec. \ref{skin}. The skin thickness, as calculated from a set of functionals, turns out to be correlated with different physical parameters, in particular the slope parameter $ L $,
see Fig. \ref{fig:LQ}. Of particular interest is the correlation with the dipole polarizability $ \alpha_D $, but in this case the correlation is rather between the skin width and the product $ \alpha_D S $, see Fig. \ref{fig:alphaskin}. Analogously the skin width is not correlated with the symmetry energy $ S $, but rather with the difference between $ S $ and the symmetry energy at a given nuclear mass A, see Fig. \ref{fig:skinS_a}. All these correlations can be explained within the DM, and they can be used to extract several physical parameters once the skin width would be measured.   
\par
Another strategy can be followed, combining the constraints coming from the IAS analysis and the condition that the functional has to reproduce the available data on the skin thickness. The data are coming from different experimental methods with hadron probes.
In this case the set of EDF is strongly restricted. Allowing for some deviations from the data, the region of the $ (S,\rho) $ plane is anyhow drastically reduced, see Fig. \ref{fig:2box}. \par 
From these considerations it is evident the relevance of an accurate experimental determination of the skin width. This is the aim of the PREX experiment, as discussed in Sect. \ref{skin}, on the asymmetry in spin polarized electron scattering. The asymmetry is directly connected with the parity violation of the weak interaction, and it is sensitive mainly only to the neutron density distribution. The phase PREX-I was already completed, but the uncertainty was too large to have an accurate enough determination of the skin width, see Fig. \ref{fig:PREX_exp}. The phase PREX-II is expected to have a substantially better accuracy, see discussion connected with Fig. \ref{fig:PREX_Xavier}. This would mark a substantial overall advancement for our picture of the symmetry energy behavior and its impact on nuclear structure. \par        
In Sec. \ref{Astro} we discussed the relevance of the nuclear symmetry energy for compact astrophysical objects. We first considered Neutron Stars, and the structure of NS crust was considered in Sec. \ref{crust}, where the beta equilibrium condition between nuclei,
eventually exotic, and the electron gas makes evident the role of the symmetry energy, see Eq. \ref{eq:chem}. At increasing density the neutrons start to drip from nuclei, and one enters the so-called inner crust. The end of this crust region as density is further increased corresponds to the transition to homogeneous catalyzed matter. The density $ \rho_t $ at which the transition occurs turns out to be linearly correlated with the slope parameter $ L $, once it is determined within a wide set of functionals. Since $ L $ is correlated with the skin width, this entails a correlation between of $ \rho_t $ with the skin width, see Fig. \ref{fig:trans_den}. However between the lattice structure of the inner crust and the homogeneous phase it could be expected the appearance of a disordered phase or of extended structures, the so-called "pasta phase". The presence or not of the pasta phase by itself can be used as a constraint on the symmetry energy, since its structure is determined by "frustration", i.e. the competition between the long range repulsive Coulomb force and the short range attractive nuclear forces. To calculate the presence and extension of the pasta phase is a quite challenging problem, that can have a definite answer only through a quantal treatment. Such a prospect is the main motivation of the "MADNESS project". When successfully realized it will give a substantial contribution to the development of the whole physics of NS. A variety of phenomena are affected by the crust structure and can be then related to the symmetry energy. The crust oscillations, if decoupled from the core, have frequency that depends on the composition of the crust, and thus on the symmetry energy. Transient phenomena, like accretion in LMXRB, are often affected by the transport properties in the crust. Improving and extending the observations of all these phenomena can have a large impact on their interpretation and on the whole crust physics. \par 
Similarly the NS core structure, discussed in Sec. \ref{core}, is directly affected by the symmetry energy, mainly above saturation density. This is apparent from the NS matter EOS, as exemplified by e.g. Eqs. (\ref{eq:xp1},\ref{eq:press}). Cooling of isolated NS is surely a process strongly affected by the symmetry energy at supra-saturation density. The direct URCA process (DURCA) is the most rapid cooling mechanism, which is the beta decay of free neutrons and its inverse, see Eq. (\ref{eq:cool}). The DURCA is active only at proton fraction above the threshold of about 12\% (or 14\% if muons are taken into account). This threshold is reached at a density that depends on the EOS and its symmetry energy. This means that larger mass NS, that have a higher central density, will cool faster than lower masses NS. It could be that this threshold is never reached in any NS. Unfortunately progress along this line are hindered by at least three factors. The first one is the large uncertainty on the estimated age of a NS and the difficulty to estimate the mass of an isolated NS. The second one is the extreme complexity of the cooling simulation code. The third one is the expected presence of superfluidity in NS matter, which makes ambiguous or not unique the interpretation of the data. Only a substantial improvement of the theoretical framework will allow a real progress in this field. An exception can be the singular case of the NS in Cassiopeia A \cite{Fesen,Shternin,Noimnras}, which however is still controversial \cite{Alfio}.\par 
The supernovae phenomena is a very rare event if we restrict the region of possible observations in the nearby of our Galaxy,
but its appearance would be a unique opportunity to study and understand a variety of fundamental processes that occur in their inside. The progress in the simulations of this extremely complex phenomenon in the last few years is impressive. The complexity of the supernovae explosion and of their simulations could be a difficulty in the interpretation of the data in a hypothetical observation. An exception is the neutrino signal, not only because neutrinos can escape from the supernova interior and pass through the external expanding mantle, but also because they could be sensitive to the EOS and the physical conditions of the matter during the first phase after bounce. As discussed in Sec. \ref{supernovae}, the neutrino luminosity as a function of time during this period could bring an imprint of the symmetry energy of the nuclear matter involved in the process. This is due to the appearance of convection in the proto-neutron star, which depends on the symmetry energy, see Eqs. (\ref{eq:convec2},\ref{eq:convec5}). It is important to notice that the matter in this case is at a temperature of few tens of MeV, and one should rather speak of symmetry free energy, which can be radically different from the (zero temperature) symmetry energy. The appearance of convection increases the neutrino luminosity, and therefore modulates the observed flux. Since the time and intensity for the onset of convection depends on the symmetry free energy, one could get a definite signal from the proto-neutron star, see Fig. \ref{fig:lumin}, which would be otherwise impossible with other probes. We are looking forward for the appearance of a nearby supernova.
\par
A new exciting development in the astrophysics of compact objects is the first observation of gravitational waves by the LIGO detector. 
The GW physics was discussed in Sec. 3.4. The first two detected events are surely coming from black hole mergers, but it is
expected that also GW from NS mergers will be observed in the future. In this case a link with the Nuclear Matter EOS is possible, 
as already shown by few simulations based on numerical relativity. Although the connection with the symmetry energy appears indirect, through the EOS properties, some sensitivity to the high density behaviour of the symmetry energy seems to be present.
Also large amplitude oscillations of NS can be sources of GW, and the corresponding frequency spectrum is surely sensitive to the EOS.
A new window in observational astrophysics has been opened  by the GW detection, and the forthcoming upgrading of the 
detectors, already active, and the new ones, planned or under construction, will bring spectacular developments  in astrophysics.    
\par
In Sec. \ref{HI} we reviewed the role of the symmetry energy in the physics of heavy ion collisions. In general the data analysis are based on extensive simulations of the collisions, either semi-classical (BUU) or quantal (QMD). The control and sensitivity of the simulations to different observables related to the symmetry energy is not yet satisfactory. For instance, the neutron/proton mass splitting derived form the comparison with the data on differential fluxes turns out to be sensitive to the adopted simulation scheme. Other physical quantities look more robust, and we considered the isospin diffusion, the isoscaling, neutron and proton energy spectra, and others. There is surely room for improvements in the theoretical data analysis based on simulations, as well as    
in the experimental apparata. Heavy ion collisions should be an ideal tool for the study of the symmetry energy at supra-saturation density, since it is the only experimental method to produce compressed nuclear matter. The projects aiming to explore the higher density regime are still very few, and it would be of great relevance to develop the experimental efforts in this direction.
In fact it has to be stressed that the theoretical predictions in this region are quite uncertain, as discussed in Sec. \ref{symsupra}.
\par
In Sec. \ref{theory} we presented a review of the microscopic EOS that are compatible with phenomenology on the saturation point of symmetric nuclear matter and incompressibility. To each one of this EOS it corresponds a well defined symmetry energy as a function of density. We compared this prediction with the constraints obtained within the EDF method from the nuclear structure data both on  the IAS and neutron skin. These constraints are necessary limited to sub-saturation density. Most of the so selected EOS stay close to the constraints, which gives an impression of the degree of reliability of the microscopic theories. Of course the agreement does not necessarily validate the microscopic theories, since the constraints were obtained from an analysis which could be biased by the particular set of EDF considered and the agreement could be fortuitous. Only a more extensive EDF analysis and the further improvement of the many-body theory will be able to give a definite answer. \par
Above saturation density the situation appears drastically different. The agreement between the different microscopic theories is rapidly lost at increasing density. The few phenomenological constraints in this density range are not able to discriminate among the different EOS. On one hand this can be disappointing, on the other hand this can be an opportunity for a more severe confrontation of the theoretical models with phenomenology. New analysis devoted to the over-saturation density region will possibly produce an extraordinary advancement in our knowledge on the structure of the nuclear medium and on the physical origin of the nuclear symmetry energy. \par
In the last section we combined together the constraints on the symmetry energy and its slope at saturation as obtained in the different fields, as described in the previous sections. In a two dimensional plot we reported in Fig. \ref{fig:L_S}
the boundaries on $ S $ and $ L $ to extract the overlap region, if any, where all these constraints are fulfilled.
The resulting plot of Fig. \ref{fig:L_S} shows that a small overlap region does exist. If we take literally all the constraints, the overlap region is very narrow, especially in the $ S $ axis, mainly due to the constraint obtained from the mass fitting within the Droplet Model. The size of the uncertainty in the DM has been obtained within the same fitting, and it can be expected to be underestimated. One can say that the extracted value for $ S $ from this analysis should be around 32 MeV, with an uncertainty of $ \pm 1 $. Much wider is the uncertainty on $ L $, which can range from 40 to 80 MeV.   
\noindent
In the same plot we reported also the predictions of the microscopic theory discussed in Sec. \ref{theory}. The EOS were selected on the basis of their satisfactory predictions of the saturation point and of the incompressibility. This restrict strongly the set of acceptable microscopic EOS. It turns out that the predictions of most of the few selected EOS  stay close the the overlap. This satisfactory agreement should be appreciated, it gives indeed further support both to the phenomenological analysis performed in different fields along the years and the microscopic many-body theories that have been developed by different groups. Again the agreement could be fortuitous,
and its confirmation by other analysis on experimental and observational data and theoretical improvements would be of great relevance for Nuclear Physics as a whole.

\section{Acknowledgements}
We acknowledge partial support from “NewCompStar", COST Action MP1304.


\newcommand{\apjl}{Astrophys. J. Lett.\ }
\newcommand{\apj}{Astrophys. J. \ }
\newcommand{\prc}{Phys. Rev. C\ }
\newcommand{\prd}{Phys. Rev. D\ }
\newcommand{\mnras}{Mon. Not. R. Astron. Soc.\ }
\newcommand{\aap}{Astron. Astrophys.\ }
\newcommand{\nphysa}{Nucl. Phys. A\ }
\newcommand{\physrep}{Phys. Rep.\ }
\newcommand{\nat}{Nature\ }

\bibliographystyle{elsarticle-num}
\bibliography{PPNP_Esym}

\begin{thebibliography}{100}
\expandafter\ifx\csname url\endcsname\relax
  \def\url#1{\texttt{#1}}\fi
\expandafter\ifx\csname urlprefix\endcsname\relax\def\urlprefix{URL }\fi
\expandafter\ifx\csname href\endcsname\relax
  \def\href#1#2{#2} \def\path#1{#1}\fi

\bibitem{EPJA2014}
{Li, Bao-An}, {Ramos, Àngels}, {Verde, Giuseppe}, {Vidaña, Isaac},
  \href{http://dx.doi.org/10.1140/epja/i2014-14009-x}{Topical issue on nuclear
  symmetry energy}, Eur. Phys. J. A 50~(2) (2014) 9.
\newblock \href {http://dx.doi.org/10.1140/epja/i2014-14009-x}
  {\path{doi:10.1140/epja/i2014-14009-x}}.
\newline\urlprefix\url{http://dx.doi.org/10.1140/epja/i2014-14009-x}

\bibitem{Tsang}
M.~B. {Tsang}, J.~R. {Stone}, F.~{Camera}, P.~{Danielewicz}, S.~{Gandolfi},
  K.~{Hebeler}, C.~J. {Horowitz}, J.~{Lee}, W.~G. {Lynch}, Z.~{Kohley},
  R.~{Lemmon}, P.~{M{\"o}ller}, T.~{Murakami}, S.~{Riordan}, X.~{Roca-Maza},
  F.~{Sammarruca}, A.~W. {Steiner}, I.~{Vida{\~n}a}, S.~J. {Yennello},
  {Constraints on the symmetry energy and neutron skins from experiments and
  theory}, \prc 86 (2012) 015803.
\newblock \href {http://arxiv.org/abs/1204.0466} {\path{arXiv:1204.0466}}.

\bibitem{Latt}
J.~M. {Lattimer}, Y.~{Lim}, {Constraining the Symmetry Parameters of the
  Nuclear Interaction}, \apj 771 (2013) 51.
\newblock \href {http://arxiv.org/abs/1203.4286} {\path{arXiv:1203.4286}}.

\bibitem{Horo}
C.~J. {Horowitz}, E.~F. {Brown}, Y.~{Kim}, W.~G. {Lynch}, R.~{Michaels},
  A.~{Ono}, J.~{Piekarewicz}, M.~B. {Tsang}, H.~H. {Wolter}, {A way forward in
  the study of the symmetry energy: experiment, theory, and observation},
  Journal of Physics G Nuclear Physics 41 (2014) 093001.
\newblock \href {http://arxiv.org/abs/1401.5839} {\path{arXiv:1401.5839}}.

\bibitem{RS}
P.~Ring, P.~Schuck, The nuclear many-body problem, Springer Science \& Business
  Media, 2004.

\bibitem{Daniel2009}
P.~{Danielewicz}, J.~{Lee}, {Symmetry energy I: Semi-infinite matter}, Nuclear
  Physics A 818 (2009) 36--96.
\newblock \href {http://arxiv.org/abs/0807.3743} {\path{arXiv:0807.3743}}.

\bibitem{lepto}
R.~C. {Nayak}, J.~M. {Pearson}, M.~{Farine}, P.~{Gleissl}, M.~{Brack.},
  {Leptodermous expansion of finite-nucleus incompressibility}, Nuclear Physics
  A 516 (1990) 62--76.

\bibitem{MS66}
W.~D. {Myers}, W.~J. {Swiatecki}, {Nuclear masses and deformations}, \nphysa 81
  (1966) 1.

\bibitem{My69}
W.~D. {Myers}, {Droplet model nuclear density distributions and single-particle
  potential wells}, Nuclear Physics A 145 (1970) 387--400.

\bibitem{MS69}
W.~D. {Myers}, W.~J. {Swiatecki}, {Average nuclear properties}, Annals of
  Physics 55 (1969) 395--505.

\bibitem{Strut}
V.~M. {Strutinsky}, {Shell effects in nuclear masses and deformation energies},
  Nuclear Physics A 95 (1967) 420--442.

\bibitem{NixPRL2012}
P.~{M{\"o}ller}, W.~D. {Myers}, H.~{Sagawa}, S.~{Yoshida}, {New Finite-Range
  Droplet Mass Model and Equation-of-State Parameters}, Physical Review Letters
  108 (2012) 052501.

\bibitem{2016ADNDT}
P.~{M{\"o}ller}, A.~J. {Sierk}, T.~{Ichikawa}, H.~{Sagawa}, {Nuclear
  ground-state masses and deformations: FRDM(2012)}, Atomic Data and Nuclear
  Data Tables 109 (2016) 1--204.
\newblock \href {http://arxiv.org/abs/1508.06294} {\path{arXiv:1508.06294}}.

\bibitem{March}
N.~March, {Self-consistent fields in atoms. Hartree and Thomas: Fermi atoms},
  Pergamon Press, Inc., New York, 1975.

\bibitem{MarioPeter}
M.~{Centelles}, P.~{Schuck}, X.~{Vi{\~n}as}, {Thomas-Fermi theory for atomic
  nuclei revisited}, Annals of Physics 322 (2007) 363--396.
\newblock \href {http://arxiv.org/abs/nucl-th/0601092}
  {\path{arXiv:nucl-th/0601092}}.

\bibitem{MS1996}
W.~D. {Myers}, W.~J. {Swiatecki}, {Nuclear properties according to the
  Thomas-Fermi model}, Nuclear Physics A 601 (1996) 141--167.

\bibitem{SB1961}
R.~G. {Seyler}, C.~H. {Blanchard}, {Classical Self-Consistent Nuclear Model},
  Physical Review 124 (1961) 227--232.

\bibitem{MS1998}
W.~D. {Myers}, W.~J. {{\'S}wia{\c T}ecki}, {Nuclear equation of state}, \prc 57
  (1998) 3020--3025.

\bibitem{BCPM}
B.~K. {Sharma}, M.~{Centelles}, X.~{Vi{\~n}as}, M.~{Baldo}, G.~F. {Burgio},
  {Unified equation of state for neutron stars on a microscopic basis}, \aap
  584 (2015) A103.
\newblock \href {http://arxiv.org/abs/1506.00375} {\path{arXiv:1506.00375}}.

\bibitem{FP}
B.~{Friedman}, V.~R. {Pandharipande}, {Hot and cold, nuclear and neutron
  matter}, Nuclear Physics A 361 (1981) 502--520.

\bibitem{HK}
P.~{Hohenberg}, W.~{Kohn}, {Inhomogeneous Electron Gas}, Physical Review 136
  (1964) 864--871.

\bibitem{Gogny}
J.~F. {Berger}, M.~{Girod}, D.~{Gogny}, {Microscopic analysis of collective
  dynamics in low energy fission}, Nuclear Physics A 428 (1984) 23--36.

\bibitem{Gogny_nl}
F.~{Chappert}, N.~{Pillet}, M.~{Girod}, J.-F. {Berger}, {Gogny force with a
  finite-range density dependence}, \prc 91 (2015) 034312.

\bibitem{RMPreview}
M.~{Bender}, P.-H. {Heenen}, P.-G. {Reinhard}, {Self-consistent mean-field
  models for nuclear structure}, Reviews of Modern Physics 75 (2003) 121--180.

\bibitem{HFB17}
N.~{Chamel}, S.~{Goriely}, J.~M. {Pearson}, {Further explorations of
  Skyrme-Hartree-Fock-Bogoliubov mass formulas. XI. Stabilizing neutron stars
  against a ferromagnetic collapse}, \prc 80 (2009) 065804.
\newblock \href {http://arxiv.org/abs/0911.3346} {\path{arXiv:0911.3346}}.

\bibitem{Agra}
B.~K. {Agrawal}, S.~K. {Dhiman}, R.~{Kumar}, {Exploring the extended
  density-dependent Skyrme effective forces for normal and isospin-rich nuclei
  to neutron stars}, \prc 73 (2006) 034319.
\newblock \href {http://arxiv.org/abs/nucl-th/0604045}
  {\path{arXiv:nucl-th/0604045}}.

\bibitem{Colo_t}
G.~{Col{\`o}}, H.~{Sagawa}, S.~{Fracasso}, P.~F. {Bortignon}, {Spin orbit
  splitting and the tensor component of the Skyrme interaction}, Physics
  Letters B 646 (2007) 227--231.
\newblock \href {http://arxiv.org/abs/nucl-th/0701015}
  {\path{arXiv:nucl-th/0701015}}.

\bibitem{Lesi}
T.~{Lesinski}, M.~{Bender}, K.~{Bennaceur}, T.~{Duguet}, J.~{Meyer}, {Tensor
  part of the Skyrme energy density functional: Spherical nuclei}, \prc 76
  (2007) 014312.
\newblock \href {http://arxiv.org/abs/0704.0731} {\path{arXiv:0704.0731}}.

\bibitem{Otsu}
T.~{Otsuka}, T.~{Matsuo}, D.~{Abe}, {Mean Field with Tensor Force and Shell
  Structure of Exotic Nuclei}, Physical Review Letters 97 (2006) 162501.

\bibitem{Egi}
J.~L. {Egido}, L.~M. {Robledo}, {Angular Momentum Projection and Quadrupole
  Correlations Effects in Atomic Nuclei}, in: G.~A. {Lalazissis}, P.~{Ring},
  D.~{Vretenar} (Eds.), Extended Density Functionals in Nuclear Structure
  Physics, Vol. 641 of Lecture Notes in Physics, Berlin Springer Verlag, 2004,
  pp. 269--302.
\newblock \href {http://arxiv.org/abs/nucl-th/0311106}
  {\path{arXiv:nucl-th/0311106}}.

\bibitem{Bend}
M.~{Bender}, T.~{Duguet}, D.~{Lacroix}, {Particle-number restoration within the
  energy density functional formalism}, \prc 79 (2009) 044319.
\newblock \href {http://arxiv.org/abs/0809.2045} {\path{arXiv:0809.2045}}.

\bibitem{Lac}
D.~{Lacroix}, T.~{Duguet}, M.~{Bender}, {Configuration mixing within the energy
  density functional formalism: Removing spurious contributions from
  nondiagonal energy kernels}, \prc 79 (2009) 044318.
\newblock \href {http://arxiv.org/abs/0809.2041} {\path{arXiv:0809.2041}}.

\bibitem{Wash}
K.~{Washiyama}, K.~{Bennaceur}, B.~{Avez}, M.~{Bender}, P.-H. {Heenen},
  V.~{Hellemans}, {New parametrization of Skyrme's interaction for regularized
  multireference energy density functional calculations}, \prc 86 (2012)
  054309.
\newblock \href {http://arxiv.org/abs/1209.5258} {\path{arXiv:1209.5258}}.

\bibitem{Satu}
W.~{Satu{\l}a}, J.~{Dobaczewski}, {Simple regularization scheme for
  multireference density functional theories}, \prc 90 (2014) 054303.
\newblock \href {http://arxiv.org/abs/1407.0857} {\path{arXiv:1407.0857}}.

\bibitem{Jirina}
M.~{Dutra}, O.~{Louren{\c c}o}, J.~S. {S{\'a} Martins}, A.~{Delfino}, J.~R.
  {Stone}, P.~D. {Stevenson}, {Skyrme interaction and nuclear matter
  constraints}, \prc 85 (2012) 035201.
\newblock \href {http://arxiv.org/abs/1202.3902} {\path{arXiv:1202.3902}}.

\bibitem{Blaizot}
J.~P. {Blaizot}, {Nuclear compressibilities}, \physrep 64 (1980) 171--248.

\bibitem{Jirina2}
M.~{Dutra}, O.~{Louren{\c c}o}, S.~S. {Avancini}, B.~V. {Carlson},
  A.~{Delfino}, D.~P. {Menezes}, C.~{Provid{\^e}ncia}, S.~{Typel}, J.~R.
  {Stone}, {Relativistic mean-field hadronic models under nuclear matter
  constraints}, \prc 90 (2014) 055203.
\newblock \href {http://arxiv.org/abs/1405.3633} {\path{arXiv:1405.3633}}.

\bibitem{KS}
W.~{Kohn}, L.~J. {Sham}, {Self-Consistent Equations Including Exchange and
  Correlation Effects}, Physical Review 140 (1965) 1133--1138.

\bibitem{BCPM_KS}
M.~{Baldo}, L.~M. {Robledo}, P.~{Schuck}, X.~{Vi{\~n}as}, {New Kohn-Sham
  density functional based on microscopic nuclear and neutron matter equations
  of state}, \prc 87 (2013) 064305.

\bibitem{Fay}
S.~A. {Fayans}, {Towards a universal nuclear density functional}, Soviet
  Journal of Experimental and Theoretical Physics Letters 68 (1998) 169--174.

\bibitem{Stei}
A.~W. {Steiner}, M.~{Prakash}, J.~M. {Lattimer}, P.~J. {Ellis}, {Isospin
  asymmetry in nuclei and neutron stars}, \physrep 411 (2005) 325--375.
\newblock \href {http://arxiv.org/abs/nucl-th/0410066}
  {\path{arXiv:nucl-th/0410066}}.

\bibitem{multipol}
G.~{Col{\`o}}, U.~{Garg}, H.~{Sagawa}, {Symmetry energy from the nuclear
  collective motion: constraints from dipole, quadrupole, monopole and
  spin-dipole resonances}, European Physical Journal A 50 (2014) 26.
\newblock \href {http://arxiv.org/abs/1309.1572} {\path{arXiv:1309.1572}}.

\bibitem{LS}
E.~{Lipparini}, S.~{Stringari}, {Sum rules and giant resonances in nuclei},
  \physrep 175 (1989) 103--261.

\bibitem{siNM}
J.~{Treiner}, H.~{Krivine}, {Semi-classical nuclear properties from effective
  interactions}, Annals of Physics 170 (1986) 406--453.

\bibitem{XavierPRL}
M.~{Centelles}, X.~{Roca-Maza}, X.~{Vi{\~n}as}, M.~{Warda}, {Nuclear Symmetry
  Energy Probed by Neutron Skin Thickness of Nuclei}, Physical Review Letters
  102 (2009) 122502.
\newblock \href {http://arxiv.org/abs/0806.2886} {\path{arXiv:0806.2886}}.

\bibitem{DanielIAS}
P.~{Danielewicz}, J.~{Lee}, {Symmetry energy II: Isobaric analog states},
  Nuclear Physics A 922 (2014) 1--70.
\newblock \href {http://arxiv.org/abs/1307.4130} {\path{arXiv:1307.4130}}.

\bibitem{Trippa}
L.~{Trippa}, G.~{Col{\`o}}, E.~{Vigezzi}, {Giant dipole resonance as a
  quantitative constraint on the symmetry energy}, \prc 77 (2008) 061304.
\newblock \href {http://arxiv.org/abs/0802.3658} {\path{arXiv:0802.3658}}.

\bibitem{Tamii}
A.~{Tamii}, I.~{Poltoratska}, P.~{von Neumann-Cosel}, Y.~{Fujita}, T.~{Adachi},
  C.~A. {Bertulani}, J.~{Carter}, M.~{Dozono}, H.~{Fujita}, K.~{Fujita},
  K.~{Hatanaka}, D.~{Ishikawa}, M.~{Itoh}, T.~{Kawabata}, Y.~{Kalmykov}, A.~M.
  {Krumbholz}, E.~{Litvinova}, H.~{Matsubara}, K.~{Nakanishi}, R.~{Neveling},
  H.~{Okamura}, H.~J. {Ong}, B.~{{\"O}zel-Tashenov}, V.~Y. {Ponomarev},
  A.~{Richter}, B.~{Rubio}, H.~{Sakaguchi}, Y.~{Sakemi}, Y.~{Sasamoto},
  Y.~{Shimbara}, Y.~{Shimizu}, F.~D. {Smit}, T.~{Suzuki}, Y.~{Tameshige},
  J.~{Wambach}, R.~{Yamada}, M.~{Yosoi}, J.~{Zenihiro}, {Complete Electric
  Dipole Response and the Neutron Skin in Pb208}, Physical Review Letters 107
  (2011) 062502.
\newblock \href {http://arxiv.org/abs/1104.5431} {\path{arXiv:1104.5431}}.

\bibitem{polar}
X.~{Roca-Maza}, X.~{Vi{\~n}as}, M.~{Centelles}, B.~K. {Agrawal}, G.~{Col{\`o}},
  N.~{Paar}, J.~{Piekarewicz}, D.~{Vretenar}, {Neutron skin thickness from the
  measured electric dipole polarizability in $^{68}$Ni, $^{120}$Sn and
  $^{208}$Pb}, \prc 92 (2015) 064304.
\newblock \href {http://arxiv.org/abs/1510.01874} {\path{arXiv:1510.01874}}.

\bibitem{dropol}
J.~{Meyer}, P.~{Quentin}, B.~K. {Jennings}, {The isovector dipole mode: A
  simple sum rule approach}, Nuclear Physics A 385 (1982) 269--284.

\bibitem{Savran}
D.~{Savran}, T.~{Aumann}, A.~{Zilges}, {Experimental studies of the Pygmy
  Dipole Resonance}, Progress in Particle and Nuclear Physics 70 (2013)
  210--245.

\bibitem{Mohan}
R.~{Mohan}, M.~{Danos}, L.~C. {Biedenharn}, {Three-Fluid Hydrodynamical Model
  of Nuclei}, \prc 3 (1971) 1740--1749.

\bibitem{Klimkiewicz}
A.~{Klimkiewicz}, N.~{Paar}, P.~{Adrich}, M.~{Fallot}, K.~{Boretzky},
  T.~{Aumann}, D.~{Cortina-Gil}, U.~D. {Pramanik}, T.~W. {Elze}, H.~{Emling},
  H.~{Geissel}, M.~{Hellstr{\"o}m}, K.~L. {Jones}, J.~V. {Kratz}, R.~{Kulessa},
  C.~{Nociforo}, R.~{Palit}, H.~{Simon}, G.~{Sur{\'o}wka}, K.~{S{\"u}mmerer},
  D.~{Vretenar}, W.~{Walu{\'s}}, {Nuclear symmetry energy and neutron skins
  derived from pygmy dipole resonances}, \prc 76 (2007) 051603.

\bibitem{Carbon}
A.~{Carbone}, G.~{Col{\`o}}, A.~{Bracco}, L.-G. {Cao}, P.~F. {Bortignon},
  F.~{Camera}, O.~{Wieland}, {Constraints on the symmetry energy and neutron
  skins from pygmy resonances in $^{68}$Ni and $^{132}$Sn}, \prc 81 (2010)
  041301.
\newblock \href {http://arxiv.org/abs/1003.3580} {\path{arXiv:1003.3580}}.

\bibitem{Daoutidis}
I.~{Daoutidis}, S.~{Goriely}, {Impact of the nuclear symmetry energy on the
  pygmy dipole resonance}, \prc 84 (2011) 027301.

\bibitem{Reinhard1}
P.-G. {Reinhard}, W.~{Nazarewicz}, {Information content of a new observable:
  The case of the nuclear neutron skin}, \prc 81 (2010) 051303.
\newblock \href {http://arxiv.org/abs/1002.4140} {\path{arXiv:1002.4140}}.

\bibitem{Reinhard2}
P.-G. {Reinhard}, W.~{Nazarewicz}, {Information content of the low-energy
  electric dipole strength: Correlation analysis}, \prc 87 (2013) 014324.
\newblock \href {http://arxiv.org/abs/1211.1649} {\path{arXiv:1211.1649}}.

\bibitem{IVGQRpre}
S.~S. {Henshaw}, M.~W. {Ahmed}, G.~{Feldman}, A.~M. {Nathan}, H.~R. {Weller},
  {New Method for Precise Determination of the Isovector Giant Quadrupole
  Resonances in Nuclei}, Physical Review Letters 107 (2011) 222501.

\bibitem{IVGQRskin}
X.~{Roca-Maza}, M.~{Brenna}, B.~K. {Agrawal}, P.~F. {Bortignon}, G.~{Col{\`o}},
  L.-G. {Cao}, N.~{Paar}, D.~{Vretenar}, {Giant quadrupole resonances in
  $^{208}$Pb, the nuclear symmetry energy, and the neutron skin thickness},
  \prc 87 (2013) 034301.
\newblock \href {http://arxiv.org/abs/1212.4377} {\path{arXiv:1212.4377}}.

\bibitem{BM}
A.~Bohr, B.~R. Mottelson, Nuclear structure, vol. ii (1975).

\bibitem{Koura}
H.~{Koura}, T.~{Tachibana}, M.~{Uno}, M.~{Yamada}, {Nuclidic Mass Formula on a
  Spherical Basis with an Improved Even-Odd Term}, Progress of Theoretical
  Physics 113 (2005) 305--325.

\bibitem{XavierEPJA}
X.~{Vi{\~n}as}, M.~{Centelles}, X.~{Roca-Maza}, M.~{Warda}, {Density dependence
  of the symmetry energy from neutron skin thickness in finite nuclei},
  European Physical Journal A 50 (2014) 27.
\newblock \href {http://arxiv.org/abs/1308.1008} {\path{arXiv:1308.1008}}.

\bibitem{XQ2009}
M.~{Warda}, X.~{Vi{\~n}as}, X.~{Roca-Maza}, M.~{Centelles}, {Neutron skin
  thickness in the droplet model with surface width dependence: Indications of
  softness of the nuclear symmetry energy}, \prc 80 (2009) 024316.
\newblock \href {http://arxiv.org/abs/0906.0932} {\path{arXiv:0906.0932}}.

\bibitem{alphaskin}
X.~{Roca-Maza}, X.~{Vi{\~n}as}, M.~{Centelles}, B.~K. {Agrawal}, G.~{Col{\`o}},
  N.~{Paar}, J.~{Piekarewicz}, D.~{Vretenar}, {Neutron skin thickness from the
  measured electric dipole polarizability in $^{68}$Ni$^{120}$Sn and
  $^{208}$Pb}, \prc 92 (2015) 064304.
\newblock \href {http://arxiv.org/abs/1510.01874} {\path{arXiv:1510.01874}}.

\bibitem{skinpyg}
A.~{Klimkiewicz}, N.~{Paar}, P.~{Adrich}, M.~{Fallot}, K.~{Boretzky},
  T.~{Aumann}, D.~{Cortina-Gil}, U.~D. {Pramanik}, T.~W. {Elze}, H.~{Emling},
  H.~{Geissel}, M.~{Hellstr{\"o}m}, K.~L. {Jones}, J.~V. {Kratz}, R.~{Kulessa},
  C.~{Nociforo}, R.~{Palit}, H.~{Simon}, G.~{Sur{\'o}wka}, K.~{S{\"u}mmerer},
  D.~{Vretenar}, W.~{Walu{\'s}}, {Nuclear symmetry energy and neutron skins
  derived from pygmy dipole resonances}, \prc 76 (2007) 051603.

\bibitem{Zuo}
J.~{Dong}, W.~{Zuo}, J.~{Gu}, {Constraints on neutron skin thickness in
  $^{208}$Pb and density-dependent symmetry energy}, \prc 91 (2015) 034315.
\newblock \href {http://arxiv.org/abs/1504.02217} {\path{arXiv:1504.02217}}.

\bibitem{general}
P.-G. {Reinhard}, W.~{Nazarewicz}, {Information content of a new observable:
  The case of the nuclear neutron skin}, \prc 81 (2010) 051303.
\newblock \href {http://arxiv.org/abs/1002.4140} {\path{arXiv:1002.4140}}.

\bibitem{Xavier_el}
X.~{Roca-Maza}, M.~{Centelles}, F.~{Salvat}, X.~{Vi{\~n}as}, {Theoretical study
  of elastic electron scattering off stable and exotic nuclei}, \prc 78 (2008)
  044332.
\newblock \href {http://arxiv.org/abs/0808.1252} {\path{arXiv:0808.1252}}.

\bibitem{Donn}
T.~W. {Donnelly}, J.~{Dubach}, I.~{Sick}, {Isospin dependences in
  parity-violating electron scattering}, Nuclear Physics A 503 (1989) 589--631.

\bibitem{Erl}
J.~{Erler}, A.~{Kurylov}, M.~J. {Ramsey-Musolf}, {Weak charge of the proton and
  new physics}, \prd 68 (2003) 016006.
\newblock \href {http://arxiv.org/abs/hep-ph/0302149}
  {\path{arXiv:hep-ph/0302149}}.

\bibitem{Data}
K.~{Nakamura}, {Particle Data Group}, {Review of Particle Physics}, Journal of
  Physics G Nuclear Physics 37 (2010) 075021.

\bibitem{Hor2001}
C.~J. {Horowitz}, S.~J. {Pollock}, P.~A. {Souder}, R.~{Michaels}, {Parity
  violating measurements of neutron densities}, \prc 63 (2001) 025501.
\newblock \href {http://arxiv.org/abs/nucl-th/9912038}
  {\path{arXiv:nucl-th/9912038}}.

\bibitem{PREXI}
S.~{Abrahamyan}, Z.~{Ahmed}, H.~{Albataineh}, K.~{Aniol}, D.~S. {Armstrong},
  W.~{Armstrong}, T.~{Averett}, B.~{Babineau}, A.~{Barbieri}, V.~{Bellini},
  R.~{Beminiwattha}, J.~{Benesch}, F.~{Benmokhtar}, T.~{Bielarski},
  W.~{Boeglin}, A.~{Camsonne}, M.~{Canan}, P.~{Carter}, G.~D. {Cates},
  C.~{Chen}, J.-P. {Chen}, O.~{Hen}, F.~{Cusanno}, M.~M. {Dalton}, R.~{de Leo},
  K.~{de Jager}, W.~{Deconinck}, P.~{Decowski}, X.~{Deng}, A.~{Deur},
  D.~{Dutta}, A.~{Etile}, D.~{Flay}, G.~B. {Franklin}, M.~{Friend},
  S.~{Frullani}, E.~{Fuchey}, F.~{Garibaldi}, E.~{Gasser}, R.~{Gilman},
  A.~{Giusa}, A.~{Glamazdin}, J.~{Gomez}, J.~{Grames}, C.~{Gu}, O.~{Hansen},
  J.~{Hansknecht}, D.~W. {Higinbotham}, R.~S. {Holmes}, T.~{Holmstrom}, C.~J.
  {Horowitz}, J.~{Hoskins}, J.~{Huang}, C.~E. {Hyde}, F.~{Itard}, C.-M. {Jen},
  E.~{Jensen}, G.~{Jin}, S.~{Johnston}, A.~{Kelleher}, K.~{Kliakhandler}, P.~M.
  {King}, S.~{Kowalski}, K.~S. {Kumar}, J.~{Leacock}, J.~{Leckey}, IV, J.~H.
  {Lee}, J.~J. {Lerose}, R.~{Lindgren}, N.~{Liyanage}, N.~{Lubinsky},
  J.~{Mammei}, F.~{Mammoliti}, D.~J. {Margaziotis}, P.~{Markowitz},
  A.~{McCreary}, D.~{McNulty}, L.~{Mercado}, Z.-E. {Meziani}, R.~W. {Michaels},
  M.~{Mihovilovic}, N.~{Muangma}, C.~{Mu{\~n}oz-Camacho}, S.~{Nanda},
  V.~{Nelyubin}, N.~{Nuruzzaman}, Y.~{Oh}, A.~{Palmer}, D.~{Parno}, K.~D.
  {Paschke}, S.~K. {Phillips}, B.~{Poelker}, R.~{Pomatsalyuk}, M.~{Posik},
  A.~J.~R. {Puckett}, B.~{Quinn}, A.~{Rakhman}, P.~E. {Reimer}, S.~{Riordan},
  P.~{Rogan}, G.~{Ron}, G.~{Russo}, K.~{Saenboonruang}, A.~{Saha},
  B.~{Sawatzky}, A.~{Shahinyan}, R.~{Silwal}, S.~{Sirca}, K.~{Slifer},
  P.~{Solvignon}, P.~A. {Souder}, M.~L. {Sperduto}, R.~{Subedi}, R.~{Suleiman},
  V.~{Sulkosky}, C.~M. {Sutera}, W.~A. {Tobias}, W.~{Troth}, G.~M. {Urciuoli},
  B.~{Waidyawansa}, D.~{Wang}, J.~{Wexler}, R.~{Wilson}, B.~{Wojtsekhowski},
  X.~{Yan}, H.~{Yao}, Y.~{Ye}, Z.~{Ye}, V.~{Yim}, L.~{Zana}, X.~{Zhan},
  J.~{Zhang}, Y.~{Zhang}, X.~{Zheng}, P.~{Zhu}, {Measurement of the Neutron
  Radius of $^{208}$Pb through Parity Violation in Electron Scattering},
  Physical Review Letters 108 (2012) 112502.
\newblock \href {http://arxiv.org/abs/1201.2568} {\path{arXiv:1201.2568}}.

\bibitem{XavierPREX}
X.~{Roca-Maza}, M.~{Centelles}, X.~{Vi{\~n}as}, M.~{Warda}, {Neutron Skin of
  $^{208}$Pb, Nuclear Symmetry Energy, and the Parity Radius Experiment},
  Physical Review Letters 106 (2011) 252501.
\newblock \href {http://arxiv.org/abs/1103.1762} {\path{arXiv:1103.1762}}.

\bibitem{Modal}
C.~{Mondal}, B.~K. {Agrawal}, J.~N. {De}, {Constraining the symmetry energy
  content of nuclear matter from nuclear masses: A covariance analysis}, \prc
  92 (2015) 024302.
\newblock \href {http://arxiv.org/abs/1507.05384} {\path{arXiv:1507.05384}}.

\bibitem{GL_cov}
X.~{Roca-Maza}, N.~{Paar}, G.~{Col{\`o}}, {Covariance analysis for energy
  density functionals and instabilities}, Journal of Physics G Nuclear Physics
  42 (2015) 034033.
\newblock \href {http://arxiv.org/abs/1406.1885} {\path{arXiv:1406.1885}}.

\bibitem{Naz}
P.-G. {Reinhard}, W.~{Nazarewicz}, {Information content of a new observable:
  The case of the nuclear neutron skin}, \prc 81 (2010) 051303.
\newblock \href {http://arxiv.org/abs/1002.4140} {\path{arXiv:1002.4140}}.

\bibitem{book_Bert}
X.~{Roca-Maza}, J.~{Piekarewicz}, T.~{Garcia-Galvez}, M.~{Centelles}, Influence
  of the nuclear symmetry energy on the structure and composition of the outer
  crust, Neutron Star Crust, eds. C. Bertulani \& J. Piekarewicz (Hauppage, NY:
  Nova Science Publishers) 265 (2012) 50.

\bibitem{Bert}
M.~{Baldo}, E.~{Saperstein}, The structure of the neutron star crust within a
  semi-microscopic energy density functional method, Neutron Star Crust, eds.
  C. Bertulani \& J. Piekarewicz (Hauppage, NY: Nova Science Publishers) 265
  (2012) 30.

\bibitem{Bao}
S.~S. {Bao}, J.~N. {Hu}, Z.~W. {Zhang}, H.~{Shen}, {Effects of the symmetry
  energy on properties of neutron star crusts near the neutron drip density},
  \prc 90 (2014) 045802.
\newblock \href {http://arxiv.org/abs/1410.0851} {\path{arXiv:1410.0851}}.

\bibitem{dripacc}
A.~F. {Fantina}, N.~{Chamel}, Y.~D. {Mutafchieva}, Z.~K. {Stoyanov}, L.~M.
  {Mihailov}, R.~L. {Pavlov}, {Role of the symmetry energy on the neutron-drip
  transition in accreting and nonaccreting neutron stars}, \prc 93 (2016)
  015801.

\bibitem{rhot}
I.~{Vida{\~n}a}, C.~{Provid{\^e}ncia}, A.~{Polls}, A.~{Rios}, {Density
  dependence of the nuclear symmetry energy: A microscopic perspective}, \prc
  80 (2009) 045806.
\newblock \href {http://arxiv.org/abs/0907.1165} {\path{arXiv:0907.1165}}.

\bibitem{Bert_t}
D.~P. {Menezes}, S.~D. {Avancini}, C.~{Provid{\^e}ncia}, M.~D. {Alloy}, The
  inner crust and its structure, Neutron Star Crust, eds. C. Bertulani \& J.
  Piekarewicz (Hauppage, NY: Nova Science Publishers) 265 (2012) 100.

\bibitem{Avan}
S.~S. {Avancini}, L.~{Brito}, P.~{Chomaz}, D.~P. {Menezes},
  C.~{Provid{\^e}ncia}, {Spinodal instabilities and the distillation effect in
  relativistic hadronic models}, \prc 74 (2006) 024317.

\bibitem{Jirina_pasta}
H.~{Pais}, J.~R. {Stone}, {Exploring the Nuclear Pasta Phase in Core-Collapse
  Supernova Matter}, Physical Review Letters 109 (2012) 151101.

\bibitem{Hor69}
C.~J. {Horowitz}, M.~A. {P{\'e}rez-Garc{\'{\i}}a}, J.~{Piekarewicz},
  {Neutrino-``pasta'' scattering: The opacity of nonuniform neutron-rich
  matter}, \prc 69 (2004) 045804.
\newblock \href {http://arxiv.org/abs/astro-ph/0401079}
  {\path{arXiv:astro-ph/0401079}}.

\bibitem{Hor70}
C.~J. {Horowitz}, M.~A. {P{\'e}rez-Garc{\'{\i}}a}, J.~{Carriere}, D.~K.
  {Berry}, J.~{Piekarewicz}, {Nonuniform neutron-rich matter and coherent
  neutrino scattering}, \prc 70 (2004) 065806.
\newblock \href {http://arxiv.org/abs/astro-ph/0409296}
  {\path{arXiv:astro-ph/0409296}}.

\bibitem{Hor78}
C.~J. {Horowitz}, D.~K. {Berry}, {Shear viscosity and thermal conductivity of
  nuclear ``pasta''}, \prc 78 (2008) 035806.
\newblock \href {http://arxiv.org/abs/0807.2603} {\path{arXiv:0807.2603}}.

\bibitem{Hor88}
A.~S. {Schneider}, C.~J. {Horowitz}, J.~{Hughto}, D.~K. {Berry}, {Nuclear
  ``pasta'' formation}, \prc 88 (2013) 065807.
\newblock \href {http://arxiv.org/abs/1307.1678} {\path{arXiv:1307.1678}}.

\bibitem{HorMAD}
I.~{Sagert}, G.~I. {Fann}, F.~J. {Fattoyev}, S.~{Postnikov}, C.~J. {Horowitz},
  {Quantum simulations of nuclei and nuclear pasta with the multiresolution
  adaptive numerical environment for scientific simulations}, \prc 93 (2016)
  055801.
\newblock \href {http://arxiv.org/abs/1509.06671} {\path{arXiv:1509.06671}}.

\bibitem{Fuchs}
C.~{Fuchs}, {Kaon production in heavy ion reactions at intermediate energies},
  Progress in Particle and Nuclear Physics 56 (2006) 1--103.
\newblock \href {http://arxiv.org/abs/nucl-th/0507017}
  {\path{arXiv:nucl-th/0507017}}.

\bibitem{Steiner}
A.~W. {Steiner}, A.~L. {Watts}, {Constraints on Neutron Star Crusts from
  Oscillations in Giant Flares}, Physical Review Letters 103 (2009) 181101.
\newblock \href {http://arxiv.org/abs/0902.1683} {\path{arXiv:0902.1683}}.

\bibitem{accr}
A.~W. {Steiner}, {Deep crustal heating in a multicomponent accreted neutron
  star crust}, \prc 85 (2012) 055804.
\newblock \href {http://arxiv.org/abs/1202.3378} {\path{arXiv:1202.3378}}.

\bibitem{thermal}
S.~{Abbar}, J.~{Carlson}, H.~{Duan}, S.~{Reddy}, {Quantum Monte Carlo
  calculations of the thermal conductivity of neutron star crusts}, \prc 92
  (2015) 045809.
\newblock \href {http://arxiv.org/abs/1503.01696} {\path{arXiv:1503.01696}}.

\bibitem{Shap}
S.~Shapiro, S.~Teukolsky, Black Holes, White Dwarfs and Neutron Stars: The
  Physics of Compact Objects, Wiley, 2008.

\bibitem{MREOS}
A.~W. {Steiner}, J.~M. {Lattimer}, E.~F. {Brown}, {The Equation of State from
  Observed Masses and Radii of Neutron Stars}, \apj 722 (2010) 33--54.
\newblock \href {http://arxiv.org/abs/1005.0811} {\path{arXiv:1005.0811}}.

\bibitem{MREOS1}
A.~W. {Steiner}, J.~M. {Lattimer}, E.~F. {Brown}, {The Neutron Star Mass-Radius
  Relation and the Equation of State of Dense Matter}, \apjl 765 (2013) L5.
\newblock \href {http://arxiv.org/abs/1205.6871} {\path{arXiv:1205.6871}}.

\bibitem{DanielHI}
P.~{Danielewicz}, R.~{Lacey}, W.~G. {Lynch}, {Determination of the Equation of
  State of Dense Matter}, Science 298 (2002) 1592--1596.
\newblock \href {http://arxiv.org/abs/nucl-th/0208016}
  {\path{arXiv:nucl-th/0208016}}.

\bibitem{n_emis}
D.~G. {Yakovlev}, A.~D. {Kaminker}, O.~Y. {Gnedin}, P.~{Haensel}, {Neutrino
  emission from neutron stars}, \physrep 354 (2001) 1--155.
\newblock \href {http://arxiv.org/abs/astro-ph/0012122}
  {\path{arXiv:astro-ph/0012122}}.

\bibitem{Kamin}
A.~D. {Kaminker}, P.~{Haensel}, D.~G. {Yakovlev}, {Nucleon superfluidity vs.
  observations of cooling neutron stars}, \aap 373 (2001) L17--L20.
\newblock \href {http://arxiv.org/abs/astro-ph/0105047}
  {\path{arXiv:astro-ph/0105047}}.

\bibitem{Usp}
D.~G. {Yakovlev}, K.~P. {Levenfish}, Y.~A. {Shibanov}, {Reviews of Topical
  Problems: Cooling of neutron stars and superfluidity in their cores}, Physics
  Uspekhi 42 (1999) 737--778.
\newblock \href {http://arxiv.org/abs/astro-ph/9906456}
  {\path{arXiv:astro-ph/9906456}}.

\bibitem{UJ}
U.~{Lombardo}, H.-J. {Schulze}, {Superfluidity in Neutron Star Matter}, in:
  D.~{Blaschke}, N.~K. {Glendenning}, A.~{Sedrakian} (Eds.), Physics of Neutron
  Star Interiors, Vol. 578 of Lecture Notes in Physics, Berlin Springer Verlag,
  2001, p.~30.
\newblock \href {http://arxiv.org/abs/astro-ph/0012209}
  {\path{arXiv:astro-ph/0012209}}.

\bibitem{BCS50}
D.~{Page}, {Pairing and the Cooling of Neutron Stars}, in: R.~A. {Broglia}, {et
  al.} (Eds.), Fifty Years of Nuclear BCS: Pairing in Finite Systems, World
  Scientific Publishing Co, 2013, pp. 324--337.

\bibitem{Noimnras}
G.~{Taranto}, G.~F. {Burgio}, H.-J. {Schulze}, {Cassiopeia A and direct Urca
  cooling}, \mnras 456 (2016) 1451--1458.
\newblock \href {http://arxiv.org/abs/1511.04243} {\path{arXiv:1511.04243}}.

\bibitem{stellar}
D.~{Page}, J.~M. {Lattimer}, M.~{Prakash}, A.~W. {Steiner}, {Stellar
  Superfluids}, ArXiv e-prints\href {http://arxiv.org/abs/1302.6626}
  {\path{arXiv:1302.6626}}.

\bibitem{Anto}
J.~{Antoniadis}, P.~C.~C. {Freire}, N.~{Wex}, T.~M. {Tauris}, R.~S. {Lynch},
  M.~H. {van Kerkwijk}, M.~{Kramer}, C.~{Bassa}, V.~S. {Dhillon}, T.~{Driebe},
  J.~W.~T. {Hessels}, V.~M. {Kaspi}, V.~I. {Kondratiev}, N.~{Langer}, T.~R.
  {Marsh}, M.~A. {McLaughlin}, T.~T. {Pennucci}, S.~M. {Ransom}, I.~H.
  {Stairs}, J.~{van Leeuwen}, J.~P.~W. {Verbiest}, D.~G. {Whelan}, {A Massive
  Pulsar in a Compact Relativistic Binary}, Science 340 (2013) 448.
\newblock \href {http://arxiv.org/abs/1304.6875} {\path{arXiv:1304.6875}}.

\bibitem{Demo}
P.~B. {Demorest}, T.~{Pennucci}, S.~M. {Ransom}, M.~S.~E. {Roberts}, J.~W.~T.
  {Hessels}, {A two-solar-mass neutron star measured using Shapiro delay}, \nat
  467 (2010) 1081--1083.
\newblock \href {http://arxiv.org/abs/1010.5788} {\path{arXiv:1010.5788}}.

\bibitem{Tara2013}
G.~{Taranto}, M.~{Baldo}, G.~F. {Burgio}, {Selecting microscopic equations of
  state}, \prc 87 (2013) 045803.
\newblock \href {http://arxiv.org/abs/1302.6882} {\path{arXiv:1302.6882}}.

\bibitem{Ken}
K.~{Fukukawa}, M.~{Baldo}, G.~F. {Burgio}, L.~{Lo Monaco}, H.-J. {Schulze},
  {Nuclear matter equation of state from a quark-model nucleon-nucleon
  interaction}, \prc 92 (2015) 065802.
\newblock \href {http://arxiv.org/abs/1507.07288} {\path{arXiv:1507.07288}}.

\bibitem{SETBF}
S.~{Gandolfi}, J.~{Carlson}, S.~{Reddy}, {Maximum mass and radius of neutron
  stars, and the nuclear symmetry energy}, \prc 85 (2012) 032801.
\newblock \href {http://arxiv.org/abs/1101.1921} {\path{arXiv:1101.1921}}.

\bibitem{Rad2001}
J.~M. {Lattimer}, M.~{Prakash}, {Nuclear matter and its role in supernovae,
  neutron stars and compact object binary mergers}, \physrep 333 (2000)
  121--146.
\newblock \href {http://arxiv.org/abs/astro-ph/0002203}
  {\path{arXiv:astro-ph/0002203}}.

\bibitem{PAL}
M.~{Prakash}, J.~M. {Lattimer}, T.~L. {Ainsworth}, {Equation of state and the
  maximum mass of neutron stars}, Physical Review Letters 61 (1988) 2518--2521.

\bibitem{RadApJ}
J.~M. {Lattimer}, M.~{Prakash}, {Neutron Star Structure and the Equation of
  State}, \apj 550 (2001) 426--442.
\newblock \href {http://arxiv.org/abs/astro-ph/0002232}
  {\path{arXiv:astro-ph/0002232}}.

\bibitem{BaoPLB}
B.-A. {Li}, A.~W. {Steiner}, {Constraining the radii of neutron stars with
  terrestrial nuclear laboratory data}, Physics Letters B 642 (2006) 436--440.
\newblock \href {http://arxiv.org/abs/nucl-th/0511064}
  {\path{arXiv:nucl-th/0511064}}.

\bibitem{Rcrust}
M.~{Baldo}, G.~F. {Burgio}, M.~{Centelles}, B.~K. {Sharma}, X.~{Vi{\~n}as},
  {From the crust to the core of neutron stars on a microscopic basis}, Physics
  of Atomic Nuclei 77 (2014) 1157--1165.
\newblock \href {http://arxiv.org/abs/1308.2304} {\path{arXiv:1308.2304}}.

\bibitem{RadGR}
S.~{Guillot}, R.~E. {Rutledge}, {Rejecting Proposed Dense Matter Equations of
  State with Quiescent Low-mass X-Ray Binaries}, \apjl 796 (2014) L3.
\newblock \href {http://arxiv.org/abs/1409.4306} {\path{arXiv:1409.4306}}.

\bibitem{RadGO}
T.~{G{\"u}ver}, F.~{{\"O}zel}, {The Mass and the Radius of the Neutron Star in
  the Transient Low-mass X-Ray Binary SAX J1748.9-2021}, \apjl 765 (2013) L1.
\newblock \href {http://arxiv.org/abs/1301.0831} {\path{arXiv:1301.0831}}.

\bibitem{RadLS2014}
J.~M. {Lattimer}, A.~W. {Steiner}, {Neutron Star Masses and Radii from
  Quiescent Low-mass X-Ray Binaries}, \apj 784 (2014) 123.
\newblock \href {http://arxiv.org/abs/1305.3242} {\path{arXiv:1305.3242}}.

\bibitem{maxmass}
N.~{Chamel}, P.~{Haensel}, J.~L. {Zdunik}, A.~F. {Fantina}, {On the Maximum
  Mass of Neutron Stars}, International Journal of Modern Physics E 22 (2013)
  1330018.
\newblock \href {http://arxiv.org/abs/1307.3995} {\path{arXiv:1307.3995}}.

\bibitem{Janka}
H.-T. {Janka}, {Explosion Mechanisms of Core-Collapse Supernovae}, Annual
  Review of Nuclear and Particle Science 62 (2012) 407--451.
\newblock \href {http://arxiv.org/abs/1206.2503} {\path{arXiv:1206.2503}}.

\bibitem{FishEPJA}
T.~{Fischer}, M.~{Hempel}, I.~{Sagert}, Y.~{Suwa}, J.~{Schaffner-Bielich},
  {Symmetry energy impact in simulations of core-collapse supernovae}, European
  Physical Journal A 50 (2014) 46.
\newblock \href {http://arxiv.org/abs/1307.6190} {\path{arXiv:1307.6190}}.

\bibitem{AAMarek}
A.~{Marek}, H.-T. {Janka}, E.~{M{\"u}ller}, {Equation-of-state dependent
  features in shock-oscillation modulated neutrino and gravitational-wave
  signals from supernovae}, \aap 496 (2009) 475--494.
\newblock \href {http://arxiv.org/abs/0808.4136} {\path{arXiv:0808.4136}}.

\bibitem{ApJSuwa}
Y.~{Suwa}, T.~{Takiwaki}, K.~{Kotake}, T.~{Fischer}, M.~{Liebend{\"o}rfer},
  K.~{Sato}, {On the Importance of the Equation of State for the
  Neutrino-driven Supernova Explosion Mechanism}, \apj 764 (2013) 99.
\newblock \href {http://arxiv.org/abs/1206.6101} {\path{arXiv:1206.6101}}.

\bibitem{ApJCouch}
S.~M. {Couch}, {The Dependence of the Neutrino Mechanism of Core-collapse
  Supernovae on the Equation of State}, \apj 765 (2013) 29.
\newblock \href {http://arxiv.org/abs/1206.4724} {\path{arXiv:1206.4724}}.

\bibitem{SNNS}
A.~W. {Steiner}, M.~{Hempel}, T.~{Fischer}, {Core-collapse Supernova Equations
  of State Based on Neutron Star Observations}, \apj 774 (2013) 17.
\newblock \href {http://arxiv.org/abs/1207.2184} {\path{arXiv:1207.2184}}.

\bibitem{Shen}
H.~{Shen}, H.~{Toki}, K.~{Oyamatsu}, K.~{Sumiyoshi}, {Relativistic equation of
  state of nuclear matter for supernova and neutron star}, Nuclear Physics A
  637 (1998) 435--450.
\newblock \href {http://arxiv.org/abs/nucl-th/9805035}
  {\path{arXiv:nucl-th/9805035}}.

\bibitem{HS}
M.~{Hempel}, J.~{Schaffner-Bielich}, {A statistical model for a complete
  supernova equation of state}, Nuclear Physics A 837 (2010) 210--254.
\newblock \href {http://arxiv.org/abs/0911.4073} {\path{arXiv:0911.4073}}.

\bibitem{nsuper}
K.~S. {Hirata}, T.~{Kajita}, M.~{Koshiba}, M.~{Nakahata}, Y.~{Oyama},
  N.~{Sato}, A.~{Suzuki}, M.~{Takita}, Y.~{Totsuka}, T.~{Kifune}, T.~{Suda},
  K.~{Takahashi}, T.~{Tanimori}, K.~{Miyano}, M.~{Yamada}, E.~W. {Beier}, L.~R.
  {Feldscher}, W.~{Frati}, S.~B. {Kim}, A.~K. {Mann}, F.~M. {Newcomer}, R.~{van
  Berg}, W.~{Zhang}, B.~G. {Cortez}, {Observation in the Kamiokande-II detector
  of the neutrino burst from supernova SN1987A}, \prd 38 (1988) 448--458.

\bibitem{nsignal}
L.~F. {Roberts}, G.~{Shen}, V.~{Cirigliano}, J.~A. {Pons}, S.~{Reddy}, S.~E.
  {Woosley}, {Protoneutron Star Cooling with Convection: The Effect of the
  Symmetry Energy}, Physical Review Letters 108 (2012) 061103.
\newblock \href {http://arxiv.org/abs/1112.0335} {\path{arXiv:1112.0335}}.

\bibitem{Wilson}
J.~R. {Wilson}, R.~W. {Mayle}, {Convection in core collapse supernovae.},
  \physrep 163 (1988) 63--77.

\bibitem{GW}
B.~P. {Abbott}, R.~{Abbott}, T.~D. {Abbott}, M.~R. {Abernathy}, F.~{Acernese},
  K.~{Ackley}, C.~{Adams}, T.~{Adams}, P.~{Addesso}, R.~X. {Adhikari}, et~al.,
  {Observation of Gravitational Waves from a Binary Black Hole Merger},
  Physical Review Letters 116 (2016) 061102.
\newblock \href {http://arxiv.org/abs/1602.03837} {\path{arXiv:1602.03837}}.

\bibitem{GW2}
B.~P. Abbott, R.~Abbott, T.~D. Abbott, M.~R. Abernathy, F.~Acernese, et~al.,
  Gw151226: Observation of gravitational waves from a 22-solar-mass binary
  black hole coalescence, Phys. Rev. Lett. 116 (2016) 241103.

\bibitem{GWPRD90}
A.~{Bauswein}, N.~{Stergioulas}, H.-T. {Janka}, {Revealing the high-density
  equation of state through binary neutron star mergers}, \prd 90 (2014)
  023002.
\newblock \href {http://arxiv.org/abs/1403.5301} {\path{arXiv:1403.5301}}.

\bibitem{2009astro}
B.~J. {Owen}, D.~H. {Reitze}, S.~E. {Whitcomb}, {Probing neutron stars with
  gravitational waves}, in: astro2010: The Astronomy and Astrophysics Decadal
  Survey, Vol. 2010 of Astronomy, 2009, pp. 1--8.
\newblock \href {http://arxiv.org/abs/0903.2603} {\path{arXiv:0903.2603}}.

\bibitem{2004GW_Roma}
O.~{Benhar}, V.~{Ferrari}, L.~{Gualtieri}, {Gravitational wave asteroseismology
  reexamined}, \prd 70 (2004) 124015.
\newblock \href {http://arxiv.org/abs/astro-ph/0407529}
  {\path{arXiv:astro-ph/0407529}}.

\bibitem{2011GW_Rev}
N.~{Andersson}, V.~{Ferrari}, D.~I. {Jones}, K.~D. {Kokkotas}, B.~{Krishnan},
  J.~S. {Read}, L.~{Rezzolla}, B.~{Zink}, {Gravitational waves from neutron
  stars: promises and challenges}, General Relativity and Gravitation 43 (2011)
  409--436.
\newblock \href {http://arxiv.org/abs/0912.0384} {\path{arXiv:0912.0384}}.

\bibitem{1998MNRAS.299}
N.~{Andersson}, K.~D. {Kokkotas}, {Towards gravitational wave
  asteroseismology}, \mnras 299 (1998) 1059--1068.
\newblock \href {http://arxiv.org/abs/gr-qc/9711088}
  {\path{arXiv:gr-qc/9711088}}.

\bibitem{apr}
A.~{Akmal}, V.~R. {Pandharipande}, D.~G. {Ravenhall}, {Equation of state of
  nucleon matter and neutron star structure}, \prc 58 (1998) 1804--1828.
\newblock \href {http://arxiv.org/abs/nucl-th/9804027}
  {\path{arXiv:nucl-th/9804027}}.

\bibitem{BBS1}
M.~{Baldo}, G.~F. {Burgio}, H.-J. {Schulze}, {Hyperon stars in the
  Brueckner-Bethe-Goldstone theory}, \prc 61 (2000) 055801.
\newblock \href {http://arxiv.org/abs/nucl-th/9912066}
  {\path{arXiv:nucl-th/9912066}}.

\bibitem{Glen}
N.~K. {Glendenning} (Ed.), {Compact stars : nuclear physics, particle physics,
  and general relativity}, 2000.

\bibitem{aprb}
A.~{Akmal}, V.~R. {Pandharipande}, {Spin-isospin structure and pion
  condensation in nucleon matter}, \prc 56 (1997) 2261--2279.
\newblock \href {http://arxiv.org/abs/nucl-th/9705013}
  {\path{arXiv:nucl-th/9705013}}.

\bibitem{Rubino}
O.~{Benhar}, R.~{Rubino}, {Stability of the mixed phase in hybrid stars}, \aap
  434 (2005) 247--256.
\newblock \href {http://arxiv.org/abs/astro-ph/0410376}
  {\path{arXiv:astro-ph/0410376}}.

\bibitem{DeyBomb}
M.~{Dey}, I.~{Bombaci}, J.~{Dey}, S.~{Ray}, B.~C. {Samanta}, {Strange stars
  with realistic quark vector interaction and phenomenological
  density-dependent scalar potential}, Physics Letters B 438 (1998) 123--128.
\newblock \href {http://arxiv.org/abs/astro-ph/9810065}
  {\path{arXiv:astro-ph/9810065}}.

\bibitem{RepPP}
M.~{Baldo}, G.~F. {Burgio}, {Properties of the nuclear medium}, Reports on
  Progress in Physics 75 (2012) 026301.
\newblock \href {http://arxiv.org/abs/1102.1364} {\path{arXiv:1102.1364}}.

\bibitem{pns_hot}
G.~F. {Burgio}, V.~{Ferrari}, L.~{Gualtieri}, H.-J. {Schulze}, {Oscillations of
  hot, young neutron stars: Gravitational wave frequencies and damping times},
  \prd 84 (2011) 044017.
\newblock \href {http://arxiv.org/abs/1106.2736} {\path{arXiv:1106.2736}}.

\bibitem{fermin}
V.~{Ferrari}, G.~{Miniutti}, J.~A. {Pons}, {Gravitational waves from newly
  born, hot neutron stars}, \mnras 342 (2003) 629--638.
\newblock \href {http://arxiv.org/abs/astro-ph/0210581}
  {\path{arXiv:astro-ph/0210581}}.

\bibitem{2015Rez}
K.~{Takami}, L.~{Rezzolla}, L.~{Baiotti}, {Spectral properties of the
  post-merger gravitational-wave signal from binary neutron stars}, \prd 91
  (2015) 064001.
\newblock \href {http://arxiv.org/abs/1412.3240} {\path{arXiv:1412.3240}}.

\bibitem{2013Fatto}
F.~J. {Fattoyev}, J.~{Carvajal}, W.~G. {Newton}, B.-A. {Li}, {Constraining the
  high-density behavior of the nuclear symmetry energy with the tidal
  polarizability of neutron stars}, \prc 87 (2013) 015806.
\newblock \href {http://arxiv.org/abs/1210.3402} {\path{arXiv:1210.3402}}.

\bibitem{2010Duez}
M.~D. {Duez}, {Numerical relativity confronts compact neutron star binaries: a
  review and status report}, Classical and Quantum Gravity 27 (2010) 114002.
\newblock \href {http://arxiv.org/abs/0912.3529} {\path{arXiv:0912.3529}}.

\bibitem{2009Faber}
J.~{Faber}, {Status of neutron star-black hole and binary neutron star
  simulations}, Classical and Quantum Gravity 26 (2009) 114004.

\bibitem{2010Abadie}
J.~{Abadie}, B.~P. {Abbott}, R.~{Abbott}, M.~{Abernathy}, T.~{Accadia},
  F.~{Acernese}, C.~{Adams}, R.~{Adhikari}, P.~{Ajith}, B.~{Allen}, et~al.,
  {TOPICAL REVIEW: Predictions for the rates of compact binary coalescences
  observable by ground-based gravitational-wave detectors}, Classical and
  Quantum Gravity 27 (2010) 173001.
\newblock \href {http://arxiv.org/abs/1003.2480} {\path{arXiv:1003.2480}}.

\bibitem{2012PRLBaus}
A.~{Bauswein}, H.-T. {Janka}, {Measuring Neutron-Star Properties via
  Gravitational Waves from Neutron-Star Mergers}, Physical Review Letters 108
  (2012) 011101.
\newblock \href {http://arxiv.org/abs/1106.1616} {\path{arXiv:1106.1616}}.

\bibitem{2012PRDBaus}
A.~{Bauswein}, H.-T. {Janka}, K.~{Hebeler}, A.~{Schwenk}, {Equation-of-state
  dependence of the gravitational-wave signal from the ring-down phase of
  neutron-star mergers}, \prd 86 (2012) 063001.
\newblock \href {http://arxiv.org/abs/1204.1888} {\path{arXiv:1204.1888}}.

\bibitem{2014Fatto}
F.~J. {Fattoyev}, W.~G. {Newton}, B.-A. {Li}, {Probing the high-density
  behavior of symmetry energy with gravitational waves}, European Physical
  Journal A 50 (2014) 45.
\newblock \href {http://arxiv.org/abs/1309.5153} {\path{arXiv:1309.5153}}.

\bibitem{2008Hinder}
{\'E}.~{\'E}. {Flanagan}, T.~{Hinderer}, {Constraining neutron-star tidal Love
  numbers with gravitational-wave detectors}, \prd 77 (2008) 021502.
\newblock \href {http://arxiv.org/abs/0709.1915} {\path{arXiv:0709.1915}}.

\bibitem{2008ApJHind}
T.~{Hinderer}, {Tidal Love Numbers of Neutron Stars}, \apj 677 (2008)
  1216--1220.
\newblock \href {http://arxiv.org/abs/0711.2420} {\path{arXiv:0711.2420}}.

\bibitem{kaos}
D.~{Mi{\'s}kowiec}, W.~{Ahner}, R.~{Barth}, M.~{Cie{\'s}lak}, M.~{D{\c
  e}bowski}, E.~{Grosse}, W.~{Henning}, P.~{Koczo{\'n}}, R.~{Schicker},
  E.~{Schwab}, P.~{Senger}, P.~{Baltes}, C.~{M{\"u}ntz}, H.~{Oeschler},
  S.~{Sartorius}, C.~{Sturm}, A.~{Wagner}, P.~{Beckerle}, C.~{Bormann},
  D.~{Brill}, Y.~{Shin}, J.~{Stein}, R.~{Stock}, H.~{Str{\"o}bele},
  B.~{Kohlmeyer}, H.~{P{\"o}ppl}, F.~{P{\"u}hlhofer}, J.~{Speer},
  K.~{V{\"o}lkel}, W.~{Walu{\'s}}, {Observation of enhanced subthreshold
  K$^{+}$ production in central collisions between heavy nuclei}, Physical
  Review Letters 72 (1994) 3650--3653.

\bibitem{fopi}
J.~L. {Ritman}, N.~{Herrmann}, D.~{Best}, J.~P. {Alard}, V.~{Amouroux},
  N.~{Bastid}, I.~{Belyaev}, L.~{Berger}, J.~{Biegansky}, A.~{Buta}, R.~{{\v
  C}aplar}, N.~{Cindro}, J.~P. {Coffin}, P.~{Crochet}, R.~{Dona}, P.~{Dupieux},
  M.~{Dzelalija}, P.~{Fintz}, Z.~{Fodor}, A.~{Genoux-Lubain}, A.~{Gobbi},
  G.~{Goebels}, G.~{Guillaume}, Y.~{Grigorian}, E.~{H{\"a}fele}, K.~D.
  {Hildenbrand}, S.~{H{\"o}lbling}, F.~{Jundt}, J.~{Kecskemeti},
  M.~{Kirejczyk}, Y.~{Korchagin}, R.~{Kotte}, C.~{Kuhn}, D.~{Lambrecht},
  A.~{Lebedev}, I.~{Legrand}, Y.~{Leifels}, C.~{Maazouzi}, V.~{Manko},
  T.~{Matulewicz}, J.~{M{\"o}sner}, S.~{Mohren}, D.~{Moisa}, W.~{Neubert},
  D.~{Pelte}, M.~{Petrovici}, C.~{Pinkenburg}, F.~{Rami}, V.~{Ramillien},
  W.~{Reisdorf}, C.~{Roy}, D.~{Sch{\"u}ll}, Z.~{Seres}, B.~{Sikora},
  V.~{Simion}, K.~{Siwek-Wilczy{\'n}ska}, V.~{Smolyankin}, U.~{Sodan},
  L.~{Tizniti}, M.~{Trzaska}, M.~A. {Vasiliev}, P.~{Wagner}, G.~S. {Wang},
  T.~{Wienold}, D.~{Wohlfarth}, A.~{Zhilin}, {On the transverse momentum
  distribution of strange hadrons produced in relativistic heavy ion
  collisions}, Zeitschrift fur Physik A Hadrons and Nuclei 352 (1995) 355--357.
\newblock \href {http://arxiv.org/abs/nucl-ex/9506002}
  {\path{arXiv:nucl-ex/9506002}}.

\bibitem{BombaciPRC55}
I.~{Bombaci}, {Observational evidence for strange matter in compact objects
  from the x-ray burster 4U 1820-30}, \prc 55 (1997) 1587--1590.

\bibitem{yoshidaPRC58}
S.~{Yoshida}, H.~{Sagawa}, N.~{Takigawa}, {Incompressibility and density
  distributions in asymmetric nuclear systems}, \prc 58 (1998) 2796--2806.

\bibitem{LeePRC57}
C.-H. {Lee}, T.~T.~S. {Kuo}, G.~Q. {Li}, G.~E. {Brown}, {Nuclear symmetry
  energy}, \prc 57 (1998) 3488--3491.

\bibitem{vau3}
D.~{Vautherin}, D.~M. {Brink}, {Hartree-Fock Calculations with Skyrme's
  Interaction. I. Spherical Nuclei}, \prc 5 (1972) 626--647.

\bibitem{Col1998}
M.~{Colonna}, M.~{di Toro}, A.~B. {Larionov}, {Collective modes in asymmetric
  nuclear matter}, Physics Letters B 428 (1998) 1--7.

\bibitem{ber_88}
G.~F. {Bertsch}, S.~{Das Gupta}, {A guide to microscopic models for
  intermediate energy heavy ion collisions}, \physrep 160 (1988) 189--233.

\bibitem{dan_00}
P.~{Danielewicz}, {Determination of the mean-field momentum-dependence using
  elliptic flow}, Nuclear Physics A 673 (2000) 375--410.
\newblock \href {http://arxiv.org/abs/nucl-th/9912027}
  {\path{arXiv:nucl-th/9912027}}.

\bibitem{fmd}
H.~{Feldmeier}, {Fermionic molecular dynamics}, Nuclear Physics A 515 (1990)
  147--172.

\bibitem{amd_1}
A.~{Ono}, H.~{Horiuchi}, {Antisymmetrized molecular dynamics of wave packets
  with stochastic incorporation of the Vlasov equation}, \prc 53 (1996)
  2958--2972.
\newblock \href {http://arxiv.org/abs/nucl-th/9601008}
  {\path{arXiv:nucl-th/9601008}}.

\bibitem{amd_2}
A.~{Ono}, {Antisymmetrized molecular dynamics with quantum branching processes
  for collisions of heavy nuclei}, \prc 59 (1999) 853--864.
\newblock \href {http://arxiv.org/abs/nucl-th/9809029}
  {\path{arXiv:nucl-th/9809029}}.

\bibitem{amd_3}
A.~{Ono}, S.~{Hudan}, A.~{Chbihi}, J.~D. {Frankland}, {Compatibility of
  localized wave packets and unrestricted single particle dynamics for cluster
  formation in nuclear collisions}, \prc 66 (2002) 014603.
\newblock \href {http://arxiv.org/abs/nucl-th/0204005}
  {\path{arXiv:nucl-th/0204005}}.

\bibitem{comd}
M.~{Papa}, T.~{Maruyama}, A.~{Bonasera}, {Constrained molecular dynamics
  approach to fermionic systems}, \prc 64 (2001) 024612.
\newblock \href {http://arxiv.org/abs/nucl-th/0012083}
  {\path{arXiv:nucl-th/0012083}}.

\bibitem{ImQMD}
N.~{Wang}, Z.~{Li}, X.~{Wu}, {Improved quantum molecular dynamics model and its
  applications to fusion reaction near barrier}, \prc 65 (2002) 064608.
\newblock \href {http://arxiv.org/abs/nucl-th/0201079}
  {\path{arXiv:nucl-th/0201079}}.

\bibitem{QMD}
J.~{Aichelin}, {"Quantum" molecular dynamics-a dynamical microscopic n-body
  approach to investigate fragment formation and the nuclear equation of state
  in heavy ion collisions}, \physrep 202 (1991) 233--360.

\bibitem{BNV}
H.~{Kruse}, B.~V. {Jacak}, J.~J. {Molitoris}, G.~D. {Westfall},
  H.~{St{\"o}cker}, {Vlasov-Uehling-Uhlenbeck theory of medium energy heavy ion
  reactions: Role of mean field dynamics and two body collisions}, \prc 31
  (1985) 1770--1774.

\bibitem{SMF}
M.~{Colonna}, M.~{Di Toro}, A.~{Guarnera}, S.~{Maccarone},
  M.~{Zielinska-Pfab{\'e}}, H.~H. {Wolter}, {Fluctuations and dynamical
  instabilities in heavy-ion reactions}, Nuclear Physics A 642 (1998) 449--460.

\bibitem{BUU}
J.~{Aichelin}, G.~{Bertsch}, {Numerical simulation of medium energy heavy ion
  reactions}, \prc 31 (1985) 1730--1738.

\bibitem{fami}
M.~A. {Famiano}, T.~{Liu}, W.~G. {Lynch}, M.~{Mocko}, A.~M. {Rogers}, M.~B.
  {Tsang}, M.~S. {Wallace}, R.~J. {Charity}, S.~{Komarov}, D.~G. {Sarantites},
  L.~G. {Sobotka}, G.~{Verde}, {Neutron and Proton Transverse Emission Ratio
  Measurements and the Density Dependence of the Asymmetry Term of the Nuclear
  Equation of State}, Physical Review Letters 97 (2006) 052701.
\newblock \href {http://arxiv.org/abs/nucl-ex/0607016}
  {\path{arXiv:nucl-ex/0607016}}.

\bibitem{tsang_2004}
M.~B. {Tsang}, T.~X. {Liu}, L.~{Shi}, P.~{Danielewicz}, C.~K. {Gelbke}, X.~D.
  {Liu}, W.~G. {Lynch}, W.~P. {Tan}, G.~{Verde}, A.~{Wagner}, H.~S. {Xu}, W.~A.
  {Friedman}, L.~{Beaulieu}, B.~{Davin}, R.~T. {de Souza}, Y.~{Larochelle},
  T.~{Lefort}, R.~{Yanez}, V.~E. {Viola}, R.~J. {Charity}, L.~G. {Sobotka},
  {Isospin Diffusion and the Nuclear Symmetry Energy in Heavy Ion Reactions},
  Physical Review Letters 92 (2004) 062701.

\bibitem{tsang_2009}
M.~B. {Tsang}, Y.~{Zhang}, P.~{Danielewicz}, M.~{Famiano}, Z.~{Li}, W.~G.
  {Lynch}, A.~W. {Steiner}, {Constraints on the Density Dependence of the
  Symmetry Energy}, Physical Review Letters 102 (2009) 122701.
\newblock \href {http://arxiv.org/abs/0811.3107} {\path{arXiv:0811.3107}}.

\bibitem{liu_2007}
T.~X. {Liu}, W.~G. {Lynch}, M.~B. {Tsang}, X.~D. {Liu}, R.~{Shomin}, W.~P.
  {Tan}, G.~{Verde}, A.~{Wagner}, H.~F. {Xi}, H.~S. {Xu}, B.~{Davin},
  Y.~{Larochelle}, R.~T.~D. {Souza}, R.~J. {Charity}, L.~G. {Sobotka}, {Isospin
  diffusion observables in heavy-ion reactions}, \prc 76 (2007) 034603.
\newblock \href {http://arxiv.org/abs/nucl-ex/0610013}
  {\path{arXiv:nucl-ex/0610013}}.

\bibitem{kohley_11}
Z.~{Kohley}, L.~W. {May}, S.~{Wuenschel}, M.~{Colonna}, M.~{di Toro},
  M.~{Zielinska-Pfabe}, K.~{Hagel}, R.~{Tripathi}, A.~{Bonasera}, G.~A.
  {Souliotis}, D.~V. {Shetty}, S.~{Galanopoulos}, M.~{Mehlman}, W.~B. {Smith},
  S.~N. {Soisson}, B.~C. {Stein}, S.~J. {Yennello}, {Transverse collective flow
  and midrapidity emission of isotopically identified light charged particles},
  \prc 83 (2011) 044601.

\bibitem{xu_00}
H.~S. {Xu}, M.~B. {Tsang}, T.~X. {Liu}, X.~D. {Liu}, W.~G. {Lynch}, W.~P.
  {Tan}, A.~{Vander Molen}, G.~{Verde}, A.~{Wagner}, H.~F. {Xi}, C.~K.
  {Gelbke}, L.~{Beaulieu}, B.~{Davin}, Y.~{Larochelle}, T.~{Lefort}, R.~T. {de
  Souza}, R.~{Yanez}, V.~E. {Viola}, R.~J. {Charity}, L.~G. {Sobotka}, {Isospin
  Fractionation in Nuclear Multifragmentation}, Physical Review Letters 85
  (2000) 716--719.
\newblock \href {http://arxiv.org/abs/nucl-ex/9910019}
  {\path{arXiv:nucl-ex/9910019}}.

\bibitem{tsang_01}
M.~B. {Tsang}, W.~A. {Friedman}, C.~K. {Gelbke}, W.~G. {Lynch}, G.~{Verde},
  H.~S. {Xu}, {Isotopic Scaling in Nuclear Reactions}, Physical Review Letters
  86 (2001) 5023--5026.
\newblock \href {http://arxiv.org/abs/nucl-ex/0103010}
  {\path{arXiv:nucl-ex/0103010}}.

\bibitem{nato_10}
J.~B. {Natowitz}, G.~{R{\"o}pke}, S.~{Typel}, D.~{Blaschke}, A.~{Bonasera},
  K.~{Hagel}, T.~{Kl{\"a}hn}, S.~{Kowalski}, L.~{Qin}, S.~{Shlomo}, R.~{Wada},
  H.~H. {Wolter}, {Symmetry Energy of Dilute Warm Nuclear Matter}, Physical
  Review Letters 104 (2010) 202501.
\newblock \href {http://arxiv.org/abs/1001.1102} {\path{arXiv:1001.1102}}.

\bibitem{wada_12}
R.~{Wada}, K.~{Hagel}, L.~{Qin}, J.~B. {Natowitz}, Y.~G. {Ma}, G.~{R{\"o}pke},
  S.~{Shlomo}, A.~{Bonasera}, S.~{Typel}, Z.~{Chen}, M.~{Huang}, J.~{Wang},
  H.~{Zheng}, S.~{Kowalski}, C.~{Bottosso}, M.~{Barbui}, M.~R.~D. {Rodrigues},
  K.~{Schmidt}, D.~{Fabris}, M.~{Lunardon}, S.~{Moretto}, G.~{Nebbia},
  S.~{Pesente}, V.~{Rizzi}, G.~{Viesti}, M.~{Cinausero}, G.~{Prete},
  T.~{Keutgen}, Y.~{El Masri}, Z.~{Majka}, {Nuclear matter symmetry energy at
  0.03 $\leq \rho/\rho_0 \leq$ 0.2}, \prc 85 (2012) 064618.
\newblock \href {http://arxiv.org/abs/1110.3341} {\path{arXiv:1110.3341}}.

\bibitem{bali_97}
B.-A. {Li}, C.~M. {Ko}, Z.~{Ren}, {Equation of State of Asymmetric Nuclear
  Matter and Collisions of Neutron-Rich Nuclei}, Physical Review Letters 78
  (1997) 1644--1647.
\newblock \href {http://arxiv.org/abs/nucl-th/9701048}
  {\path{arXiv:nucl-th/9701048}}.

\bibitem{coup}
D.~D.~S. {Coupland}, M.~{Youngs}, W.~G. {Lynch}, M.~B. {Tsang}, Z.~{Chajecki},
  Y.~X. {Zhang}, M.~A. {Famiano}, T.~K. {Ghosh}, B.~{Giacherio}, M.~A.
  {Kilburn}, J.~{Lee}, F.~{Lu}, P.~{Russotto}, A.~{Sanetullaev}, R.~H.
  {Showalter}, G.~{Verde}, J.~{Winkelbauer}, {Effective Nucleon Masses from
  Heavy Ion Collisions}\href {http://arxiv.org/abs/1406.4546}
  {\path{arXiv:1406.4546}}.

\bibitem{SkM*}
J.~{Bartel}, P.~{Quentin}, M.~{Brack}, C.~{Guet}, H.-B. {H{\aa}kansson},
  {Towards a better parametrisation of Skyrme-like effective forces: A critical
  study of the SkM force}, Nuclear Physics A 386 (1982) 79--100.

\bibitem{SLy4}
F.~{Douchin}, P.~{Haensel}, {A unified equation of state of dense matter and
  neutron star structure}, \aap 380 (2001) 151--167.
\newblock \href {http://arxiv.org/abs/astro-ph/0111092}
  {\path{arXiv:astro-ph/0111092}}.

\bibitem{zuo_1999}
W.~{Zuo}, I.~{Bombaci}, U.~{Lombardo}, {Asymmetric nuclear matter from an
  extended Brueckner-Hartree-Fock approach}, \prc 60 (1999) 024605.
\newblock \href {http://arxiv.org/abs/nucl-th/0102035}
  {\path{arXiv:nucl-th/0102035}}.

\bibitem{zuo_2002}
W.~{Zuo}, A.~{Lejeune}, U.~{Lombardo}, J.~F. {Mathiot}, {Microscopic three-body
  force for asymmetric nuclear matter}, European Physical Journal A 14 (2002)
  469--475.
\newblock \href {http://arxiv.org/abs/nucl-th/0202077}
  {\path{arXiv:nucl-th/0202077}}.

\bibitem{chen_07}
L.-W. {Chen}, C.~M. {Ko}, B.-A. {Li}, {Isospin-dependent properties of
  asymmetric nuclear matter in relativistic mean field models}, \prc 76 (2007)
  054316.
\newblock \href {http://arxiv.org/abs/0709.0900} {\path{arXiv:0709.0900}}.

\bibitem{baran_05}
V.~{Baran}, M.~{Colonna}, M.~D. {Toro}, M.~{Zielinska-Pfab{\'e}}, H.~H.
  {Wolter}, {Isospin transport at Fermi energies}, \prc 72 (2005) 064620.
\newblock \href {http://arxiv.org/abs/nucl-th/0506078}
  {\path{arXiv:nucl-th/0506078}}.

\bibitem{Dalen_2005}
E.~N. {van Dalen}, C.~{Fuchs}, A.~{Faessler}, {Effective Nucleon Masses in
  Symmetric and Asymmetric Nuclear Matter}, Physical Review Letters 95 (2005)
  022302.
\newblock \href {http://arxiv.org/abs/nucl-th/0502064}
  {\path{arXiv:nucl-th/0502064}}.

\bibitem{Rong_2006}
J.~{Rong}, Z.-Y. {Ma}, N.~V. {Giai}, {Isospin-dependent optical potentials in
  Dirac-Brueckner-Hartree-Fock approach}, \prc 73 (2006) 014614.

\bibitem{Ma_2004}
Z.-Y. {Ma}, J.~{Rong}, B.-Q. {Chen}, Z.-Y. {Zhu}, H.-Q. {Song}, {Isospin
  dependence of nucleon effective mass in Dirac Brueckner Hartree Fock
  approach}, Physics Letters B 604 (2004) 170--174.
\newblock \href {http://arxiv.org/abs/nucl-th/0412030}
  {\path{arXiv:nucl-th/0412030}}.

\bibitem{Sam_2005}
F.~{Sammarruca}, W.~{Barredo}, P.~{Krastev}, {Predicting the single-proton and
  single-neutron potentials in asymmetric nuclear matter}, \prc 71 (2005)
  064306.
\newblock \href {http://arxiv.org/abs/nucl-th/0411053}
  {\path{arXiv:nucl-th/0411053}}.

\bibitem{Ulr_97}
S.~{Ulrych}, H.~{M{\"u}ther}, {Relativistic structure of the nucleon
  self-energy in asymmetric nuclei}, \prc 56 (1997) 1788--1794.
\newblock \href {http://arxiv.org/abs/nucl-th/9706030}
  {\path{arXiv:nucl-th/9706030}}.

\bibitem{bauge_2001}
E.~{Bauge}, J.~P. {Delaroche}, M.~{Girod}, {Lane-consistent, semimicroscopic
  nucleon-nucleus optical model}, \prc 63 (2001) 024607.

\bibitem{zhang_2014}
Y.~{Zhang}, M.~B. {Tsang}, Z.~{Li}, H.~{Liu}, {Constraints on nucleon effective
  mass splitting with heavy ion collisions}, Physics Letters B 732 (2014)
  186--190.
\newblock \href {http://arxiv.org/abs/1402.3790} {\path{arXiv:1402.3790}}.

\bibitem{bao_2013}
B.-A. {Li}, X.~{Han}, {Constraining the neutron-proton effective mass splitting
  using empirical constraints on the density dependence of nuclear symmetry
  energy around normal density}, Physics Letters B 727 (2013) 276--281.
\newblock \href {http://arxiv.org/abs/1304.3368} {\path{arXiv:1304.3368}}.

\bibitem{kong_2015}
H.-Y. {Kong}, Y.~{Xia}, J.~{Xu}, L.-W. {Chen}, B.-A. {Li}, Y.-G. {Ma},
  {Reexamination of the neutron-to-proton-ratio puzzle in intermediate-energy
  heavy-ion collisions}, \prc 91 (2015) 047601.
\newblock \href {http://arxiv.org/abs/1502.00778} {\path{arXiv:1502.00778}}.

\bibitem{bao_2015_1}
B.-A. {Li}, W.-J. {Guo}, Z.~{Shi}, {Effects of the kinetic symmetry energy
  reduced by short-range correlations in heavy-ion collisions at intermediate
  energies}, \prc 91 (2015) 044601.

\bibitem{bao_2015_2}
B.-A. {Li}, L.-W. {Chen}, {Neutron-proton effective mass splitting in
  neutron-rich matter and its impacts on nuclear reactions}, Modern Physics
  Letters A 30 (2015) 1530010.
\newblock \href {http://arxiv.org/abs/1503.00370} {\path{arXiv:1503.00370}}.

\bibitem{shi_2003}
L.~{Shi}, P.~{Danielewicz}, {Nuclear isospin diffusivity}, \prc 68 (2003)
  064604.
\newblock \href {http://arxiv.org/abs/nucl-th/0304030}
  {\path{arXiv:nucl-th/0304030}}.

\bibitem{wang_15}
Y.~{Wang}, C.~{Guo}, Q.~{Li}, H.~{Zhang}, {$^{3}$H/$^{3}$He ratio as a probe of
  the nuclear symmetry energy at sub-saturation densities}, European Physical
  Journal A 51 (2015) 37.
\newblock \href {http://arxiv.org/abs/1407.7625} {\path{arXiv:1407.7625}}.

\bibitem{kowa_07}
S.~{Kowalski}, J.~B. {Natowitz}, S.~{Shlomo}, R.~{Wada}, K.~{Hagel}, J.~{Wang},
  T.~{Materna}, Z.~{Chen}, Y.~G. {Ma}, L.~{Qin}, A.~S. {Botvina}, D.~{Fabris},
  M.~{Lunardon}, S.~{Moretto}, G.~{Nebbia}, S.~{Pesente}, V.~{Rizzi},
  G.~{Viesti}, M.~{Cinausero}, G.~{Prete}, T.~{Keutgen}, Y.~E. {Masri},
  Z.~{Majka}, A.~{Ono}, {Experimental determination of the symmetry energy of a
  low density nuclear gas}, \prc 75 (2007) 014601.
\newblock \href {http://arxiv.org/abs/nucl-ex/0602023}
  {\path{arXiv:nucl-ex/0602023}}.

\bibitem{hagel_14}
K.~{Hagel}, J.~B. {Natowitz}, G.~{R{\"o}pke}, {The equation of state and
  symmetry energy of low-density nuclear matter}, European Physical Journal A
  50 (2014) 39.
\newblock \href {http://arxiv.org/abs/1401.2074} {\path{arXiv:1401.2074}}.

\bibitem{urqmd}
S.~A. {Bass}, M.~{Belkacem}, M.~{Bleicher}, M.~{Brandstetter}, L.~{Bravina},
  C.~{Ernst}, L.~{Gerland}, M.~{Hofmann}, S.~{Hofmann}, J.~{Konopka}, G.~{Mao},
  L.~{Neise}, S.~{Soff}, C.~{Spieles}, H.~{Weber}, L.~A. {Winckelmann},
  H.~{St{\"o}cker}, W.~{Greiner}, C.~{Hartnack}, J.~{Aichelin}, N.~{Amelin},
  {Microscopic models for ultrarelativistic heavy ion collisions}, Progress in
  Particle and Nuclear Physics 41 (1998) 255--369.
\newblock \href {http://arxiv.org/abs/nucl-th/9803035}
  {\path{arXiv:nucl-th/9803035}}.

\bibitem{fopi_1}
W.~{Reisdorf}, A.~{Andronic}, R.~{Averbeck}, M.~L. {Benabderrahmane}, O.~N.
  {Hartmann}, N.~{Herrmann}, K.~D. {Hildenbrand}, T.~I. {Kang}, Y.~J. {Kim},
  M.~{Ki{\v s}}, P.~{Koczo{\'n}}, T.~{Kress}, Y.~{Leifels}, M.~{Merschmeyer},
  K.~{Piasecki}, A.~{Sch{\"u}ttauf}, M.~{Stockmeier}, V.~{Barret}, Z.~{Basrak},
  N.~{Bastid}, R.~{{\v C}aplar}, P.~{Crochet}, P.~{Dupieux}, M.~{D{\v
  z}elalija}, Z.~{Fodor}, P.~{Gasik}, Y.~{Grishkin}, B.~{Hong},
  J.~{Kecskemeti}, M.~{Kirejczyk}, M.~{Korolija}, R.~{Kotte}, A.~{Lebedev},
  X.~{Lopez}, T.~{Matulewicz}, W.~{Neubert}, M.~{Petrovici}, F.~{Rami}, M.~S.
  {Ryu}, Z.~{Seres}, B.~{Sikora}, K.~S. {Sim}, V.~{Simion},
  K.~{Siwek-Wilczy{\'n}ska}, V.~{Smolyankin}, G.~{Stoicea}, Z.~{Tymi{\'n}ski},
  K.~{Wi{\'s}niewski}, D.~{Wohlfarth}, Z.~G. {Xiao}, H.~S. {Xu},
  I.~{Yushmanov}, A.~{Zhilin}, {FOPI Collaboration}, {Systematics of central
  heavy ion collisions in the 1A GeV regime}, Nuclear Physics A 848 (2010)
  366--427.

\bibitem{fopi_2}
W.~{Reisdorf}, Y.~{Leifels}, A.~{Andronic}, R.~{Averbeck}, V.~{Barret},
  Z.~{Basrak}, N.~{Bastid}, M.~L. {Benabderrahmane}, R.~{{\v C}aplar},
  P.~{Crochet}, P.~{Dupieux}, M.~{D{\v z}elalija}, Z.~{Fodor}, P.~{Gasik},
  Y.~{Grishkin}, O.~N. {Hartmann}, N.~{Herrmann}, K.~D. {Hildenbrand},
  B.~{Hong}, T.~I. {Kang}, J.~{Kecskemeti}, Y.~J. {Kim}, M.~{Kirejczyk},
  M.~{Ki{\v s}}, P.~{Koczo{\'n}}, M.~{Korolija}, R.~{Kotte}, T.~{Kress},
  A.~{Lebedev}, X.~{Lopez}, T.~{Matulewicz}, M.~{Merschmeyer}, W.~{Neubert},
  M.~{Petrovici}, K.~{Piasecki}, F.~{Rami}, M.~S. {Ryu}, A.~{Sch{\"u}ttauf},
  Z.~{Seres}, B.~{Sikora}, K.~S. {Sim}, V.~{Simion}, K.~{Siwek-Wilczy{\'n}ska},
  V.~{Smolyankin}, M.~{Stockmeier}, G.~{Stoicea}, Z.~{Tymi{\'n}ski},
  K.~{Wi{\'s}niewski}, D.~{Wohlfarth}, Z.~G. {Xiao}, H.~S. {Xu},
  I.~{Yushmanov}, A.~{Zhilin}, {FOPI Collaboration}, {Systematics of azimuthal
  asymmetries in heavy ion collisions in the 1A GeV regime}, Nuclear Physics A
  876 (2012) 1--60.

\bibitem{wang_14}
Y.~{Wang}, C.~{Guo}, Q.~{Li}, H.~{Zhang}, Y.~{Leifels}, W.~{Trautmann},
  {Constraining the high-density nuclear symmetry energy with the
  transverse-momentum-dependent elliptic flow}, \prc 89 (2014) 044603.
\newblock \href {http://arxiv.org/abs/1403.7041} {\path{arXiv:1403.7041}}.

\bibitem{russ_11}
P.~{Russotto}, P.~Z. {Wu}, M.~{Zoric}, M.~{Chartier}, Y.~{Leifels}, R.~C.
  {Lemmon}, Q.~{Li}, J.~{{\L}ukasik}, A.~{Pagano}, P.~{Paw{\l}owski},
  W.~{Trautmann}, {Symmetry energy from elliptic flow in $^{197}$Au +
  $^{197}$Au}, Physics Letters B 697 (2011) 471--476.
\newblock \href {http://arxiv.org/abs/1101.2361} {\path{arXiv:1101.2361}}.

\bibitem{cozma_13}
M.~D. {Cozma}, Y.~{Leifels}, W.~{Trautmann}, Q.~{Li}, P.~{Russotto}, {Toward a
  model-independent constraint of the high-density dependence of the symmetry
  energy}, \prc 88 (2013) 044912.
\newblock \href {http://arxiv.org/abs/1305.5417} {\path{arXiv:1305.5417}}.

\bibitem{li_05}
Q.~{Li}, Z.~{Li}, S.~{Soff}, M.~{Bleicher}, H.~{St{\"o}cker}, {Probing the
  density dependence of the symmetry potential at low and high densities}, \prc
  72 (2005) 034613.
\newblock \href {http://arxiv.org/abs/nucl-th/0506030}
  {\path{arXiv:nucl-th/0506030}}.

\bibitem{zhang_05}
Y.~{Zhang}, Z.~{Li}, {Probing the density dependence of the symmetry potential
  with peripheral heavy-ion collisions}, \prc 71 (2005) 024604.
\newblock \href {http://arxiv.org/abs/nucl-th/0501039}
  {\path{arXiv:nucl-th/0501039}}.

\bibitem{chen_03_1}
L.-W. {Chen}, C.~M. {Ko}, B.-A. {Li}, {Light clusters production as a probe to
  nuclear symmetry energy}, \prc 68 (2003) 017601.
\newblock \href {http://arxiv.org/abs/nucl-th/0302068}
  {\path{arXiv:nucl-th/0302068}}.

\bibitem{chen_03_2}
L.-W. {Chen}, C.~M. {Ko}, B.-A. {Li}, {Light cluster production in intermediate
  energy heavy-ion collisions induced by neutron-rich nuclei}, Nuclear Physics
  A 729 (2003) 809--834.
\newblock \href {http://arxiv.org/abs/nucl-th/0306032}
  {\path{arXiv:nucl-th/0306032}}.

\bibitem{chen_04}
L.-W. {Chen}, C.~M. {Ko}, B.-A. {Li}, {Effects of momentum-dependent nuclear
  potential on two-nucleon correlation functions and light cluster production
  in intermediate energy heavy-ion collisions}, \prc 69 (2004) 054606.
\newblock \href {http://arxiv.org/abs/nucl-th/0403049}
  {\path{arXiv:nucl-th/0403049}}.

\bibitem{chim1}
E.~{De Filippo}, A.~{Pagano}, P.~{Russotto}, F.~{Amorini}, A.~{Anzalone},
  L.~{Auditore}, V.~{Baran}, I.~{Berceanu}, B.~{Borderie}, R.~{Bougault},
  M.~{Bruno}, T.~{Cap}, G.~{Cardella}, S.~{Cavallaro}, M.~B. {Chatterjee},
  A.~{Chbihi}, M.~{Colonna}, M.~{D'Agostino}, R.~{Dayras}, M.~{Di Toro},
  J.~{Frankland}, E.~{Galichet}, W.~{Gawlikowicz}, E.~{Geraci},
  A.~{Grzeszczuk}, P.~{Guazzoni}, S.~{Kowalski}, E.~{La Guidara},
  G.~{Lanzalone}, G.~{Lanzan{\`o}}, N.~{Le Neindre}, I.~{Lombardo},
  C.~{Maiolino}, M.~{Papa}, E.~{Piasecki}, S.~{Pirrone}, R.~{P{\l}aneta},
  G.~{Politi}, A.~{Pop}, F.~{Porto}, M.~F. {Rivet}, F.~{Rizzo}, E.~{Rosato},
  K.~{Schmidt}, K.~{Siwek-Wilczy{\'n}ska}, I.~{Skwira-Chalot},
  A.~{Trifir{\`o}}, M.~{Trimarchi}, G.~{Verde}, M.~{Vigilante}, J.~P.
  {Wieleczko}, J.~{Wilczy{\'n}ski}, L.~{Zetta}, W.~{Zipper}, {Correlations
  between emission timescale of fragments and isospin dynamics in
  $^{124}$Sn+$^{64}$Ni and $^{112}$Sn+$^{58}$Ni reactions at 35A MeV}, \prc 86
  (2012) 014610.
\newblock \href {http://arxiv.org/abs/1206.0697} {\path{arXiv:1206.0697}}.

\bibitem{chim2}
Z.~Y. {Sun}, M.~B. {Tsang}, W.~G. {Lynch}, G.~{Verde}, F.~{Amorini},
  L.~{Andronenko}, M.~{Andronenko}, G.~{Cardella}, M.~{Chatterje},
  P.~{Danielewicz}, E.~{de Filippo}, P.~{Dinh}, E.~{Galichet}, E.~{Geraci},
  H.~{Hua}, E.~{La Guidara}, G.~{Lanzalone}, H.~{Liu}, F.~{Lu}, S.~{Lukyanov},
  C.~{Maiolino}, A.~{Pagano}, S.~{Piantelli}, M.~{Papa}, S.~{Pirrone},
  G.~{Politi}, F.~{Porto}, F.~{Rizzo}, P.~{Russotto}, D.~{Santonocito}, Y.~X.
  {Zhang}, {Isospin diffusion and equilibration for Sn+Sn collisions at E/A=35
  MeV}, \prc 82 (2010) 051603.
\newblock \href {http://arxiv.org/abs/1009.1669} {\path{arXiv:1009.1669}}.

\bibitem{sumy_08}
K.~{Sumiyoshi}, G.~{R{\"o}pke}, {Appearance of light clusters in post-bounce
  evolution of core-collapse supernovae}, \prc 77 (2008) 055804.
\newblock \href {http://arxiv.org/abs/0801.0110} {\path{arXiv:0801.0110}}.

\bibitem{Oco_10}
E.~{O'Connor}, C.~D. {Ott}, {A new open-source code for spherically symmetric
  stellar collapse to neutron stars and black holes}, Classical and Quantum
  Gravity 27 (2010) 114103.
\newblock \href {http://arxiv.org/abs/0912.2393} {\path{arXiv:0912.2393}}.

\bibitem{Lentz_12}
E.~J. {Lentz}, A.~{Mezzacappa}, O.~E.~B. {Messer}, W.~R. {Hix}, S.~W. {Bruenn},
  {Interplay of Neutrino Opacities in Core-collapse Supernova Simulations},
  \apj 760 (2012) 94.
\newblock \href {http://arxiv.org/abs/1206.1086} {\path{arXiv:1206.1086}}.

\bibitem{wang_07}
J.~{Wang}, T.~{Keutgen}, R.~{Wada}, K.~{Hagel}, S.~{Kowalski}, T.~{Materna},
  L.~{Qin}, Z.~{Chen}, J.~B. {Natowitz}, Y.~G. {Ma}, M.~{Murray}, A.~{Keksis},
  E.~{Martin}, A.~{Ruangma}, D.~V. {Shetty}, G.~{Souliotis}, M.~{Veselsky},
  E.~M. {Winchester}, S.~J. {Yennello}, D.~{Fabris}, M.~{Lunardon},
  S.~{Moretto}, G.~{Nebbia}, S.~{Pesente}, V.~{Rizzi}, G.~{Viesti},
  M.~{Cinausero}, G.~{Prete}, J.~{Cibor}, Z.~{Majka}, P.~{Staszel},
  W.~{Zipper}, Y.~E. {Masri}, R.~{Alfaro}, A.~{Martinez-Davalos},
  A.~{Menchaca-Rocha}, A.~{Ono}, {Properties of the initial participant matter
  interaction zone in near-Fermi-energy heavy-ion collisions}, \prc 75 (2007)
  014604.
\newblock \href {http://arxiv.org/abs/nucl-ex/0603009}
  {\path{arXiv:nucl-ex/0603009}}.

\bibitem{qin_12}
L.~{Qin}, K.~{Hagel}, R.~{Wada}, J.~B. {Natowitz}, S.~{Shlomo}, A.~{Bonasera},
  G.~{R{\"o}pke}, S.~{Typel}, Z.~{Chen}, M.~{Huang}, J.~{Wang}, H.~{Zheng},
  S.~{Kowalski}, M.~{Barbui}, M.~R.~D. {Rodrigues}, K.~{Schmidt}, D.~{Fabris},
  M.~{Lunardon}, S.~{Moretto}, G.~{Nebbia}, S.~{Pesente}, V.~{Rizzi},
  G.~{Viesti}, M.~{Cinausero}, G.~{Prete}, T.~{Keutgen}, Y.~{El Masri},
  Z.~{Majka}, Y.~G. {Ma}, {Laboratory Tests of Low Density Astrophysical
  Nuclear Equations of State}, Physical Review Letters 108 (2012) 172701.
\newblock \href {http://arxiv.org/abs/1110.3345} {\path{arXiv:1110.3345}}.

\bibitem{roep_82}
G.~{R{\"o}pke}, L.~{M{\"u}nchow}, H.~{Schulz}, {Particle clustering and Mott
  transitions in nuclear matter at finite temperature (I). Method and general
  aspects}, Nuclear Physics A 379 (1982) 536--552.

\bibitem{roep_84}
G.~{R{\"o}pke}, M.~{Schmidt}, H.~{Schulz}, {Particle clustering and Mott
  transitions in hot nuclear matter at finite temperature (III). Heavy cluster
  abundances and coulomb interaction}, Nuclear Physics A 424 (1984) 594--604.

\bibitem{roep_90}
M.~{Schmidt}, G.~{R{\"o}pke}, H.~{Schulz}, {Generalized Bethe-Uhlenbeck
  approach for hot nuclear matter}, Annals of Physics 202 (1990) 57--99.

\bibitem{mabi_15}
J.~{Mabiala}, H.~{Zheng}, A.~{Bonasera}, P.~{Cammarata}, K.~{Hagel},
  L.~{Heilborn}, Z.~{Kohley}, L.~W. {May}, A.~B. {McIntosh}, M.~D. {Youngs},
  A.~{Zarrella}, S.~J. {Yennello}, {Novel technique to extract experimental
  symmetry free energy information for nuclear matter}, \prc 92 (2015) 024605.
\newblock \href {http://arxiv.org/abs/1507.06895} {\path{arXiv:1507.06895}}.

\bibitem{ABrown}
B.~Alex~Brown, Neutron radii in nuclei and the neutron equation of state, Phys.
  Rev. Lett. 85 (2000) 5296--5299.

\bibitem{bao_02}
B.-A. {Li}, {Probing the High Density Behavior of the Nuclear Symmetry Energy
  with High Energy Heavy-Ion Collisions}, Physical Review Letters 88 (2002)
  192701.
\newblock \href {http://arxiv.org/abs/nucl-th/0205002}
  {\path{arXiv:nucl-th/0205002}}.

\bibitem{lipr08}
B.-A. {Li}, L.-W. {Chen}, C.~M. {Ko}, {Recent progress and new challenges in
  isospin physics with heavy-ion reactions}, \physrep 464 (2008) 113--281.
\newblock \href {http://arxiv.org/abs/0804.3580} {\path{arXiv:0804.3580}}.

\bibitem{lisustich01}
B.-A. {Li}, A.~T. {Sustich}, B.~{Zhang}, {Proton differential elliptic flow and
  the isospin dependence of the nuclear equation of state}, \prc 64 (2001)
  054604.
\newblock \href {http://arxiv.org/abs/nucl-th/0108047}
  {\path{arXiv:nucl-th/0108047}}.

\bibitem{baran05}
V.~{Baran}, M.~{Colonna}, V.~{Greco}, M.~{Di Toro}, {Reaction dynamics with
  exotic nuclei}, \physrep 410 (2005) 335--466.
\newblock \href {http://arxiv.org/abs/nucl-th/0412060}
  {\path{arXiv:nucl-th/0412060}}.

\bibitem{abelev07}
B.~I. {Abelev}, M.~M. {Aggarwal}, Z.~{Ahammed}, B.~D. {Anderson},
  D.~{Arkhipkin}, G.~S. {Averichev}, Y.~{Bai}, J.~{Balewski}, O.~{Barannikova},
  L.~S. {Barnby}, et~al., {Partonic Flow and $\phi$-meson Production in Au+Au
  Collisions at $s_{NN}$=200GeV}, Physical Review Letters 99 (2007) 112301.
\newblock \href {http://arxiv.org/abs/nucl-ex/0703033}
  {\path{arXiv:nucl-ex/0703033}}.

\bibitem{aamodt10}
K.~{Aamodt}, B.~{Abelev}, A.~{Abrahantes Quintana}, D.~{Adamov{\'a}}, A.~M.
  {Adare}, M.~M. {Aggarwal}, G.~{Aglieri Rinella}, A.~G. {Agocs}, S.~{Aguilar
  Salazar}, Z.~{Ahammed}, et~al., {Elliptic Flow of Charged Particles in Pb-Pb
  Collisions at s$_{NN}$=2.76TeV}, Physical Review Letters 105 (2010) 252302.
\newblock \href {http://arxiv.org/abs/1011.3914} {\path{arXiv:1011.3914}}.

\bibitem{adare12}
A.~{Adare}, S.~{Afanasiev}, C.~{Aidala}, N.~N. {Ajitanand}, Y.~{Akiba},
  H.~{Al-Bataineh}, J.~{Alexander}, K.~{Aoki}, Y.~{Aramaki}, E.~T. {Atomssa},
  et~al., {Deviation from quark number scaling of the anisotropy parameter
  v$_{2}$ of pions, kaons, and protons in Au+Au collisions at s$_{NN}$=200
  GeV}, \prc 85 (2012) 064914.
\newblock \href {http://arxiv.org/abs/1203.2644} {\path{arXiv:1203.2644}}.

\bibitem{aad12}
G.~{Aad}, B.~{Abbott}, J.~{Abdallah}, A.~A. {Abdelalim}, A.~{Abdesselam},
  O.~{Abdinov}, B.~{Abi}, M.~{Abolins}, H.~{Abramowicz}, H.~{Abreu}, et~al.,
  {Measurement of the pseudorapidity and transverse momentum dependence of the
  elliptic flow of charged particles in lead-lead collisions at 2.76 TeV with
  the ATLAS detector}, Physics Letters B 707 (2012) 330--348.
\newblock \href {http://arxiv.org/abs/1108.6018} {\path{arXiv:1108.6018}}.

\bibitem{gutbrod90}
H.~H. {Gutbrod}, K.~H. {Kampert}, B.~{Kolb}, A.~M. {Poskanzer}, H.~G. {Ritter},
  R.~{Schicker}, H.~R. {Schmidt}, {Squeeze-out of nuclear matter as a function
  of projectile energy and mass}, \prc 42 (1990) 640--651.

\bibitem{leif93}
Y.~{Leifels}, T.~{Blaich}, T.~W. {Elze}, H.~{Emling}, H.~{Freiesleben},
  K.~{Grimm}, W.~{Henning}, R.~{Holzmann}, J.~G. {Keller}, H.~{Klingler}, J.~V.
  {Kratz}, R.~{Kulessa}, D.~{Lambrecht}, S.~{Lange}, E.~{Lubkiewicz}, E.~F.
  {Moore}, W.~{Prokopowicz}, R.~{Schmidt}, C.~{Sch{\"u}tter}, H.~{Spies},
  K.~{Stelzer}, J.~{Stroth}, E.~{Wajda}, W.~{Walu{\'s}}, M.~{Zinser},
  E.~{Zude}, {Exclusive studies of neutron and charged particle emission in
  collisions of $^{197}$Au +$^{197}$Au at 400 MeV/nucleon}, Physical Review
  Letters 71 (1993) 963--966.

\bibitem{lamb94}
D.~{Lambrecht}, T.~{Blaich}, T.~W. {Elze}, H.~{Emling}, H.~{Freiesleben},
  K.~{Grimm}, W.~{Henning}, R.~{Holzmann}, J.~G. {Keller}, H.~{Klingler}, J.~V.
  {Kratz}, R.~{Kulessa}, S.~{Lange}, Y.~{Leifels}, E.~{Lubkiewicz}, E.~F.
  {Moore}, W.~{Prokopowicz}, R.~{Schmidt}, C.~{Sch{\"u}tter}, H.~{Spies},
  K.~{Stelzer}, J.~{Stroth}, E.~{Wajda}, W.~{Walu{\'s}}, M.~{Zinser},
  E.~{Zude}, {Energy dependence of collective flow of neutrons and protons
  in$^{197}$Au+$^{197}$Au collisions}, Zeitschrift fur Physik A Hadrons and
  Nuclei 350 (1994) 115--120.

\bibitem{LAND}
T.~{Blaich}, T.~W. {Elze}, H.~{Emling}, H.~{Freiesleben}, K.~{Grimm},
  W.~{Henning}, R.~{Holzmann}, G.~{Ickert}, J.~G. {Keller}, H.~{Klingler},
  W.~{Kneissl}, R.~{K{\"o}nig}, R.~{Kulessa}, J.~V. {Kratz}, D.~{Lambrecht},
  J.~S. {Lange}, Y.~{Leifels}, E.~{Lubkiewicz}, M.~{Proft}, W.~{Prokopowicz},
  C.~{Sch{\"u}tter}, R.~{Schmidt}, H.~{Spies}, K.~{Stelzer}, J.~{Stroth},
  W.~{Walus}, E.~{Wajda}, H.~J. {Wollersheim}, M.~{Zinser}, E.~{Zude}, {LAND
  Collaboration}, {A large area detector for high-energy neutrons}, Nuclear
  Instruments and Methods in Physics Research A 314 (1992) 136--154.

\bibitem{russotto11}
P.~{Russotto}, P.~Z. {Wu}, M.~{Zoric}, M.~{Chartier}, Y.~{Leifels}, R.~C.
  {Lemmon}, Q.~{Li}, J.~{{\L}ukasik}, A.~{Pagano}, P.~{Paw{\l}owski},
  W.~{Trautmann}, {Symmetry energy from elliptic flow in $^{197}$Au +
  $^{197}$Au}, Physics Letters B 697 (2011) 471--476.
\newblock \href {http://arxiv.org/abs/1101.2361} {\path{arXiv:1101.2361}}.

\bibitem{cozma11}
M.~D. {Cozma}, {Neutron-proton elliptic flow difference as a probe for the high
  density dependence of the symmetry energy}, Physics Letters B 700 (2011)
  139--144.
\newblock \href {http://arxiv.org/abs/1102.2728} {\path{arXiv:1102.2728}}.

\bibitem{hi_96}
X.~{Lopez}, Y.~J. {Kim}, N.~{Herrmann}, A.~{Andronic}, V.~{Barret},
  Z.~{Basrak}, N.~{Bastid}, M.~L. {Benabderrahmane}, R.~{{\v C}aplar},
  E.~{Cordier}, P.~{Crochet}, P.~{Dupieux}, M.~{D{\v z}elalija}, Z.~{Fodor},
  I.~{Ga{\v s}pari{\'c}}, Y.~{Grishkin}, O.~N. {Hartmann}, K.~D. {Hildenbrand},
  B.~{Hong}, T.~I. {Kang}, J.~{Kecskemeti}, M.~{Kirejczyk}, M.~{Ki{\v s}},
  P.~{Koczon}, M.~{Korolija}, R.~{Kotte}, A.~{Lebedev}, Y.~{Leifels},
  M.~{Merschmeyer}, W.~{Neubert}, D.~{Pelte}, M.~{Petrovici}, F.~{Rami},
  W.~{Reisdorf}, M.~S. {Ryu}, A.~{Sch{\"u}ttauf}, Z.~{Seres}, B.~{Sikora},
  K.~S. {Sim}, V.~{Simion}, K.~{Siwek-Wilczy{\'n}ska}, V.~{Smolyankin},
  G.~{Stoicea}, Z.~{Tyminski}, P.~{Wagner}, K.~{Wi{\'s}niewski},
  D.~{Wohlfarth}, Z.~G. {Xiao}, I.~{Yushmanov}, X.~Y. {Zhang}, A.~{Zhilin},
  G.~{Ferini}, T.~{Gaitanos}, {Isospin dependence of relative yields of K$^{+}$
  and K$^{0}$ mesons at 1.528A GeV}, \prc 75 (2007) 011901.
\newblock \href {http://arxiv.org/abs/nucl-ex/0701006}
  {\path{arXiv:nucl-ex/0701006}}.

\bibitem{xlopez07}
X.~{Lopez}, Y.~J. {Kim}, N.~{Herrmann}, A.~{Andronic}, V.~{Barret},
  Z.~{Basrak}, N.~{Bastid}, M.~L. {Benabderrahmane}, R.~{{\v C}aplar},
  E.~{Cordier}, P.~{Crochet}, P.~{Dupieux}, M.~{D{\v z}elalija}, Z.~{Fodor},
  I.~{Ga{\v s}pari{\'c}}, Y.~{Grishkin}, O.~N. {Hartmann}, K.~D. {Hildenbrand},
  B.~{Hong}, T.~I. {Kang}, J.~{Kecskemeti}, M.~{Kirejczyk}, M.~{Ki{\v s}},
  P.~{Koczon}, M.~{Korolija}, R.~{Kotte}, A.~{Lebedev}, Y.~{Leifels},
  M.~{Merschmeyer}, W.~{Neubert}, D.~{Pelte}, M.~{Petrovici}, F.~{Rami},
  W.~{Reisdorf}, M.~S. {Ryu}, A.~{Sch{\"u}ttauf}, Z.~{Seres}, B.~{Sikora},
  K.~S. {Sim}, V.~{Simion}, K.~{Siwek-Wilczy{\'n}ska}, V.~{Smolyankin},
  G.~{Stoicea}, Z.~{Tyminski}, P.~{Wagner}, K.~{Wi{\'s}niewski},
  D.~{Wohlfarth}, Z.~G. {Xiao}, I.~{Yushmanov}, X.~Y. {Zhang}, A.~{Zhilin},
  G.~{Ferini}, T.~{Gaitanos}, {Isospin dependence of relative yields of K$^{+}$
  and K$^{0}$ mesons at 1.528A GeV}, \prc 75 (2007) 011901.
\newblock \href {http://arxiv.org/abs/nucl-ex/0701006}
  {\path{arXiv:nucl-ex/0701006}}.

\bibitem{reisdorf07}
W.~{Reisdorf}, M.~{Stockmeier}, A.~{Andronic}, M.~L. {Benabderrahmane}, O.~N.
  {Hartmann}, N.~{Herrmann}, K.~D. {Hildenbrand}, Y.~J. {Kim}, M.~{Ki{\v s}},
  P.~{Koczo{\'n}}, T.~{Kress}, Y.~{Leifels}, X.~{Lopez}, M.~{Merschmeyer},
  A.~{Sch{\"u}ttauf}, V.~{Barret}, Z.~{Basrak}, N.~{Bastid}, R.~{{\v C}aplar},
  P.~{Crochet}, P.~{Dupieux}, M.~{D{\v z}elalija}, Z.~{Fodor}, Y.~{Grishkin},
  B.~{Hong}, T.~I. {Kang}, J.~{Kecskemeti}, M.~{Kirejczyk}, M.~{Korolija},
  R.~{Kotte}, A.~{Lebedev}, T.~{Matulewicz}, W.~{Neubert}, M.~{Petrovici},
  F.~{Rami}, M.~S. {Ryu}, Z.~{Seres}, B.~{Sikora}, K.~S. {Sim}, V.~{Simion},
  K.~{Siwek-Wilczy{\'n}ska}, V.~{Smolyankin}, G.~{Stoicea}, Z.~{Tymi{\'n}ski},
  K.~{Wi{\'s}niewski}, D.~{Wohlfarth}, Z.~G. {Xiao}, H.~S. {Xu},
  I.~{Yushmanov}, A.~{Zhilin}, {FOPI Collaboration}, {Systematics of pion
  emission in heavy ion collisions in the 1 A GeV regime}, Nuclear Physics A
  781 (2007) 459--508.

\bibitem{xiao09}
Z.~{Xiao}, B.-A. {Li}, L.-W. {Chen}, G.-C. {Yong}, M.~{Zhang}, {Circumstantial
  Evidence for a Soft Nuclear Symmetry Energy at Suprasaturation Densities},
  Physical Review Letters 102 (2009) 062502.
\newblock \href {http://arxiv.org/abs/0808.0186} {\path{arXiv:0808.0186}}.

\bibitem{feng10}
Z.-Q. {Feng}, G.-M. {Jin}, {Probing high-density behavior of symmetry energy
  from pion emission in heavy-ion collisions}, Physics Letters B 683 (2010)
  140--144.
\newblock \href {http://arxiv.org/abs/0904.2990} {\path{arXiv:0904.2990}}.

\bibitem{xie13}
W.-J. {Xie}, J.~{Su}, L.~{Zhu}, F.-S. {Zhang}, {Symmetry energy and pion
  production in the Boltzmann-Langevin approach}, Physics Letters B 718 (2013)
  1510--1514.

\bibitem{russo_14}
P.~{Russotto}, M.~D. {Cozma}, A.~{Le F{\`e}vre}, Y.~{Leifels}, R.~{Lemmon},
  Q.~{Li}, J.~{{\L}ukasik}, W.~{Trautmann}, {Flow probe of symmetry energy in
  relativistic heavy-ion reactions}, European Physical Journal A 50 (2014) 38.
\newblock \href {http://arxiv.org/abs/1310.2896} {\path{arXiv:1310.2896}}.

\bibitem{chimera}
A.~{Pagano}, M.~{Alderighi}, F.~{Amorini}, A.~{Anzalone}, L.~{Arena},
  L.~{Auditore}, V.~{Baran}, M.~{Bartolucci}, I.~{Berceanu}, J.~{Blicharska},
  J.~{Brzychczyk}, A.~{Bonasera}, B.~{Borderie}, R.~{Bougault}, M.~{Bruno},
  G.~{Cardella}, S.~{Cavallaro}, M.~B. {Chatterjee}, A.~{Chbihi}, J.~{Cibor},
  M.~{Colonna}, M.~{D'Agostino}, R.~{Dayras}, E.~{De Filippo}, M.~{Di Toro},
  W.~{Gawlikowicz}, E.~{Geraci}, F.~{Giustolisi}, A.~{Grzeszczuk},
  P.~{Guazzoni}, D.~{Guinet}, M.~{Iacono-Manno}, S.~{Kowalski}, E.~{La
  Guidara}, G.~{Lanzano}, G.~{Lanzalone}, N.~{Le Neindre}, S.~{Li}, S.~{Lo
  Nigro}, C.~{Maiolino}, Z.~{Majka}, G.~{Manfredi}, T.~{Paduszynski},
  M.~{Papa}, M.~{Petrovici}, E.~{Piasecki}, S.~{Pirrone}, R.~{Planeta},
  G.~{Politi}, A.~{Pop}, F.~{Porto}, M.~F. {Rivet}, E.~{Rosato}, F.~{Rizzo},
  S.~{Russo}, P.~{Russotto}, M.~{Sassi}, G.~{Sechi}, V.~{Simion},
  K.~{Siwek-Wilczynska}, I.~{Skwira}, M.~L. {Sperduto}, J.~C. {Steckmeyer},
  L.~{Swiderski}, A.~{Trifiro`}, M.~{Trimarchi}, G.~{Vannini}, M.~{Vigilante},
  J.~P. {Wieleczko}, J.~{Wilczynski}, H.~{Wu}, Z.~{Xiao}, L.~{Zetta},
  W.~{Zipper}, {Fragmentation studies with the CHIMERA detector at LNS in
  Catania: recent progress}, Nuclear Physics A 734 (2004) 504--511.

\bibitem{kratta}
J.~{{\L}ukasik}, P.~{Paw{\l}owski}, A.~{Budzanowski}, B.~{Czech},
  I.~{Skwirczy{\'n}ska}, J.~{Brzychczyk}, M.~{Adamczyk}, S.~{Kupny},
  P.~{Lasko}, Z.~{Sosin}, A.~{Wieloch}, M.~{Ki{\v s}}, Y.~{Leifels},
  W.~{Trautmann}, {KRATTA, a versatile triple telescope array for charged
  reaction products}, Nuclear Instruments and Methods in Physics Research A 709
  (2013) 120--128.
\newblock \href {http://arxiv.org/abs/1301.2127} {\path{arXiv:1301.2127}}.

\bibitem{book}
M.~Baldo (Ed.), Nuclear Methods and The Nuclear Equation of State, World
  Scientific, Singapore, 1999.

\bibitem{Martino}
G.~{Hagen}, T.~{Papenbrock}, A.~{Ekstr{\"o}m}, K.~A. {Wendt}, G.~{Baardsen},
  S.~{Gandolfi}, M.~{Hjorth-Jensen}, C.~J. {Horowitz}, {Coupled-cluster
  calculations of nucleonic matter}, \prc 89 (2014) 014319.
\newblock \href {http://arxiv.org/abs/1311.2925} {\path{arXiv:1311.2925}}.

\bibitem{Machlei1989}
R.~Machleidt, Advances in Nuclear Physics, Springer US, 1989, Ch. The Meson
  Theory of Nuclear Forces and Nuclear Structure, pp. 189--376.

\bibitem{OM1}
O.~{Benhar}, M.~{Valli}, {Shear Viscosity of Neutron Matter from Realistic
  Nucleon-Nucleon Interactions}, Physical Review Letters 99 (2007) 232501.
\newblock \href {http://arxiv.org/abs/0707.2681} {\path{arXiv:0707.2681}}.

\bibitem{OM2}
A.~{Lovato}, O.~{Benhar}, S.~{Fantoni}, A.~Y. {Illarionov}, K.~E. {Schmidt},
  {Density-dependent nucleon-nucleon interaction from three-nucleon forces},
  \prc 83 (2011) 054003.
\newblock \href {http://arxiv.org/abs/1011.3784} {\path{arXiv:1011.3784}}.

\bibitem{RG}
S.~K. {Bogner}, R.~J. {Furnstahl}, A.~{Schwenk}, {From low-momentum
  interactions to nuclear structure}, Progress in Particle and Nuclear Physics
  65 (2010) 94--147.
\newblock \href {http://arxiv.org/abs/0912.3688} {\path{arXiv:0912.3688}}.

\bibitem{compr_Stone}
J.~R. {Stone}, N.~J. {Stone}, S.~A. {Moszkowski}, {Incompressibility in finite
  nuclei and nuclear matter}, \prc 89 (2014) 044316.
\newblock \href {http://arxiv.org/abs/1404.0744} {\path{arXiv:1404.0744}}.

\bibitem{compr1}
G.~{Col{\`o}}, N.~{van Giai}, J.~{Meyer}, K.~{Bennaceur}, P.~{Bonche},
  {Microscopic determination of the nuclear incompressibility within the
  nonrelativistic framework}, \prc 70 (2004) 024307.
\newblock \href {http://arxiv.org/abs/nucl-th/0403086}
  {\path{arXiv:nucl-th/0403086}}.

\bibitem{compr2}
G.~{Colo`}, N.~{Van Giai}, {Theoretical understanding of the nuclear
  incompressibility: where do we stand?}, Nuclear Physics A 731 (2004) 15--27.
\newblock \href {http://arxiv.org/abs/nucl-th/0309002}
  {\path{arXiv:nucl-th/0309002}}.

\bibitem{v18}
R.~B. {Wiringa}, V.~G.~J. {Stoks}, R.~{Schiavilla}, {Accurate nucleon-nucleon
  potential with charge-independence breaking}, \prc 51 (1995) 38--51.
\newblock \href {http://arxiv.org/abs/nucl-th/9408016}
  {\path{arXiv:nucl-th/9408016}}.

\bibitem{UIX}
J.~{Carlson}, V.~R. {Pandharipande}, R.~B. {Wiringa}, {Three-nucleon
  interaction in 3-, 4- and {$\infty$}-body systems}, Nuclear Physics A 401
  (1983) 59--85.

\bibitem{UIX_1}
R.~{Schiavilla}, V.~R. {Pandharipande}, R.~B. {Wiringa}, {Momentum
  distributions in A = 3 and 4 nuclei}, Nuclear Physics A 449 (1986) 219--242.

\bibitem{UIX_2}
B.~S. {Pudliner}, V.~R. {Pandharipande}, J.~{Carlson}, S.~C. {Pieper}, R.~B.
  {Wiringa}, {Quantum Monte Carlo calculations of nuclei with A$\lt$7}, \prc 56
  (1997) 1720--1750.
\newblock \href {http://arxiv.org/abs/nucl-th/9705009}
  {\path{arXiv:nucl-th/9705009}}.

\bibitem{Li1}
Z.~H. {Li}, H.-J. {Schulze}, {Neutron star structure with modern nucleonic
  three-body forces}, \prc 78 (2008) 028801.

\bibitem{Li2}
Z.~H. {Li}, U.~{Lombardo}, H.-J. {Schulze}, W.~{Zuo}, {Consistent
  nucleon-nucleon potentials and three-body forces}, \prc 77 (2008) 034316.

\bibitem{Bonn}
R.~{Brockmann}, R.~{Machleidt}, {Relativistic nuclear structure. I. Nuclear
  matter}, \prc 42 (1990) 1965--1980.

\bibitem{Dalen}
E.~N. {van Dalen}, C.~{Fuchs}, A.~{Faessler}, {Momentum, density, and isospin
  dependence of symmetric and asymmetric nuclear matter properties}, \prc 72
  (2005) 065803.
\newblock \href {http://arxiv.org/abs/nucl-th/0511040}
  {\path{arXiv:nucl-th/0511040}}.

\bibitem{Drisch}
C.~{Drischler}, V.~{Som{\`a}}, A.~{Schwenk}, {Microscopic calculations and
  energy expansions for neutron-rich matter}, \prc 89 (2014) 025806.
\newblock \href {http://arxiv.org/abs/1310.5627} {\path{arXiv:1310.5627}}.

\bibitem{Kenji1}
M.~{Baldo}, K.~{Fukukawa}, {Nuclear Matter from Effective Quark-Quark
  Interaction}, Physical Review Letters 113 (2014) 242501.
\newblock \href {http://arxiv.org/abs/1409.7206} {\path{arXiv:1409.7206}}.

\bibitem{DBHF_TBF}
G.~Brown, W.~Weise, {Relativistic Effects in Nuclear Physics}, Commun.
  Nucl.Part.Phys. 17 (1987) 39.

\bibitem{Chen_2010}
L.-W. {Chen}, C.~M. {Ko}, B.-A. {Li}, J.~{Xu}, {Density slope of the nuclear
  symmetry energy from the neutron skin thickness of heavy nuclei}, \prc 82
  (2010) 024321.
\newblock \href {http://arxiv.org/abs/1004.4672} {\path{arXiv:1004.4672}}.

\bibitem{Fesen}
R.~A. {Fesen}, M.~C. {Hammell}, J.~{Morse}, R.~A. {Chevalier}, K.~J.
  {Borkowski}, M.~A. {Dopita}, C.~L. {Gerardy}, S.~S. {Lawrence}, J.~C.
  {Raymond}, S.~{van den Bergh}, {The Expansion Asymmetry and Age of the
  Cassiopeia A Supernova Remnant}, \apj 645 (2006) 283--292.
\newblock \href {http://arxiv.org/abs/astro-ph/0603371}
  {\path{arXiv:astro-ph/0603371}}.

\bibitem{Shternin}
P.~S. {Shternin}, D.~G. {Yakovlev}, C.~O. {Heinke}, W.~C.~G. {Ho}, D.~J.
  {Patnaude}, {Cooling neutron star in the Cassiopeia A supernova remnant:
  evidence for superfluidity in the core}, \mnras 412 (2011) L108--L112.
\newblock \href {http://arxiv.org/abs/1012.0045} {\path{arXiv:1012.0045}}.

\bibitem{Alfio}
A.~{Bonanno}, M.~{Baldo}, G.~F. {Burgio}, V.~{Urpin}, {The neutron star in
  Cassiopeia A: equation of state, superfluidity, and Joule heating}, \aap 561
  (2014) L5.
\newblock \href {http://arxiv.org/abs/1311.2153} {\path{arXiv:1311.2153}}.

\end{thebibliography}

\end{document}